\documentclass[natbib,smallextended]{svjour3}       
\smartqed  
\usepackage{graphicx}
\usepackage{amsmath,amssymb,amsfonts}
\usepackage{mathtools}
\usepackage{color}
\usepackage{soul}
\usepackage[shortcuts]{extdash}
\usepackage{tcolorbox}
\usepackage[colorlinks=true,citecolor=blue,urlcolor=blue,breaklinks]{hyperref}

\numberwithin{equation}{section}

\def\e{{\rm e}}

\def\x{{\mathrm{x}}}
\def\y{{\mathrm{y}}}

\newcommand{\s}{\mathrm{s}}
\newcommand{\n}{\mathrm{n}}
\newcommand{\p}{\mathrm{p}}
\newcommand{\X}{\mathrm{x}}
\newcommand{\Y}{\mathrm{y}}

\newcommand{\A}{\mathcal{A}}
\newcommand{\Ans}{\mathcal{A}_\mathrm{ns}}
\newcommand{\B}{\mathcal{B}}

\def\N{{\mathrm{N}}}
\def\S{{\mathrm{S}}}
\newcommand{\R}{{\cal R}}

\newcommand{\pc}{\check{p}}

\newcommand{\betac}{\check\beta}
\newcommand{\muc}{\check\mu}

\newcommand{\ec}{\check\varepsilon}

\newcommand{\be}{\begin{equation}}
\newcommand{\ee}{\end{equation}}
\newcommand{\beq}{\begin{equation}}
\newcommand{\eeq}{\end{equation}}
\newcommand{\bea}{\begin{eqnarray}}
\newcommand{\eea}{\end{eqnarray}}
\newcommand{\bear}{\begin{eqnarray}}
\newcommand{\eear}{\end{eqnarray}}

\newcommand{\ny}{n_\y^2}
\newcommand{\nxy}{n_{\x\y}^2}

\newcommand{\Bn}{\mathcal{B}^\n}
\newcommand{\Bs}{\mathcal{B}^\s}
\newcommand{\BX}{\mathcal{B}^\x}
\newcommand{\BY}{\mathcal{B}^\y}
\newcommand{\AXY}{\mathcal{A}^{\x\y}}

\newcommand{\BXab}{\BX_{a b}}
\newcommand{\Acalxab}{{\mathcal{A}}^\x_{a b}}
\newcommand{\Acalxyab}{{\mathcal{A}}^{\x \y}_{a b}}

\newcommand{\Xcalab}{\Xcal^{\x \y}_{a b}}
\newcommand{\Ccc}{\mathcal{C}_{cc}}

\newcommand{\Xcal}{\mathcal{X}}

\def\ax{A_\x}
\def\bx{B_\x}

\def\dx{D_\x}
\def\ex{E_\x}

\def\hxab{g^{\ax \bx}}

\def\hxba{g^{\bx \ax}}

\def\hxde{g^{\dx \ex}}

\def\fd{\xi}
\def\lefx{{\cal L}_{\fd_\x}}

\def\RXYa{ R^{\x \y}_a}
\def\RYXa{ R^{\y \x}_a}
\def\rtotxa{R_a^\x}
\def\DXab{{S}^\x_{a b}}
\def\DXba{{S}^\x_{b a}}
\def\dtotxba{D^\x_{b a}}
\def\dtotxab{D^\x_{a b}}
\def\dtotxabu{D_\x^{a b}}

\begin{document}

\title{Relativistic fluid dynamics: physics for many different scales}

\author{Nils Andersson \and Gregory L.\ Comer%
  \thanks{This article is a revised version of \url{https://doi.org/10.12942/lrr-2007-1}.\\
    \textbf{Change summary:} Major revision, updated and expanded. \\
    \textbf{Change details:} This revision represents a significant revision/update of the review. We have expanded the discussion to make it more pedagogical (introducing new sections on thermodynamics and matter equations of state as well as general variational principles), added sections to account for more physics (electromagnetism, elasticity and heat conductivity) and made contact with state of the art numerical relativity simulations and a range of issues (from cosmology to AdS/CFT and field theory models). The discussion of the relevant literature has been expanded by the addition of more than 300 new references.}
}

\institute{N. Andersson \at 
         Mathematical Sciences and STAG Research Centre\\
          University of Southampton \\
          Southampton SO17 1BJ, U.K.
         \email {na@maths.soton.ac.uk}
        \and
        G. L. Comer \at
         Department of Physics \& Center for Fluids at All Scales \\
          Saint Louis University \\
          St.\ Louis, MO, 63156-0907, U.S.A.
         \email{comergl@slu.edu}
}

\date{Received: date / Accepted: date}

\maketitle


\begin{abstract}
  The relativistic fluid is a highly successful model used to describe
  the dynamics of many-particle systems moving at high velocities and/or in strong gravity. It takes as
  input physics from microscopic scales and yields as output
  predictions of bulk, macroscopic motion. By inverting the process--e.g., drawing on astrophysical observations---an understanding of relativistic features can lead to insight into physics
  on the microscopic scale. Relativistic fluids have been used to
  model systems as ``small'' as colliding heavy ions in laboratory experiments, and as large
  as the Universe itself, with ``intermediate'' sized objects like
  neutron stars being considered along the way. The purpose of this
  review is to discuss the mathematical and theoretical physics
  underpinnings of the relativistic (multi-) fluid model. We focus
  on the variational principle approach championed by Brandon Carter
  and collaborators, in which a crucial element is to distinguish
  the momenta that are conjugate to the particle number density
  currents. This approach differs from the ``standard'' text-book
  derivation of the equations of motion from the divergence of the
  stress-energy tensor in that one explicitly obtains the relativistic
  Euler equation as an ``integrability'' condition on the relativistic
  vorticity. We discuss the conservation laws and the equations of
  motion in detail, and provide a number of (in our opinion)
  interesting and relevant applications of the general theory. The formalism provides a foundation for complex models, e.g., including electromagnetism, superfluidity and elasticity---all of which are relevant for state of the art neutron-star modelling. 
\end{abstract}

\newpage
\setcounter{tocdepth}{3}
\tableofcontents

\newpage

\section{Setting the stage}
\label{sec:intro}

If one performs a search on the topic of relativistic fluids on any of the major
physics article databases one is overwhelmed by the number of ``hits''. This
reflects the importance that the fluid model has long had for
physics and engineering. For relativistic physics, in particular, the
fluid model is essential. After all, many-particle astrophysical and
cosmological systems are the best sources of detectable effects associated
with General Relativity. Two obvious examples, the expansion of the Universe
and oscillations (or, indeed, mergers) of neutron stars, indicate the vast range of scales
on which relativistic fluids are relevant. A particularly topical
context for general relativistic fluids is their use in the modeling of
gravitational-wave sources. This includes the compact binary inspiral
problem, either of two neutron stars or a neutron star and a black hole, the
collapse of stellar cores during supernovae, or various neutron star
instabilities. One should also not forget the use of (special) relativistic
fluids in modeling collisions of heavy nuclei, astrophysical jets, and
gamma-ray burst emission. 

This review provides an introduction to the modeling of fluids in General
Relativity. As the (main) target audience is graduate students with a need for an
understanding of relativistic fluid dynamics we have made an effort
to keep the presentation pedagogical, carefully introducing the central concepts. The discussion will (hopefully) also be
useful to researchers who work in areas outside of General Relativity and
gravitation per se (e.g., a nuclear physicist who develops neutron star
equations of state), but who require a working knowledge of relativistic fluid
dynamics.

Throughout (most of) the discussion we will assume that General Relativity is
the proper description of gravity. From a conservative point of view, this restriction is not too severe. Einstein's theory is extremely well tested and it is natural to focus our attention on it. At the same time, it is important to realize that the problem of fluids in other theories of gravity has
interesting aspects. And perhaps more importantly, we know that General Relativity cannot be the ultimate theory of gravity---it absolutely breaks on the quantum scale and may also have trouble on the large scales of cosmology (taking the presence of the mysterious dark energy as evidence that something is missing in our understanding). As we hope that the review will be used by students
and researchers who are not necessarily experts in General Relativity and the
techniques of differential geometry, we have included an introduction to the
mathematical tools required to build relativistic models. Our summary is not a proper introduction to General Relativity, but we have made an effort to define all the tools we need for the discussion that
follows. Hopefully, our description is sufficiently self-contained to
provide a less experienced reader with a working understanding of (at least
some of) the mathematics involved. In particular, the reader will find
an extended discussion of the covariant and Lie derivatives. This is natural
since many important properties of fluids, both relativistic and
non-relativistic, can be established and understood by the use of parallel
transport and Lie-dragging, and it is vital to appreciate the
distinctions between the two. As we do not want to make the initial learning too steep, we have tried to avoid the language of differential geometry. This makes the discussion  less ``elegant'' in places, but we feel that this is a price worth paying if the aim is to make the material more generally accessible. 

Ideally, the reader should have some familiarity with standard fluid
dynamics, e.g., at the level of the discussion in \cite{landau59:_fluid_mech}, basic
thermodynamics \citep{reichl98:_book}, and the mathematics of action
principles and how they are used to generate equations of
motion \citep{lanczos49:_var_mechs}. Having stated this, it is clear
that we are facing a  challenge. We are trying to
introduce a topic on which numerous books have been written
(e.g., \citealt{tolman34:_book, landau59:_fluid_mech,lichnerowica67:_book,
  anile90:_relfl_book, wilson03:_book,2013rehy.book.....R}), and which requires an
understanding of a significant fraction of modern theoretical physics. This does not, however, mean that there is no place for this kind of survey.  We continue to see exciting developments for multi-constituent systems, such as
superfluid/superconducting neutron star cores\footnote{We use ``superfluid'' to refer to any system which has the
  ability to flow without friction. In this sense, superfluids and
  superconductors are viewed in the same way. When we wish to
  distinguish charge-carrying superfluids, we will call them
  superconductors.}. Much of the recent theory work has been guided by the geometric
approach to fluid dynamics championed by
Carter \citeyearpar{carter83:_in_random_walk, carter89:_covar_theor_conduc,
  carter92:_brane}, which provides a powerful framework that makes
extensions to multi-fluid situations intuitive. A typical
example of a phenomenon that arises naturally is the so-called
entrainment effect, which plays a crucial role in a superfluid neutron
star core. Given the flexible nature of the formalism, its natural connection with General Relativity and the potential for future applications, we have opted to base much of our description on the work
of Carter and colleagues.

It is important to appreciate that, even though the subject of relativistic fluids is far from new, issues still remain to be resolved. The most obvious shortcoming of the
available theory concerns dissipative effects. As we will see, different
dissipation channels are (at least in principle) easy to incorporate in
Newtonian theory but the extension to General Relativity remains ``problematic''. This is an issue---with a number of notable recent 
efforts---of key 
importance for future gravitational-wave source modelling  (e.g., in numerical relativity) as well as the description of laboratory systems (like heavy-ion collisions). In order to develop the required framework, we need to make progress on both the underpinning theory and implementations (e.g., computationally ``affordable'' simulations)---a real, but at the same time inspiring, challenge.


\subsection{A brief history of fluids}

The two fluids air and water are essential to human survival. This obvious
fact implies a basic need to divine their innermost secrets. Homo
Sapiens have always needed to anticipate air and water behaviour under a myriad of circumstances, such as those that concern water supply, weather, and
travel. The essential importance of fluids for survival---and how they can
be exploited to enhance survival---implies that the study of fluids likely
reaches as far back into antiquity as the human race
itself. Unfortunately, our historical records of this ever-ongoing
study are not so great that we can reach very far accurately.

A wonderful account (now in affordable Dover print) is ``A History and Philosophy of
Fluid Mechanics'' by \cite{tokaty}. He points out that while
early cultures may not have had universities, government sponsored laboratories, or
privately funded  centers pursuing fluids research (nor a Living Reviews
archive on which to communicate results!), there was certainly some
collective understanding. After all, there is a clear connection between the
viability of early civilizations and their access to water. For example, we
have the societies associated with the Yellow and Yangtze rivers in China, the
Ganges in India, the Volga in Russia, the Thames in England, and the Seine in
France, to name just a few. We must also not forget the Babylonians and
their amazing technological (irrigation) achievements in the land between the
Tigris and Euphrates, and  the Egyptians, whose intimacy with the flooding
of the Nile is well documented. In North America, we have the so-called
Mississippians, who left behind their mound-building
accomplishments. For example, the Cahokians (in Collinsville,
Illinois) constructed Monk's Mound\footnote{\url{http://en.wikipedia.org/wiki/Monk's\_Mound}}, the largest pre-Columbian earthen
structure in existence that is ``\dots over
100 feet tall, 1000 feet long, and 800 feet wide (larger at its base
than the Great Pyramid of Giza)''.

In terms of ocean and sea travel, we know that the maritime ability of the
Mediterranean people was the key to ensuring cultural and economic
growth and societal stability. The finely-tuned skills of the Polynesians
in the South Pacific allowed them to travel great distances, perhaps reaching
as far as South America, and certainly making it to the ``most remote
spot on the Earth'', Easter Island. Apparently, they were 
adept at reading the smallest of signs---water colour, views of weather on
the horizon, subtleties of wind patterns, floating objects, birds,
etc.---as indications of nearby land masses. Finally, the harsh
climate of the North Atlantic was overcome by the highly accomplished
Nordic sailors, whose skills allowed them to reach 
North America. Perhaps it would be appropriate to think of these early
explorers as adept geophysical fluid dynamicists/oceanographers?

\begin{figure}[htb]
    \centerline{\includegraphics[scale = 0.4]{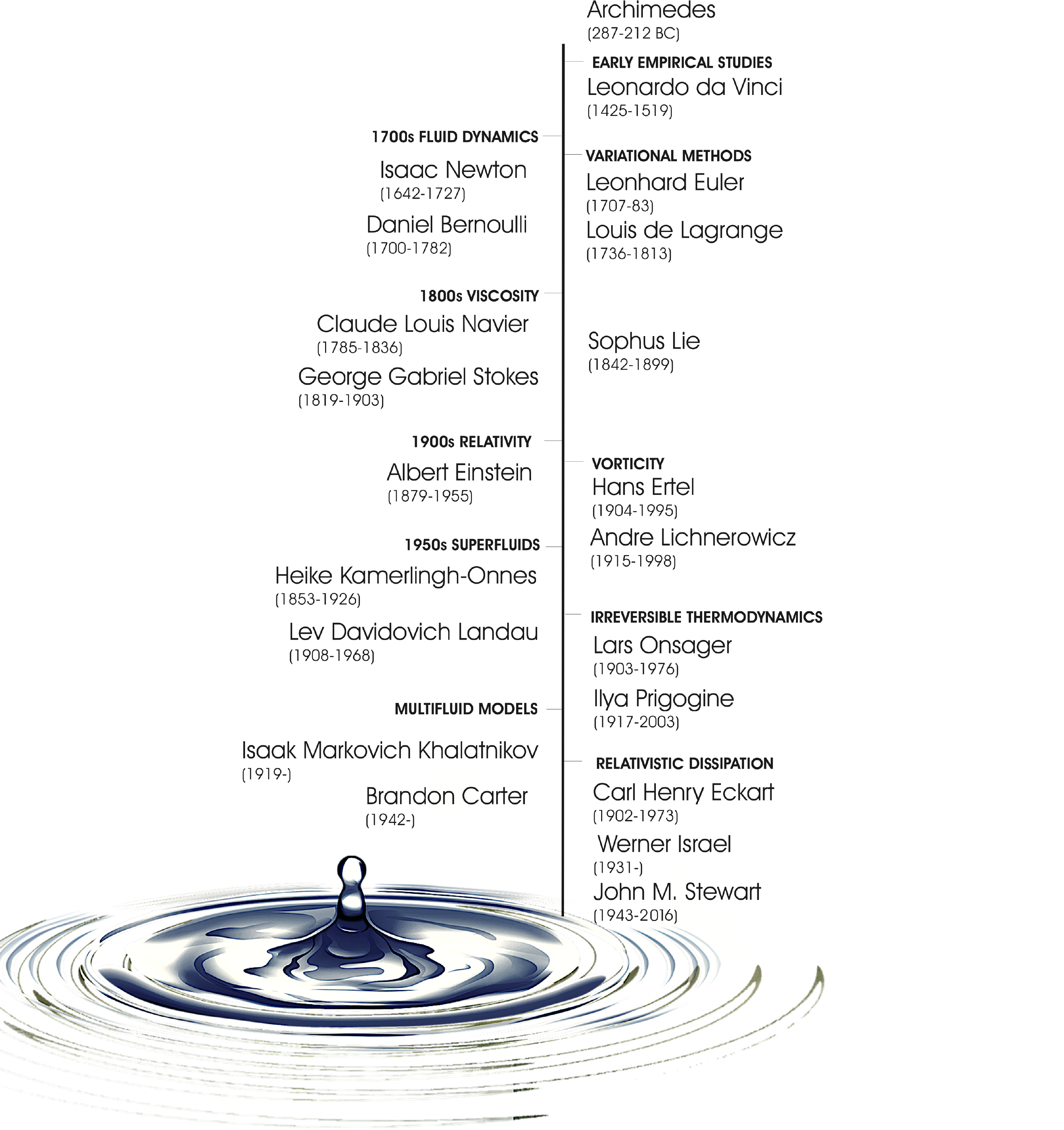}}
    \caption{A ``timeline'' focussed on the topics covered in this
        review, including chemists, engineers, mathematicians,
        philosophers, and physicists who have contributed to the
        development of non-relativistic fluids, their relativistic
        counterparts, multi-fluid versions of both, and exotic
        phenomena like superfluidity.}
    \label{time}
\end{figure}

Many great scientists are associated with the study of fluids. Lost are the
names of the individuals who, almost 400,000 years ago, carved
``aerodynamically correct'' \citep{gadelhak} wooden spears. Also lost are
those who developed boomerangs and fin-stabilized arrows. Among those not
lost is Archimedes, the Greek mathematician (287\,--\,212 BC), who provided a
mathematical expression for the buoyant force on bodies. Earlier, Thales of
Miletus (624\,--\,546 BC) asked the simple question: What \emph{is} air and
water? His question is profound as it represents a  departure
from the main, myth-based modes of inquiry at that time. Tokaty ranks Hero
of Alexandria as one of the great, early contributors. Hero (c.~10\,--\,70) was a
Greek scientist and engineer, who left behind  writings and drawings that,
from today's perspective, indicate a good grasp of basic fluid mechanics.
To make a complete account of individual contributions to our present
understanding of fluid dynamics is, of course, impossible. Yet, it is
useful to list some of the contributors to the field. We provide a highly
subjective ``timeline'' in Fig.~\ref{time}. The list is to a large extent
focussed on the topics covered in this review, and includes chemists,
engineers, mathematicians, philosophers, and physicists. It recognizes those
that have contributed to the development of non-relativistic fluids, their
relativistic counterparts, multi-fluid versions of both, and exotic phenomena like superfluidity. The list provides context---both historical and scientific---and also serves as an informal table of contents for this survey.

\cite{tokaty} discusses the human propensity for destruction when it
comes to  water resources. Depletion and pollution are the main
offenders. He refers to a ``Battle of the Fluids'' as a struggle between
their destruction and protection. His context for this discussion was the
Cold War. He rightly points out the failure to protect our water and air
resources by the two dominant powers---the USA and USSR. In an
ironic twist, modern study of the relativistic properties of fluids has its own ``Battle of the Fluids''. A self-gravitating mass can become absolutely
unstable and collapse to a black hole, the ultimate destruction of any form of
matter.

\subsection{Why are fluid models useful?}
\label{sec:fluidsuseful}

The {\it Merriam-Webster} online dictionary\footnote{\url{http://www.m-w.com/}}
defines a fluid as ``\dots a substance (as a liquid or gas) tending to
flow or conform to the outline of its container'' when taken as a noun
and ``\dots having particles that easily move and change their
relative position without a separation of the mass and that easily
yield to pressure: capable of flowing'' when taken as an
adjective. The best model of physics is the Standard Model which is
ultimately the description of the ``substance'' that  makes up our
fluids. The substance of the Standard Model consists of a remarkably small set of elementary particles: leptons, quarks, and the so-called
``force'' carriers (gauge-vector bosons). Each elementary particle is
quantum mechanical, but the Einstein equations require explicit
trajectories. Effectively, there is a disconnect between the quantum scale and our classical description of gravity. Moreover, cosmology and neutron stars are (essentially) many
particle systems and---even forgetting about quantum mechanics---it is
not possible to track each and every ``particle'' that makes them up,
regardless of whether these are elementary (leptons, quarks, etc.) or
collections of elementary particles (e.g., individual stars in galaxies and
galaxies in cosmology). The fluid model is such that the inherent
quantum mechanical behaviour, and the existence of many particles are
averaged over in such a way that it can be implemented consistently in
the Einstein equations.

\begin{figure}[htb]
    \centerline{\includegraphics[width=0.7\textwidth]{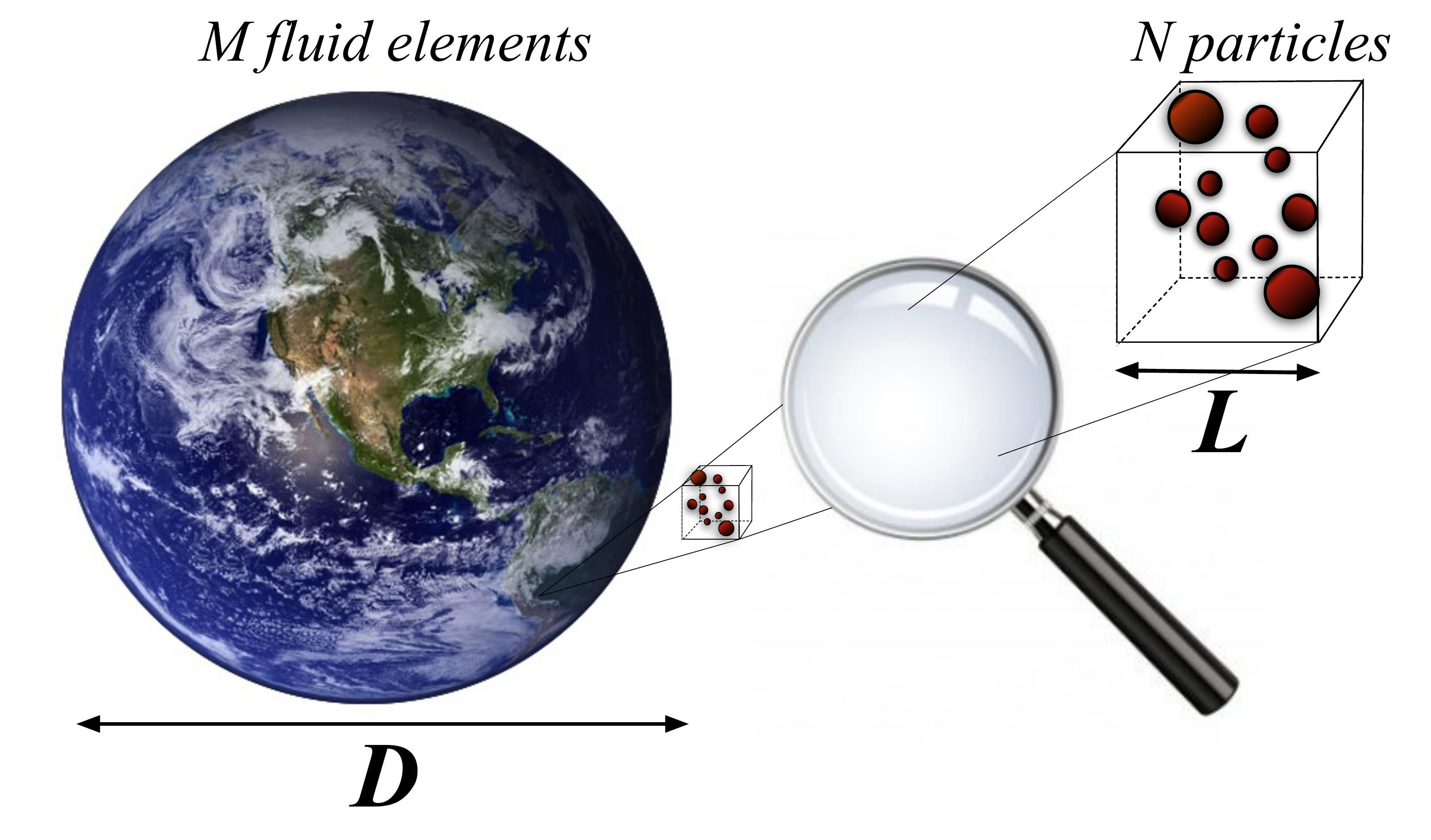}}
    \caption{An object with a characteristic size $D$ is modeled
    as a fluid that contains $M$ fluid elements. From inside the
    object we magnify a generic fluid element of characteristic size
    $L$. In order for the fluid model to work we require $M \gg N \gg
    1$ and $D \gg L$.}
    \label{fluidparticle}
\end{figure}

Central to the model is the notion of a ``fluid element'', also known as a
``fluid partlicle'' or ``material particle'' \citep{lautrup05:_book}. This is an
imagined, local ``box'' that is infinitesimal with respect to the system
\emph{en masse} and yet large enough to contain a large number of particles
(e.g., an Avogadro's number of particles). The idea is illustrated in
Fig.~\ref{fluidparticle}. We consider an object with characteristic size
$D$ that is modeled as a fluid
that contains $M$ fluid elements. From inside the object we magnify a
generic fluid element of characteristic size $L$. In order for the
fluid model to work we require
$M \gg N \gg 1$ and $D \gg L$. Strictly speaking, the model has $L$
infinitesimal, $M \to \infty$, but with the total number of particles
remaining finite. An operational point of view is that discussed by
Lautrup in his fine text ``Physics of Continuous
Matter'' \citeyearpar{lautrup05:_book}. He rightly points out the implicit
 connection to the intended precision. At some
level, any real system will be discrete and no longer represented by a
continuum. As long as the scale where the discreteness of matter and
fluctuations are important is much smaller than the desired precision,
the continuum approximation is valid.  The key point is that the fluid model allows us to consider complex dynamical phenomena in terms of a (relatively) small number of variables. We do not have to keep track of individual particles. The connection between the different scales (macroscopic and microscopic) plays a role, but  many of the tricky issues are assumed to be ``known''  (read: encoded in the matter equation of state, the determination of which may be someone else's ``problem'').

The aim of this review is to describe how the fluid model can be used (and understood) in the context of Einstein's curved spacetime theory for gravity. As will become clear, 
this necessarily involves attention to detail. For example, we need to consider how the coordinate invariance of General Relativity (with no preferred observers) impacts on (by necessity) observer-dependent notions from thermodynamics and the underlying microphysics. We also need to explore to what extent the dynamics of spacetime enters the problem. This is particularly relevant in the context of numerical simulations of energetic gravitational-wave sources (like merging neutron stars or massive stars collapsing under their own weight). The first step we have to take is natural---we need to consider how a given fluid element moves through spacetime and how this fluid motion enters the Einstein field equations. To some extent, this is a text-book problem with a well-known solution (=\,the perfect fluid model). However, as we will learn along the way, more realistic matter descriptions (including for example superfluidity, as expected in the core of a mature neutron star, or the elasticity of the star's crust) require a more sophisticated approach.  Nevertheless, the first step we have to take is natural.  

\begin{figure}[htb]
    \centerline{\includegraphics[width=0.6\textwidth]{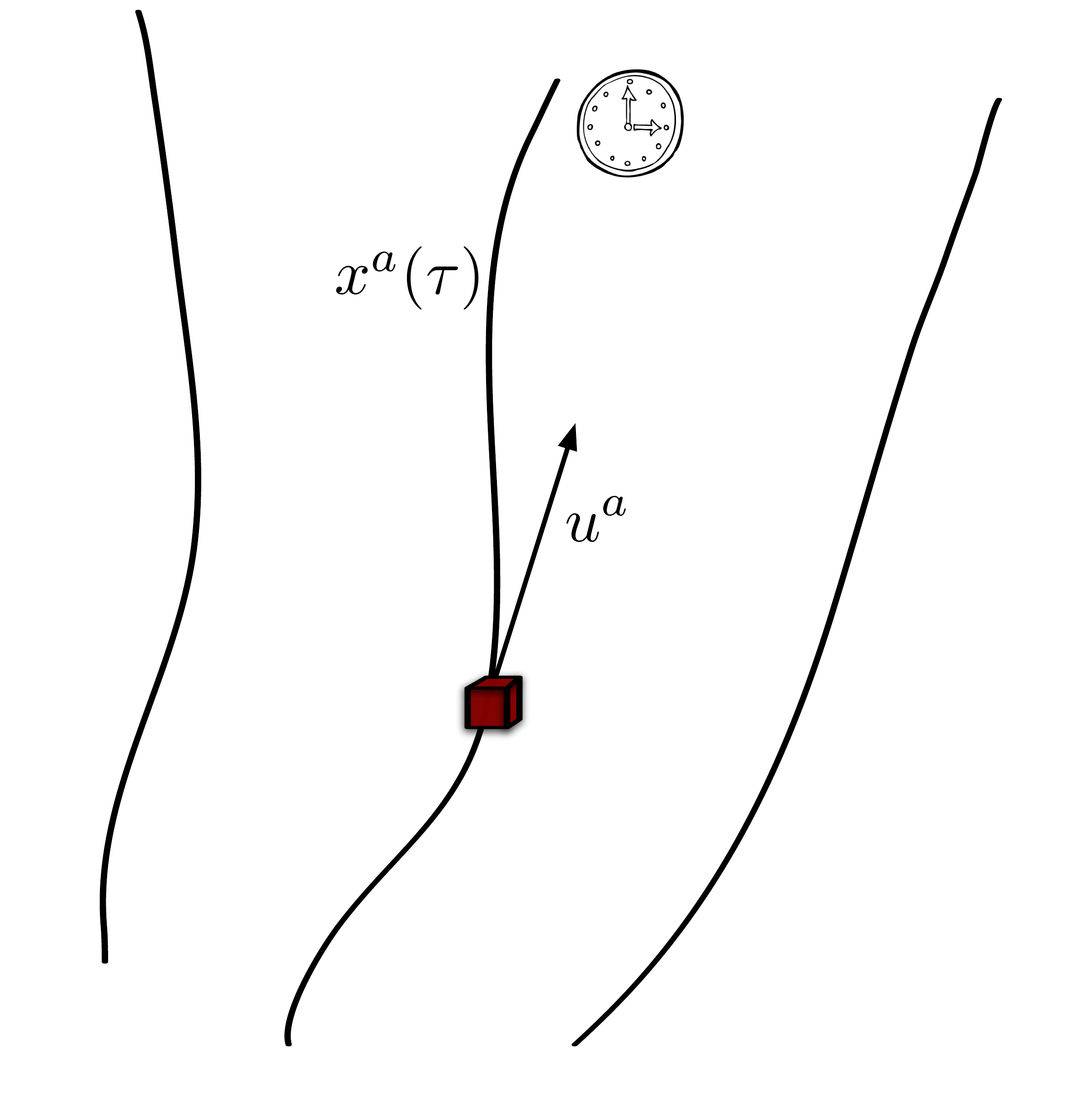}}
    \caption{An illustration of the fibration of spacetime associated with a set of fluid ``observers'', each with their own four velocity $u^a$ and notion of time (the proper time measured on a co-moving clock). In the fluid model, individual worldlines are assigned to specific fluid elements (which involve averages over the  large number of constituent particles).}
    \label{fibrate}
\end{figure}

The explicit trajectories that enter the Einstein equations are those of the fluid elements, 
\emph{not} the much smaller (generally fundamental) particles that are ``confined'' (on average) 
to the elements. Hence, when we talk about the fluid velocity, we mean the velocity of fluid 
elements. In this sense, the use of the phrase ``fluid particle'' is very apt. For instance, each fluid 
element  traces out a timelike trajectory in spacetime $x^a(\tau)$, such that the unit tangent vector 
\begin{equation}
u^a = {dx^a \over d\tau} \ , \quad \mbox{with} \quad u_a u^a = -1
\end{equation}
where $\tau$ is time measured on a co-moving clock (proper time), provides the four velocity of the particle. The idea is illustrated in Fig.~\ref{fibrate}. 

The fundamental variable that enters the fluid equations is the particle flux density, in the following given by $n^a = n u^a$, where $n \approx N/L^3$ is the particle number density 
of the fluid element whose worldline is given by $u^a$. An object like a neutron star is then modelled 
as a  collection of particle flux density worldlines that continuously fill a portion of spacetime. 
In fact, we will see later that the relativistic Euler equation is little more than an ``integrability'' 
condition that guarantees that this filling (or fibration) of spacetime can be performed.  

Equivalently, we may consider the family of three-dimensional hypersurfaces that are 
pierced by the worldlines at given instants of time, as illustrated in Fig.~\ref{pullback}.  The 
integrability condition in this case  guarantees that the family of hypersurfaces continuously fill  a portion of spacetime. In this view, a fluid is a so-called three-brane 
(see \citealt{carter92:_brane} for a general discussion of branes). In fact, the strategy adopted in 
Sect.~\ref{sec:pullback} to derive the relativistic fluid equations is based on thinking of a fluid 
as living in a three-dimensional ``matter'' space (i.e., the left-hand-side of Fig.~\ref{pullback}). At first sight, this approach may seem confusing. However, as we will demonstrate,  it allows us to develop a versatile framework for complicated  systems which (in turn) enables progress on a number of relevant problems in astrophysics and cosmology. 

Once we understand how to build a fluid model using the matter space,
it is straight-forward to extend the technique to single fluids with
several constituents, as in Sect.~\ref{pbonemc}, and multiple fluid
systems, as in Sect.~\ref{sec:twofluids}. An example of the former would be
a fluid with one species of particles at a non-zero temperature,
i.e., non-zero entropy, that does not allow for heat conduction relative
to the particles. (Of course, entropy still flows through spacetime.)
The latter example can be obtained by relaxing the constraint of no
heat conduction. In this case the particles and the entropy are both
considered to be fluids\footnote{The notion that heat can be considered a ``fluid'' may seem somewhat heretical, but we will demonstrate that it allows us to explain aspects that otherwise remain somewhat ad hoc.} that are dynamically independent, meaning that
the entropy will have a four-velocity that is generally different from
that of the particles. There is thus an associated collection of fluid
elements for the particles and another for the entropy. At each point
of spacetime that the system occupies there will be two fluid
elements, in other words, there are two matter spaces
(cf.\ Sect.~\ref{sec:twofluids}). Perhaps the most important consequence of
this is that there can be a relative flow of the entropy with respect
to the particles. In general, relative flows lead to the so-called
entrainment effect, i.e., the momentum of one fluid in a multiple
fluid system is in principle a linear combination of all the fluid
velocities \citep{andersson05:_flux_con}. The canonical examples of two
fluid models with entrainment are superfluid
$\mathrm{He}^4$ \citep{putterman74:_sfhydro} at non-zero temperature
  and a mixture of superfluid $\mathrm{He}^4$ and
  $\mathrm{He}^3$ \citep{andreev75:_entrain}. We will develop a detailed understanding of all these concepts in due course, but as it is important to proceed with care we will first focus on the physics that provide input for the fluid model.


\subsection{Notation and conventions}

Throughout the article we assume the ``MTW'' \citep{mtw73} conventions.  We also generally 
assume geometrized units $c=G=1$, unless specifically noted otherwise, and set the Boltzmann constant $k_B = 1$. A coordinate basis will 
always be used, with spacetime indices denoted by lowercase Latin 
letters $\{a,b,...\}$ etc.~that range over $\{0,1,2,3\}$ (time being the zeroth coordinate), and 
purely spatial indices denoted by lowercase Latin letters $\{i,j,...\}$ etc.~that range over 
$\{1,2,3\}$. Unless otherwise noted, we assume that the Einstein summation convention applies. Finally, we adopt the convention that $u^\x_a=g_{ab}u_\x^a$ where $\x$ is a fluid constituent label. These are never summed over when repeated. Also note that, while it is possible to build a chemically covariant formalism (with the $\x$ treated on a par with spacetime indices) we will not do so here. Our approach has the ``advantage'' that the constituent labels can be placed up or down, without this having any particular meaning, which  helps keep many of the  expressions tidy. We will also regularly have to deal with expressions where more than two of these labels are repeated and this complicates a fully covariant approach.



\section{Thermodynamics and equations of state}
\label{sec:thermo}

As fluids consists of many fluid elements---and each fluid element consists of
many particles---the state of matter in a given fluid element is (inevitably)
determined thermodynamically \citep{reichl98:_book}. This means that only a
few parameters are tracked as the fluid element evolves. In a typical situation, not
all the thermodynamic variables are independent---they are connected through
the so-called equation of state. Moreover, the number of independent variables
may be reduced if the system has an overall additivity property. As
this is a very instructive example, we will illustrate this point
in detail.


\subsection{Fundamental, or Euler, relation}

Consider the standard form of the combined First and Second Laws\footnote{We say 
``combined'' here because the First Law is a statement about heat and work, and says nothing 
about the entropy, which enters through the Second Law. Heat is not strictly equal to 
$T \, \mathrm{d} S$ for all processes; they are equal for quasistatic processes, but not for free 
expansion of a gas into vacuum \citep{schroeder00:_thermobook}} for a simple, single-species
system:
\begin{equation}
  d E =
  T \, d S - p \, d V + \mu \, d N.
\end{equation}
This follows because there is an equation of state, meaning that $E =
E(S,V,N)$ where
\begin{equation}
  T = \left. \frac{\partial E}{\partial S} \right|_{V,N}\ ,
  \quad \quad
  p = - \left. \frac{\partial E}{\partial V} \right|_{S,N}\ ,
  \quad \quad
  \mu = \left. \frac{\partial E}{\partial N} \right|_{S,V}\ .
\end{equation}
The total energy $E$, entropy $S$, volume $V$, and particle number $N$
are said to be \emph{extensive} if when $S$, $V$, and $N$ are doubled, say, then
$E$ will also double. Conversely, the temperature $T$, pressure $p$,
and chemical potential $\mu$ are called \emph{intensive} if they do not change
their values when $V$, $N$, and $S$ are doubled. This is the additivity
property and we will now show why it implies an Euler relation (also
known as the ``fundamental relation''; \citealt{reichl98:_book}) among the
thermodynamic variables. This relation is essential for any effort to connect the microphysics and thermodynamics to the fluid dynamics.

Let a tilde represent the change in thermodynamic variables when $S$, $V$,
and $N$ are all increased by the same amount $\lambda$, i.e.,
\begin{equation}
  \tilde{S} = \lambda S\ ,
  \qquad
  \tilde{V} = \lambda V\ ,
  \qquad
  \tilde{N} = \lambda N\ .
\end{equation}
Taking $E$ to be extensive then means
\begin{equation}
  \tilde{E}(\tilde{S},\tilde{V},\tilde{N}) = \lambda E(S,V,N).
\end{equation}
Of course, we have for the intensive variables
\begin{equation}
  \tilde{T} = T,
  \qquad
  \tilde{p} = p,
  \qquad
  \tilde{\mu} = \mu.
\end{equation}
Now,
\begin{multline}
  d \tilde{E} =
  \lambda \, d E + E \, d \lambda = \tilde{T} \, d \tilde{S} -
  \tilde{p} \, d \tilde{V} +
  \tilde{\mu} \, d \tilde{N}
  \\
  = \lambda
  \left( T d S - p d V + \mu d N \right) +
  \left( T S - p V + \mu N \right) d \lambda,
\end{multline}%
and (since the change in the energy should be proportional to $\lambda$) we find the Euler relation
\begin{equation}
  E = T S - p V + \mu N.
\end{equation}
If we let $\varepsilon = E / V$ denote the total energy density, $s = S / V$
the total entropy density, and $n = N / V$ the total particle number
density, then
\begin{equation}
  p + \varepsilon = T s + \mu n.
  \label{funrel}
\end{equation}

The nicest feature of an extensive system is that the number of
parameters required for a complete specification of the thermodynamic
state can be reduced by one,  in such a way that only intensive variables remain. To see this, let $\lambda = 1/V$, in
which case
\begin{equation}
  \tilde{S} = s,
  \qquad
  \tilde{V} = 1,
  \qquad
  \tilde{N} = n.
\end{equation}
The re-scaled energy becomes just the total energy density,
i.e., $\tilde{E} = E / V = \varepsilon$, and moreover $\varepsilon = \varepsilon(s,n)$
since
\begin{equation}
  \varepsilon = \tilde{E}(\tilde{S},\tilde{V},\tilde{N}) =
  \tilde{E}(S/V,1,N/V) = \tilde{E}(s,n).
\end{equation}
The first law thus becomes
\begin{equation}
  d \tilde{E} =
  \tilde{T} \, d \tilde{S} -
  \tilde{p} \, d \tilde{V} +
  \tilde{\mu} \, d \tilde{N} =
  T \, d s + \mu \, d n,
\end{equation}
or
\begin{equation}
  d \varepsilon = T \, d s + \mu \, d n.
  \label{1stlaw}
\end{equation}
This implies
\begin{equation}
  T = \left. \frac{\partial \varepsilon}{\partial s} \right|_n\!\!\!,
  \qquad
  \mu = \left. \frac{\partial \varepsilon}{\partial n} \right|_s\!\!.
\end{equation}
That is, $\mu$ and $T$ are the chemical potentials\footnote{Loosely speaking, the ``energy'' associated with adding or removing one particle of the given species from the system.} associated with the particles and entropy, respectively.
The Euler relation~(\ref{funrel}) then yields the pressure as
\begin{equation}
  p = - \varepsilon +
  s \left. \frac{\partial \varepsilon}{\partial s} \right|_n \!\!\! +
  n \left. \frac{\partial \varepsilon}{\partial n} \right|_s\!\!.
  \label{eulerrel}
\end{equation}

In essence, we can think of a given relation  $\varepsilon(s,n)$ as the equation of state, to
be determined in the flat, tangent space at each point of spacetime, or,
physically, small enough patches across which the changes in the
gravitational field are negligible, but also large enough to contain
a large number of particles. For example, for a neutron star,
\cite{glendenning97:_compact_stars} argues that the
relative change in the metric over the size of a nucleon with respect to
the change over the entire star is about $10^{- 19}$, and thus one must
consider many inter-nucleon spacings before a substantial change in the
metric occurs. In other words, it is sufficient to determine the properties
of matter in special relativity, neglecting effects due to the spacetime
curvature.\footnote{This is fortunate, as we may otherwise have to face the thorny issue of quantum gravity head-on.} The equation of state is the key link between the
microphysics that governs the local fluid behaviour and global
quantities (such as the mass and radius of a star).

In what follows we will use a thermodynamic formulation that satisfies the
fundamental scaling relation, meaning that the local thermodynamic state
(modulo entrainment, see later) is a function of the variables $N/V$, $S/V$,
and so on. This is in contrast to the discussion in, for example, 
``MTW'' \citep{mtw73}. In their approach one fixes from the outset the
total number of particles $N$, meaning that one simply sets $d N
= 0$ in the first law of thermodynamics. Thus, without imposing any
scaling relation, one can write
\begin{equation}
  d \varepsilon =
  d \left( E/V \right) =
  T \, d s +
  \frac{1}{n} \left( p + \varepsilon - T s \right) d n.
\end{equation}
This is consistent with our starting point, because we assume
that the extensive variables associated with a fluid element do not
change as the fluid element moves through spacetime. However, we feel
that the scaling is necessary in that the fully conservative (read:
non-dissipative) fluid formalism presented below can be adapted to
non-conservative, or dissipative, situations where $d N = 0$ cannot be
imposed.

\subsection{Case study: neutron stars}

\begin{figure}[htb]
    \centerline{\includegraphics[width=0.7\textwidth]{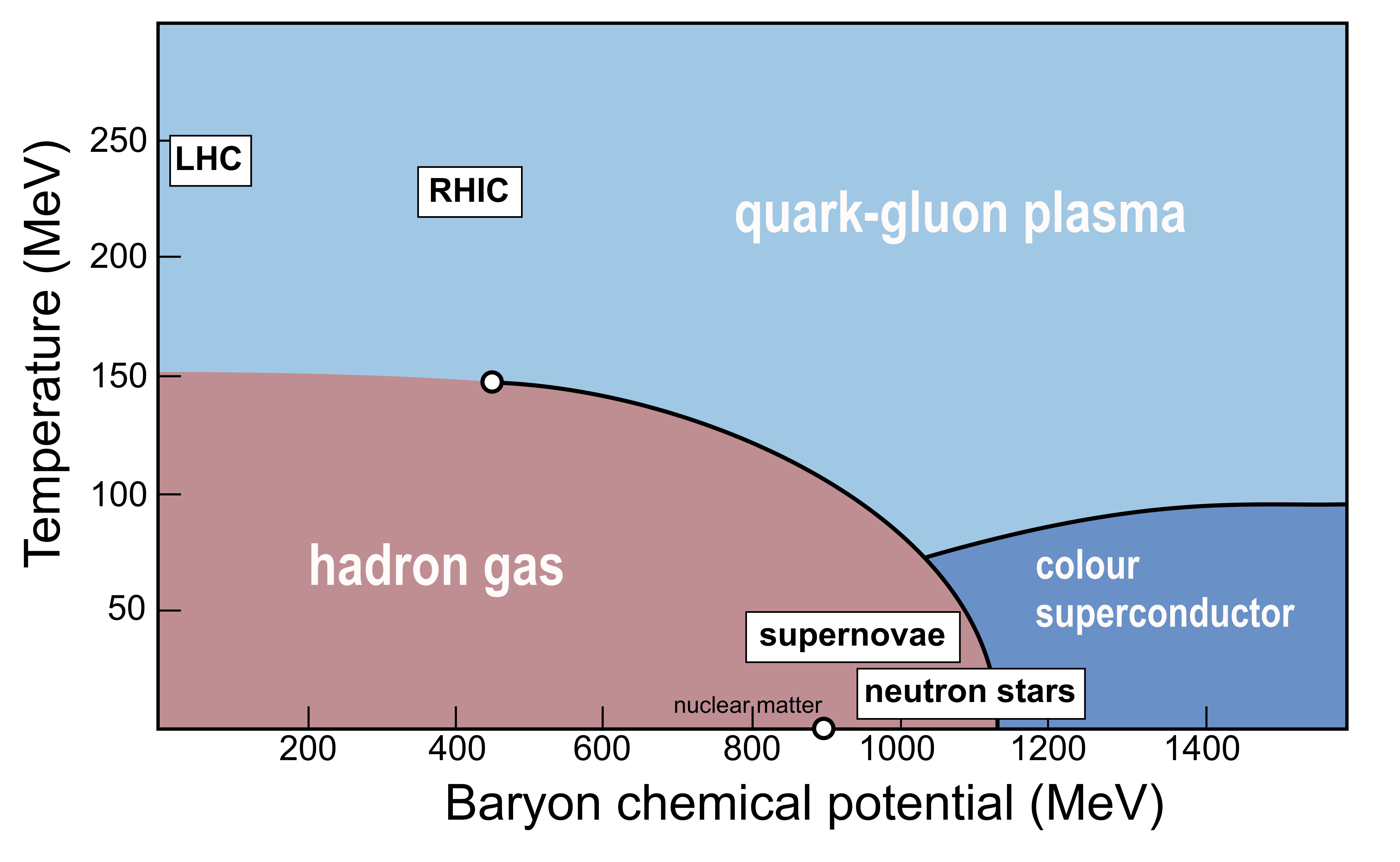}}
    \caption{A broad-brush illustration of the phase space for dense matter physics, represented by the baryon chemical potential ($\mu_{\mathrm b}$) (horizontal axis) and the temperature (vertical axis). Experiments carried out using high-energy colliders, like the LHC and RHIC, aim to explore the nature of the quark-gluon plasma and the conditions of the early Universe---hot matter at relatively low densities. In contrast,an understanding of relativistic stars depends on the dense-low temperature regime, which unlikely to be within reach of laboratory efforts. First principles calculation in the $\mu_{\mathrm b}\to \infty$ limit of QCD suggests that the core of a mature neutron star may contain a colour superconductor, but the exact nature of the quark pairing at the relevant densitites is not (particularly) well understood \citep{2008RvMP...80.1455A}.  }
    \label{phase}
\end{figure}

With a mass of more than that of the Sun squeezed inside a radius of about 10~km, a neutron star represents many extremes of physics. The relevant matter description involves issues that cannot be explored in terrestrial laboratories, yet relies on aspects similar to those  probed by high-energy colliders. However, while the LHC at CERN and RHIC at Brookhaven (among others) probe low density matter at high temperatures, neutron stars are cold (on the nuclear physics temperature scale) and reach significantly higher densities. In effect, the problems are complementary, see Fig.~\ref{phase} for a schematic illustration. Moreover, atrophysical modelling of neutron star dynamics (e.g., the global oscillations of the star) typically involves large enough scales that a fluid description is an absolute necessity.  Yet, such models must build on appropriate microphysics input (encoded in the equation of state). This is problematic because first principle calculations of the interactions for many-body QCD systems are not yet within reach (due to the fermion sign problem). In essence, we do not know the composition of matter. There may be a large population of hyperons present at densities relevant for neutron star cores. Perhaps the quarks are deconfined to form a quark-gluon plasma? Our models needs to be flexible enough to account for different possibilities, and the problem is further complicated by the state of matter. At the relevant temperatures, many of the particle constituents (neutrons, protons, hyperons, etc.) are expected to exhibit Cooper pairing to form superfluid/superconducting condensates.  This brings in aspects from low-temperature physics and a realistic neutron-star model must recognize this. In short, the problem is overwhelming and one would typically (at some point) have to  resort to phenomenology, using experiments and observations to test predictions as new models become available \citep{2016RvMP...88b1001W}. 

The details may be blurry but (at least) the rules that guide the exercise are fairly clear. We need to build models that allow for a complex matter composition and account for different states of matter (from solids to superfluids). This involves going beyond the single-fluid setting and considering systems with distinct components exhibiting relative flows. In short, we need to model {\it multi-constituent multi-fluid systems}. As both concepts will be central to the discussion, let us introduce the main ideas already at this point.

It is natural to start by considering the matter in the outer core of a neutron star, dominated by neutrons with a small fraction of protons and electrons. Assuming that the different constituents flow together (we will  relax this assumption later), we have the thermodynamic relation (assuming matter at zero temperature, for simplicity)
\begin{equation}
p+\varepsilon = \sum_{\x} n_\x \mu_\x \ , \quad \mbox{with} \quad \x = \n,\p,\e \ ,
\end{equation}
where $n_\x$ are the respective number densities and $\mu_\x$ the corresponding chemical potentials. This is a straightforward extension of \eqref{eulerrel}. At the microscopic scale (e.g., the level of the equation of state), it is usually assumed that the matter is charge neutral. This implies that the number of electrons must balance that of the protons. We have $n_\p=n_\e$ and it follows that
\begin{equation}
p+\varepsilon =  n_\n \mu_\n + n_\p (\mu_\p+\mu_\e)
\end{equation}
 
 Next, we need to consider the issue of chemical equilibrium. For the case under consideration this would involve the system being such that the Urca reactions are in balance. In essence, this means that we have
 \begin{equation}
\beta \equiv \mu_\n - (\mu_\p + \mu_\e) = 0 \ .
 \end{equation}
 This condition determines how many neutrons we need per proton, which means that the composition is specified. In general, we can rewrite the thermodynamical relation as\footnote{In general, one may have to worry about neutrinos here.}
 \begin{equation}
p+\varepsilon =  n \mu_\n - n_\p \beta \ ,
\end{equation}
where we have introduced the baryon number density $n= n_\n +n_\p$. Assuming equilibrium, this leads to
\begin{equation}
p = n \mu_\n (n) - \varepsilon (n) \ ;
\end{equation}
that is, we have a one-parameter equation of state. It is common to think of the equation of state in this way---the pressure is provided as a function of the (baryon number) density.

Many formulations for numerical simulations take this ``barotropic'' model as the starting point. The usual logic works (in some sense) ``backwards'' by focussing on the mass density and separating out the mass density contribution to the chemical potential by introducing $\rho = mn$ where $m$ is the baryon mass. That is, we use
\begin{equation}
\mu_\n = m+ \tilde\mu
\end{equation}
This expression reflects that simple fact that the (rest) mass of a particle in isolation should be $mc^2$, leaving the (to some extent) unknown aspects of the many-body interactions to be encoded in $\tilde \mu$. This allows us to write
\begin{equation}
p =  \rho +  (n\tilde\mu- \varepsilon) =  \rho(1- \epsilon) 
\label{peps}
\end{equation}
where $\epsilon$ represents the (specific) internal energy. Numerical efforts often focus on $\epsilon$. The reason for this will become shortly. First, it is easy to see that we also have
\begin{equation}
\varepsilon = \rho(1+\epsilon)
\label{reps}
\end{equation}
since
\begin{equation}
\tilde \mu = {d (\rho \epsilon) \over dn}
\end{equation}
It is also useful to note that 
\begin{equation}
d\epsilon = {p\over \rho^2}  d\rho
\end{equation}
 
Let us now see what happens when we try to account for additional aspects, like the effects due to a finite temperature.  Assuming that we are comfortable working with the chemical potential (as we will do throughout much of this review) the natural starting point would  be 
\eqref{1stlaw}. However, it could be that we would prefer to extend the discussion using the internal energy. In that case, we first of all need to convince ourselves that \eqref{peps} and \eqref{reps} remain valid when $\varepsilon = \varepsilon (n,s)$. We then have $\epsilon=\epsilon(\rho,s)$, which leads to 
\begin{equation}
{\partial \epsilon \over \partial s} = {T \over \rho}
\end{equation}
and we find that 
\begin{equation}
d \epsilon = {p\over \rho^2} d\rho + {T\over \rho } ds - {sT \over \rho^2} d \rho = {p\over \rho^2} d\rho +  T d \hat s
\end{equation}
where we have introduced the specific entropy
\begin{equation}
\hat s = {s\over \rho} \ .
\end{equation}
If we want to progress beyond this point, we need to provide the form for the internal energy. This requires a finite temperature treatment on the microphysics level, as discussed in (for example) \cite{2015PhRvC..92b5801C,2016PhR...621..127L}.

Before we move on, it is useful to note that many numerical simulations have been based on implementing a pragmatic result drawn from the ideal gas law
\begin{equation}
p = n k_B T 
\label{ideal}
\end{equation}
where $k_B$ is Boltzmann's constant. Noting that this model leads to $\epsilon = C_v T$, with $C_v$ the heat capacity (at fixed volume) while 
\begin{equation}
{k_B\over C_v} = m(\Gamma-1)
\end{equation}
we readily arrive at
\begin{equation}
p = \rho \epsilon(\Gamma-1)
\label{gamma}
\end{equation}
For obvious reasons this is commonly referred to as the Gamma-law equation of state. It may not be particularly realistic---at least not for neutron stars---but it is simple (and relatively easy to implement). It also provides a straightforward measure of the temperature. Combining \eqref{ideal} and \eqref{gamma} we arrive at
\begin{equation}
T = {m\epsilon\over k_B} (\Gamma-1) = {m\over k_B} {p\over \rho}
\end{equation}
This is useful, but we need to be careful with this result. In a more general setting---like a multi-constituent system for which the ideal gas law argument is dubious---we are not quantifying the actual temperature. This would require use of the relevant physics from the beginning of the argument rather than at the end. However,  sometimes you have to accept a bit of pragmatism as the price of progress.

 Up to this point, we have separated the microphysics (determining the equation of state)
from the hydrodynamics (governing stellar oscillations and the like). Let us now consider the scale associated with fluid dynamics. 
For ordinary matter, the relevant scale is set by interparticle collisions.
Collisions tend to dissipate relative motion, leading to the system reaching (local dynamical and thermodynamical) equilibrium.
Since we want to associate a single ``velocity'' with each fluid element, the particles
must be able to equilibrate in a meaningful sense (e.g., have a velocity distribution with a well defined peak, allowing us to average over the system).
The relevant length-scale is the mean-free path. This concept is closely related to the
shear viscosity of matter (which arises due to particle scattering). In the case of neutrons (which dominate the outer core of a typical neutron star)
we would have
\be
\lambda \approx { \eta \over \rho v_F} \approx 10^{-4} \left( {\rho \over 10^{14}\ \mbox{g/cm}^3} \right)^{11/12} \left( {10^8 \ \mbox{K} \over T}\right)^2 \ \mbox{cm} \ ,
\ee
where $v_F$ is the relevant Fermi velocity and we have used the estimate for the neutron-neutron scattering shear viscosity $\eta$ from \cite{2005NuPhA.763..212A}. This estimate gives us an idea of the {\it smallest} scale on which it makes sense to consider the
system as a fluid. Notably, the mean-free path is many orders of magnitude larger than the interparticle separation (typically, the Fermi scale). The actual scale assumed in a fluid model typically depends on the problem one wants to study and tends to be limited by computational resources. For example, in current state of the art simulations of neutron star mergers, the computational fluid elements tend to be of order a few tens to perhaps a hundred meters across. They are in no sense microscopic entities. It is important to appreciate that these models involve a significant amount of ``extrapolation''. 

Assuming that the averaging procedure makes sense (we will have more to say about this later), 
the equations of hydrodynamics can be obtained from a set of (more or less) phenomenological 
balance laws representing the conservation (or not...) of the key quantities. The possibility that different fluid components may be able to flow (or perhaps rather ``drift'') relative to one another, leads to a {\it multi-fluid system}. In order to model such systems we assume that the system contains a number of distinguishable 
components, the dynamics of which are coupled. The formalism that we will develop draws on experience from 
chemistry, where one regularly has to consider the mechanics of mixtures, but is adapted to the kind of systems that 
are relevant for General Relativity. The archetypal such system is (again) represented by the neutron 
star core, where we expect different components (neutrons, protons, hyperons) to be in a superfluid state. However, 
the formalism is general enough that it can be applied in a variety of contexts, including (as we shall see later) the 
problem of heat conduction and the charged flows relevant for electromagnetism.

As the concept may not be familiar, it is worth considering the notion of a multi-fluid system in a bit more detail before we move on.
In principle, it is easy to see how such a system may arise. Recall the discussion of the mean-free path, but consider a
system with two distinct particle species. Suppose that the mean-free path associated with  scattering of particles of the same kind
is (for some reason) significantly shorter than the scale for inter-species collisions. Then we have two clearly defined ``fluids''. In fact, any system where it is meaningful to consider one component drifting (on average) relative to another one can be considered from this point-of-view (a liquid with gas bubbles would be an obvious example). 

Another relevant context involves systems that exhibit superfluidity.
At the most basic level, superfluidity implies that no friction impedes the flow. Technically,  the previous argument leading to a scale for averaging does not work anymore. However, a superfluid system has a different scale associated with it;
the so-called coherence length. The coherence length arises from the fact that a superfluid is a ``macroscopic'' quantum state,
the flow of which depends on the gradient of the phase of the wave-function (the so-called order parameter, see Sect.~\ref{sec:bec}).
On some small scale, the superfluidity breaks down due to quantum fluctations. This defines the coherence
length. It can be taken as the typical ``size'' of a Cooper pair in a fermionic system.
On any larger scale the system exhibits collective (fluid) behaviour.

For neutron-star superfluids, the coherence length is of the order of tens of Fermi; evidently, much smaller than the mean-free path
in the normal fluid case. This means that superfluids can exhibit extremely small scale dynamics. Since a superfluid is inviscid,  superfluid neutrons and superconducting protons (say) do not scatter (at least not at as long as thermal excitations can be ignored) and hence the outer core of a neutron star demands a multi-fluid treatment \citep{2011MNRAS.410..805G}. One can meaningfully take
the fluid elements to have a size of the order of the coherence length, i.e. they are tiny.
However, in reality the problem is more complicated, as yet another length-scale needs to be considered. First of all, on scales larger than the Debye screening length, the electrons will be electromagnetically locked to the protons, forming a charge-neutral conglomerate that
{\it does} exhibit friction (due to electron-electron scattering). This brings us back to the mean-free path argument. At finite temperatures we  also need to consider thermal excitations for both neutrons and protons (which may scatter and dissipate), making the problem rather complex.
Finally, ideal superfluids are irrotational and neutron stars are not. In order to mimic bulk rotation the neutron
superfluid must form a dense array of vortices (locally breaking the superfluidity). This brings yet another length  scale
into the picture. In order to develop a useful fluid model, we need to average over the vortices, as well. This makes the
effective fluid elements much larger. The typical vortex spacing in a neutron star is of the order;
\be
d_\n \approx 4\times10^{-4} \left({P \over 1\ \mbox{ms}} \right)^{1/2} \ \mbox{cm} \ ,
\ee
where $P$ is the star's spin period.
In other words, the fluid elements we consider may (at the end of the day)  be quite large also in a superfluid system.



\section{Physics in a curved spacetime}
\label{sec:gr}

There is an extensive literature on Special and General Relativity and
the spacetime-based view\footnote{There are three space and one time
dimensions that form a type of topological space known as a
manifold \citep{wald84:_book}. Local, suitably small
patches of a curved spacetime are practically the same as patches of
flat, Minkowski spacetime. Moreover, where two patches overlap, the
identification of points in one patch with those in the other is
smooth.} of the laws of physics, providing historical context, technical insight and topical updates. For a student at any level
interested in developing a working understanding we recommend \cite{taylorwheeler92:_book} for an introduction, followed
by Hartle's excellent text \citeyearpar{hartle03:_book} designed for
students at the undergraduate level. The recent contribution from \cite{2014grav.book.....P} provides a detailed discussion of the link between Newtonian gravity and Einstein's four dimensional picture. For more advanced students,
we suggest two of the classics, ``MTW'' \citep{mtw73} and
\cite{weinberg72:_book}, or the more contemporary book by
\cite{wald84:_book}. Finally, let us not forget the \textit{Living
Reviews} archive as a premier online source of up-to-date information!

In terms of the experimental and/or observational support for Special and
General Relativity, we recommend two articles by Will that were written
for the 2005 World Year of Physics
celebration \citeyearpar{will05:_sr,will05:_gr}. They summarize a variety of
tests that have been designed to expose breakdowns in both
theories. (We also recommend Will's popular book \textit{Was Einstein
  Right?} \citeyearpar{will86:_rightbook} and his technical exposition \textit{Theory and Experiment in Gravitational Physics} \citeyearpar{will93:_theorybook}.) Updates including the breakthrough observations of gravitational waves can be found in recent monographs \citep{mmbook,nabook} . There have been significant recent developments, but... to date, Einstein's theoretical
edifice is still standing! 

For Special Relativity, this is not surprising, given its long list of
successes: explanation of the Michelson--Morley result, the prediction
and subsequent discovery of anti-matter, and the standard model of
particle physics, to name a few. \cite{will05:_sr} offers the
observation that genetic mutations via cosmic rays require Special
Relativity, since otherwise muons would decay before making it to the
surface of the Earth. On a more somber note, we may consider the Trinity
site in New Mexico, and the tragedies of Hiroshima and Nagasaki, as
reminders of $E = m c^2$.

In support of General Relativity, there are E\"otv\"os-type experiments
testing the equivalence of inertial and gravitational mass, detection of
gravitational red-shifts of photons, the passing of the solar system
tests, confirmation of energy loss via gravitational radiation in the
Hulse--Taylor binary pulsar---and eventually the first direct detection of these faint whispers from the Universe in 2015---and the expansion of the
Universe. Incredibly, General Relativity even finds  a practical
application in the GPS system. In fact, we need both of Einstein's theories. The speed of the moving clock leads to it slowing down by 7 micro-seconds every day, while the fact that a clock in a gravitational field runs slow, leads to the orbiting clock appearing to speed up by 45 micro-seconds each day. All in all, if we ignore relativity position errors accumulate at a rate of about 10~km every day \citep{will05:_sr}. This would make reliable navigation impossible. 

The evidence is  overwhelming that General Relativity, or at least
some closely related theory that passes the entire collection of tests, is the proper description of
gravity. Given this, we assume the Einstein Equivalence Principle,
i.e., that \citep{will05:_sr,will05:_gr,will93:_theorybook}
\begin{itemize}
\item[-] test bodies fall with the same acceleration independently of
  their internal structure or composition;
\item[-] the outcome of any local non-gravitational experiment is
  independent of the velocity of the freely-falling reference frame in
  which it is performed;
\item[-] the outcome of any local non-gravitational experiment is
  independent of where and when in the Universe it is performed.
\end{itemize}
If the Equivalence Principle holds, then gravitation must be described by a
metric-based theory \citep{will05:_sr}. This means that
\begin{enumerate}
\item spacetime is endowed with a symmetric metric,
\item the trajectories of freely falling bodies are geodesics of that
  metric, and
\item in local freely falling reference frames, the non-gravitational
  laws of physics are those of Special Relativity.
\end{enumerate}
For our present purposes this is very good news. The availability of a
metric\footnote{The metric has a lot of ``heavy lifting'' to do. It allows us to measure spacetime intervals, provides  a causal structure---the local meaning of past and future---introduces the notions of proper time and local inertial frames and dictates the motion of test particles.} means that we can develop the theory without requiring much of the
differential geometry edifice that would be needed in a more general case.
We will develop the description of relativistic fluids with this in mind.
Readers that find our approach too ``pedestrian'' may want to consult
the article by \cite{gour06}, which serves as a useful
complement to our description.


\subsection{The metric and spacetime curvature}

Our strategy is to provide a ``working understanding'' of the
mathematical objects that enter the Einstein equations of General
Relativity. We assume that the metric is the fundamental ``field'' of
gravity. For a four-dimensional spacetime the metric determines the distance
between two spacetime points along a given curve, which can generally be
written as a one parameter function with, say, components
$x^a(\tau)$. For a material body, it is natural to take the parameter to be proper time, but we may opt to make a different choice. As we will see, once a notion of parallel transport
is established, the metric also encodes  information about the
curvature of spacetime, which is taken to be pseudo-Riemannian, meaning that the signature\footnote{It is worth noting that much work originating from  particle physics assumes a metric signature  $-2$. The main impact of this difference as far as fluids are concerned is that it changes the normalization of the four velocity.} of the metric is $-+++$
(cf.\ Eq.~(\ref{sig}) below). 

In a coordinate basis, which we will assume throughout this review, the metric
is denoted by $g_{a b} = g_{b a}$. The symmetry implies that
there are in general ten independent components (modulo the freedom to set
arbitrarily four components that is inherited from coordinate transformations;
cf.\ Eqs.~(\ref{ct1}) and~(\ref{ct2}) below). The spacetime version of the
Pythagorean theorem takes the form
\begin{equation}
  d s^2 = g_{a b} \, d x^a \, d x^b \ ,
\end{equation}
and in a local set of Minkowski coordinates $\{t,x,y,z\}$ (i.e., in a local
inertial frame, or small patch of the manifold) it looks like
\begin{equation}
 d s^2 =
  - \left( d t \right)^2 +
  \left( d x \right)^2 +
  \left( dy \right)^2 +
  \left( dz \right)^2.
  \label{sig}
\end{equation}
This illustrates the $-+++$ signature. The inverse metric $g^{a b}$ is
such that
\begin{equation}
  g^{a c} g_{c b} = \delta^a{}_b,
\end{equation}
where $\delta^a{}_b$ is the unit tensor. The metric is also used to
raise and lower spacetime indices, i.e., if we let $V^a$ denote a
contravariant vector, then its associated covariant vector (also known as a
covector or one-form) $V_a$ is obtained as
\begin{equation}
  V_a = g_{a b} V^b 
  \qquad \Leftrightarrow \qquad
  V^a = g^{a b} V_b \ .
\end{equation}

We can now consider three different classes of curves: timelike, null, and spacelike. A vector is 
said to be timelike if $g_{a b} V^a V^b < 0$, null if $g_{a b} V^a V^b = 0$, and spacelike if 
$g_{a b} V^a V^b > 0$. We can naturally define timelike, null, and spacelike curves in terms of 
the congruence of tangent vectors that they generate. A particularly useful timelike curve for 
fluids is one that is parameterized by the so-called proper time, i.e., $x^a(\tau)$ where
\begin{equation}
 d \tau^2 = - d s^2.
\end{equation}
The tangent $u^a$ to such a curve has unit magnitude; specifically,
\begin{equation}
  u^a \equiv \frac{dx^a}{d\tau},
  \label{uvec}
\end{equation}
and thus
\begin{equation}
  g_{a b} u^a u^b = g_{a b}
  \frac{d x^a}{d \tau}
  \frac{d x^b}{d\tau} =
  \frac{d s^2}{d \tau^2} = - 1.
\end{equation}

Under a coordinate transformation $x^a \to \overline{x}^a$, contravariant vectors transform as
\begin{equation}
  \overline{V}^a =
  \frac{\partial \overline{x}^a}{\partial x^b} V^b
  \label{ct1}
\end{equation}
and covariant vectors as
\begin{equation}
  \overline{V}_a =
  \frac{\partial x^b}{\partial \overline{x}^a} V_b \ .
  \label{ct2}
\end{equation}
Tensors with a greater rank (i.e., a greater number of indices), transform
similarly by acting linearly on each index using the above two rules.

When integrating, as we have to when we discuss conservation laws for fluids,
we must make use of an appropriate measure that ensures the coordinate
invariance of the integration. In the context of three-dimensional
Euclidean space this measure is referred to as the Jacobian. For spacetime,
we use the so-called volume form $\epsilon_{abcd}$. It is
completely antisymmetric, and for four-dimensional spacetime, it has only one
independent component, which is
\begin{equation}
  \epsilon_{0 1 2 3} = \sqrt{- g}
  \qquad \mbox{and} \qquad
  \epsilon^{0 1 2 3} = \frac{1}{\sqrt{- g}},
\end{equation}
where $g$ is the determinant of the metric (cf.\ Appendix~\ref{appendix} for
details). The minus sign is required under the square root because of the
metric signature. By contrast, for three-dimensional Euclidean space
(i.e., when considering the fluid equations in the Newtonian limit) we
have
\begin{equation}
  \epsilon_{1 2 3} = \sqrt{g}
  \qquad \mbox{and} \qquad
  \epsilon^{1 2 3} = \frac{1}{\sqrt{g}},
\end{equation}
but now $g$ is the determinant of the three-dimensional space metric. A
general identity that is extremely useful for writing the fluid vorticity in
three-dimensional, Euclidean space---using lower-case Latin indices and
setting $s = 0$, $n = 3$ and $j = 1$ in Eq.~(\ref{epsum2}) of
Appendix~\ref{appendix}---is
\begin{equation}
  \epsilon^{m i j} \epsilon_{m k l} =
  \delta^i{}_k \delta^j{}_l - \delta^j{}_k \delta^i{}_l.
\end{equation}
The general identities in Eqs.~(\ref{epsum1}, \ref{epsum2}, \ref{epsum3}) of
Appendix~\ref{appendix} will be frequently used in the following.


\subsection{Parallel transport and the covariant derivative}
\label{section_2.2}

In order to have a generally covariant prescription for fluids---in terms of spacetime tensors---we must have a notion of derivative
$\nabla_a$ that is itself covariant. For example, when $\nabla_a$ acts
on a vector $V^a$ a rank-two tensor of mixed indices must result:
\begin{equation}
  \overline{\nabla}_b \overline{V}^a =
  \frac{\partial x^c}{\partial \overline{x}^b}
  \frac{\partial \overline{x}^a}{\partial x^d}
  \nabla_c V^d \ .
\end{equation}
The ordinary partial derivative does not work because under a general
coordinate transformation
\begin{equation}
  \frac{\partial \overline{V}^a}{\partial \overline{x}^b} =
  \frac{\partial x^c}{\partial \overline{x}^b}
  \frac{\partial \overline{x}^a}{\partial x^d}
  \frac{\partial V^d}{\partial x^c} +
  \frac{\partial x^c}{\partial \overline{x}^b}
  \frac{\partial^2 \overline{x}^a}{\partial x^c \partial x^d}
  V^d \ .
\end{equation}
The second term spoils the general covariance, since it vanishes only for the
restricted set of rectilinear transformations
\begin{equation}
  \overline{x}^a = a^a{}_b x^b + b^a \ ,
\end{equation}
where $a^a{}_b$ and $b^a$ are constants. Note that this includes the Lorentz transformation of Special Relativity.

For both physical and mathematical reasons, one expects a covariant derivative
to be defined in terms of a limit. This is, however, a bit problematic. In
three-dimensional Euclidean space limits can be defined uniquely as
vectors can be moved around without their length and direction changing, for
instance, via the use of Cartesian coordinates (the
$\{{\boldsymbol i},{\boldsymbol j},{\boldsymbol k}\}$ set of basis vectors) and the usual dot
product. Given these limits, those corresponding to more general
curvilinear coordinates can be established. The same is not true for
curved spaces and/or spacetimes because they do \emph{not} have an a
priori notion of parallel transport.

\begin{figure}[htb]
    \centerline{\includegraphics[scale = 0.5]{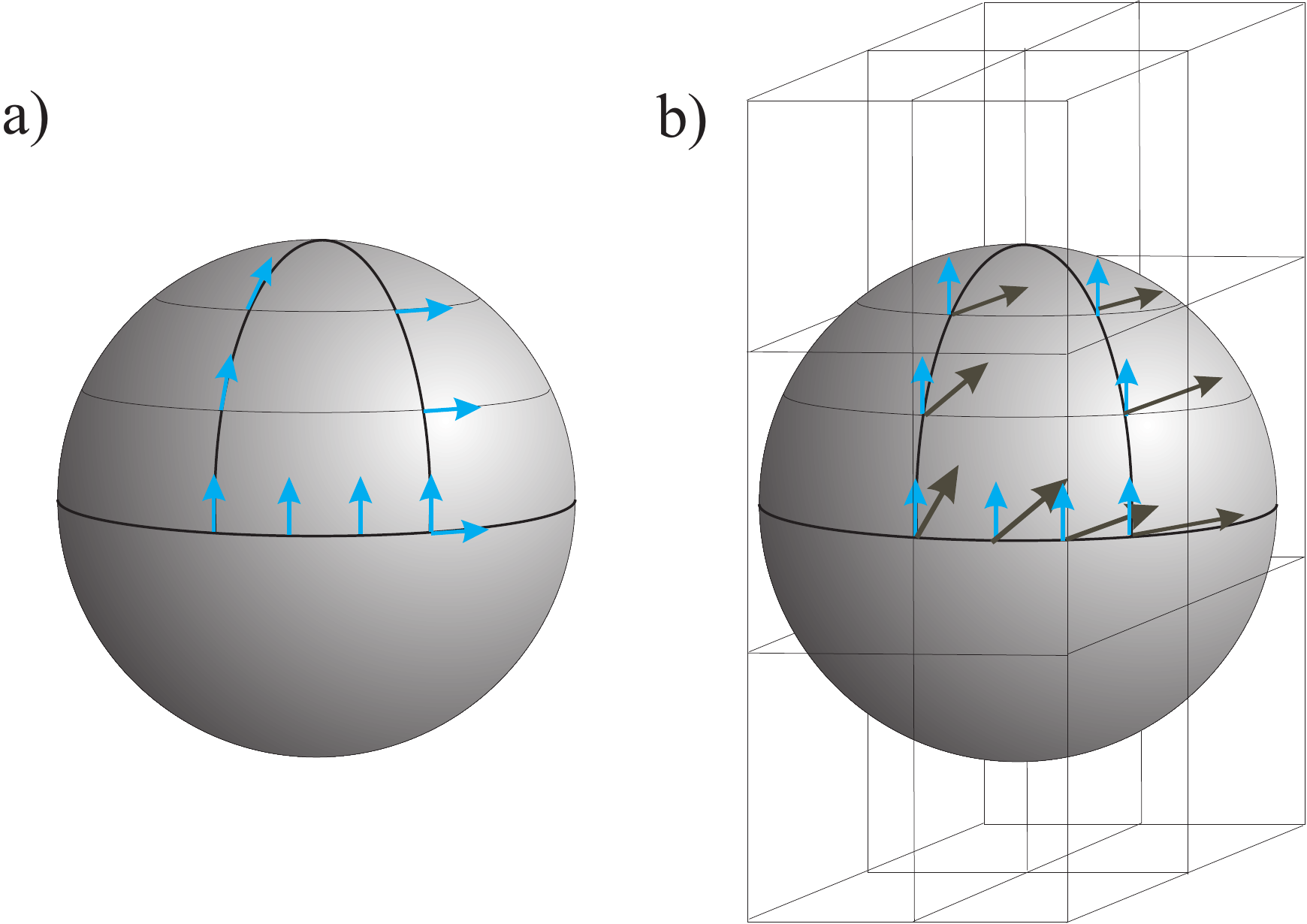}}
    \caption{A schematic illustration of two possible versions of
    parallel transport. In the first case (a) a vector is transported
    along great circles on the sphere locally maintaining the same
    angle with the path. If the contour is closed, the final
    orientation of the vector will differ from the original one. In
    case (b) the sphere is considered to be embedded in a
    three-dimensional Euclidean space, and the vector on the sphere
    results from projection. In this case, the vector returns to the
    original orientation for a closed contour.}
    \label{sphere}
\end{figure}

Consider the classic example of a vector on the surface of a sphere
(illustrated in Fig.~\ref{sphere}). Take this vector and move it along
some great circle from the equator to the North pole in such a way as to
always keep the vector pointing along the circle. Pick a different great
circle, and without allowing the vector to rotate, by forcing it to maintain
the same angle with the locally straight portion of the great circle that it
happens to be on, move it back to the equator. Finally, move the vector in
a similar way along the equator until it gets back to its starting
point. The vector's spatial orientation will be different from its
original direction, and the difference is directly related to the
particular path that the vector followed.

On the other hand, we could consider  the sphere to be embedded in a
three-dimensional Euclidean space, and let the two-dimensional vector on
the sphere result from projection of a three-dimensional vector. Then we move
the projection so that its higher-dimensional counterpart always maintains
the same orientation with respect to its original direction in the embedding
space. When the projection returns to its starting place it will have
exactly the same orientation as it started out with (see
Fig.~\ref{sphere}). It is now clear that a derivative operation that
depends on comparing a vector at one point to that of a nearby point
is not unique, because it depends on the choice of parallel transport.

\cite{pauli81:_rel_book} notes that
\cite{levicivita03:_partrans} is the first to have
formulated the concept of parallel ``displacement'', with
\cite{weyl52:_rel_book} generalizing it to manifolds that do not
have a metric. The point of view expounded in the books of Weyl and
Pauli is that parallel transport is best defined as a mapping of the
``totality of all vectors'' that ``originate'' at one point of a
manifold with the totality at another point. (In modern texts, this discussion tends to be based on fiber bundles.) Pauli points out that
we cannot simply require equality of vector components as the
mapping.

Let us examine the parallel transport of the force-free, point particle
velocity in Euclidean three-dimensional space as a means for motivating
the form of the mapping. As the velocity is constant, we know that the
curve traced out by the particle will be a straight line. In fact, we can
turn this around and say that the velocity parallel transports itself
because the path traced out is a geodesic (i.e., the straightest possible
curve allowed by Euclidean space). In our analysis we will borrow
liberally from the excellent discussion of 
\cite{lovelock89:_tensor_book}. Their text is comprehensive yet
readable for anyone  not well-versed with differential
geometry. Finally, we note that this analysis will be relevant later
when we consider the Newtonian limit of the  relativistic
equations, in an arbitrary coordinate basis.

We are all well aware that the points on the curve traced out by the
particle can be described, in Cartesian coordinates, by three functions
$x^i(t)$ where $t$ is the universal Newtonian time. Likewise, we know
that the tangent vector at each point of the curve is given by the velocity
components $v^i(t) = d x^i/d t$, and that the
force-free condition is equivalent to
\begin{equation}
  a^i(t) = \frac{dv^i}{d t} = 0
  \qquad \Rightarrow \qquad
  v^i(t) = \mathrm{const}.
\end{equation}
Hence, the velocity components $v^i(0)$ at the point $x^i(0)$ are equal
to those at any other point along the curve, say $v^i(T)$ at $x^i(T)$,
and so we could simply take $v^i(0) = v^i(T)$ as the mapping. But as
Pauli warns, we only need to reconsider this example using spherical
coordinates to see that the velocity components
$\{\dot{r},\dot{\theta},\dot{\phi}\}$ must  change as they
undergo parallel transport along a straight-line path (assuming the
particle does not pass through the origin). The question is what should be
used in place of component equality? The answer follows once we find a
curvilinear coordinate version of $dv^i/dt = 0$.

What we need is a new ``time'' derivative
$\overline{D}/d t$, that yields a generally
covariant statement
\begin{equation}
  \frac{\overline{D} \overline{v}^i}{d t} = 0,
\end{equation}
where the $\overline{v}^i(t) = d \overline{x}^i/d t$ are the velocity components in a curvilinear system of coordinates. Consider
now a coordinate transformation to the new coordinate system
$\overline{x}^i$, the inverse being $x^i = x^i(\overline{x}^j)$. Given
that
\begin{equation}
  v^i = \frac{\partial x^i}{\partial \overline{x}^j} \overline{v}^j
\end{equation}
we can write
\begin{equation}
  \frac{d v^i}{d t} =
  \left( \frac{\partial x^i}{\partial \overline{x}^j}
  \frac{\partial \overline{v}^j}{\partial \overline{x}^k} +
  \frac{\partial^2 x^i}{\partial \overline{x}^k \partial \overline{x}^j}
  \overline{v}^j \right) \overline{v}^k,
  \label{accinter}
\end{equation}
where
\begin{equation}
  \frac{d \overline{v}^i}{d t} =
  \frac{\partial \overline{v}^i}{\partial \overline{x}^j} \overline{v}^j.
\end{equation}
Again, we have an ``offending''  term that vanishes only for
rectilinear coordinate transformations. However, we are now in a position to
show the importance of this term to the definition of the covariant derivative.

First note that the metric $\overline{g}_{i j}$ for our curvilinear
coordinate system is obtained from
\begin{equation}
  \overline{g}_{i j} =
  \frac{\partial x^k}{\partial \overline{x}^i}
  \frac{\partial x^l}{\partial \overline{x}^j}
  \delta_{k l},
  \label{mettrans}
\end{equation}
where
\begin{equation}
  \delta_{i j} = \left\{
    \begin{array}{ll}
      1 & \qquad \mathrm{for\ } i = j, \\
      0 & \qquad \mathrm{for\ } i \neq j.
    \end{array}
  \right.
  \label{flatmet}
\end{equation}
Differentiating Eq.~(\ref{mettrans}) with respect to $\overline{x}$, and
permutating indices, we can show that
\begin{equation}
  \frac{\partial^2 x^h}{\partial \overline{x}^i \partial \overline{x}^j}
  \frac{\partial x^l}{\partial \overline{x}^k} \delta_{h l} =
  \frac{1}{2} \left( \overline{g}_{i k,j} + \overline{g}_{j k,i} -
  \overline{g}_{i j,k} \right) \equiv \overline{g}_{i l}
  \overline{ \left\{ \scriptstyle{l \atop j~k} \right\}},
\end{equation}
where we use commas to indicate partial derivatives:
\begin{equation}
  \overline{g}_{i j, k} \equiv
  \frac{\partial \overline{g}_{i j}}{\partial \overline{x}^k}.
\end{equation}
Using the inverse transformation of $\overline{g}_{i j}$ to $\delta_{i j}$
implied by Eq.~(\ref{mettrans}), and the fact that
\begin{equation}
  \delta^i{}_j =
  \frac{\partial \overline{x}^k}{\partial x^j}
  \frac{\partial x^i}{\partial \overline{x}^k},
\end{equation}
we get
\begin{equation}
  \frac{\partial^2 x^i}{\partial \overline{x}^j \partial \overline{x}^k} =
  \overline{\left\{\scriptstyle{l \atop j~k}\right\}}
  \frac{\partial x^i}{\partial \overline{x}^l}.
  \label{2part}
\end{equation}
Now we substitute Eq.~(\ref{2part}) into Eq.~(\ref{accinter})
and find
\begin{equation}
  \frac{d v^i}{d t} =
  \frac{\partial x^i}{\partial \overline{x}^j}
  \frac{\overline{{D}} \overline{v}^j}{d t},
\end{equation}
where
\begin{equation}
  \frac{\overline{{D}} \overline{v}^i}{{d} t} =
  \overline{v}^j \left( \frac{\partial \overline{v}^i}{\partial \overline{x}^j} +
  \overline{\left\{ \scriptstyle{i \atop k~j} \right\}}
  \overline{v}^k \right).
  \label{geodesic}
\end{equation}
The operator $\overline{{D}}/{d} t$ is easily seen to be
covariant with respect to general transformations of curvilinear
coordinates.

We now identify the generally covariant derivative (dropping the
overline) as
\begin{equation}
    \nabla_j v^i = \frac{\partial v^i}{\partial x^j} +
                   \left\{\scriptstyle{i \atop k~j}\right\} v^k \equiv
                   v^i{}_{; j}. \label{covdevcon}
\end{equation}
Similarly, the covariant derivative of a covector is
\begin{equation}
    \nabla_j v_i = \frac{\partial v_i}{\partial x^j} -
                   \left\{\scriptstyle{k \atop i~j}\right\} v_k \equiv
                   v_{i ; j}. \label{covdevcov}
\end{equation}
One extends the covariant derivative to higher rank tensors by adding to
the partial derivative each term that results by acting linearly on each
index with $\left\{\scriptstyle{i \atop j~k}\right\}$ using the two rules
given above.

Relying on our understanding of the force-free point particle, we have
built a notion of parallel transport that is consistent with our
intuition based on equality of components in Cartesian coordinates. We
can now expand this intuition to see how the vector components in a
curvilinear coordinate system must change under an infinitesimal,
parallel displacement from $x^i(t)$ to $x^i(t + \delta t)$. Setting
Eq.~(\ref{geodesic}) to zero, and noting that $v^i \delta t = \delta
x^i$, implies
\begin{equation}
  \delta v^i \equiv \frac{\partial v^i}{\partial x^j} \delta x^j =
  - \left\{ \scriptstyle{i \atop k~j} \right\} v^k \delta x^j.
\end{equation}
In General Relativity we assume that under an infinitesimal parallel
transport from a spacetime point $x^a(\tau)$ on a given curve to a
nearby point $x^a(\tau + \delta \tau)$ on the same curve, the
components of a vector $V^a$ will change in an analogous way, namely
\begin{equation}
  \delta V^a_\parallel \equiv
  \frac{\partial V^a}{\partial x^b} \delta x^b =
  - \Gamma^a_{c b} V^c \delta x^b \ ,
  \label{partransgr}
\end{equation}
where
\begin{equation}
  \delta x^a \equiv
  \frac{d x^a}{{d} \tau} \delta \tau \ .
\end{equation}
\cite{weyl52:_rel_book} refers to the symbol $\Gamma^a_{b c}$
as the ``components of the affine relationship'', but we will use the
modern terminology and call it the connection. In the language of Weyl
and Pauli, this is the mapping that we were looking for.

For Euclidean space, we can verify that the metric satisfies
\begin{equation}
  \nabla_i g_{j k} = 0
\end{equation}
for a general, curvilinear coordinate system. The metric is thus said
to be ``compatible'' with the covariant derivative. Metric
compatibility is imposed as an assumption in General Relativity. This results in the so-called
Christoffel symbol for the connection, defined as
\begin{equation}
  \Gamma^a_{b c} = \frac{1}{2} g^{a d}
  \left( g_{b d, c} + g_{c d, b} - g_{b c, d}\right).
  \label{gabc}
\end{equation}
The rules for the covariant derivative of a contravariant vector and a
covector are the same as in Eqs.~(\ref{covdevcon}) and~(\ref{covdevcov}),
except that all indices are spacetime ones.

\vspace*{0.1cm}
\begin{tcolorbox}
\textbf{Comment:} In addition to covariant derivative, we will need to draw on some aspects of differential geometry. In particular, it is useful to understand the \emph{wedge product} and the \emph{exterior derivative}. The wedge produce is (simply) an antisymmetrized tensor product. In the particular case of two one-forms $\boldsymbol A$ and $\boldsymbol B$, we have
$$
(\boldsymbol A\wedge \boldsymbol B)_{ab} = 2! A_{[a} B_{b]}
$$
In general, we can get away with suppressing the indices when we use forms because we know that we are dealing with forms (all indices downstairs) and the tensors are anti-symmetric.

Meanwhile, the exterior derivative is defined as a (normalized) anti-symmetric partial derivative:
$$
(d\boldsymbol A)_{ab} = 2\partial_{[a} A_{b]}
$$
The advantage of this definition is that the exterior derivative is a tensor, even though the partial derivative is not. From the definition---and the fact that partial derivatives commute---it follows that (for any form $\boldsymbol A$) we have
$$
d(d\boldsymbol A) = 0 
$$
This leads to the notion that a form is \emph{closed} if $d\boldsymbol A = 0 $ and \emph{exact} if $\boldsymbol A = d\boldsymbol B$ for some form $\boldsymbol B$.
\end{tcolorbox}
\vspace*{0.1cm}


\subsection{The Lie derivative and spacetime symmetries}
\label{section_2.3}

From the above discussion it should be evident that there are other ways to
take derivatives in a curved spacetime. A particularly important tool for
measuring changes in tensors from point to point
in spacetime is the Lie derivative. It requires a vector field, but no
connection, and is a more natural definition in the sense that it does not
even require a metric. The Lie derivative yields a tensor of the same type and rank as
the tensor on which the derivative operated (unlike the covariant
derivative, which increases the rank by one). It is as
important for Newtonian, non-relativistic fluids as for relativistic ones
(a fact which needs to be continually emphasized as it has not yet
permeated the fluid literature for chemists, engineers, and physicists).
For instance, the classic papers on the gravitational-wave driven Chandrasekhar--Friedman--Schutz
instability \citep{friedman78:_lagran,friedman78:_secul_instab} in rotating
stars are great illustrations of the use of the Lie derivative in Newtonian
physics. We recommend the book by \cite{schutz80:_geomet} for a
complete discussion and derivation of the Lie derivative and its role in
Newtonian fluid dynamics (see also the series of papers by \citealt{Carter03:_newtI,Carter03:_newtII,Carter04:_newtIII}). Here, we
will adapt  the coordinate-based discussion of
\cite{schouten89:_tenanal}, as it may be more readily
understood by readers not well-versed in differential geometry.

In a first course on classical mechanics, when students encounter rotations,
they are introduced to the idea of active and passive transformations. An
\emph{active} transformation would be to fix the origin and
axis-orientations of a given coordinate system with respect to some external
observer, and then move an object from one point to another point of the
same coordinate system. A \emph{passive} transformation would be to
place an object so that it remains fixed with respect to some external
observer, and then induce a rotation of the object with respect to a given
coordinate system, rotating the coordinate system itself with respect to
the external observer. We will derive the Lie derivative of a vector by
first performing an active transformation and then following it with a
passive transformation to determine how the final vector differs from its
original form. In the language of differential geometry, we will first
``push-forward'' the vector, and then subject it to a ``pull-back''. 

\vspace*{0.1cm}
\begin{tcolorbox}
\textbf{Comment:} In the following we will make regular use of maps between different manifolds. The basic idea is that, given two manifolds, $M$, and $N$ (say), possibly of different dimension and with coordinates $x^a$ and $X^A$, we imagine a map $\phi:M\to N$ and a function $f:N\to \textbf{R}$, in turn, a function on $M$. This set-up allows us to construct a map $(f\circ \phi):M\to \textbf{R}$, giving a function on $M$. This is referred to as the \emph{pull-back} of $f$ by $\phi$, the idea being that we are \emph{pulling back} the function from $N$ to $M$.
\end{tcolorbox}
\begin{tcolorbox}
The inverse of this does not work---we cannot push a function ``forward''. However, we know that we can think of a vector as a derivative that maps smooth functions into numbers. This then allows us to define the \emph{push-forward} of a vector. The idea may seem somewhat abstract at this point, but should become clear later. 
The Lie derivative provides the first example of the procedure.
\end{tcolorbox}
\vspace*{0.1cm}

In the active (push-forward) sense we imagine that there are two
spacetime points connected by a smooth curve $x^a(\lambda)$. Let the
first point be at $\lambda = 0$, and the second, nearby point at $\lambda
= \epsilon$, i.e., $x^a(\epsilon)$; that is,
\begin{equation}
  x^a_\epsilon \equiv
  x^a(\epsilon) \approx x^a_0 + \epsilon \, \xi^a \ ,
  \label{push}
\end{equation}
where $x^a_0 \equiv x^a(0)$ and
\begin{equation}
  \xi^a =
  \left.\frac{dx^a}{{d} \lambda} \right|_{\lambda = 0}
\end{equation}
is the tangent to the curve at $\lambda = 0$. In the passive
(pull-back) sense we imagine that the coordinate system itself is
changed to $\overline{x}{}^a =\overline{x}{}^a(x^b)$, but in the
very special form
\begin{equation}
  \overline{x}{}^a = x^a - \epsilon \, \xi^a \ .
  \label{pull}
\end{equation}
In this second step  the Lie derivative differs from the
covariant derivative. If we insert Eq.~(\ref{push}) into
Eq.~(\ref{pull}) we find the result $\overline{x}{}^a_\epsilon =
x^a_0$. This is called ``Lie-dragging'' of the coordinate frame, 
meaning that the coordinates at $\lambda = 0$ are carried along so
that at $\lambda = \epsilon$ (and in the new coordinate system) the
coordinate labels take the same numerical values.

\begin{figure}[htb]
    \centerline{\includegraphics[scale = 0.5]{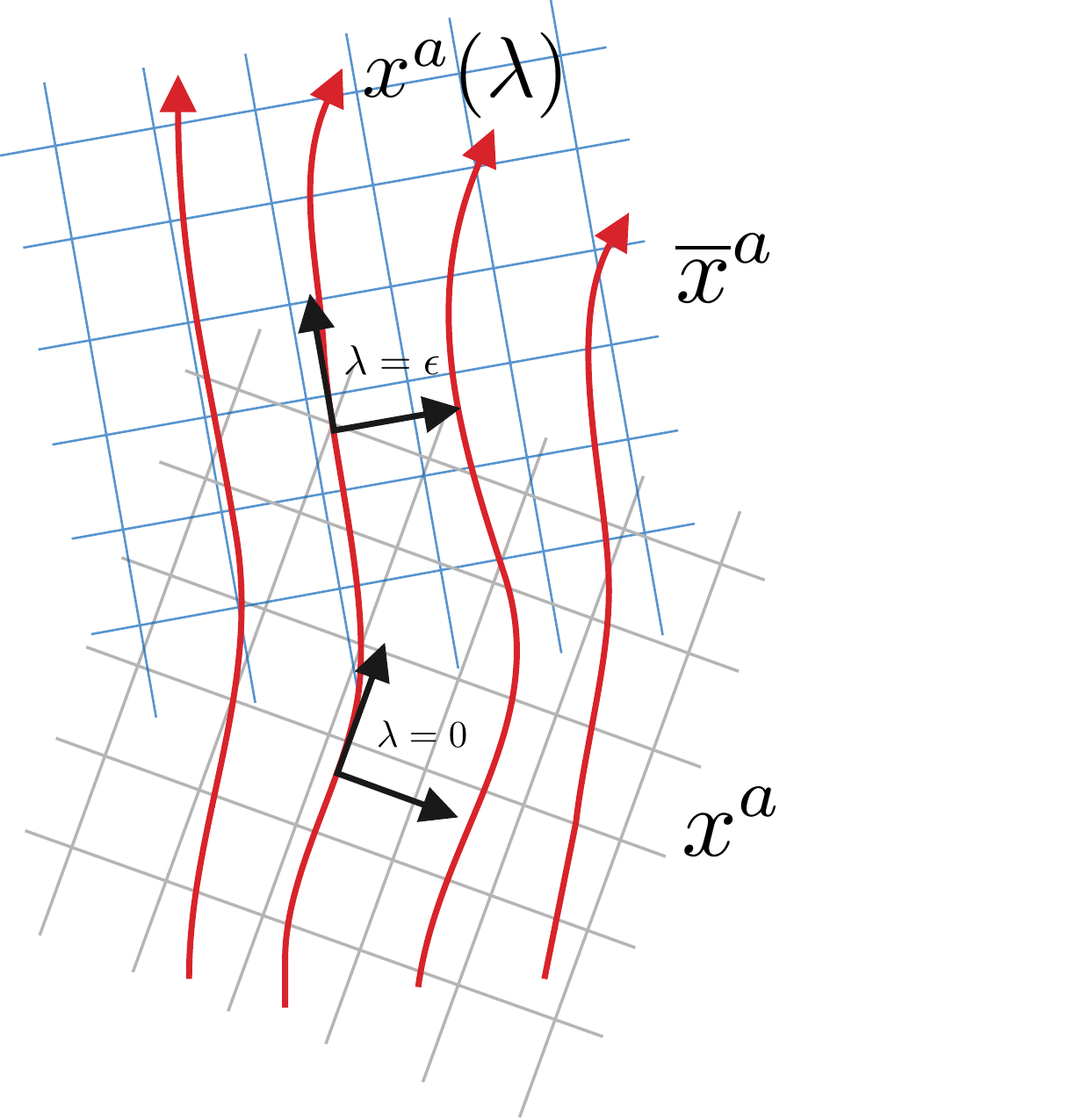}}
    \caption{A schematic illustration of the Lie derivative. The
    coordinate system is dragged along with the flow, and one can
    imagine an observer ``taking derivatives'' as he/she moves with
    the flow (see the discussion in the text).}
    \label{transport}
\end{figure}

As an interesting aside it is worth noting that
\cite{arnold95:_mathbook}---only a little whimsically---refers to
this construction as the ``fisherman's derivative''. He imagines a
fisherman sitting in a boat on a river, ``taking derivatives'' as the
boat moves along with the current. Let us now see how Lie-dragging reels in vectors.

For some given vector field that takes values $V^a(\lambda)$, say,
along the curve, we write
\begin{equation}
  V^a_0 = V^a(0)
\end{equation}
for the value of $V^a$ at $\lambda = 0$ and
\begin{equation}
  V^a_\epsilon = V^a(\epsilon)
\end{equation}
for the value at $\lambda = \epsilon$. Because the two points $x^a_0$
and $x^a_\epsilon$ are infinitesimally close ($\epsilon \ll 1$)  we have
\begin{equation}
  V^a_\epsilon \approx V^a_0 + \epsilon \, \xi^b
  \left. \frac{\partial V^a}{\partial x^b} \right|_{\lambda = 0}
  \label{nearbyvec}
\end{equation}
for the value of $V^a$ at the nearby point and in the \emph{same}
coordinate system. However, in the new coordinate system (at the nearby
point) we find
\begin{equation}
  \overline{V}{}^a_\epsilon = \left.
  \left( \frac{\partial \overline{x}{}^a}{\partial x^b} V^b\right)
  \right|_{\lambda = \epsilon} \!\!\!\!\!\! \approx
  V^a_\epsilon - \epsilon\, V^b_0
  \left. \frac{\partial \xi^a}{\partial x^b} \right|_{\lambda = 0}\!\!\!\!\!\!.
\end{equation}
The Lie derivative now is defined to be
\begin{eqnarray}
  {\cal L}_\xi V^a &=&
  \lim_{\epsilon \to 0} \frac{\overline{V}{}^a_\epsilon - V^a}{\epsilon}
  \nonumber
  \\
  &=& \xi^b \frac{\partial V^a}{\partial x^b} -
  V^b \frac{\partial \xi^a}{\partial x^b}
  \nonumber
  \\
  &=& \xi^b \nabla_b V^a - V^b \nabla_b \xi^a \ ,
  \label{liedev}
\end{eqnarray}%
where we have dropped the ``$0$'' subscript and the last equality follows
easily by noting $\Gamma^c_{a b} = \Gamma^c_{b a}$.

The Lie derivative of a covector $A_a$ is easily obtained by acting on the
scalar $A_a V^a$ for an arbitrary vector $V^a$:
\begin{eqnarray}
  {\cal L}_\xi A_a V^a &=&
  V^a {\cal L}_\xi A_a + A_a {\cal L}_\xi V^a
  \nonumber
  \\
  &=& V^a {\cal L}_\xi A_a +
  A_a \left( \xi^b \nabla_b V^a - V^b \nabla_b \xi^a \right).
\end{eqnarray}%
But, because $A_a V^a$ is a scalar,
\begin{eqnarray}
  {\cal L}_\xi A_a V^a &=& \xi^b \nabla_b A_a V^a
  \nonumber
  \\
  &=& \xi^b \left( V^a \nabla_b A_a + A_a \nabla_b V^a \right) \ ,
\end{eqnarray}%
and thus
\begin{equation}
  V^a \left( {\cal L}_\xi A_a - \xi^b \nabla_b A_a -
  A_b \nabla_a \xi^b \right) = 0.
\end{equation}
Since $V^a$ is arbitrary we have
\begin{equation}
  {\cal L}_\xi A_a =
  \xi^b \nabla_b A_a + A_b \nabla_a \xi^b \ .
\end{equation}

Eq.~(\ref{partransgr}) introduced the effect of parallel transport
on vector components. By contrast, the Lie-dragging of a vector causes
its components to change as
\begin{equation}
  \delta V^a_{\cal L} = {\cal L}_\xi V^a \, \epsilon \ .
\end{equation}
We see that if ${\cal L}_\xi V^a = 0$, then the components of the
vector do not change as the vector is Lie-dragged. Suppose now that
$V^a$ represents a vector field and that there exists a corresponding
congruence of curves with tangent given by $\xi^a$. If the components
of the vector field do not change under Lie-dragging, we can show that
this implies a symmetry, meaning that a coordinate system can be found such
that the vector components do not depend on one of the
coordinates. This is a potentially very powerful statement.

Let $\xi^a$ represent the tangent to the curves drawn out by, say, the
$a = \phi$ coordinate. Then we can write $x^a(\lambda) = \lambda$ which
means
\begin{equation}
  \xi^a = \delta^a{}_\phi \ .
\end{equation}
If the Lie derivative of $V^a$ with respect to $\xi^b$ vanishes we
find
\begin{equation}
  \xi^b \frac{\partial V^a}{\partial x^b} =
  V^b \frac{\partial \xi^a}{\partial x^b} = 0 \ .
\end{equation}
Using this in Eq.~(\ref{nearbyvec}) implies $V^a_\epsilon = V^a_0$,
that is to say, the vector field $V^a(x^b)$ does not depend on the
$x^a$ coordinate. Generally speaking, every $\xi^a$ that exists that
causes the Lie derivative of a vector (or higher rank tensors) to vanish
represents a symmetry.

Let us take the spacetime metric $g_{a b}$ as an example. A spacetime
symmetry can be represented by a generating vector field $\xi^a$
such that
\begin{equation}
  {\cal L}_\xi g_{a b} = \nabla_a \xi_b + \nabla_b \xi_a = 0 \ .
  \label{kvec}
\end{equation}
This is known as Killing's equation, and solutions to this equation are
naturally referred to as Killing vectors. It is now fairly easy to demonstrate the claim that the existence of a Killing vector relates to an underlying symmetry of the spacetime metric. First we expand \eqref{kvec} to get
\begin{equation}
    g_{bc} \partial_a \xi^c + g_{ac} \partial_b \xi^c + \xi^d \partial_d g_{ab} = 0 \ .
    \label{xiexp}
\end{equation}
Then we assume that the Killing vector is associated with one of the coordinates, e.g., by letting $\xi^a = \delta_0^a$. The first two terms in \eqref{xiexp} then vanish by definition, and we are left with
\begin{equation}
    \xi^d \partial_d g_{ab} = \partial_0 g_{ab} = 0 \ , 
\end{equation}
demonstrating that the metric does not depend on the $x^0$ coordinate.

An important application of this idea is provided by stationary, axisymmetric, and asymptotically flat
spacetimes---highly relevant in the present context as they
capture the physics of rotating, equilibrium configurations. The associated geometries are   fundamental for the relativistic astrophysics of spinning black holes and neutron
stars. Stationary, axisymmetric, and asymptotically flat spacetimes are
such that \citep{bonazzola93:_rotbodies}
\begin{enumerate}
\item there exists a Killing vector $t^a$ that is timelike at
  spatial infinity, and the independence of the metric on the associated time coordinate leads to the solution being stationary;
\item there exists a Killing vector $\phi^a$ that vanishes on a
  timelike 2-surface---the axis of symmetry---is spacelike
  everywhere else, and whose orbits are closed curves; and
\item asymptotic flatness means the scalar products $t_a t^a$,
  $\phi_a \phi^a$, and $t_a \phi^a$ tend to, respectively,
  $- 1$, $+ \infty$, and $0$ at spatial infinity.
\end{enumerate}
%


\subsection{Spacetime curvature}

The main message of the previous two Sections~\ref{section_2.2}
and~\ref{section_2.3} is that one must have
an a priori idea of how vectors and higher rank tensors are moved from
point to point in spacetime. An immediate manifestation of the complexity
associated with carrying tensors about in spacetime is that the covariant
derivative does not commute. For a vector we find 
\begin{equation}
  \nabla_b \nabla_c V^a - \nabla_c \nabla_b V^a =
  R^a{}_{d b c} V^d \ ,
  \label{covcom}
\end{equation}
where $R^a{}_{d b c}$ is the Riemann tensor. It is obtained from
\begin{equation}
  R^a{}_{d b c} =
  \Gamma^a_{d c, b} - \Gamma^a_{d b, c} +
  \Gamma^a_{e b} \Gamma^e_{d c} -
  \Gamma^a_{e c} \Gamma^e_{d b} \ .
\end{equation}
Closely associated are the Ricci tensor $R_{ab} = R_{ba}$
and scalar $R$ that are defined by the contractions
\begin{equation}
  R_{a b} = R^c{}_{a c b} \ ,
  \qquad
  R = g^{a b} R_{a b} \ .
\end{equation}
We will also need the Einstein tensor, which is given
by
\begin{equation}
  G_{a b} = R_{a b} - \frac{1}{2} R g_{a b} \ .
\end{equation}
It is such that $\nabla_b G^b{}_a$ vanishes identically. This is known
as the Bianchi identity.

A more intuitive understanding of the Riemann tensor is obtained by seeing
how its presence leads to a path-dependence in the changes that a vector
experiences as it moves from point to point in spacetime. Such a situation
is known as a ``non-integrability'' condition, because the result depends on
the whole path and not just the initial and final points. That is, it is
not like a total derivative which can be integrated and depends on only the
 limits of integration. Geometrically we say that the
spacetime is curved, which is why the Riemann tensor is also known as the
curvature tensor.

To illustrate the meaning of the curvature tensor, let us suppose that we are
given a surface that is parameterized by the two parameters $\lambda$ and
$\eta$. Points that live on this surface will have coordinate labels
$x^a(\lambda,\eta)$. We want to consider an infinitesimally small
``parallelogram'' whose four corners (moving counterclockwise with the first
corner at the lower left) are given by $x^a(\lambda,\eta)$,
$x^a(\lambda,\eta + \delta \eta)$, $x^a(\lambda + \delta \lambda,\eta +
\delta \eta)$, and $x^a(\lambda + \delta \lambda,\eta)$. Generally
speaking, any ``movement'' towards the right of the parallelogram is effected
by varying $\eta$, and ones towards the top results by varying $\lambda$.
The plan is to take a vector $V^a(\lambda,\eta)$ at the lower-left
corner $x^a(\lambda,\eta)$, parallel transport it along a $\lambda =
\mathrm{const}$ curve to the lower-right corner at $x^a(\lambda,\eta + \delta
\eta)$ where it will have the components
$V^a(\lambda,\eta + \delta \eta)$, and end up by parallel transporting
$V^a$ at $x^a(\lambda,\eta + \delta \eta)$ along an $\eta = \mathrm{const}$
curve to the upper-right corner at $x^a(\lambda + \delta \lambda,\eta +
\delta \eta)$. We will call this path I and denote the final component
values of the vector as $V^a_\mathrm{I}$. We then repeat the  process
except that the path will go from the lower-left to the upper-left and
then on to the upper-right corner. We will call this path II and denote
the final component values as $V^a_\mathrm{II}$.

Recalling Eq.~(\ref{partransgr}) as the definition of parallel transport,
we first of all have
\begin{equation}
  V^a(\lambda,\eta + \delta \eta) \approx V^a(\lambda,\eta) +
  \delta_\eta V^a_\parallel (\lambda,\eta) =
  V^a(\lambda,\eta) - \Gamma^a_{b c} V^b \delta_\eta x^c
\end{equation}
and
\begin{equation}
  V^a(\lambda + \delta \lambda,\eta) \approx V^a(\lambda,\eta) +
  \delta_\lambda V^a_\parallel (\lambda,\eta) =
  V^a(\lambda,\eta) - \Gamma^a_{b c} V^b
  \delta_\lambda x^c \ ,
\end{equation}
where
\begin{equation}
  \delta_\eta x^a \approx
  x^a(\lambda,\eta + \delta \eta) - x^a(\lambda,\eta) \ ,
  \qquad
  \delta_\lambda x^a \approx
  x^a(\lambda + \delta \lambda,\eta) - x^a(\lambda,\eta) \ .
\end{equation}
Next, we need
\begin{eqnarray}
  V^a_\mathrm{I} &\approx& V^a(\lambda,\eta + \delta \eta) +
  \delta_\lambda V^a_\parallel(\lambda,\eta + \delta \eta),
  \\
  V^a_\mathrm{II} &\approx& V^a(\lambda + \delta \lambda,\eta) +
  \delta_\eta V^a_\parallel(\lambda + \delta \lambda,\eta).
\end{eqnarray}%
Working things out, we find that the difference between the two paths
is
\begin{equation}
  \Delta V^a \equiv V^a_\mathrm{I} - V^a_\mathrm{II} =
  R^a{}_{d b c} V^d \delta_\lambda x^c \delta_\eta x^b \ ,
\end{equation}
which follows because $\delta_\lambda \delta_\eta x^a = \delta_\eta
\delta_\lambda x^a$, i.e., we have closed the parallelogram.

\subsection{The Einstein field equations}

We now have  the  tools we need to outline the argument that leads to the field equations of General Relativity. This sketch will be complemented by a variational derivation in Sect.~\ref{sec:efevar}. 

Consider 
two freely falling particles moving along neighbouring geodesics with a vector $\xi^a$ measuring the separation.
Assuming that this vector is purely spatial according to the trajectory of one of the bodies, who we also assign to measure time (such that the corresponding four-velocity only has a time-component), we have 
\begin{equation}
u^a \xi_a = 0 \ .
\end{equation}
The second derivative of the separation vector will be affected by the spacetime curvature. With this set-up it follows that
\begin{equation}
u^a \nabla_a \xi^b - \xi^a \nabla_a u^b = 0
\label{uxirel}
\end{equation}
and we find that  
\begin{equation}
u^c \nabla_c (u^b\nabla_b \xi^a) = u^c \xi^b ( \nabla_c \nabla_b - \nabla_b \nabla_c) u^a =  -
R^a_{\ d b c} u^d \xi^b u^c \ ,
\label{geodev}
\end{equation}
where we have used the fact that the Riemann
tensor encoded the failure of second covariant derivatives to commute.
This is the equation of {\em geodesic deviation}. 

At this point it is useful to introduce a total time derivative, such that
\begin{equation}
{D \over D\tau} = u^a \nabla_a
\end{equation}
which means that \eqref{geodev} becomes
\begin{equation}
{D^2 \xi^a \over D\tau^2}  =   -
R^a_{\ d b c} u^d \xi^b u^c \ .
\end{equation}
This provides us with an expression for the
relative acceleration caused by the spacetime curvature. As gravity is a tidal interaction, we can meaningfully compare our relation to the corresponding relation in Newtonian gravity. This leads to the identification
\begin{equation}
R^j_{\ 0k0} = {\mathcal E}^j_{\ k} = \delta^{jl} \left(
{ \partial^2 \Phi \over \partial x^l \partial x^k} \right) \ ,
\end{equation}
where $\mathcal E_j^{\ k}$ is the tidal tensor and $\Phi$ is the gravitational potential. 
This provides a constraint that
the curved spacetime theory must satisfy (in the limit
of weak gravity and low velocities). 

After some deliberation, including a careful counting of the dynamical degrees of freedom (noting the freedom to introduce coordinates), one arrives at the field equations for General Relativity:
\begin{equation}
  G_{a b} = {8 \pi G \over c^4} T_{a b} \ ,
  \label{ein_eq}
\end{equation}where $G$ is Newton's constant and $c$ is the speed of light.

At this point it is evident that any discussion of relativistic physics (involving matter) must include the energy-momentum-stress tensor\footnote{Even though it is less descriptive, and even somewhat deceiving when there are 
multiple flows, we will adopt the convention that $T_{a b}$ is referred to as the 
``stress-energy'' tensor from now on.
}, 
$T_{a b}$. This is where the messy physics of reality enter the problem. \cite{mtw73} refer to $T_{a b}$ as ``\dots a machine that contains 
a knowledge of the energy density, momentum density, and stress as measured by any and all 
observers at that event.''  Encoding this is a severe challenge.  However, 
we need to understand how this works---both phenomenologically (allowing us to move swiftly to the challenge of solving the equations) and from a  detailed microphysics point of view (as required in order for our models to be realistic). We will develop this understanding step by step, starting with the simple perfect fluid model and proceeding towards more complex settings including distinct components exhibiting relative flows and dissipation. However, before we take the next step in this direction we need to introduce the main technical machinery that forms the basis for much of the discussion.



\section{Variational analysis}
\label{sec:variational}

The key geometric difference between  generally covariant Newtonian fluids and
their general relativistic counterparts is that the former have an a priori
notion of time \citep{Carter03:_newtI, Carter03:_newtII, Carter04:_newtIII}.
Newtonian fluids also have an a priori notion of space (cf.\ the discussion in \citealt{Carter03:_newtI}). Such a structure has clear advantages
for evolution problems, where one needs to be unambiguous about the
rate-of-change of a given system. However, once a problem requires, say,
electromagnetism, then the a priori Newtonian time is at odds with the
spacetime covariance of the electromagnetic fields (as the Lorentz invariance of Maxwell's equations dictates that the problem is considered in---at least---Special Relativity). Fortunately, for spacetime
covariant theories there is the so-called ``3\,+\,1'' formalism (see, for
instance, \citealt{smarr78:_kinematic} and the discussion in Sect.~\ref{sec:numsim}) which allows one to define
``rates-of-change'' in an unambiguous manner, by introducing a family
of spacelike hypersurfaces (the ``3'') given as the level surfaces of
a spacetime scalar (the ``1'') associated with a timelike progression.

Something that Newtonian and relativistic fluids have in common is that there
are preferred frames for measuring changes---those that are attached to the
fluid elements. In the parlance of hydrodynamics, one refers to Lagrangian
and Eulerian frames, or observers. In Newtonian theory, an Eulerian observer is one who
sits at a fixed point in space, and watches fluid elements pass by, all the
while taking measurements of their densities, velocities, etc.\ at the given
location. In contrast, a Lagrangian observer rides along with a particular
fluid element and records changes of that element as it moves through space
and time. A relativistic Lagrangian observer is the same, but the
relativistic Eulerian observer is more complicated to define (as we have to explain what we mean by a ''fixed point'' in space). One way to do this, see \cite{smarr78:_kinematic}, is to define such an observer as one who moves along a worldline that remains everywhere orthogonal to
the family of spacelike hypersurfaces.

The existence of a preferred frame for a fluid system can be a great
advantage. In Sect.~\ref{shelve} we will use an ``off-the-shelf''
approach that exploits a preferred frame to derive the standard
perfect fluid equations. Later, we will use Eulerian and Lagrangian
variations to build an action principle for both single and multiple fluid
systems. In this problem the Lagrangian displacements play a central role, as they allow us to introduce the constraints that are required in order to arrive at the desired results. Moreover, these types of variations turn out to be useful for many applications, e.g.,  they can be used as the foundation for a
linearized perturbation analysis of neutron stars \citep{kkbs}. As we will
see, the use of Lagrangian variations is  essential for establishing
instabilities in rotating fluids \citep{friedman78:_lagran,
  friedman78:_secul_instab}. However, it is worth noting already at this relatively early stage that systems with several distinct flows are more complex as they can have as many notions of Lagrangian observers as there are
fluids in the system.

\subsection{A simple starting point: The point particle}
\label{sec:point}

The simplest physics problem, i.e.~the motion of a point particle, serves
as a guide to deep principles  used in much harder problems. We 
have used it already to motivate parallel transport as the foundation for the 
covariant derivative. Let us call upon the point particle again to set the 
context for the action-based derivation of the fluid equations. We 
will simplify the discussion by considering only motion in one dimension---assuring the reader that we have good reasons for this, and asking for patience while 
we remind him/her of what may be very basic facts. 

\begin{figure}[htb]
    \centerline{\includegraphics[width=0.7\textwidth]{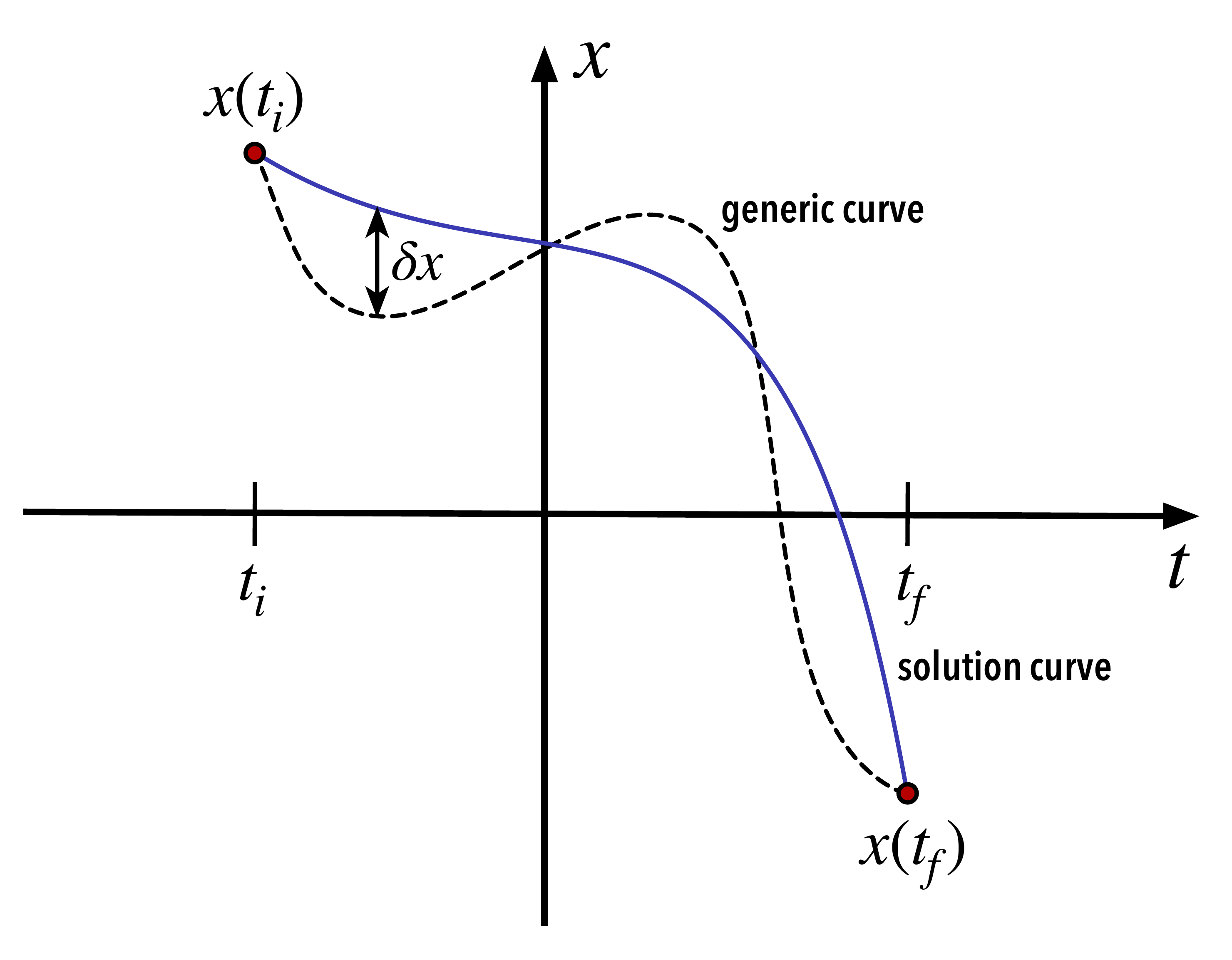}}
    \caption{A simple illustration of the variation that leads to the point particle equations of motion. The solid line in this parameter space represents a curve which is understood to be a solution to the equations of motion, while the dashed line is some arbitrarily specified curve. At a given value of time, the variation $\delta x$ represents the vertical displacement between the curves; obviously, at the endpoints $t = t_1$ and $t = t_2$, the two curves meet and the displacement vanishes. Keeping the endpoints fixed, the equations of motion are obtained from the extrema of the action, as demonstrated in the main text. The same idea applies in the more complicated cases of field theories that we consider later;  the fields have actions, and the field equations of motion are obtained by locating the extrema. The field values at the extrema are often referred to as being ``on shell' (or ``on the mass shell'') for reasons we do not really have to elaborate on here.}
    \label{vary}
\end{figure}

Early on in life (relatively!) we learn that an action appropriate for the point particle is 
\begin{equation} 
    I = \int^{t_f}_{t_i} T dt = \int^{t_f}_{t_i} 
    \left(\frac{1}{2} m \dot{x}^2\right) dt\ , 
\end{equation} 
where $m$ is the mass and $T$ the kinetic energy. A first-order variation 
of the action with respect to $x(t)$ yields 
\begin{equation} 
   \delta I = - \int^{t_f}_{t_i}   \left(m \ddot{x}\right)
              \delta x  dt+ \left.\left(m \dot{x} \delta x\right) 
              \right|^{t_f}_{t_i} \ , 
\end{equation} 
see Fig.~\ref{vary}. If this is all the physics to be incorporated, i.e.~if there are no forces 
acting on the particle, then we impose d'Alembert's principle of 
least action, which states that the
trajectories $x(t)$ that make the action stationary, i.e.~$\delta I = 0$, 
yield the true motion. We then see  that functions 
$x(t)$ that satisfy the boundary conditions 
\begin{equation} 
    \delta x(t_i) = 0 = \delta x(t_f) \ , 
\end{equation} 
and the equation of motion 
\begin{equation} 
    m \ddot{x} = 0 \ , 
\end{equation} 
will indeed make $\delta I = 0$. The same logic applies in the 
substantially more difficult variational problems that will be considered later. 

\vspace*{0.1cm}
\begin{tcolorbox}
\textbf{Comment:} The simple text-book variational derivation of Newton's second law \eqref{newteq} may seem somewhat out of place in a discussion of general relativistic fluids. However, as we proceed it is useful to keep this problem is mind. It provides an intuitive understanding of the more complicated settings we will explore. The general aim is to use a variation of an action---involving (off-shell) deviations away from the solution curve in the relevant parameter space. The steps generally involve ``integration by parts'' (as in the derivation of \eqref{newteq}) and an assumption of fixed ``boundary conditions''. The boundary terms---in general representing the behaviour on a surface in spacetime---can be ignored, as long as we are mainly focussed on the equation of motion. We will make this assumption throughout the discussion, often without spelling it out.   
\end{tcolorbox}
\vspace*{0.1cm}

In general we need to account for forces acting on the particle. First on the list are the 
so-called conservative forces, describable by a potential $V(x)$, which are 
placed into the action according to: 
\begin{equation} 
    I = \int^{t_f}_{t_i} L(x,\dot{x}) dt = \int^{t_f}_{t_i} 
        \left[\frac{1}{2} m \dot{x}^2 - V(x)\right] dt \ , 
\end{equation} 
where $L = T - V$ is known as the Lagrangian. The variation now leads to
\begin{equation} 
    \delta I = - \int^{t_f}_{t_i}  \left(m \ddot{x} + 
              \frac{\partial V}{\partial x}\right) \delta x dt
              + \left.\left(m \dot{x} \delta x\right) 
              \right|^{t_f}_{t_i} \ . 
\end{equation} 
Assuming no externally applied forces, d'Alembert's principle yields the 
equation of motion 
\begin{equation} 
    m \ddot{x} + \frac{\partial V}{\partial x} = 0 \ . 
\end{equation} 
An alternative way to write this is to introduce the momentum $p$ (not to 
be confused with the fluid pressure introduced earlier) defined as 
\begin{equation} 
     p = \frac{\partial L}{\partial \dot{x}} = m \dot{x} \ , 
\end{equation} 
in which case 
\begin{equation} 
     \dot{p} + \frac{\partial V}{\partial x} = 0 \ . 
\end{equation} 
 
In the most honest applications, one has the obligation to incorporate 
dissipative, i.e., non-conservative, forces. Unfortunately, dissipative 
forces $F_d$ cannot be put into action principles (at least not directly, see the discussion in Sect.~\ref{sec:viscosity} where we discuss recent progress towards dissipative variational models). Fortunately, 
Newton's second law is great guidance, since it states 
\begin{equation} 
    m \ddot{x} + \frac{\partial V}{\partial x} = F_d \ , \label{newteq} 
\end{equation} 
when both conservative and dissipative forces act. A crucial observation of 
Eq.~(\ref{newteq}) is that the ``kinetic'' ($m \ddot{x} = \dot{p}$) 
and conservative ($\partial V/\partial x$) forces, which enter the 
left-hand side, still follow from the action, i.e., 
\begin{equation} 
    \frac{\delta I}{\delta x} = - \left(m \ddot{x} + 
    \frac{\partial V}{\partial x}\right) \ ,
\end{equation} 
where we have introduced the ``variational derivative'' ${\delta I}/{\delta x}$.
When there are \emph{no} dissipative forces acting, the action principle 
gives us the appropriate equation of motion. When there \emph{are} 
dissipative forces, the action \emph{defines} the kinetic and 
conservative force terms that are to be balanced by the dissipative 
contribution. It also defines the momentum. These are the key lessons from this 
toy-problem.

We should  emphasize that this way of using the action to define the 
kinetic and conservative pieces of the equation of motion, as well as the 
momentum, can also be used in situations when a  system experiences 
an externally applied force $F_\mathrm{ext}$. The force can be conservative or 
dissipative (see, e.g., \citealt{2013PhRvL.110q4301G}), and will enter the equation of motion in the same way as 
$F_d$ did above. That is 
\begin{equation} 
    - \frac{\delta I}{\delta x} = F_d + F_\mathrm{ext} \ . 
\end{equation}
Like a dissipative force, the main effect of the external force can be to 
siphon kinetic energy from the system. Of course, whether a force is 
considered to be external or not depends on the {\em a priori} definition 
of the system. 

\subsection{More general Lagrangians}

Returning to the discussion of the variational approach for obtaining the dynamical equations that govern a given system, 
let us consider a generalised version of the problem. Basically, we want to extend the idea to the case of a field theory in 
spacetime. To do this, we assume that the system is described by a set of fields $\Phi^A$ defined on spacetime, i.e., depending 
on the coordinates $x^a$. At this level, we can keep the discussion abstract and consider any number of fields, labelled by $A$.
This set can (in principle) contain any number of scalar, vector or tensor fields.  If we are interested in models containing vector fields, then the label $A$ runs over all four components of each of the relevant fields. In that situation, the label $A$ essentially becomes a 
spacetime index, like $a$. Tensor fields are treated in a similar way. As an example, discussed in more detail later, consider electromagnetism, for which the set of fields would be the vector potential $A^a$ and the spacetime metric $g_{a b}$, so that we have $\Phi^A = \{A^a, g_{a b}\}$.

The action for the system should now take the form of an integral of a Lagrangian (density) $\mathcal L$, which depends on the fields $\Phi^A$ and 
their various derivatives (as ``appropriate''). Integrating over a spacetime region $R$ we would have
\begin{equation}
I= \int_R \mathcal L \left( \Phi^A, \partial_a \Phi^A, \partial_a \partial_b \Phi^A, ...\right) d^4 x
\end{equation}
Since we expect the theory to be covariant, we need the action to transform as a scalar under a general coordinate transformation. To ensure this, we need to involve the invariant volume element $\sqrt{-g}d^4x$, where $g$ is the determinant of the metric, as before. Defining the scalar Lagrangian $L$ we then have
\be
I= \int_R L \sqrt{-g}\  d^4 x
\ee
(which is a scalar by construction).

\vspace*{0.1cm}
\begin{tcolorbox}
\textbf{Comment:} This is the first time that we come across the volume element in spacetime. This notion requires some care and involves the Levi-Civita tensor which we will make frequent use of later. The connection is quite intuitive. Consider the well-known fact (from linear algebra) that the volume of the parallelepiped spanned by three vectors $\vec A$, $\vec B$ and $
    \vec C$ is given by the triple product
    $$
    V = | \vec A \cdot (\vec B \times \vec C) | = |\epsilon_{ijk}A^i B^j C^k|\ .
    $$
    Taking the vectors to represent the edges of a volume element, that shears and stretches as it moves, we see that the volume element may be associated with an anti-symmetric tensor. In flat space and Cartesian coordinates, we have
    $$
    \epsilon_{ijk} = [i, j, k]  = \left\{ \begin{array}{lll} +1 \  \mbox{if}\  ijk=123\ \mbox{or a cyclic permutation}\ ,  \\
    -1\  \mbox{if}\ ijk=213\ \mbox{or a cyclic permutation}\ ,  \\
    0\  \mbox{otherwise}\end{array}\ . \right.
    $$
    This should be quite familiar. Inspired by this, we identify the volume element with the antisymmetric tensor density (using the wedge product from differential geometry)
    $$
    d^4x = dx^0\wedge dx^1 \wedge dx^2 \wedge dx^3\ .
    $$
    However, in this expression the right-hand side is coordinate dependent, so we replace it by
    $$
    d^4x = dx^0\wedge dx^1 \wedge dx^2 \wedge dx^3 = {1\over 4!} [a ,b,c,d]\ dx^a \wedge dx^b \wedge dx^c \wedge dx^d \ . 
    $$
    However, the symbol $[a,b,c,d]$ is (by definition) the same in all coordinate systems, so the object we have written down transforms as a density, not a tensor. We have
    $$
    d^4x' = \left| {\partial x^{a'} \over \partial x^a}\right| d^4x\ .
    $$
    This is problematic, but there is a simple solution. Noting that the determinant of the spacetime metric ($=g$) also transforms as a density;
    $$
    g \left( x^{a'}\right) = \left| {\partial x^{a'} \over \partial x^a}\right|^2 g\left( x^a\right) \ ,
    $$
    we simply multiply by $\sqrt{-g}$ to get the invariant volume element
    $$
    \sqrt{-g} dx^0\wedge dx^1 \wedge dx^2 \wedge dx^3 \equiv \sqrt{-g} d^4x \ .
    $$
    The argument also leads us to introduce the Levi-Civita tensor
    $$
    \epsilon_{abcd} = \sqrt{-g}\ [a, b, c, d] \ .
    $$
    As it is a tensor object, we can raise the indices with the metric, as we have become accustomed to. The logic is, of course, equally relevant in three dimensions and flat space. As soon as we move away from Cartesian coordinates, we must include the metric determinant in the definition of the $\epsilon_{ijk}$ tensor.
\end{tcolorbox}
\vspace*{0.1cm}

As in the case of a point particle, we can derive the field equations by demanding that the action is stationary under variations in the fields. 
Letting 
\begin{equation}
\Phi^A \to \Phi^A + \delta \Phi^A
\end{equation}
and assuming, for simplicity, that the theory is ``local'' (meaning that only first derivatives of the fields appear in the action) we need
also
\begin{equation}
\partial_a \Phi^A \to \partial_a \Phi^A + \partial_a \left(\delta \Phi^A\right) = \partial_a \Phi^A + \delta \left(\partial_a \Phi^A\right)
\end{equation}
Given these relation, the variation in the action is $I+\delta I$, where
\begin{equation}
\delta I  = \int_R \delta \mathcal L d^4 x = \int_R \left[ {\partial \mathcal L \over \partial \Phi^A} \delta \Phi^A +
{\partial \mathcal L \over \partial \left(\partial_a \Phi^A\right) } \delta \left( \partial_a \Phi^A\right) \right] d^4x
\end{equation}
To make progress we need to factor out $\delta \Phi^A$ from the second term in the integrand. This is achieved by integrating by parts;
\begin{equation}
 \int_R
{\partial \mathcal L \over \partial \left(\partial_a \Phi^A\right) } \delta \left( \partial_a \Phi^A\right)\ d^4x
=  \int_R \partial_a \left[ 
{\partial \mathcal L \over \partial \left(\partial_a \Phi^A\right) } \delta \Phi^A \right] d^4x -  \int_R \partial _a 
\left[ 
{\partial \mathcal L \over \partial \left(\partial_a \Phi^A\right) } \right] \delta \Phi^A \ d^4x
\end{equation}
At this point we make use of the fact that the first term is a total derivative, which can be turned into a integral over the bounding surface
(in the usual way). Inspired by the boundary conditions imposed on the variations in the point-particle case, we then restrict ourselves to 
variations $\delta \Phi^A$ that vanish on the boundary.  Thus, we can neglect the first integral (later referred to as the ``surface terms''), ending up with 
\begin{equation}
\delta I  =  \int_R \left\{ {\partial \mathcal L \over \partial \Phi^A}-
\partial _a 
\left[ 
{\partial \mathcal L \over \partial \left(\partial_a \Phi^A\right) } \right] \delta \Phi^A
 \right\} \delta \Phi^Ad^4x
\end{equation}
Demanding that $\delta I=0$ we see that the variational derivative satisfies
\begin{equation}
{\delta \mathcal L \over \delta \Phi^A } =  {\partial \mathcal L \over \partial \Phi^A}-
\partial _a 
\left[ 
{\partial \mathcal L \over \partial \left(\partial_a \Phi^A\right) } \right] = 0 \ .
\label{ELpart}
\end{equation}
These are the Euler-Lagrange equations that govern the evolution of the fields $\Phi^A$.

So far, we have developed the theory for the Lagrangian density $\mathcal L$, rather than the Lagrangian $L$ itself. 
This is not a problem, we can simply consider the components of the metric as belonging to the set of fields that we vary.
However, the added complication (due to the presence of $\sqrt{-g}$ and the derivatives that need to be evaluated) 
may be unnecessary in many cases. In such situations one can often express the Lagrangian in terms of the covariant 
derivative $\nabla_a$ instead of the partial $\partial_a$. Essentially, this involves reworking the algebra taking as starting 
point an action of form
\begin{equation}
I = \int_R L \left( \Phi^A, \nabla_a \Phi^A, \ldots, g_{a b}, \partial_c g_{a b}, \ldots\right) \sqrt{-g}\ d^4x
\end{equation}
where the fields $\Phi^A$ are now independent of the metric, although the Lagrangian may still contain $g_{a b}$ in contractions of 
spacetime indices to construct the required scalar. After some algebra, we  find that
\begin{equation}
{\delta  L \over \delta \Phi^A } =  {\partial  L \over \partial \Phi^A} -
\nabla _a \left[ 
{\partial  L \over \partial \left(\nabla_a \Phi^A\right) } \right] = 0
\label{ELeq}\end{equation}
This is the form of the Euler-Lagrange equations that we will be using in the following.

\subsection{Electromagnetism}
\label{sec:emvar}

As a first ``explicit'' example of the variational approach, let us derive the field equations for electromagnetism \cite{efsth}. In this case, the starting point is the electromagnetic vector potential $A^a$, which (in turn) leads to the Faraday tensor
\begin{equation}
F_{ab}=\nabla_a A_b - \nabla_b A_a
\end{equation}
Because of the anti-symmetry, this object has 6 components which can (as we will see later) be associated with the electric and magnetic fields, leading to a (presumably) more familiar picture. However, these fields are manifestly observer dependent (a moving charge leads to a magnetic field etc.) so, from a formal point of view, it is better to develop the theory in terms of $F_{ab}$. Making contact with the previous discussion and the variational approach, the fields
$\Phi^A$ to be varied will be the four components of $A^a$. The first step of the derivation is to construct a suitable scalar Lagrangian 
from $A^a$ and its first derivatives. However, already at this point do we run into ``trouble''. We know that the theory is gauge-invariant, 
since we can add $\nabla_a \psi = \partial_a \psi$ (where $\psi$ is an arbitrary scalar) to the vector potential without altering the physics (read: $F_{ab}$). The upshot of this is that we need to ensure that the electromagnetic action is invariant under the transformation
\begin{equation}
A_a \to A_a + \nabla_a \psi
\end{equation}
This constrains the permissible Lagrangians. For example, we cannot use the contraction $A^a A_a=g_{ab}A^a A^b$ since this combination
is not gauge invariant. However, it is easy to see that $F_{ab}$ exhibits the required invariance, so we can use it as our main building block. The obvious thing to do would be to try to use
the scalar $F_{a b} F^{a b}$ to build the Lagrangian. However, this would not account for the fact that the charge current 
$j^a$ acts as source of the electromagnetic field. To reflect this, we add an ``interaction term'' $-j^a A_a$ to the Lagrangian (leaving the details of this for later). 
At the end of the day, the Lagrangian takes the form
\begin{equation}
 L= - {1 \over 4 \mu_0} F_{a b} F^{a b} + j^a A_a
 \label{EMlag}
\end{equation}
where $\mu_0$ is a constant (describing the strength of the coupling).

At this point, we realize that the current term is not gauge-invariant. It would transform as
\begin{equation}
j^a A_a \to j^a A_a + j^a \nabla_a \psi = j^a A_a + \nabla_a \left( \psi j^a\right) - \psi \left( \nabla_a j^a\right)
\end{equation}
We  already know that the second term contributes a surface term to the action integral, and hence can be ``ignored''.
The third term is different. In order to ensure that the action is gauge-invariant, we must demand that the current is 
conserved, i.e.
\begin{equation}
\nabla_a j^a = 0 \ . 
\end{equation}
The field equations that we derive require  this constraint to be satisfied. Later, when we consider the fluid problem, we will see that the
conservation of the matter flux plays a similar role.

Having established an invariant scalar Lagrangian, we determine the Euler-Lagrange equations by varying the fields $A_a$ (keeping the 
source $j^a$ fixed). From \eqref{ELeq} we then have
\begin{equation}
{\partial  L \over \partial A_a} - \nabla_b \left[ {\partial  L \over \partial \left( \partial_b A_a\right)} \right] = 0 \ .
\end{equation}
From the stated form of the action (and recalling the discussion of the point particle) we see that 
\begin{equation}
{\partial  L \over \partial A_a} =  j^a
\end{equation}
The second term is messier, but after a bit of work we arrive at;
\begin{equation}
{\partial  L \over \partial \left( \partial_b A_a\right)}  = - {1 \over \mu_0} F^{a b}
\end{equation}
which leads to the final field equation
\begin{equation}
\nabla_b F^{a b} = \mu_0 j^a \ .
\end{equation}

The relativistic Maxwell equations are completed by 
\begin{equation}
\nabla_{[c} F_{a b]} = 0 \quad \Longrightarrow \quad \nabla_c F_{a b} + \nabla_b F_{c a} + \nabla_a F_{b c} = 0 
\label{maxw2}
\end{equation}
which is automatically satisfied for our definition of $F^{a b}$, as it is anti-symmetric.

\vspace*{0.1cm}
\begin{tcolorbox}
\textbf{Comment:}
At this point we have an opportunity to comment on the connection with differential geometry and also introduce the Hodge dual, which will play a role later. The Hodge dual 
of the electromagnetic field tensor is defined by 
$$
^\star F_{ab} = {1\over 2!} \epsilon^{cd}_{\ \ ab}F_{cd}
$$
where we adopt the convention that the contraction always involves the first indices of the Levi-Civita tensor. A different choice may affect the overall sign.
The generalization to other tensor objects is natural. 

It is also worth noting that, in terms of the exterior derivative, \eqref{maxw2} represents the fact that the two-form $F_{ab}$ is closed:
$$
d \boldsymbol{F} = 0 
$$
This means that there must exist a one-form, $A_a$ such that
$$
\boldsymbol{F} = d \boldsymbol{A} \ \Longrightarrow\ F_{ab} = \partial_a A_b - \partial_b A_a = \nabla_a A_b - \nabla_b A_a
$$
This is, of course, the vector potential.

\end{tcolorbox}
\vspace*{0.1cm}

\subsection{The Einstein field equations}
\label{sec:efevar}

Having discussed the underlying principles and considered the explicit example of electromagnetism, we have reached the level of confidence required to
derive the field equations of General Relativity. We know that the metric $g_{a b}$ is the central object of the theory (essentially, because we
are looking for a theory where the geometry plays a key role). To build the Lagrangian we therefore want to construct a simple (for elegance) scalar from the metric and its derivatives. The simplest object we can think of is the Ricci scalar, $R$. This is, in fact, the \emph{only} scalar 
that contains only the metric and its first two derivatives. Moreover, it is natural that the Lagrangian involves a quantity which is directly 
linked to the spacetime curvature, and the Ricci scalar fits this bill, as well.  

This argument leads to the celebrated Einstein-Hilbert action
\begin{equation}
I_\mathrm{EH} = \int_R R \sqrt{-g}\ d^4x \ .
\end{equation}
In this case, where the Lagrangian depends on the metric, it is natural to work directly with the density $\mathcal L = R \sqrt{-g}$.
From \eqref{ELpart} we then see that 
\begin{equation}
{\partial \mathcal L\over \partial g_{a b} } - \partial_c \left[ { \partial \mathcal L \over \partial \left( \partial_c g_{a b}\right) }\right]
+\partial_d  \partial_c \left[ { \partial \mathcal L \over \partial \left( \partial_d \partial_c g_{a b}\right) }\right] = 0 \ ,
\end{equation}
where we have allowed for the fact that the Lagrangian also depends on the second derivatives of the metric (the extension of the analysis to allow for this is straightforward). Having a go at evaluating the required derivatives, we soon appreciate that this task is formidable. Luckily, 
there is an easier way to arrive at the answer. 

Let us consider the variation in the action that results from a metric variation $g_{a b} \to g_{a b} + \delta g_{a b}$.  Carrying out this analysis we need the variation of the covariant metric, 
which follows readily:
\begin{equation}
g^{a b} g_{b c} = \delta^a_c \qquad \Longrightarrow \qquad 
\delta g^{a b} = - g^{a c} g^{b d} \delta g_{c d} \ .
\end{equation}
Making use of the fact that $R = g^{a b} R_{a b}$, we then have
\begin{equation}
\delta I_\mathrm{EH} = \int_R \left[ \delta g^{a b} R_{a b} + g^{a b} \delta R_{a b} \right] \sqrt{-g}\ d^4x +
\int_R g^{a b} R_{a b} \delta \sqrt{-g}\ d^4x \ .
\label{einint}
\end{equation}
Since the metric is the fundamental variable, we need to factor out $\delta g^{ab}$ (somehow). The terms in the second integral 
are easiest to deal with. Given that $g$ is the determinant of the metric, the expression we need follows from  \eqref{deltadet}. That is, we have
\begin{equation}
\delta\sqrt{-g} = -{1 \over 2} \sqrt{-g}\ g_{a b} \delta g^{a b} \ .
\end{equation}
Turning to the second term in the first bracket of \eqref{einint}, the easiest way to progress is to consider the variation of the Riemann tensor
and then constructing the expression for the Ricci tensor by contraction. Moreover, noting that the Riemann tensor variation is expressed
in terms of variations of the connection, $\delta \Gamma^c_{\ a b}$, which is a tensor, we can simplify the analysis by working in a local inertial frame (where $\Gamma^c_{\ a b}=0$). Thus, we have
\begin{equation}
\delta R^d_{\ a b c} = \nabla_b \left( \delta \Gamma^d_{\ a c}\right) -  \nabla_c \left( \delta \Gamma^d_{\ a b}\right) \ .
\end{equation}
As this is also a tensor expression it is valid in any coordinate system. Carrying out the required contraction, we find that
\begin{equation}
\delta R_{a b} = \nabla_b \left( \delta \Gamma^c_{\ a c}\right) -  \nabla_c \left( \delta \Gamma^c_{\ a b}\right) \ .
\end{equation}
Using this expression we see that 
\begin{equation}
 g^{a b} \delta R_{a b} = \nabla_b \left( g^{a b} \delta \Gamma^{c}_{\ a c} - g^{a c} \delta\Gamma^b_{\ a c} 
 \right) \ .
\end{equation}
In other words, the term that we need in \eqref{einint} can be written as a total derivative. Given that this leads to a surface term, we duly 
neglect it and arrive at the final result:
\begin{equation}
\delta I_\mathrm{EH} = \int_R \left( R_{a b} -{1 \over 2} g_{a b} R \right) \delta g^{a b} \sqrt{-g} \ d^4x \ .
\end{equation}
The vanishing of the variation leads to the vacuum Einstein equations
\begin{equation}
G_{a b} = R_{a b} - {1 \over 2} g_{a b} R = 0 \ .
\end{equation}
The derivation highlights the fact that Einstein's theory is one of the 
most elegant constructions of modern physics. 

\subsection{The stress-energy tensor as obtained from the action principle}

However aesthetically pleasing the theory may be, 
our main interest here is not in the vacuum dynamics of Einstein's theory. Rather, we want to explore the matter sector. In Einstein's Universe, matter plays a dual role---it (actively) provides the origin of the spacetime curvature and the gravitational field and (perhaps not quite passively) adjusts its motion according to this curvature. 

In particular, we want to explore systems of astrophysical relevance for which general relativistic aspects
are crucial. Inevitably, this involves some rather complex physics. However, the coupling to the spacetime 
curvature remains relatively straightforward as it is encoded in a single object; the stress-energy tensor
$T_{a b}$. This object is as important for 
General Relativity as the Einstein tensor $G_{a b}$ in that it enters the 
Einstein equations in as direct a way as possible, i.e. (in geometric units)
\begin{equation}
    G_{a b} = 8 \pi T_{a b} \ . 
\end{equation}
From a conceptual point-of-view it is relatively easy to incorporate matter in the 
variational derivation from the previous section. Essentially, we add a matter component such that (cf. the argument for electromagnetism)
\begin{equation}
I = I_\mathrm{EH} + I_\mathrm{M} = \int_R \left( {1\over 2\kappa} R + L \right)\sqrt{-g}\ d^4x
\end{equation}
where $\kappa = 8\pi G/c^4$ is a coupling constant fixed by Newtonian correspondence in the weak-field limit.
Given the results for the vacuum gravity problem, it is easy to see that the matter contribution to the 
field equations follow from the variation of the matter action with respect to the metric. This insight will be very important later. In essence, the Einstein equations
take the form
\begin{equation}
G_{a b} = \kappa T_{a b} 
\end{equation}
provided that 
\begin{equation} 
     T_{a b} = - \frac{2}{\sqrt{- g}}
   { \delta  \mathcal{L}_\mathrm{M} \over  \delta g^{a b}}=  - \frac{2}{\sqrt{- g}} 
    { \delta \left(\sqrt{- g} L \right) \over  \delta g^{a b}}
     \ , \label{seten} 
\end{equation} 
or, equivalently,
\begin{equation} 
     T^{a b}= \frac{2}{\sqrt{- g}} {\delta 
     \left(\sqrt{- g} L \right) \over  \delta g_{a b}}
     \ . 
\end{equation} 
Applying this result to the case of electromagnetism and \eqref{EMlag}, we see that the relevant stress-energy tensor takes the form 
\begin{equation}
T_{a b}^\mathrm{EM} = - {1\over \mu_0} \left[g^{c d}F_{a c}F_{b d}-{1\over 4}g_{a b} \left(F_{c d}F^{c d}\right) \right] \ .
\end{equation}



\section{Case study: single fluids}
\label{tab1}

Without an a priori, physics-based specification for $T_{a b}$, solutions to the Einstein 
equations are void of physical content, a point which has been emphasized, for instance, by 
Geroch and Horowitz (in \citealt{hawk1979:_gr}). Unfortunately, the following algorithm for
producing ``solutions'' has been much abused: (i) specify the form of the metric, typically by 
imposing some type of symmetry (or symmetries), (ii) work out the components of $G_{a b}$ 
based on this metric, (iii) define the energy density to be $G_{0 0}$ and the pressure
to be $G_{1 1}$, say, and thereby ``solve'' those two equations, and (iv) based on the ``solutions'' 
for the energy density and pressure solve the remaining Einstein equations. The problem is that 
this algorithm is little more than a mathematical parlour game. It is only by sheer luck that it will
generate a physically relevant solution for a non-vacuum spacetime. As such, the strategy is 
antithetical to the \emph{raison d'\^etre} of, say, gravitational-wave astrophysics, which is to use 
observed data as a probe of the microphysics, say, in the cores of neutron
stars. Much effort is currently going into taking given microphysics and combining it with the 
Einstein equations to model gravitational-wave emission from astrophysical scenarios, like binary neutron star mergers \cite{2017RPPh...80i6901B}.
To achieve this aim, we need an appreciation of the 
stress-energy tensor and how it is encodes the physics.

\subsection{General stress decomposition}

Readers familiar with Newtonian fluids will be aware of the roles that the internal energy (recall the discussion in Sect.~\ref{sec:thermo}), 
the particle flux, and the stress tensor play in the fluid equations. In special relativity we learn that, in 
order to have spacetime covariant theories (e.g., well-behaved with respect to the Lorentz 
transformation) energy and momentum must be combined into a spacetime vector, whose 
zeroth component is the energy while the spatial components give the momentum (as measured by a given observer). The fluid stress 
must also be incorporated into a spacetime object, hence the necessity for
$T_{a b}$. Because the Einstein tensor's covariant divergence vanishes identically, we must have 
 \begin{equation}
\nabla_b T^b{}_a = 0 \ .
\label{divT1}
\end{equation}
This provides us with four equations, often interpreted as the equations for relativistic fluid dynamics. As we will soon see, this interpretation makes ``sense'' (as the equations we arrive at reduce to the familiar Newtonian ones in the appropriate limit). However, from a formal point of view the argument is somewhat misleading. It leaves us with the impression that the job is done, but this is not (quite) the case. Sure, we are able to speedily write down the equations for a perfect fluid. But, we still have work to do if we want to consider more complex settings (e.g., including relative flows). This requires additional assumptions or a different approach altogether. One of the main aims with this review is to develop such an alternative and explore the results in a variety of settings. Having done this, we will see that  \eqref{divT1} follows automatically once the ``fluid equations''
are satisfied. This may seem like splitting hairs at the moment, but the point we are trying to make should become clear as we progress.

The fact that we advocate a different strategy does not mean that the importance of the stress-energy tensor is (somehow) reduced. Not at all. We still need $T_{ab}$ to provide the matter input for the Einstein equations and we may opt to use \eqref{divT1} to get (some of) the dynamical equations we need. Given this, it is important to  understand  the physical meaning of the components of $T_{a b}$. In order to do this, we need to introduce a suitable observer (someone has to measure energy etc. for us).  This then allows us to express the tensor components in terms of projections into the timelike and spacelike directions associated with this observer, in essence providing a fibration of spacetime as illustrated in Fig.~\ref{fibrate}. 

\begin{figure}[htb]
    \centerline{\includegraphics[width=0.7\textwidth]{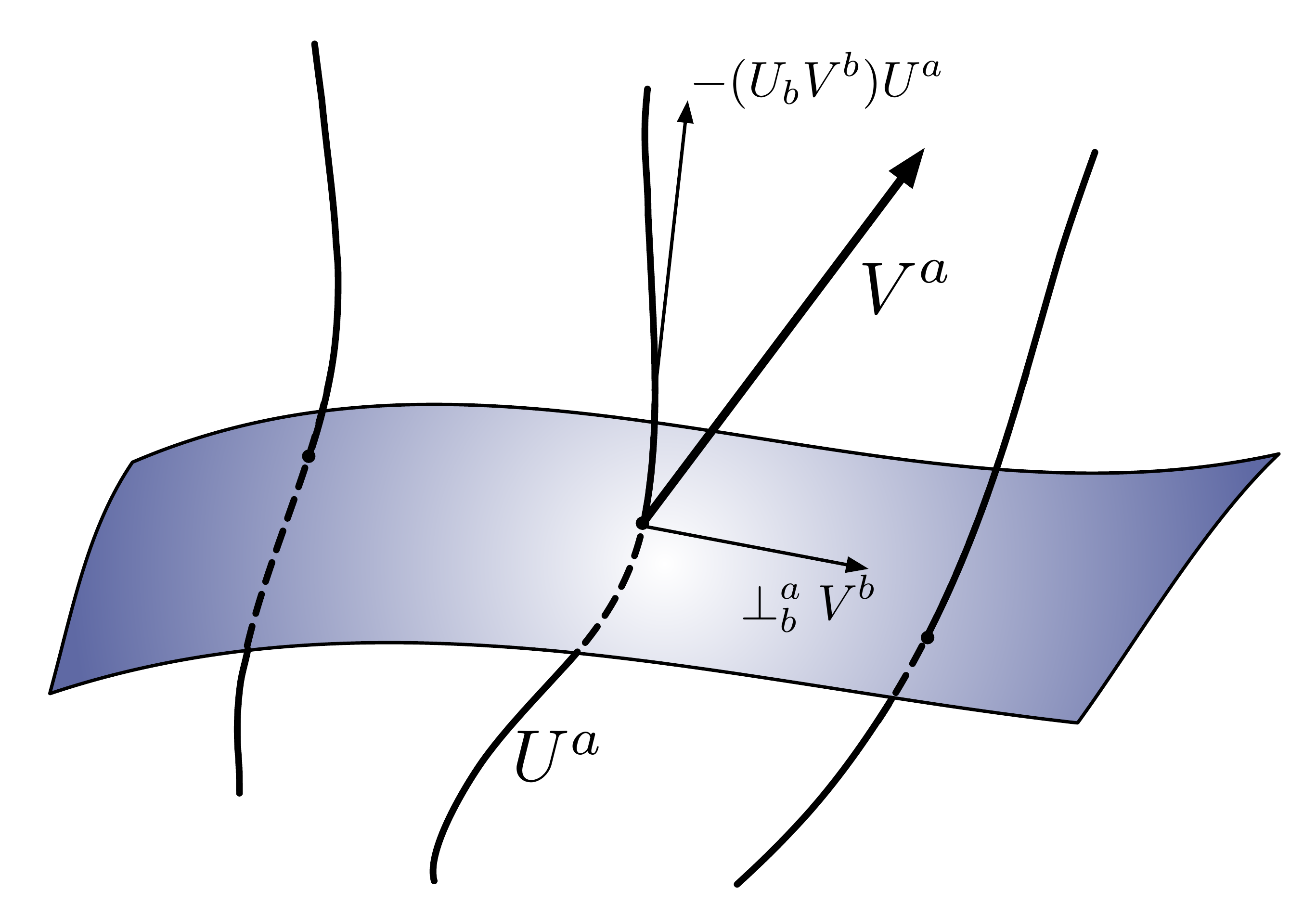}}
    \caption{The projections of a vector $V^a$ onto
    the worldline defined by $U^a$ (providing a fibration of spacetime) into the perpendicular
    hypersurface (obtained from a projection with $\perp^a_b$).}
    \label{projection}
\end{figure}

In order to project a tensor along an observer's timelike direction we contract that index 
with the observer's four-velocity, $U^a$. The required projection of a tensor into spacelike directions 
perpendicular to the timelike direction defined by $U^a$ is 
effected via the operator $\perp^a_b$, defined as
\begin{equation}
  \perp^a_b = \delta^a{}_b + U^a U_b \ ,
  \qquad
  U^a U_a = - 1 \quad \Longrightarrow \quad \perp^a_b U^b = 0 
\end{equation}
Any tensor index that has been ``hit'' with the projection operator will be perpendicular to the 
timelike direction defined (locally) by $U^a$.  It is then easy to see that any vector can be expressed in terms of its component along a given $U^a$ and components orthogonal (in the spacetime sense) to it. That is, we have 
\begin{equation}
V^a = \delta^a_b V^b + \underbrace{(U^a U_b V^b - U^a U_b V^b)}_{=0} = -(U_bV^b) U^a + \perp^a_b V^b
\end{equation}
The two projections (of a vector $V^a$ for an observer with 
unit four-velocity $U^a$) are illustrated in figure~\ref{projection}. More general tensors are projected by acting with $U^a$ or $\perp^a_b$ 
on each index separately (i.e., multi-linearly).

Let us now see how we can use the projection to give physical ``meaning'' to the components of the stress-energy tensor.
The energy density $\varepsilon$ as perceived by the observer is (see
Eckart~\cite{eckart40:_rel_diss_fluid} for one of the earliest discussions)
\begin{equation}
  \varepsilon = U^a U^b T_{a b}  \ ,
\end{equation}
while
\begin{equation}
  {\cal P}_a = - \perp^b_a U^c T_{b c}
\end{equation}
is the spatial momentum density (as it does not have a contribution along $U^a$ it is a three vector), and the spatial stresses are encoded in
\begin{equation}
  {\cal S}_{a b} = \perp^c_a \perp^d_b T_{c d} \ .
\end{equation}
As usual, the manifestly spatial component $\mathcal S_{i j}$ is understood to be the $i^\mathrm{th}$-component of 
the force across a unit area perpendicular to the $j^\mathrm{th}$-direction. With respect to 
the observer, the stress-energy tensor can now be written (in complete generality) as 
\begin{equation}
  T_{a b} =
  \varepsilon \, U_a U_b + 2 U_{(a} {\cal P}_{b)} + {\cal S}_{a b},
\end{equation}
where $2 U_{(a} {\cal P}_{b)} \equiv U_a {\cal P}_b + U_b {\cal P}_a$. Because 
$U^a {\cal P}_a = 0$, we see that the trace $T = T^a{}_a$ is
\begin{equation}
  T = {\cal S} - \varepsilon,
\end{equation}
where ${\cal S} = {\cal S}^a{}_a$. 

It is important at this stage to appreciate that we are discussing a mathematical construction. We  need to take further steps to connect the phenomenology to the underlying physics.

\subsection{``Off-the-shelf'' analysis}
\label{shelve}

As we have already suggested, there are different ways of deriving the general relativistic fluid
equations. Our purpose here is not to review all possible approaches, but
rather to focus on a couple: (i) an ``off-the-shelf'' consistency analysis
for the simplest fluid a la Eckart~\cite{eckart40:_rel_diss_fluid}, to
establish some of the key ideas, and then (ii) a more powerful method based on
an action principle that varies fluid element world lines. We now consider  the first of these. The second avenue will be explored in Sect.~\ref{sec:pullback}.

We have seen how the components of a general stress-energy tensor can be projected 
 onto a coordinate system carried by an observer moving with four-velocity 
$U^a$. Let us now connect this with the motion of a fluid. The simplest fluid is one for which there is only one four-velocity 
$u^a$.  As both four velocities are normalized (to unity) we must have
\begin{equation}
u^a = \gamma(U^a + v^a) \ , \quad \mbox{with} \quad U_a v^a = 0 \quad \mbox{and} \quad \gamma = (1-v^2)^{-1/2}
\label{varel}
\end{equation}
the familiar redshift factor from special relativity. Clearly, the problem simplifies if we assume that the observer ride along with the fluid. That is, we introduce a
 preferred frame defined by $u^a$, and then simply take $U^a = u^a$. With respect to the fluid there will then (by definition) be no momentum 
flux, i.e., ${\cal P}_a = 0$. Moreover, since we use a fully spacetime covariant formulation, i.e., there are 
only spacetime indices, the resulting stress-energy tensor will transform properly 
under general coordinate transformations, and hence can be used for any observer.

In general, the spatial stresses are given by a two-index, symmetric tensor, and the only objects that can be used to 
carry the indices (in the simple model we are considering at this point) are the four-velocity $u^a$ and the metric $g_{a b}$. Furthermore, because the 
spatial stress must also be symmetric, the only possibility is a linear combination of $g_{a b}$ 
and $u^a u^b$. Given that $u^b {\cal S}_{b a} = 0$, we must have
\begin{equation}
  {\cal S}_{a b} = \frac{1}{3} {\cal S} (g_{a b} + u_a u_b).
\end{equation}
As the system is assumed to be locally isotropic, it is possible to diagonalize the spatial stress 
tensor. This also implies that its three independent diagonal elements should actually be equal to 
the same quantity, which turns out to be the local pressure. Hence we have $p = {\cal S}/3$  and
\begin{equation}
     T_{a b} = \left(\varepsilon + p\right) u_a u_b + p g_{a b} = \varepsilon u_a u_b + p \perp_{ab} \ .
\end{equation}
This is the well-established result for a perfect fluid.

Given a relation $p = p(\varepsilon)$ (an equation of state), there are four independent fluid variables. Because of this the 
equations of motion are often understood to be given by \eqref{divT1}.  Let us proceed along these lines, but first  simplify
matters by assuming that the equation of state is given by a relation of the form $\varepsilon = \varepsilon(n)$ where $n$ is the 
particle number density. As discussed in Sect.~\ref{sec:thermo}, the chemical potential $\mu$ is then given by
\begin{equation}
  {d} \varepsilon =
  \frac{d \varepsilon}{d n} {d} n \equiv
  \mu \, {d} n \ , \label{mudef1f}
\end{equation}
and we know from the Euler relation~(\ref{funrel}) that
\begin{equation}
  \mu n = p + \varepsilon.
  \label{funrel1}
\end{equation}
In essence, we have connected the model to the thermodynamics. This is an important step.

Let us now get rid of the free index of $\nabla_b T^b{}_a = 0$ in two ways: first, by contracting  
with $u^a$ and second, by projecting  with $\perp^a_b$ (recalling that $U^a = u^a$). Given that
that $u^a u_a = - 1$ we have the identity
\begin{equation}
  \nabla_a \left(u^b u_b\right) = 0
  \qquad \Longrightarrow \qquad
  u_b \nabla_a u^b = 0.
\end{equation}
Contracting \eqref{divT1} with $u^a$ and using this identity gives
\begin{equation}
  u^a \nabla_a \varepsilon + (\varepsilon + p) \nabla_a u^a = 0 \ . \label{fstlaw1f}
\end{equation}
The definition of the chemical potential $\mu$ and the Euler relation allow us to rewrite this as
\begin{equation}
  \mu u^a \nabla_a n + \mu n \nabla_a u^a = 0
  \qquad \Longrightarrow \qquad
  \nabla_a n^a = 0 \ ,
\end{equation}
where we have introduced the particle flux, $n^a \equiv n u^a$. This result simply represents the fact that the particles are conserved. 

Meanwhile, projection of the free index in \eqref{divT1} using $\perp^b_a$ leads to
\begin{equation}
  (\varepsilon + p) a_a = -  \perp^b_a \nabla_b p \ , \label{euler}
\end{equation}
where $a_a \equiv u^b \nabla_b u_a$ is the fluid (four) acceleration.  This is reminiscent of the Euler 
equation for Newtonian fluids. In fact, we demonstrate in  Sect.~\ref{sec:newton} that the non-relativistic limit of  \eqref{euler} this leads to the Newtonian result.

However, we should not be too quick to think that this is the only way to understand 
\eqref{divT1}! There is an alternative form that makes the perfect fluid have more in 
common with vacuum electromagnetism. If we define
\begin{equation}
  \mu_a = \mu u_a \ ,
\end{equation}
then the stress-energy tensor can be written in the form
\begin{equation}
  T^a{}_b = p \delta^a{}_b + n^a \mu_b \ .
\end{equation}
We have here our first encounter with the fluid element momentum $\mu_a$ that is conjugate to 
the particle flux, the number density current $n^a$. 
Its importance will become clearer as this review 
develops, particularly when we discuss the multi-fluid problem. For now, we simply note  that $u_a {d} u^a = 0$, implies that we will have
\begin{equation}
  {d} \varepsilon = - \mu_a \, {d} n^a \ .  \label{frstlw1}
\end{equation}
 This relation will serve as the starting point for the fluid action principle in Sect.~\ref{sec:pullback}, where $- \varepsilon$ will be taken to be the fluid Lagrangian.

If we project onto the free
index of \eqref{divT1} using $\perp^b_a$, as before, we arrive at
\begin{equation}
  f_a + \left( \nabla_b n^b \right) \mu_a = 0 \ ,
  \label{diveq}
\end{equation}
where the force density $f_a$ is
\begin{equation}
  f_a = n^b \omega_{b a} \ ,
  \label{fdens}
\end{equation}
and the vorticity $\omega_{a b}$ is defined as
\begin{equation}
  \omega_{a b} \equiv 2 \nabla_{[ a} \mu_{b ]} =
  \nabla_a \mu_b - \nabla_b \mu_a \ .
\end{equation}
Contracting Eq.~(\ref{diveq}) with $n^a$ we see (since $\omega_{a b} = - \omega_{b a}$) 
that
\begin{equation}
  \nabla_a n^a = 0
  \label{cons1f}
\end{equation}
and, as a consequence, the equations of motion take the form
\begin{equation}
  f_a = n^b \omega_{b a} = 0 \ .
  \label{euler1f}
\end{equation}

The vorticity two-form $\omega_{a b}$ has emerged quite naturally as an essential ingredient of 
the fluid dynamics \citep{lichnerowica67:_book, carter89:_covar_theor_conduc,bekenstein87:_helicity, katz84:_vorticity}. This is a key result. Readers familiar
with Newtonian fluids should be inspired by this, as the vorticity is used to establish 
theorems on fluid behaviour (for instance the Kelvin--Helmholtz 
theorem; \citealt{landau59:_fluid_mech}) and is at the heart of turbulence 
modeling \citep{pullin98:_vortex_turb}.

\vspace*{0.1cm}
\begin{tcolorbox}
\textbf{Comment:} While we have inferred the equations of motion from the identity 
$\nabla_b T^b{}_a = 0$, we now emphatically state that---while the equations are correct---the 
logic is limited. In fact, from a field theory point of view it is completely wrong! The 
proper way to think about the identity is that the equations of motion are satisfied first, which then 
guarantees that $\nabla_b T^b{}_a = 0$. There is no clearer way to understand this than to study 
the multi-fluid case. The vanishing of the covariant divergence represents only four 
equations, whereas the multi-fluid problem clearly requires more information (as there are additional
fluxes that need to be determined). 
\end{tcolorbox}
\vspace*{0.1cm}


To demonstrate the role of $\omega_{a b}$ as the vorticity, consider a small region of the fluid 
where the time direction $t^a$, in local Minkowski coordinates, is adjusted to be the same as that 
of the fluid four-velocity so that $u^a = t^a = (1,0,0,0)$. Eq.~(\ref{euler1f}) and the 
antisymmetry then imply that $\omega_{a b}$ can only have purely spatial components. Because 
the rank of $\omega_{a b}$ is two, there are two ``nulling'' vectors, meaning their contraction with 
either index of $\omega_{a b}$ yields zero (a condition which is true also for vacuum
electromagnetism). We have arranged already that $t^a$ be one such vector. By a suitable 
rotation of the coordinate system the other one can be taken to be $z^a = (0,0,0,1)$,  implying 
that the only non-zero component of $\omega_{a b}$ is $\omega_{x y}$. 

\begin{figure}[htb]
    \centerline{\includegraphics[scale = 0.7]{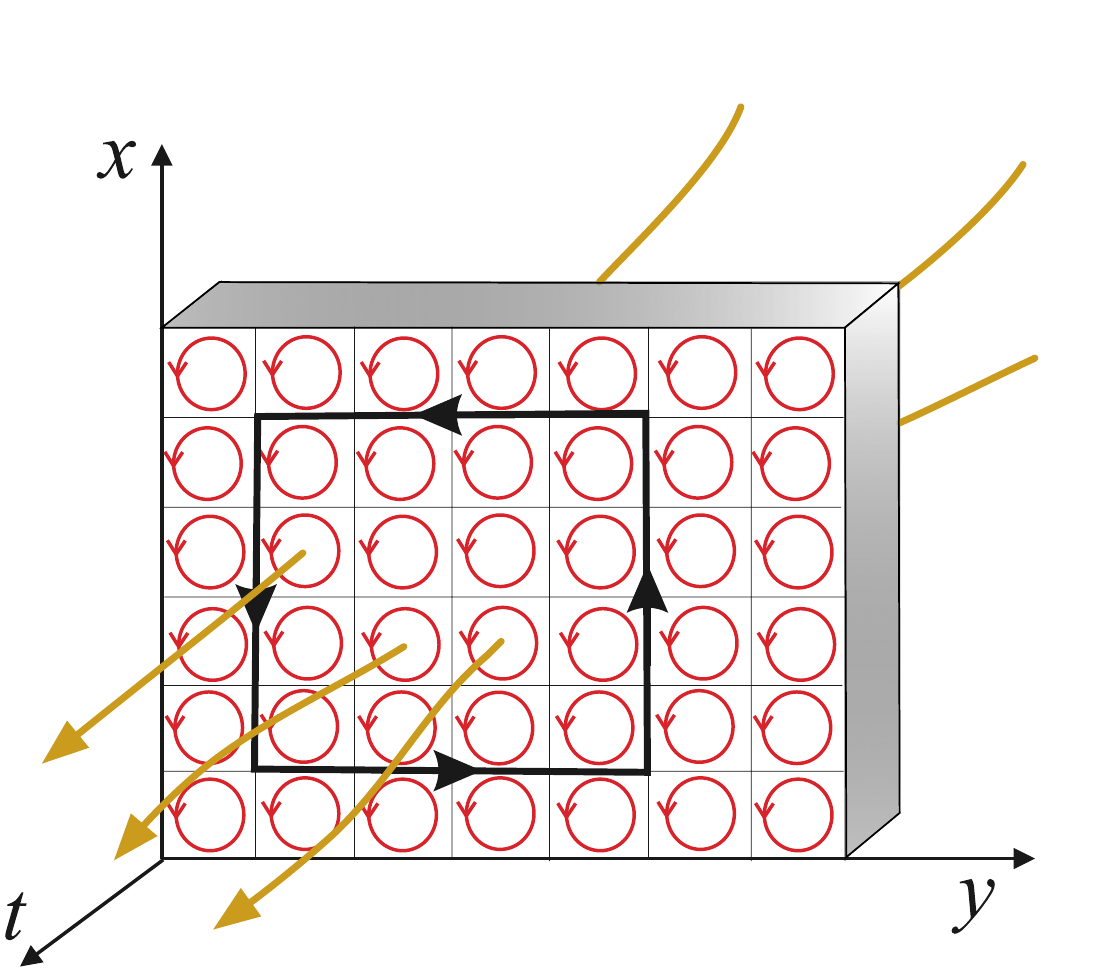}}
    \caption{A local, geometrical view of the Euler equation as an
    integrability condition of the vorticity for a single-constituent
    perfect fluid.}
    \label{vorticity}
\end{figure}

Geometrically, this kind of two-form can be  pictured as a 
collection of oriented worldtubes, whose walls lie in the $x = \mathrm{const}$ and 
$y = \mathrm{const}$ planes \cite{mtw73}. Any contraction of a vector with a two-form that does not yield zero 
implies that the vector pierces the walls of the worldtubes. But when the contraction \emph{is} 
zero, as in Eq.~(\ref{euler1f}), the vector \emph{does not} pierce the walls. This is illustrated 
in Fig.~\ref{vorticity}, where the red circles indicate the orientation of each world-tube. The 
individual fluid element four-velocities lie in the centers of the world-tubes. Finally, consider the 
closed contour in Fig.~\ref{vorticity}. If that contour is attached to fluid-element worldlines, then 
the number of worldtubes contained within the contour will not change because the worldlines 
cannot pierce the walls of the worldtubes. This is essentially the Kelvin--Helmholtz theorem on the
conservation of vorticity. From this we learn that the Euler equation is (in fact) an integrability condition 
which ensures that the vorticity two-surfaces mesh together to fill spacetime.

\vspace*{0.1cm}
\begin{tcolorbox}
\textbf{Comment}:
We get a different perspective on Eq.~\eqref{euler1f} if we 
view it as a matrix equation. Basically, the result implies that $n^a$
is an eigenvector associated with a zero eigenvalue; specifically, 
\begin{eqnarray}
    \begin{bmatrix} 
    0 & \omega_{0 1} & \omega_{0 2} & \omega_{0 3} \\ 
    - \omega_{0 1} & 0 & \omega_{1 2} & \omega_{1 3} \\
    - \omega_{0 2} & - \omega_{1 2} & 0 & \omega_{2 3} \\
    - \omega_{0 3} & - \omega_{1 3} & - \omega_{2 3} & 0
    \end{bmatrix} 
    \begin{bmatrix} 
    n^0 \\ 
    n^1 \\
    n^2 \\
    n^3
    \end{bmatrix}
    =
    \begin{bmatrix} 
    0 \\ 
    0 \\
    0 \\
    0
    \end{bmatrix}
\end{eqnarray}
Of course, a solution then exists only if the determinant of the $4 \times 4$ matrix vanishes; i.e.
\begin{eqnarray}
\det \begin{bmatrix} 
    0 & \omega_{0 1} & \omega_{0 2} & \omega_{0 3} \\ 
    - \omega_{0 1} & 0 & \omega_{1 2} & \omega_{1 3} \\
    - \omega_{0 2} & - \omega_{1 2} & 0 & \omega_{2 3} \\
    - \omega_{0 3} & - \omega_{1 3} & - \omega_{2 3} & 0
    \end{bmatrix} 
    = \left(\omega_{0 1} \omega_{2 3} - \omega_{0 2} \omega_{1 3} + \omega_{0 3} \omega_{1 2}\right)^2 = 0 \ .
\end{eqnarray}
\end{tcolorbox}
\begin{tcolorbox}
This arguments relates directly to
the wedge product $\boldsymbol{\omega} \wedge \boldsymbol{\omega}$ of the two-form 
$\boldsymbol{\omega}$ with itself is a four-form, and thus its components must be proportional 
to $\epsilon^{a b c d}$ meaning the only independent component is 
$\epsilon^{a b c d} \omega_{a b} \omega_{c d}$ (up to normalization). An explicit 
calculation shows
\begin{eqnarray}
\epsilon^{a b c d} \omega_{a b} \omega_{c d} = \omega_{0 1} \omega_{2 3} - \omega_{0 2} 
\omega_{1 3} + \omega_{0 3} \omega_{1 2} = 0
\end{eqnarray}
and therefore  $\boldsymbol{\omega} \wedge \boldsymbol{\omega}$ vanishes. The geometric meaning of this is discussed in more detail in the lead-up to Eq.~\eqref{simple1}.
\end{tcolorbox}
\vspace*{0.1cm}

\subsection{Conservation laws}

The variational model we will develop contains the same information as the standard approach (a point that is emphasized by the Newtonian limit in Sect.~\ref{sec:newton})---as it must if we want it to be useful---but it is more directly linked to the conservation of vorticity. In fact, the definition of the vorticity implies that its exterior derivative vanishes. This means that
\begin{equation}
\nabla_{[a} \omega_{bc]} = 0\ .
\label{symm}\end{equation}
Whenever the Euler equation \eqref{euler1f} holds, this leads to the vorticity being conserved along the flow. That is, we have
\begin{equation}
\label{vortcon}
\mathcal L_u \omega_{ab} = 0\ .
\end{equation}
The upshot of this is that,  Eq.~(\ref{euler1f}) can be used to 
discuss the conservation of vorticity in an elegant way. It can also be used as the basis for a 
derivation of other theorems in fluid mechanics.

As is well-known, constants of motion are often associated with symmetries of the problem under consideration.
In General Relativity, spacetime symmetries can be expressed in terms of Killing vectors, $\hat \xi^a$ (the hat is used to make a distinction from the Lagrangian displacement later). As an example, 
let us assume that the spacetime does not depend on the coordinate $a = X$. The corresponding Killing vector
would be 
\be
\hat \xi^a = \delta^a_X {\partial \over \partial X}\ ,
\ee
and the  symmetry leads to Killing's equation
\be
\mathcal L_{\hat\xi} g_{a b} = 0 \qquad \Longrightarrow \qquad \nabla_a \hat \xi_b + \nabla_b \hat \xi_a = 0 \ .
\ee 
Associated with each such Killing vector will be a conserved quantity. In the vacuum case, it is easy to combine the geodesic equation
\be
u^b \nabla_b u_a = 0 \ ,
\ee
with Killing's equation to show that 
\be
u^b \nabla_b  \left( \hat \xi^a u_a\right) = {d \over d\tau}  \left( \hat \xi^a u_a\right)  = 0 \ .
\ee
In other words, the combination $\hat \xi^a u_a$ remains constant along each geodesic.

Let us now consider how this argument extends to the fluid case. 
Assuming that the flow is invariant with respect to transport by the vector
field $\hat \xi^a$, we have 
\begin{equation} 
     \mathcal{L}_{\hat\xi} \mu_a = 0 \ , \qquad \Longrightarrow 
     \qquad 
     \hat \xi^b \nabla_b \mu_a + \mu_b \nabla_a\hat \xi^b = 0 \ . 
\label{inva}\end{equation}
Now combine this with the equation of motion in the form \eqref{euler1f} to find
\be
\hat \xi^a n^b \left( \nabla_b \mu_a - \nabla_a \mu_b\right) = n^b \nabla_b \left( \hat \xi^a \mu_a\right) = 0 \ . 
\ee
Since $n^a = n u^a$ we see that the quantity $\hat \xi^a \mu_a$ is conserved along the fluid world lines, reminding us of the vacuum result. The difference is due to the fact that pressure gradients in the fluid leads to the flow
no longer being along geodesics. One may consider two specific situations. If $\hat \xi^a$ is taken to 
be the four-velocity, then the scalar $\hat \xi^a \mu_a$ represents the 
``energy per particle''. If instead $\hat \xi^a$ represents an axial generator 
of rotation, then the scalar will correspond to an angular momentum. 
For the 
purposes of the present discussion we can leave $\hat \xi^a$ unspecified, but it 
is still useful to keep these possibilities in mind.  

Given that the flux is conserved, i.e. \eqref{consv2} holds, we can take one further step to show that we have 
\begin{equation}
    n^a \nabla_a \left(\mu_b \hat \xi^b\right) = \nabla_a 
    \left(n^a \mu_b\hat  \xi^b\right) = 0 \ , 
\end{equation} 
and we have shown that $n^a \mu_b \hat \xi^b$ is a conserved 
quantity. 

In many cases one can also obtain integrals of the motion, analogous to the Bernoulli equation for  stationary 
rotating Newtonian fluids. Quite generally, the derivation proceeds as follows. Assume that $\hat \xi^a$ is such that 
\be
\hat \xi^b \omega_{b a} = 0 \ .
\ee
This condition can be written
\be
\mathcal L_{\hat \xi} \mu_a - \nabla_a \left( \hat \xi^b \mu_b \right) = 0  
\ee
where the first term vanishes as long as \eqref{inva} holds. Hence, we arrive at the first integral
\be
 \nabla_a \left( \hat \xi^b \mu_b \right) = 0 \qquad\Longrightarrow \qquad\hat \xi^b \mu_b = \mathrm{constant} \ .
\ee

An obvious version of this analysis is an irrotational flow, when $\omega_{a b} = 0$. Another situation of direct astrophysical interest is
``rigid'' flow---when $\hat \xi^a = \lambda u^a$ for some scalar field $\lambda$. Rotating compact stars, in equilibrium, belong to 
this category.  In that case, one would have $\hat \xi^a =t^a + \Omega \phi^a$, where
$\Omega$ is the rotation frequency and $t^a$ and $\phi^a$ represent the timelike Killing vector and the spatial Killing vector associated with axisymmetry, respectively (the system permits a helical Killing vector). 

\subsection{A couple of steps towards relative flows}

With the comments at the close of the previous section, we have reached the end of the road as far as the 
``off-the-shelf'' strategy is concerned. We will now move towards an action-based derivation of the fluid 
equations of motion.  
As a first step, let us look ahead to see what is coming and why we need to go in this direction. 

Return to the perfect fluid stress-energy tensor but now let us not associate the observer with the fluid flow.  The thermodynamical relations still hold in the co-moving (fluid) frame associated with $u^a$, but the observer sees the fluid flow by with the relative velocity $v^a$ from \eqref{varel}. In essence, we then have
\begin{multline}
T_{ab} = (p+\varepsilon)\gamma^2 (U_a+v_a)(U_b +v_b)+pg_{ab} \\
= \varepsilon\gamma^2 U_a U_b +p(U_aU_b+g_{ab})+2 (p+\varepsilon)\gamma^2 U_{(a}v_{b)} + (p+\varepsilon)\gamma^2 v_a v_b
\end{multline}
We learn several important lessons from this. The perfect fluid does not seem quite so simple in the frame of a general observer. First of all, the different thermodynamical quantities will be redshifted (as expected from Special Relativity) so we need to keep track of the $\gamma$ factors. Secondly, we now appear to have both a momentum flux and anisotropic spatial stresses. In order to arrive at the main point we want to make, let us assume that the relative velocity is small enough that we can linearize the problem. As we will see later, this should be an adequate assumption in many situations of interest. Leaving out terms quadratic in $v^a$ we lose the spatial stresses and $\gamma\to 1$ (which is convenient as the thermodynamics then remains as before).
We are left with
\begin{equation}
T_{ab} \approx  \varepsilon U_a U_b +p(U_aU_b+g_{ab})+2 (p+\varepsilon) U_{(a}v_{b)} \ .
\label{Tlin1}
\end{equation}
At this point, we can make use of the freedom to choose the observer. We may return to the case where the observer rides along with the fluid by setting ($v^a=0$). This choice is commonly called the Eckart frame, as it was first introduced in the discussion of relativistic heat flow (see Sect.~\ref{sec:heat}). This is the obvious choice for a single fluid problem, but when we are dealing with multiple flows there are alternatives. 

As an illustration, in the case of a problem with both matter and heat flowing, we have to replace the stress energy tensor  by (don't worry, we will derive this later)
\begin{equation}
T_{ab} \approx p g_{ab} + n\mu u_a u_b + sT u^\s_a u^\s_b \ ,
\end{equation}
where $s$ and $T$ are the entropy (density) and temperature, respectively, and $u_\s^a$ accounts for the heat flux. We have assumed that both flows may be linearized relative to the observer so 
\begin{equation}
u_\s^a \approx U^a + q^a \ , \quad \mbox{with} \quad U^aq_a = 0 \ ,
\end{equation}
where $q^a$ is the heat flux.
This means that we have
\begin{multline}
T_{ab} \approx p g_{ab} + (n\mu+sT) U_a U_b + 2 n\mu U_{(a}v_{b)}  + 2 sT  U_{(a}q_{b)} \\
=   \varepsilon U_a U_b +p(U_aU_b+g_{ab}) + 2 n\mu U_{(a}v_{b)}  + 2 sT  U_{(a}q_{b)} \ .
\end{multline}
 In this case,  the momentum flux relative to the observer will be 
\begin{equation}
\mathcal P_a = - \perp^b_a U^c T_{b c} =  n\mu  v_a + sT q_a \ .
\end{equation}
Basically, an observer riding along with the matter will experience heat flowing. We may, however,  work with a different observer according to whom no energy flows. It is easy to see that this involves setting
\begin{equation}
v_a  = {sT \over n\mu} q_a = {sT \over p+\varepsilon-sT} q_a \ .
\end{equation}
With this choice we are left with 
\begin{equation}
T_{ab} \approx \varepsilon U_a U_b + (p+\varepsilon) ( g_{ab} +  U_a U_b ) \ ,
\end{equation}
reminding us of the perfect fluid situation, even though we are considering a more complicated problem. It follows that
\begin{equation}
U^a T^b_{\ a} = - \varepsilon U^b \ .
\label{llframe}
\end{equation}
Formally, the energy density $\varepsilon$ is an eigenvalue of the stress-energy tensor (with the observer four velocity $U^a$ the corresponding eigenvector). This choice of observer is usually referred to as the Landau-Lifschitz frame \citep{landau59:_fluid_mech}. 

We are free to work with whatever observer we like---different options have different advantages---but there is no free lunch. For example, with the Landau-Lifschitz choice the fluid equations simplify, but the particle conservation law becomes more involved. We now have
\begin{equation}
\nabla_a n^a \approx \nabla_a ( nU^a + nv^a) = \nabla_a \left(  nU^a + {n sT \over p+\varepsilon-sT} q^a \right) = 0 \ .
\end{equation} 
The contribution from the heat flux is not particularly intuitive.

The main lesson we learn from this exercise is that any situation with relative flows involves making choices, and we have to keep careful track of how these choices impact on the connection with the underlying physics. This motivates the formal development of  the variational approach for general relativistic multifluid systems, to be described in Sections~\ref{sec:twofluids}.


\subsection{From microscopic models to the equation of state}
\label{micro}

We have discussed how the equations for relativistic fluid dynamics relate to a given stress-energy tensor, involving as set of suitably averaged variables (energy, pressure, four-velocity etc.). We have also seen how one can obtain the equations of motion from 
\begin{equation}
    \nabla_a T^{ab} = 0 \ , 
    \label{fleom}
\end{equation}
as required by the Einstein field equations (by virtue of the Bianchi identities).
Moreover, in Sect.~\ref{sec:variational} we showed how the stress-energy tensor can be obtained via a variation of the Lagrangian with respect to the spacetime metric. This description is neatly self-consistent---and we will make frequent use of it later---but it is helpful to pause and consider the logic. In principle, the relation \eqref{fleom} follows from the fact that the Einstein tensor $G_{ab}$ is divergence free, which in turn represents the fact that the problem involves four ``unphysical'' degrees of freedom, usually taken to mean that we have the freedom to choose the four spacetime coordinates. However, by turning \eqref{fleom} into the equations for fluid dynamics we are changing the perspective. The four degrees of freedom now represent the conservation of energy and momentum. Why are we allowed to do this? Is it simply a fluke that the four degrees of freedom involved can be suitably interpreted in a manner that fits out purpose? One can argue that this is, indeed, the case and we will discuss this later.

For the moment, we want to consider a different aspect of the problem. If it is the case that \eqref{fleom} encodes the fluid equations of motion, then there ought to be a way to derive the stress-energy tensor from some underlying microscopical theory (presumably involving quantum physics). This issue turns out to be somewhat involved. As a starting point, suppose we focus on a 
one-parameter system, with the parameter being the particle number density. The equation of 
state will then be of the form $\varepsilon = \varepsilon(n)$, representing the energy per particle. In many-body physics (as studied in 
condensed matter, nuclear, and particle physics) one can then in principle construct the quantum 
mechanical particle number density $n_{\mathrm{QM}}$, stress-energy tensor 
$T^{\mathrm{QM}}_{a b}$, and associated conserved particle number density current 
$n^a_{\mathrm{QM}}$ (starting from some fundamental Lagrangian, say;
cf.\ \citealt{Walecka:1995mi,glendenning97:_compact_stars,Weber:1999qn}). But unlike in quantum 
field theory in a curved spacetime \citep{Birrell:1982ix}, one typically assumes that the matter exists in an 
infinite Minkowski spacetime. 


Once $T^{\mathrm{QM}}_{a b}$ is obtained, and after (quantum mechanical and statistical) 
expectation values with respect to the system's (quantum and statistical) states are taken, one 
defines the energy density as
\begin{equation}
  \varepsilon = u^a u^b \langle T^{\mathrm{QM}}_{a b} \rangle,
\end{equation}
where
\begin{equation}
  u^a \equiv \frac{1}{n} \langle n^a_{\mathrm{QM}} \rangle,
  \qquad
  n = \langle n_{\mathrm{QM}} \rangle.
\end{equation}
Similarly, the pressure is obtained as
\begin{equation}
  p = \frac{1}{3} \left( \langle T^{\mathrm{QM} a}{}_a \rangle + \varepsilon \right)
\end{equation}
and it will also be a function of $n$.

One must be very careful to distinguish $T^{\mathrm{QM}}_{a b}$ from $T_{a b}$. The former 
describes the states of elementary particles with respect to a fluid element, whereas the latter 
describes the states of fluid elements with respect to the system. Comer and
Joynt~\citeyearpar{comer03:_rel_ent} have shown how this line of reasoning applies to the two-fluid 
case. 

This outline description stays close to the fluid picture, but it does not shed much light on the origin of $T^{\mathrm{QM}}_{a b}$. This is where we run into ``trouble''. A typical field theory description would take a given symmetry of the system as its starting point, and then obtain equations of motion for conserved quantities associated with this symmetry. Let us consider this problem in flat space and use a scalar field with Lagrangian $L=L(\phi, \partial_a \phi)$ as our example. Assuming that the system is symmetric under spacetime translations, we have four conserved (Noether) currents given by
\begin{equation}
  \tau^a_{\ b} = {\partial L \over \partial(  \partial_a \phi)} \partial_b \phi - \delta ^a_b L  \ .
\end{equation}
That is, we have
\begin{equation}
\partial_a \tau^a_{\ b} = 0 \ , 
\label{conse1}
\end{equation}
which follows by virtue of the Euler-Lagrange equations:
\begin{equation}
    \partial_a \left( {\partial L \over \partial(  \partial_a \phi)} \right) - {\partial L \over \partial \phi} = 0 \ ,
\end{equation}
and the fact that we are working in flat space (so partial derivatives commute). It may seem tempting to take $\tau^a_{\ b}$ to be the stress-energy tensor---intuitively, we can change partial derivatives to covariant ones, introduce the spacetime metric (instead of $\eta^{ab}$, as appropriate), to arrive at an expression similar to \eqref{fleom}. However, the Devil is in the detail. The flat-space field equations represent a true conservation law (with four conserved currents, one for each value of $b$ in \eqref{conse1}), which is what we expect, but $\tau^a_{\ b}$ is (in general) not symmetric. Since symmetry is required for the \emph{gravitational} stress-energy tensor $T^{ab}$ (as long as we do not deviate from Einstein's theory) we have a problem. The issue is resolved by invoking the Belinfante-Robinson ``correction'' to $\tau^a_{\ b}$ (see for example \cite{paston} for a recent discussion). This is a uniquely defined object which effects the change from a flat to a curved spacetime. While we will not need to understand the details of this procedure to make progress, it is important to be aware of it. 


\section{Variational approach for a single-fluid system}
\label{sec:pullback}

Let us now consider the single-fluid problem from a different perspective and derive the equations of motion and the stress-energy tensor from an action principle. The ideas behind
this variational approach can be traced back to \cite{taub54:_gr_variat_princ} (see also~\cite{schutz_var}). Our
approach relies heavily on the work of Brandon
Carter, his students, and
collaborators~\citep{carter89:_covar_theor_conduc,
  comer93:_hamil_multi_con, comer94:_hamil_sf, carter95:_kalb_ramond,
  carter98:_relat_supercond_superfl,
  langlois98:_differ_rotat_superfl_ns, prix00:_these,
  prix04:_multi_fluid}. This strategy is attractive as it  makes maximal use of the tools of the trade of relativistic fields, i.e., no
special tricks or devices will be required (unlike even the case of
the ``off-the-shelf'' approach). Our footing is  made
sure by well-grounded, action-based arguments. As Carter has 
made clear: When there are multiple fluids, of both the charged and
uncharged variety, it is essential to distinguish the fluid momenta
from the velocities, in  order to make the geometrical
and physical content of the equations transparent. A well-posed action
is, of course, perfect for systematically constructing the momenta.

Specifically, we will make use of a  ``pull-back'' approach (see, e.g., \citealt{comer93:_hamil_multi_con, comer94:_hamil_sf,comer02:_zero_freq_subspace})  to construct a Lagrangian
displacement of the particle number density flux $n^a$, whose magnitude $n$ is the particle 
number density.  This will form the basis for the variations of the fundamental fluid variables in 
the action principle.

\subsection{The action principle}

It is useful to begin by explaining why we need to develop a constrained action principle. The argument is quite simple. Consider a single matter component, represented by a flux $n^a$.
For an isotropic system the matter Lagrangian, which we will call $\Lambda$ (taking over the role of $L$ from Sect.~\ref{sec:variational}), should be a relativistic invariant and hence 
depend only on $n^2 = -g_{ab} n^a n^b$. In effect, this means that it depends on both the flux and the spacetime metric. This is, of course, important as the dependence on the metric leads to the stress-energy tensor (again, as is Sect.~\ref{sec:variational}).
An arbitrary variation of $\Lambda=\Lambda(n^2)=\Lambda(n^a,g_{ab})$ now leads to (ignoring terms that can be written as total derivatives representing``surface terms'', as in the point-particle discussion) 
\begin{equation}
    \delta \left(\sqrt{- g} \Lambda\right) = \sqrt{- g} \left[\
    \mu_a \delta n^a + \frac{1}{2} \left(\Lambda g^{a b} +  
    n^a \mu^b\right) \delta g_{a b}\right] \ , \label{dlamb}
\end{equation}
where $\mu_a$ is the canonical  momentum, which is given by
\begin{equation}
     \mu_{a} = {\partial \Lambda \over \partial n^a} = -2 {\partial \Lambda \over \partial n^2} g_{ab} n^b \ .
     \label{momdef}
                   \end{equation} 
We have also used (see Sect.~\ref{sec:efevar})
\begin{equation}
\delta \sqrt{-g} =  {1\over 2} g^{ab} \delta g_{ab}  \ .
\end{equation}

Here is the problem: As it stands, Eq.~\eqref{dlamb} suggests that the equations of motion would simply be $\mu_a=0$, which means that the fluid carries neither energy nor momentum. This is obviously not what we are looking for,

In order to make progress, we impose the constraint that the flux is conserved\footnote{It is worth
  pointing out that we are restricting the problem somewhat by
  imposing particle conservation already from the outset. As we will
  see later, one can make good progress on less constrained problems,
  e.g., related to dissipation, using an extended variational
  approach (inspired by the point particle example from
  Sect.~\ref{sec:point}). However, we feel that it is useful to first
  understand the simpler, fully conservative,
  situation.}. That is, we insist that
\begin{equation}
\nabla_ a n^a = 0 \ .
\end{equation}

From a strict field theory point of view, it makes sense to introduce this constraint. The conservation of the particle flux (the number density current)
should not be a part of the equations of motion, but rather should be automatically
satisfied when evaluated on a solution of the ``true'' equations.

For reasons that will become clear shortly, it is useful to rewrite the conservation law in terms of the dual three-form\footnote{In order to be fully consistent we should really introduce notation to identify the dual here, but as we will keep the indices explicit there is little risk of confusion.}
\begin{equation}
n_{abc} = \epsilon_{dabc} n^d\ , 
\label{ndual}
\end{equation}
such that 
\begin{equation}
n^a = {1\over 3!} \epsilon^{bcda} n_{bcd} \ .
\end{equation}
It also follows that 
\begin{equation}
n^2 = - g_{ab} n^a n^b = {1\over 3!} n_{abc}n^{abc} \ , 
\end{equation}
which shows that $n_{abc}$ acts as a volume measure which allows us to ``count'' the number of fluid elements. In
Fig.~\ref{vorticity} we have seen that a two-form is associated with worldtubes. A
three-form is the next higher-ranked object and it can be thought of, in an
analogous way, as leading to boxes \citep{mtw73}.
This is quite intuitive, and we will comment on it again later.

\vspace*{0.1cm}
\begin{tcolorbox}
As we develop the variational approach, we need to be comfortable with volume forms. This, in particular, involves working with contraction of  $\epsilon_{abcd}$. The general relations we need are provided in Appendix~\ref{appendix}, but let us note a couple of particularly pertinent ones here. First of all, we have already used
$$
\epsilon^{dabc} \epsilon_{defg} = - 3! \delta^{[a}_e \delta^b_f \delta^{c]}_g \ ,
$$
where the signs comes from $(-1)^s$ where $s$ is the number of minus signs in the spacetime metric (e.g., $s = 1$ in our case). Meanwhile, when the work in three dimensions (as in the case of the spatial part of  3+1 decomposition later), we have the familiar relation
$$
\epsilon^{ijk}\epsilon_{ilm} = 2! \delta^{[j}_l \delta^{k]}_m = \delta^j_l \delta ^k_m - \delta^j_m \delta^k_l \ .
$$
\end{tcolorbox}
\vspace*{0.1cm}

With this set-up, the conservation of the matter flux is ensured provided that the three-form $n_{abc}$ is closed. It is easy to see that
\begin{equation} 
     \partial_{[a} n_{bcd]}=\nabla_{[a} n_{bcd]} = 0\quad   \Longrightarrow \quad 
  \nabla_{a} n^{a} = 0 \ . \label{consv2} 
\end{equation}

\begin{figure}[htb]
    \centerline{\includegraphics[scale = 0.5]{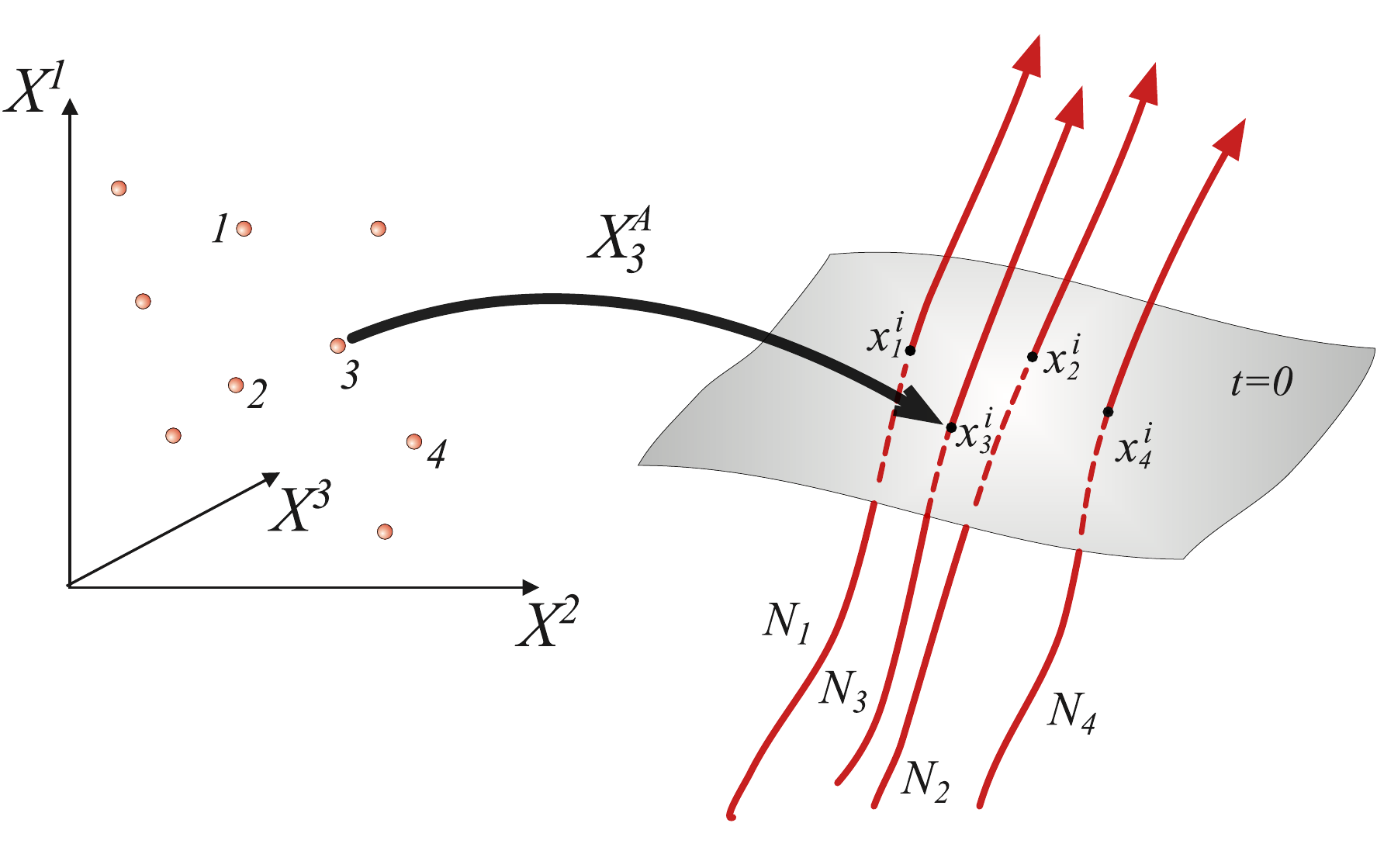}}
    \caption{The pull-back from ``fluid-particle'' points in
    the three-dimensional matter space, labelled by the coordinates
    $\{X^1,X^2,X^3\}$, to fluid-element worldlines in spacetime. Here,
    the pull-back of the ``$I^{\mathit{th}}$'' ($I = 1,2,\dots,n$)
    fluid-particle to, say, an initial point on a worldline in
    spacetime can be taken as $X^A_I = X^A(0,x^i_I)$ where $x^i_I$ is
    the spatial position of the intersection of the worldline with the
    $t = 0$ time slice.}
    \label{pullback}
\end{figure}

The main reason for introducing the dual is that it is straightforward to
construct a particle number density three-form that is automatically
closed. We achieve this by introducing a three-dimensional ``matter''
space---the left-hand part of Fig.~\ref{pullback}---which is
labelled by coordinates $X^A$, where $A,B,C, \ldots = 1,2,3$. For each
time slice in spacetime, we have the same configuration in the matter
space. That is, as time moves forward, the fluid particle positions in
the matter space remain fixed---even through the worldlines weave through
spacetime. In this sense we are ``pulling back'' from the matter space
to spacetime (cf.\ the discussion of the Lie derivative). The
 $n_{abc}$ three-form can then be ``pushed forward'' to the three-dimensional matter space by using
the map associated with the coordinates $X^A$ (which represent scalar fields on spacetime):
\begin{equation}
\psi^A_a = \partial_a X^A \ .
\end{equation}
This construction leads to a matter-space three form $N_{ABC}$, 
\begin{equation}
  n_{a b c} = \psi_a^A \psi_b^B \psi^C_c
  N_{A B C} \ ,
  \label{pb3form}
\end{equation}
which is completely anti-symmetric in its indices. The final step involves noting that 
\beq
\partial_{[a}n_{bcd]} = \psi^A_a \psi^B_b \psi^C_c\psi^D_d \partial_{[A} n_{BCD]} = 0 \ ,
\eeq
is automatically satisfied if 
\beq
\partial_{[A} n_{BCD]} = 0 \ ,
\eeq
which, in turn, follows if $n_{ABC}$ is taken to be a function only of the $X^A$ coordinates. This completes the argument.

Now we need to connect this idea to the variational principle. The key step 
involves introducing the 
Lagrangian displacement $\xi^a$, tracking the motion of a given fluid element. From the standard definition
of  Lagrangian variations,  we have 
\beq
\Delta X^A = \delta X^A + \mathcal{L}_{\xi} X^A = 0 \ , 
\label{DelX}
\eeq
where $\delta X^A$ is the Eulerian variation and $\mathcal{L}_{\xi}$ is the Lie derivative along $\xi^a$. This means that we have
\beq
    \delta X^A = -  \mathcal{L}_{\xi} X^A = -  \xi^a {\partial X^A \over \partial x^a} = -\xi^a \psi^A_a\ . 
    \label{xlagfl}
\end{equation} 
It also follows that
\begin{multline}
\Delta \psi^A_a = \delta \psi^A_a + \xi^b \partial_b \psi^A_a + \psi^A_b \partial_a \xi^b = \partial_a \delta X^a + \xi^b \partial_b \psi^A_a + \psi^A_b \partial_a \xi^b \\
= \partial_a \left( \Delta X^a - \xi^b \partial_b X^a \right) + \xi^b \partial_b \psi^A_a + \psi^A_b \partial_a \xi^b=  0 \ , 
\end{multline}
since partial derivatives commute.
Given these results, it is easy to show that 
\beq
\Delta n_{abc} = \psi^A_a \psi^B_b\psi^C_c \partial_D N_{ABC} \Delta X^D = 0 \ .
\eeq
This implies that 
\begin{equation}
\delta n_{abc} = - \mathcal L_\xi n_{abc} \ , 
\end{equation}
and hence
\begin{equation}
 \delta n^a = {1\over 3!} \delta \left( \epsilon^{bcda} n_{bcd} \right) =  
 {1\over 3!} \left( \delta \epsilon^{bcda} n_{bcd} - \epsilon^{bcda} \mathcal L_\xi n_{abc} \right) \ .
\end{equation}

Making use of a little bit of elbow grease and  the standard relations
\beq
\delta g_{db} =  - g_{da} g_{bc} \delta g^{ac} \ ,
\eeq
and
\beq
 \delta \epsilon^{abcd} = {1\over 2}  \epsilon^{abcd} g_{ef} \delta g^{ef}  \ ,
 \label{epsvar0}
\eeq
we arrive at
\begin{multline}
\delta n^a  = {1\over 3!} \delta ( \epsilon^{ b c d a} n_{b c d} )  = n^b \nabla_b \xi^a - \xi^b 
                   \nabla_b n^a - n^a \left(\nabla_b 
                   \xi^b -  \frac{1}{2} g_{b c} \delta 
                   g^{b c}\right)\\
                   = - \mathcal L_\xi n^a - n^a \left(\nabla_b 
                   \xi^b  -  \frac{1}{2} g_{b c} \delta 
                   g^{b c}\right)                   
                   \ ,
                     \label{delnvec0} 
\end{multline} 
or 
\begin{equation}
\Delta n^a = - n^a \left( \nabla_b \xi^b + { 1 \over 2} g^{bd} \delta g_{bd} \right) = - {1 \over 2} n^a \left( g^{bd} \Delta g_{bd}\right)\ ,
\label{dna0}
\end{equation}
where
\begin{equation}
\Delta g_{ab} = \delta g_{a b} + 2 \nabla_{(a} \xi_{b)} \ , \label{lagvarmet}
\end{equation}
(the parentheses indicate symmetrization, as usual).  Eq.~\eqref{dna0} has a natural 
interpretation: The variation of a fluid worldline with respect to its own Lagrangian displacement 
has to be along the worldline and can only measure the changes of the volume of its own fluid 
element. This is one of the advantages of the Lagrangian variation approach. 

\vspace*{0.1cm}
\begin{tcolorbox}
\textbf{Comment:}
At first glance, there appears to be a glaring inconsistency between the
pull-back construction and the Lagrangian variation, since the latter seems
to have four independent components, but the former clearly has three. However, there is a gauge freedom in the Lagrangian variation that can be used
to reduce the number of independent components. Take Eq.~(\ref{delnvec0})
and substitute
\begin{equation}
  \xi^a = \overline{\xi}^a + {\cal G}^a \ ,
\end{equation}
%
%
%
%
to get
\begin{equation}
  \delta n^a = \delta \overline{n}^a +
  \nabla_b
  \left( n^b {\cal G}^a - n^a {\cal G}^b \right) - {\cal G}^a \nabla_b n^b \ ,
\end{equation}
where $\delta \overline{n}^a$ is as in Eq.~(\ref{delnvec0}) except $\xi^a$
is replaced with $\overline{\xi}^a$. Using the fact that $\nabla_a n^a = 0$, and setting
\begin{equation}
  {\cal G}^a = {\cal G} n^a \ ,
\end{equation}
 the last two terms vanish and $\delta n^a = \delta \overline{n}^a$.
Thus, we can use the arbitrary function ${\cal G}^a$ (the gauge freedom) to reduce the number of
independent components of $\xi^a$ to three.
\end{tcolorbox}
\vspace*{0.1cm}

Expressing the variations of the matter Lagrangian in terms of the displacement $\xi^a$, rather than the perturbed flux, we 
ensure that the flux conservation is accounted for in the equations of motion.
The variation of 
$\Lambda$ now leads to 
\begin{multline}
    \delta \left(\sqrt{- g} \Lambda\right) = \sqrt{- g} \left\{  f_a \xi^a -  \frac{1}{2}\left[\left( \Lambda - n^c\mu_c\right) g_{a b} +  
    n_a \mu_b \right] \delta g^{a b}  \right\} \\
    + \nabla_a \left(\frac{1}{2} \sqrt{- g} \mu^{abc} n_{bcd} \xi^d\right) \ , \label{variable}
\end{multline}
and the fluid equations of motion are given by
\begin{equation} 
     f_b \equiv 2 n^a \nabla_{[a} \mu_{b]}   = 0 \ ,
     \label{force}
\eeq
 (where the square brackets indicate anti-symmetrization, as usual). Finally, introducing the vorticity two-form
 \beq
 \omega_{ab} = 2\nabla_{[a} \mu_{b]}  \ ,
 \label{omdef}
 \eeq
 we have the simple relation
 \beq
 n^a \omega_{ab} = 0 \ ,
 \eeq
 which should be familiar (see Sect.~\ref{shelve}).
 
 We can also read off the stress-energy tensor from \eqref{variable}. We need (see Sect.~\ref{sec:variational})
\begin{equation}
T_{ab} = - {2 \over \sqrt{-g}} {\delta \left( \sqrt{-g}\Lambda\right) \over \delta g^{ab}}  =  \Lambda g_{ab} - 2 {\delta \Lambda \over \delta g^{ab}} \ .
\end{equation}
Finally, introducing the matter four-velocity, such that $n^a=nu^a$ and $\mu_a = \mu u_a$, where $\mu$ is the chemical potential (as before), we see that the energy is
\begin{equation}
\varepsilon = u_a u_b T^{ab} = - \Lambda \ .
\end{equation}
Moreover, we identify the pressure from the thermodynamical relation:
\begin{equation}
p = -\varepsilon + n\mu  =  \Lambda - n^c\mu_c  \ .
\end{equation}
This means that we have
\begin{equation}
T^{ab} = pg^{ab} + n^a \mu^b = \varepsilon u^a u^b + p \perp^{ab} \ ,
\label{stressen0}
\end{equation}
and it is straightforward to confirm that
\begin{equation} 
\nabla_a T^{ab} = - f^b + \nabla^b \Lambda - \mu^b \nabla_a n^a = - f^b = 0 \ ,
\label{divT}
\end{equation}
since (i) $\Lambda$ is a function only of $n^a$ and $g_{ab}$, and (ii) the definition of the momentum $\mu_a$.

\vspace*{0.1cm}
\begin{tcolorbox}
Let us pause to recall the discussion of the point particle, where we
pointed out that only the fully conservative form of Newton's Second Law
follows from the action. External or dissipative forces are
excluded. However, we argued that a well-established form of Newton's
second law is known that allows for external and/or dissipative
forces (cf.\ Eq.~(\ref{newteq})). This lends meaning to the use of  $f_a$ in Eq.~(\ref{euler1f}). We may take the
$f_a$ to be the relativistic analogue of the left-hand-side of
Eq.~(\ref{newteq}) in every sense. In particular, when dissipation
and/or external ``forces'' act in a general relativistic setting, they may be 
 introduced as in the right-hand-side of Eq.~(\ref{euler1f}). However, it is natural to wonder if it is possible to do better than this somewhat  phenomenological approach. Is it possible to incorporate dissipative aspects in the action? Later, in Sect.~\ref{sec:viscosity}, we will argue that this can, indeed, be done.  
\end{tcolorbox}
\vspace*{0.1cm}

\subsection{Lagrangian perturbations}

Later, we will consider linear dynamics of different systems---both at the local level and for macroscopic bodies like rotating stars. This inevitably draws on an understanding of perturbation theory, which (in turn) makes contact with the variational argument we have just completed. Given this, it is worth making a few additional remarks before we move on.

First of all, 
an unconstrained variation of $\Lambda(n^2)$ is with respect to $n^a$
and the metric $g_{a b}$, and allows the four components of
$n^a$ to be varied independently. It takes the form
\begin{equation}
  \delta \Lambda =
  \mu_a \delta n^a + \frac{1}{2} n^a \mu^b \delta g_{a b} \ ,
\end{equation}
where
\begin{equation}
  \mu_a = \B n_a \ ,
  \qquad
  \B \equiv - 2 \frac{\partial \Lambda}{\partial n^2}.
  \label{mudef}
\end{equation}
The use of the letter $\B$ is to remind us that this is a bulk fluid
effect, which is present regardless of the number of fluids and
constituents. The momentum covector $\mu_a$ is (as we have seen) dynamically, and thermodynamically, conjugate to
$n^a$, and its magnitude is the chemical potential of
the particles (recalling that $\Lambda = - \varepsilon$). 

Next, by introducing the displacement $\xi^a$, effectively tracking the fluid elements, we have 
prepared the ground for a study of general Lagrangian perturbations (as relevant for, for example, a relativistic study of neutron-star instabilities \citep{jf}, see Sect.~\ref{sec:cfs}). In fact, given the results from the variational derivation it is straightforward to write down the perturbed fluid equations.

By introducing the
decomposition $n^a = n u^a$ we can show that the argument that led to \eqref{dna0} also provides\footnote{This step may lead to conceptual confusion as we (deliberately) represent the displacement vector by $\xi^a$. The mathematics for (say) the perturbed flux $\delta n^a$ is the same as in the variation derivation of the fluid equation, but the meaning of the variation is different. In the fluid derivation we consider variations away from the actual solution curve in parameter space, as illustrated in Fig.~\ref{vary}. In the case of Lagrangian perturbations, the displacement relate different configurations \emph{within} the solution space, i.e. that satisfy the equations of motion. }
\begin{equation}
  \delta n = - \nabla_a \left( n \xi^a \right) -
  n \left( u_a u^b \nabla_b \xi^a +
  \frac{1}{2} \perp^{ab}
  \delta g_{a b}\right)\ ,
  \label{dens_perb}
\end{equation}
and
\begin{equation}
  \delta u^a = \left( \delta^a{}_b + u^a u_b \right)
  \left( u^c \nabla_c \xi^b -
  \xi^c \nabla_c u^b \right) +
  \frac{1}{2} u^a u^b u^c \delta g_{b c}\ .
  \label{vel_perbs}
\end{equation}

Similar arguments lead to
\begin{eqnarray}
  \Delta u^a &=& \frac{1}{2} u^a u^b u^c \Delta g_{b c},
  \label{lagradelu}
  \\
  \Delta \epsilon_{a b c d} &=&
  \frac{1}{2} \epsilon_{a b c d} g^{e f}
  \Delta g_{e f},
  \label{lagradeleps}
  \\
  \Delta n &=&
  - \frac{n}{2} \perp^{ab}
  \Delta g_{a b}.
  \label{lagradeln}
\end{eqnarray}%
These results and their Newtonian analogues were used by
Friedman and Schutz in establishing the so-called
Chandrasekhar--Friedman--Schutz (CFS)
instability \citep{chandrasekhar70:_grav_instab, friedman78:_lagran,
  friedman78:_secul_instab} (see Sect.~\ref{sec:cfs}).

\subsection{Working with the matter space}

The derivation of the Euler equations \eqref{force} made ``implicit'' use of the matter space as a device to ensure the conservation of the particle flux. In many ways it makes sense to introduce the argument this way, but---as we will see when we consider elasticity---it can be useful to work more explicitly with the matter space quantities. 

Let us first note that, 
as implied by Fig.~\ref{pullback}, the $X^A$ coordinates
are comoving with their respective worldlines, meaning that they are
independent of the proper time $\tau$, say, that parameterizes each
curve. This is easy to demonstrate. Introducing the four velocity associated with the world line through $n^a = n u^a$, we have 
\begin{multline}
n{dX^A \over d\tau} = n {dx^a \over d\tau} \partial_a X^A = n^a \partial_a X^A = {\cal L}_n X^A \\
= -
  \frac{1}{3!} \epsilon^{b c d a}
\psi^A_a \psi^B_b \psi^C_c \psi^D_d   N_{B C D} = 0.
\end{multline}
We see that the time part of the spacetime dependence of the $X^A$ is somewhat ad
hoc. If we take the flow of time $t^a$ to be the proper time of the
worldlines ($t^a$ is parallel to $n^a$ and hence $u^a$), the $X^A$ do not change. An
apparent time dependence in spacetime means that $t^a$ is such as to cut
across fluid worldlines ($t^a$ is not parallel to $n^a$), which of course
have different values for the $X^A$.

It is also worth noting the (closely related) fact that $n_{abc}$ is  a ``fixed'' tensor, in the sense that
\beq
u^a n_{abc} = n u^a u^d \epsilon_{dabc} = 0  \ ,
\eeq
(i.e. the three-form is spatial) and
\beq
\mathcal L_u n_{abc} = 0 \ ,
\label{Lien}
\eeq
(it does not change along the flow).
The latter is equivalent to requiring that the three-form $n_{abc}$ be closed; i.e.,
\beq
\nabla_{[a}n_{bcd]} = \partial_{[a}n_{bcd]} = 0\ ,
\eeq
which, of course, holds by construction.

From a formal point of view, we have changed perspective by taking the (scalar fields) $X^A$   to be the fundamental 
variables.  
The construction also provides matter space with a geometric structure. As a first example of this note that, if integrated over a volume in matter space, $n_{ABC}$ provides a measure of the number of particles in that volume. To see this, simple introduce a matter space three form $\epsilon_{ABC}$ such that
\begin{equation}
n_{ABC} = n\epsilon_{ABC} \ ,
\end{equation}
and recall that such an object represents a volume. Since $n$ is the number density, it follows immediately that $n_{ABC}$ represents the number of particles in the volume. This object is directly linked to the spacetime version; 
\begin{equation}
n_{abc} = n u^d \epsilon_{dabc} \equiv n \epsilon_{abc}
\end{equation}
where $\epsilon_{abc}$ is associated with a right-handed tetrad moving along $u^a$. It then follows immediately that
\begin{equation}
\epsilon_{abc} = \psi^A_a \psi^B_b \psi^C_c \epsilon_{ABC} \ .
\end{equation}
Inspired by this, we may also introduce
\begin{equation}
      g^{A B} = \psi^A_a \psi^B_b g^{a b} = \psi^A_a \psi^B_b \perp^{a b} \ ,
\end{equation}
representing the induced metric on matter space. 

Equipped with these matter space quantities, it is fairly natural to ask; is it possible to express the Lagrangian $\Lambda(n^2)$ in terms of matter space quantities? The answer will soon be relevant, so let us consider it now. 
It is straightforward to show that we may consider $\Lambda$ to be a function of $g^{A B}$ and $n_{ABC}$:
\begin{multline}
n^2 = - g_{ab} n^a n^b =  {1\over 3!} n_{abc} n^{abc} \\
= {1\over 3!}  \left(\psi^A_a g^{ad} \psi^D_d\right) \left(\psi^B_b g^{be} \psi^E_e\right) \left(\psi^C_c g^{cf} \psi^F_f\right) n_{ABC} n_{DEF} \\
= {1\over 3!} g^{AD} g^{BE} g^{CF} n_{ABC} n_{DEF} \ .
\end{multline}
It follows that, if we introduce
\beq
\gamma_{AB} = \left(\sqrt{\det \left(g_{GH}\right)} n\right)^{2/3} g_{AB} \ , 
\eeq
then (using Eq.~\eqref{lcM} from Appendix B)
\begin{equation}
n^2 = \frac{1}{3!} \gamma^{AD} \gamma^{BE} \gamma^{CF} [ABC] [DEF] = \det \left(\gamma_{AB}\right) \ .
\end{equation}
and
\beq
\Lambda (n^2)\quad \Leftrightarrow\quad \Lambda ( \mathrm{det} \left( \gamma_{AB}\right))  \ .
\eeq

Finally, it is worth noting that, alongside the number three-form we may introduce the analogous object for the momentum:
\begin{equation}
\mu^{abc} = \epsilon^{dabc} \mu_d \ , \quad \mu_a = {1\over 3!} \epsilon_{bcda} \mu^{bcd} \ . 
\end{equation}
This then leads to 
\begin{equation}
n\mu = -n^a \mu_a = n_{abc}\mu^{abc} = n_{ABC} \mu^{ABC} \ , 
\end{equation}
where
\begin{equation}
\mu^{ABC} = \psi^A_a\psi^B_b\psi^C_c \mu^{abc} \ .
\end{equation}


\subsection{A step towards field theory}
\label{sec:ftheory}

 The quantities we introduced in the previous section may seem somewhat abstract at this point, but their meaning will (hopefully) become clearer later. As a first exercise in working with them, let us ask what happens if we consider the matter space ``fields'' as the fundamental variables of the theory.

In general, we might take the  Lagrangian to be $\Lambda = \Lambda(X^A, \psi^A_a, g^{ab})$ (as in, for example, \citealt{jeze}). This leads to 
\beq
 \delta \left(\sqrt{- g} \Lambda\right) = \sqrt{- g} \left\{ {\partial \Lambda \over \partial X^A} \delta X^A+  {\partial \Lambda \over \partial \psi^A_a} \delta \psi^A_a + \left[ {\partial \Lambda \over \partial g^{ab} }- {\Lambda \over 2} g_{ab}\right] \delta g^{ab} \right\} \ . 
\eeq
If we introduce the Lagrangian displacement, as before, we already know that
\beq
\Delta X^A = 0 \ , 
\end{equation}
and
\begin{equation}
 \Delta \psi^A_a = 0 \quad \Longrightarrow \quad \delta \psi^A_a = - \xi^c \nabla_c \psi^A_a - \psi^A_c \nabla_a \xi^c = - \nabla_a \left( \xi^c \psi^A_c\right)\ ,
\eeq
where we have used the fact that partial derivatives commute.
It then follows that
\beq
{\partial \Lambda \over \partial X^A} \delta X^A+  {\partial \Lambda \over \partial \psi^A_a} \delta \psi^A_a = -\xi^c \psi^A_c \left[ {\partial \Lambda \over \partial X^A} - \nabla_a \left( {\partial \Lambda \over \partial \psi^A_a} \right) \right] \ ,
\eeq
and we see that the Euler-Lagrange equations are
\beq
\psi^A_c \left[ {\partial \Lambda \over \partial X^A} - \nabla_a \left( {\partial \Lambda \over \partial \psi^A_a} \right) \right]=0 \ .
\label{eulag}\eeq
We also see that the stress-energy tensor is
\beq
T_{ab} = - {2 \over \sqrt{-g}} {\delta \left( \sqrt{-g} \Lambda \right) \over \delta g^{ab} } = \Lambda  g_{ab} - 2{\partial \Lambda \over \partial g^{ab} } \ .
\eeq
It is easy to see that these results lead us back to \eqref{seten}.

In order to compare the Euler--Lagrange equations for the fields to the Euler equations \eqref{euler1f}, we need two intermediate results. First of all, 
\begin{multline}
{\partial \Lambda \over \partial \psi^A_a} = \mu_b {\partial n^b \over \partial \psi^A_a} = {1\over 3!} \mu_b \epsilon^{cdeb} n_{CDE} {\partial \over \partial \psi^A_a} \left(  \psi^C_c \psi^D_d \psi^E_e \right) \\
= {1\over 2} \mu_b \epsilon^{adeb} \psi^D_d \psi^E_e n_{ADE} 
= - {1\over 2} \mu^{ade} \psi^D_d \psi^E_e n_{ADE} \\
= 
-  {1\over 2} \mu^{ade} \delta_A^B \psi^D_d \psi^E_e n_{BDE} = -  {1\over 2} \mu^{ade} \left( \psi_A^b \psi^B_b \right) \psi^D_d \psi^E_e n_{BDE} \\
= 
 -  {1\over 2}  \psi_A^b\mu^{ade} n_{bde}
 =
 \psi^b_A\left[ \delta^a_b \left( \mu_c n^c\right) - \mu_b n^a\right] \ .
\end{multline}
This is true because (i) the metric is held fixed in the partial derivative, and (ii) $n_{ABC}$ depends only on the matter space coordinates $X^A$. We then see that
\beq
 \psi^A_b {\partial \Lambda \over \partial \psi^A_a}  = h_b^c \left[ \delta^a_c \left( \mu_d n^d\right) - \mu_c n^a\right] = - n\mu \perp_b^a \ , 
\eeq
since $n^a=nu^a$,  $\mu_a= \mu u_a$ and $\perp^c_b u_c=0$. Secondly, we need
\beq
\psi^A_c {\partial \Lambda \over \partial X^A} = \nabla_c \Lambda - {\partial \Lambda \over \partial \psi^A_b} \nabla_c \psi^A_b  \ , 
\eeq
Making use of these results, we get
\begin{multline}
\psi^A_b \left[ {\partial \Lambda \over \partial X^A} - \nabla_a \left( {\partial \Lambda \over \partial \psi^A_a} \right) \right] 
= \nabla_a \left[ \delta^a_b \Lambda - \psi^A_b {\partial \Lambda \over \partial \psi^A_a} \right] = 
\nabla_a \left[ \delta^a_b \Lambda + n\mu \perp^a_b\right] \\
= \nabla_a \left[ \delta^a_b (\Lambda -n^c \mu_c) + n^a \mu_b\right] = \nabla_aT^a_{\ b} = 0 \ .
\end{multline}
In essence, the two descriptions are consistent---as they had to be. 

What we have outlined is a field-theory approach to the problem, based on the idea that the matter space variables can be viewed as fields in spacetime \citep{nicol1}. It is, of course, not a truly independent variational approach, and (as we have seen) the equations of motion one obtains need to be massaged into a more intuitive form. However, this does not mean that the argument is without merit. Looking at a problem from different perspectives tends to help understanding. In this particular instance, we may explore the connection between the symmetries of the problem and the matter space variables. By changing the focus from the familiar macroscopic fluid degrees of freedom to three scalar functions $X^A$ it is easy to keep track of the expected Poincar\'e
invariance. First of all, if we expect the system to be homogeneous and isotropic we have to require the fields to be invariant under internal translations and rotations. This means that 
\begin{equation}
X^A \rightarrow X^A + a^A \ , 
\end{equation}
for  constant $a^A$, and
\beq
X^A \rightarrow O^A_{\ B} X^B \ , 
\eeq
where $O^A_{\ B}$ is an SO(3) matrix (associated with rotation). These conditions do not restrict us to fluids, however, as they will also hold for isotropic solids. The final condition we need relates to invariance under volume-preserving diffeomorphisms, leading to
\beq
X^A \to \xi^A(X^B) \ , \ \mbox{with} \ \mathrm{det} {\partial \xi^A \over \partial X^B} = 1 \ .
\eeq
In practice, this corresponds to the dynamics being invariant as the fluid elements move around without expansion or contraction.

What are the implications of these conditions? First of all, we  need each of the $X^A$ fields to be acted on by at least one derivative (although see \citealt{2017CQGra..34l5001A} for a discussion on how this assumption can be relaxed for dissipative systems). This means that the Lagrangian cannot depend on $X^A$ directly (as we assumed). Moreover, taking a field-theory view of the problem (see the discussion of the fluid-gravity correspondence in Sect.~\ref{sec:flugrav}) we may focus on low momenta/low frequencies, for which the most relevant terms are those with the fewest derivatives. In effect, the lowest order Lagrangian will involve exactly
one derivative acting on each $X^A$. The focus then shifts to the map, $\psi^A_a$.  As we expect to work with Lorentz scalars, it would be natural to assume that the Lagrangian must involve the contraction
\beq
g^{AB}= g^{ab} \psi^A_a \psi^B_b \ , 
\eeq
from before (i.e., the induced metric on the matter space). Moreover, we have already seen that the symmetries require us to work with invariant functions of $g^{AB}$ and the volume preserving argument picks out the determinant as the key combination.

The connection with quantum field theory is explored by \cite{nicol1}, with particularly interesting developments relating to symmetry breaking and the emergence of superfluidity \citep{nicol0,nicol4} and extensions to incorporate quantum anomalies\footnote{The idea is that the fluid dynamics is modified in the presence of an external (gauge) field, leading to the current no longer being conserved.} in the  field theory \citep{nicol2}. And example of the latter is the Wess--Zumino anomaly, which leads to terms that remain only after integration by parts. In effect, the action is invariant, but the Lagrangian is not. Somewhat simplistically, one may associate such terms with the surface terms we neglected in the variational argument. There has also been some effort to extend the approach to dissipative systems \citep{nicol3}.


\section{Newtonian limit and Lagrangian perturbations}

\subsection{The Newtonian limit}
\label{sec:newton}

Having written down the equations that govern a single (barotropic) relativistic fluid, it is natural to consider the 
connection between the final expressions and standard Newtonian fluid dynamics. In order to make this connection, we need to establish how 
one arrives at the Newtonian limit of the relativistic equations. It is useful to work this out because---even though the framework 
we are developing is intended to describe relativistic systems---modelling often draws on intuition gained from good old Newtonian physics.
This is especially the case when one considers ``new'' applications. Useful qualitative understanding can often be 
obtained from a Newtonian analysis, but we need  relativistic models for precision and in order to explore unique aspects, like rotational frame-dragging and gravitational radiation.

There has been much progress on the analysis of Newtonian 
multifluid systems.  \cite{prix04:_multi_fluid} has developed 
an action-based formalism, analogous to the model we consider here (based on the notion of time-shifts, closely related to the Lagrangian variations in spacetime). \cite{Carter03:_newtI,Carter03:_newtII,Carter04:_newtIII}  have 
done the same, except that they use a fully spacetime covariant formalism (taking the work of Milne and Cartan as starting points), taking full account of the fact that the Newtonian limit is singular. Our aim here is less ambitious. 
We simply want to demonstrate how the Newtonian fluid equations can be extracted 
as the non-relativistic limit of the  relativistic model. 

We take as the starting point the leading order line element in the weak-field limit; 
\begin{equation} 
    {d} s^2 = - c^2 d\tau^2 = - c^2 \left(1 + \frac{2 \Phi}{c^2}\right)  {d} t^2 + 
           \eta_{i j} { d} x^i  { d} x^j \ , 
\label{wfmet}\end{equation}
where $x^i\ (i=1-3)$ are Cartesian coordinates, $\eta_{i j}$ is the flat three-dimensional metric and $\Phi$ is the gravitational potential.
The Newtonian limit then follows by writing the  
equations to leading order  in an expansion in powers of  the speed of light $c$. Formally, the Newtonian 
results are obtained in the limit where $c \to \infty$. 

Let us apply this strategy to the equations of fluid dynamics.
With $\tau$  the proper time  measured along a fluid element's worldline, 
 the curve it traces out can be written 
\begin{equation} 
    x^{a}(\tau) = \{c t(\tau),x^i(\tau)\} \ . \label{xatau}
\end{equation} 
In order to work out the  four-velocity,  
\begin{equation} 
    u^{a} = \frac{{ d} x^a}{{ d} \tau} \ , \label{uatau}
\end{equation} 
we note that  \eqref{wfmet} leads to
\begin{equation} 
  { d} \tau^2 =  \left(1 + \frac{2 \Phi}{c^2} 
                  - \frac{\eta_{ij} v^i v^j}{c^2}\right) {d} t^2 \ , 
\end{equation} 
with $v^i = { d}x^i/{d}t$ the Newtonian three-velocity of the 
fluid. Since the velocity is assumed to  be small, in the sense that 
\begin{equation} 
     {\left|v^i\right| \over c} \ll 1 \ , 
\end{equation} 
this leads to 
\begin{equation}
{dt\over d\tau} \approx 1- {\Phi\over c^2} + {v^2 \over 2 c^2} \ , 
\end{equation}
where $v^2 = \eta_{i j} v^i v^j$, 
and 
\begin{equation}
u^0 = {dx^0 \over d\tau} = c {dt \over d\tau} \approx c\left( 1- {\Phi\over c^2} + {v^2 \over 2 c^2} \right) \ .
\end{equation}

It is also easy to see that 
\begin{equation}
u^i = {d x^i \over d\tau} = v^i {dt \over d\tau} \approx v^i  \ .
\end{equation}

In order to obtain the covariant components, we use the metric (which is manifestly diagonal). Thus, we find that 
\begin{equation}
u_0 = g_{00}u^0 = - c \left(1 + \frac{2 \Phi}{c^2}\right) \left( 1- {\Phi\over c^2} + {v^2 \over 2 c^2} \right) \approx -c \left( 1+ {\Phi\over c^2} + {v^2 \over 2 c^2} \right) \ , 
\end{equation}
and
\begin{equation}
u_i = v_i \ .
\end{equation}
Note that these relations lead to
\begin{equation}
u^a u_a = -c^2 \left( 1- {\Phi\over c^2} + {v^2 \over 2 c^2} \right)  \left( 1+ {\Phi\over c^2} + {v^2 \over  2 c^2}\right) + v^2 \approx
- c^2  \ ,
\end{equation}
as expected.

We can now work out the Newtonian limit for the conserved particle flux
\begin{multline}
\nabla_a ( n u^a)  = 0 
\quad \Longrightarrow
\quad
{1\over c} \partial_t \left( n u^0 \right) + \nabla_i \left( nv^i  \right) = 0 \\
\Longrightarrow \quad 
\partial_t n  + \nabla_i \left( nv^i \right) =  \mathcal O\left(c^{-1}\right)
\end{multline}
To leading order we retain the expected result
\begin{equation}
\partial_t n  + \nabla_i \left( nv^i  \right) = 0 \ ,
\end{equation}
recovering the usual continuity equation by introducing the mass density $\rho = mn$, with $m$ the mass per particle.

 In order to work out the corresponding limit of the Euler equations, we need the curvature contributions to the covariant derivative. However, from the definition \eqref{gabc} and the weak-field metric, we see that only $g_{00}$ gives a non-vanishing contribution. Moreover, it is clear that 
\begin{equation}
\Gamma^a_{b c} = \mathcal O(1/c^2) \ , 
\end{equation}
which is why we did not need to worry about this in the case of the flux conservation. The curvature contributes at higher orders.

Explicitly, we have
\begin{equation}
u^a \nabla_a u^b = u^a \partial_a u^b + \Gamma^b_{ca} u^a u^c \\
= {1\over c} u^0 \partial_t u^b + u^i \partial_i u^b + \Gamma^b_{ca} u^a u^c \ .
\end{equation}
We only need the spatial components, so we set $b=j$ to get
\begin{multline}
u^a \nabla_a u^j = {1\over c} u^0 \partial_t u^j + u^i \partial_i u^j + \Gamma^j_{ca} u^a u^c \\
=   \partial_t v^j + v^i \partial_i v^j + c^2 \Gamma^j_{00}  + \mbox{higher order terms}  \\
= \partial_t v^j + v^i \partial_i v^j + {1\over 2} \eta^{jk} \partial_k \left( {2\Phi \over c^2} \right) \\
= 
\partial_t v^j + v^i \partial_i v^j +  \eta^{jk} \partial_k \Phi \ .
\end{multline}

Finally, we need the pressure contribution. For this we note that the projection becomes
\begin{equation}
\perp^{ab} = g^{ab} + {1\over c^2} u^a u^b \ , 
\end{equation}
in order to be dimensionally consistent. We also need $ \varepsilon \gg p $. This means that we have
\begin{equation}
\perp^{ba} \nabla_a p \quad \Longrightarrow \quad  \eta^{jk} \partial_k p  \ , 
\end{equation}
and we (finally) arrive at the Euler equations 
\begin{equation}
\partial_t v^j + v^i \partial_i v^j = -   \eta^{jk} \left(  { 1\over \rho} \partial_k p +  \partial_k \Phi \right) \ , 
\label{eulereq1}
\end{equation}
which represent momentum conservation.

\subsection{Local dynamics}
\label{ldyn}

In principle, the fluid equations (from  Sect.~\ref{shelve} or above) completely specify the problem for a 
single-component barotropic
flow (once an equation of state has been provided, of course). In general, the problem is 
nonlinear and difficult to solve analytically.  Once we couple the fluid motion to the 
dynamic spacetime of the Einstein equations, it becomes exceedingly so.  However, if we want to 
understand the behaviour of a given system we can make progress using linearized theory. 
This approach would be suitable whenever the dynamics only deviates slightly from a known 
background/equilibrium state.  The deviations should be small enough that we can neglect 
nonlinearities. This is a very common strategy, for example, to study the oscillations of neutron 
stars. Moreover, it is a good strategy if we want to explore the local dynamics of a given system.

Consider the case where the length and time scales of the deviations are such that the 
spacetime curvature can be ignored; then, we can work in the local inertial frame associated with 
the flow---i.e.~use Minkowski coordinates $x^a = [t,x^i]$ and assume that the spacetime 
curvature is flat. Letting $\tau$ be the proper time associated with a given fluid worldline,  we 
see from Eqs.~\eqref{xatau} and \eqref{uatau} and the normalization of the four-velocity 
$u^a$ (i.e.~$u^a u_a = -1$) that---in the local inertial frame---the particle flux density takes the form 
\beq
     n^a = n u^a = n \left(1 - v^2\right)^{- 1/2} [1,v^i] \ , \label{locna}
\eeq
where $v^i = dx^i/dt$ is the local three-velocity and $v^2 = \eta_{i j} v^i v^j$.  In the linearized case, 
the three-velocity $v^i$ is small and therefore a deviation. The background four-velocity is thus 
uniform, taking the form $u^a = [1,0,0,0]$, and it is obviously the case that $\nabla_b u^a = 0$.  
As long as the associated scales of the deviations are sufficiently small, we should be able to 
take the background particle number density $n$ to be uniform both temporally and spatially so 
that $\nabla_a n = 0$. Therefore, it is easy to see that the background/equilibrium state trivially 
satisfies the dynamical equations.  

Now consider (Eulerian) variations, such that $n \to n + \delta n$ and $v^i \to \delta v^i$ and let the 
deviations  be expressed as plane waves (making use of a Fourier decomposition). The 
normalization of the four-velocity $u^a$ demands that the perturbed velocity is spatial 
($u^a \delta u_a = 0$), which is consistent with the linearization of Eq.~\eqref{locna}:
\be
      \delta n^a = [\delta n ,n \delta v^i] \ .
\ee  
A standard sound speed derivation, however, takes the point of view that the energy density and 
four-velocity are the fundamental variables.  For now, we  adopt this approach in order to make 
contact with the well-known results.

From Eq.~\eqref{mudef1f}, we see a perturbation in $n$ leads to a perturbation in $\rho$ (recall $\varepsilon\approx\rho = mn$ in the weak-field limit); 
namely,
\be
      \delta \rho = \mu \delta n \ .
\ee
Likewise, Eq.~\eqref{funrel1} shows that there are corresponding perturbations in the pressure 
{\em and} chemical potential. With that in mind, we linearize  Eqs.~\eqref{fstlaw1f} and 
\eqref{euler}, and find that the perturbation problem becomes
\be
\partial_t \delta \rho + \left(p + \rho\right) \nabla_i \delta v^i = 0 \ , 
\label{p1}
\ee
and
\be
\left(p + \rho\right) \partial_t \delta v_i + \nabla_i \delta p = 0 \ .
\label{p2}
\ee
To close the system, we introduce a barotropic equation of state:
\be
p = p(\rho) \quad \longrightarrow \quad \delta p = \left( {dp \over d\rho} \right) \delta \rho \equiv 
C_s^2 \delta \rho \ . \label{barotrope}
\ee

The plane-wave Ansatz means that we have
\be
\delta p = A_{p} e^{i k (- \sigma t + \hat{k}_j x^j)}
\ee
\be
\delta \rho = A_{\rho} e^{i k (- \sigma t + \hat{k}_j x^j)}
\ee
and
\be
\delta v^i = A^i_v e^{i k (- \sigma t + \hat{k}_j x^j)} \ .
\label{ppwave}
\ee
In these expressions, the constant $\sigma$ is the wave-speed, the constant $k_i$ is the 
(spatial)  wave-vector, such that $k^2 = k_i k^i$ ($k^i = g^{i j} k_j$) and $\hat{k}_i = k_i/k$.  We see from 
Eq.~\eqref{barotrope} that the pressure amplitude $A_p$ must satisfy (assuming that the perturbations are described by the same equation of state as the background)
\be
 A_p = C_s^2 A_{\rho} \ .
\ee
Inserting the plane-wave decompositions for $\delta \rho$ and $\delta v^i$ into \eqref{p1} and 
\eqref{p2} we find 
\be
 \sigma A_{\rho} + (p + \rho) \hat{k}_i A^i_v = 0 
\ee
and
\be
(p + \rho) \sigma A^i_v + C^2_s A _{\rho} \hat{k}^i = 0 \ .
\ee

It is easy to see that we cannot have non-trivial transverse waves; i.e., if $\hat{k} _iA^i_v = 0$ 
then we must have $A_{\rho} = 0$ as well.  Focussing on the longitudinal case, we can 
contract the second equation with $\hat{k}_i$ to obtain a scalar equation. Making use of this 
equation, we obtain the dispersion relation
\be
\sigma^2 - C_s^2 = 0 \quad \Longrightarrow \quad \sigma = \pm C_s \ .
\ee
In this simple situation it is obvious that we should identify $C_s$ as the speed of sound.

It is worth noting that we can go back to the case where the particle flux $n^a$ is taken to be 
fundamental and the equation of state has the form $\rho = \rho(n)$. If we do that, then we have
\be
d\rho = \mu dn \qquad \mbox{and} \qquad dp = n d\mu
\ee
and it follows that the speed of sound is given by
\be
C_s^2 = {dp \over d \rho} = {n \over \mu} {d \mu \over dn} \ . \label{csdmudn}
\ee

\subsection{Newtonian fluid perturbations}
\label{sec:newtper}

Studies of the stability properties of rotating self-gravitating
bodies are of obvious relevance to astrophysics. By improving our
understanding of the relevant issues we can hope to shed light on
the nature of the various dynamical and secular instabilities that may
govern the spin-evolution of rotating stars. The relevance of such
knowledge for neutron star astrophysics may be highly significant, especially
since instabilities may lead to detectable gravitational-wave signals.
In this section we will outline the Lagrangian perturbation framework developed
by \cite{friedman78:_lagran, friedman78:_secul_instab} for
rotating non-relativistic stars, leading to criteria that can be
used to decide when the oscillations of a rotating neutron star are
unstable. We also provide an explicit example proving the instability of
the so-called r-modes at all rotation rates in a perfect fluid star.

Following \cite{friedman78:_lagran, friedman78:_secul_instab}, we work with
Lagrangian variations. We have already seen that the Lagrangian perturbation
$\Delta Q$ of a quantity $Q$ is related to the Eulerian variation $\delta Q$
by
\begin{equation}
  \Delta Q = \delta Q + \mathcal{L}_\xi Q,
\end{equation}
where (as before) $\mathcal{L}_\xi$ is the Lie derivative (introduced
in Sect.~\ref{sec:gr}). The Lagrangian change in the fluid velocity now follows
from the Newtonian limit of Eq.~(\ref{vel_perbs}):
\begin{equation}
  \Delta v^i = \partial_t \xi^i,
\end{equation}
where $\xi^i$ is the Lagrangian displacement. Given this, and
\begin{equation}
  \Delta g_{ij} = \nabla_i \xi_j + \nabla_j \xi_i,
\end{equation}
where $g_{ij}$ is the flat three-dimensional metric, we have
\begin{equation}
  \Delta v_i = \partial_t \xi_i + v^j\nabla_i \xi_j + v^j \nabla_j \xi_i.
\end{equation}

Let us consider the simplest case, namely a barotropic ordinary fluid for
which $\varepsilon=\varepsilon(n)$. Then we want to perturb the continuity
and Euler equations. The
conservation of mass for
the perturbations follows immediately from the Newtonian limits of
Eqs.~(\ref{dens_perb}) and~(\ref{lagradelu}) (which as we recall
automatically satisfy the continuity equation):
\begin{equation}
  \Delta n = - n \nabla_i \xi^i,
  \qquad
  \delta n = - \nabla_i (n \xi^i).
\end{equation}
Consequently, the perturbed gravitational potential follows from
\begin{equation}
  \nabla^2 \delta \Phi = 4\pi G \delta \rho = 4 \pi G m \, \delta n = - 4\pi G m \nabla_i(n \xi^i).
\end{equation}
In order to perturb the Euler equations we first rewrite
Eq.~(\ref{eulereq1}) as
\begin{equation}
  (\partial_t +\mathcal{L}_v) v_i + \nabla_i
  \left( \tilde{\mu} + \Phi - \frac{1}{2}  v^2 \right) = 0,
  \label{euleq2}
\end{equation}
where $\tilde{\mu}= \mu/m$. This form is particularly useful since the
Lagrangian variation commutes with the operator $\partial_t + \mathcal{L}_v$.
Perturbing Eq.~(\ref{euleq2}) we thus have
\begin{equation}
  (\partial_t +\mathcal{L}_v) \Delta v_i + \nabla_i
  \left(\Delta \tilde{\mu} + \Delta \Phi - \frac{1}{2} \Delta( v^2) \right) = 0.
  \label{peuls}
\end{equation}

We want to rewrite this equation in terms of the displacement vector $\xi$.
After some algebra we arrive at
\begin{multline}
  \partial_t^2 \xi_i + 2 v^j \nabla_j \partial_t \xi_i +
  (v^j \nabla_j)^2 \xi_i + \nabla_i \delta \Phi +
  \xi^j \nabla_i \nabla_j \Phi \\
  - (\nabla_i \xi^j) \nabla_j \tilde{\mu} +
  \nabla_i \Delta \tilde{\mu} = 0.
  \label{peul2}
\end{multline}
Finally, we need
\begin{equation}
  \Delta \tilde{\mu} = \delta \tilde{\mu} + \xi^i\nabla_i \tilde{\mu} =
  \left( \frac{\partial \tilde{\mu}}{\partial n} \right) \delta n +
  \xi^i\nabla_i \tilde{\mu} =
  - \left( \frac{\partial \tilde{\mu}}{\partial n} \right) \nabla_i
  (n \xi^i) + \xi^i\nabla_i \tilde{\mu}.
\end{equation}
Given this, we have arrived at the following form for the perturbed
Euler equation:
\begin{multline}
  \partial_t^2 \xi_i + 2 v^j \nabla_j \partial_t \xi_i +
  (v^j \nabla_j)^2 \xi_i + \nabla_i \delta \Phi +
  \xi^j \nabla_i \nabla_j \left( \Phi + \tilde{\mu} \right) \\
  -
  \nabla_i \left[ \left( \frac{\partial \tilde{\mu}}{\partial n} \right)
  \nabla_j (n \xi^j) \right] = 0.
  \label{peul3}
\end{multline}
This equation should be compared to Eq.~(15) of \cite{friedman78:_lagran}.

\subsection{The CFS instability}
\label{sec:cfs}

Having derived the perturbed Euler equations, we are interested in
constructing conserved quantities that can be used to assess the stability
of the system. To do this, we first multiply Eq.~(\ref{peul3}) by the number
density $n$, and then write the result (schematically) as
\begin{equation}
  A \partial_t^2 \xi + B \partial_t \xi + C \xi  = 0,
\end{equation}
omitting the indices since there is little risk of confusion. Defining the
inner product
\begin{equation}
  \left< \eta^i,\xi_i \right> = \int \eta^{i*} \xi_i \, \mathrm{d} V,
\end{equation}
where $\eta$ and $\xi$ both solve the perturbed Euler equation, and
the asterisk denotes complex conjugation (and we integrate over the volume of the body), one can now show that
\begin{equation}
  \left< \eta, A\xi \right> =
  \left< \xi,A\eta \right>^*
  \qquad \mathrm{and}
  \qquad
  \left< \eta,B\xi \right> =
  - \left< \xi,B\eta \right>^*.
\end{equation}
The latter requires the background relation $\nabla_i (n v^i) = 0$, and holds
as long as $n \to 0$ at the surface of the star. A slightly more
involved calculation leads to
\begin{equation}
  \left< \eta, C\xi \right> = \left< \xi, C\eta \right>^*.
\end{equation}
Inspired by the fact that the momentum conjugate to $\xi^i$ is
$\rho(\partial_t + v^j \nabla_j)\xi_i$, we now consider the symplectic
structure
\begin{equation}
  W(\eta,\xi) =
  \left<\eta, A\partial_t \xi + \frac{1}{2} B \xi\right> -
  \left< A\partial_t \eta + \frac{1}{2} B \eta, \xi\right>.
  \label{Wdef}
\end{equation}
It is straightforward to show that $W(\eta,\xi)$ is conserved,
i.e., $\partial_t W = 0$. This leads us to define the \emph{canonical energy}
of the system as
\begin{equation}
  E_\mathrm{c} = \frac{m}{2} W (\partial_t \xi,\xi) =
  \frac{m}{2} \left\{ \left< \partial_t \xi, A \partial_t \xi \right> +
  \left< \xi, C \xi \right> \right\}.
\end{equation}
After some manipulations, we arrive at the explicit
expression:
\begin{multline}
  E_\mathrm{c} = \frac{1}{2} \int \left\{ \rho |\partial_t \xi|^2 -
  \rho | v^j \nabla_j \xi_i|^2 + \rho \xi^i \xi^{j*}\nabla_i \nabla_j
  (\tilde{\mu} + \Phi) \right. \\
 \left.  + \left( \frac{\partial \mu}{\partial n} \right)
  |\delta n|^2 - \frac{1}{4 \pi G} |\nabla_i \delta \Phi|^2 \right\}
  \mathrm{d} V \ , 
\end{multline}
which can be compared to Eq.~(45) of \cite{friedman78:_lagran}. In the
case of an axisymmetric system, e.g., a rotating star, we can also define a
\emph{canonical angular momentum} as
\begin{equation}
  J_\mathrm{c} = - \frac{m}{2} W (\partial_\varphi \xi, \xi) =
  - \mathrm{Re} \left< \partial_\varphi \xi, A\partial_t \xi +
  \frac{1}{2} B\xi \right>.
\end{equation}
The proof that this quantity is conserved relies on the fact that (i)
$W(\eta, \xi)$ is conserved for any two solutions to the perturbed Euler
equations, and (ii) $\partial_\varphi$ commutes with $\rho v^j \nabla_j$
in axisymmetry, which means that if $\xi$ solves the Euler equations then
so does $\partial_\varphi \xi$.

As discussed in \cite{friedman78:_lagran, friedman78:_secul_instab}, the
stability analysis is complicated by the presence of so-called ``trivial''
displacements. These trivials can be thought of as representing a
relabeling of the physical fluid elements. A trivial displacement $\zeta^i$
leaves the physical quantities unchanged, i.e., is such that $\delta n =
\delta v^i = 0$. This means that we must have
\begin{eqnarray}
  \nabla_i (\rho \zeta^i) &=& 0,
  \\
  \left( \partial_t + \mathcal{L}_v \right) \zeta^i &=& 0.
\end{eqnarray}%
The solution to the first of these equations can be written 
\begin{equation}
  \rho \zeta^i = \epsilon^{ijk} \nabla_j \chi_k \ , 
\end{equation}
where, in order to satisfy the second equations, the vector $\chi_k$ must
have time-dependence such that
\begin{equation}
  ( \partial_t + \mathcal{L}_v) \chi_k = 0.
\end{equation}
This means that the trivial displacement will remain constant along the
background fluid trajectories. Or, as 
\cite{friedman78:_lagran} put it, the ``initial relabeling is
carried along with the unperturbed motion''.

The trivials  cause trouble because they affect the
canonical energy. Before one can use the canonical energy to assess the
stability of a rotating configuration one must deal with this ``gauge
problem''. To do this one should ensure that the displacement vector
$\xi$ is orthogonal to all trivials. A prescription for this is provided
by \cite{friedman78:_lagran}. In particular, they
show that the required canonical perturbations preserve the vorticity of
the individual fluid elements. Most importantly, one can also prove that
a normal mode solution is orthogonal to the trivials. Thus,  mode
solutions can serve as canonical initial data, and be used to assess
stability.

The importance of the canonical energy stems from the fact that
it can be used to test the stability of the system. In particular:
\begin{itemize}
\item[-] Dynamical instabilities are only possible for motions such that
  $E_\mathrm{c}=0$. This makes intuitive sense since the amplitude of
  a mode for which $E_\mathrm{c}$ vanishes can grow without bound and
  still obey the conservation laws.
\item[-] If the system is coupled to radiation (e.g., gravitational
  waves) which carries positive energy away from the system (which
  should be taken to mean that $\partial_t E_\mathrm{c} < 0$) then any
  initial data for which $E_\mathrm{c}<0$ will lead to an unstable
  evolution.
\end{itemize}

Consider a real frequency normal-mode solution to the perturbation equations,
a solution of form $\xi = \hat{\xi} e^{i(\omega t+m\varphi)}$. One can
readily show that the associated canonical energy becomes
\begin{equation}
  E_\mathrm{c} = \omega \left[ \omega \left<{\xi}, A {\xi}\right> -
  \frac{i}{2} \left<{\xi}, B{\xi}\right> \right],
  \label{Ec}
\end{equation}
where the expression in the bracket is real. Similarly, for the canonical angular
momentum, we get
\begin{equation}
  J_\mathrm{c} = -m \left[ \omega \left<{\xi}, A {\xi} \right> -
  \frac{i}{2} \left< {\xi}, B{\xi} \right> \right].
  \label{Jc}
\end{equation}
Combining Eq.~(\ref{Ec}) and Eq.~(\ref{Jc}) we see that, for
real frequency modes, we  have
\begin{equation}
  E_\mathrm{c} = - \frac{\omega}{m} J_\mathrm{c} = \sigma_\mathrm{p} J_\mathrm{c},
  \label{EJrel}
\end{equation}
where $\sigma_\mathrm{p}$ is the pattern speed of the mode.

Now note that Eq.~(\ref{Jc}) can be rewritten as
\begin{equation}
  \frac{J_\mathrm{c}}{\left< \hat{\xi}, \rho\hat{\xi} \right>} =
  - m\omega + m \frac{\left< {\xi}, i \rho v^j \nabla_j {\xi} \right>}
  {\left< \hat{\xi}, \rho\hat{\xi} \right>}.
\end{equation}
Using cylindrical coordinates, and $v^j = \Omega \varphi^j $, one can
show that
\begin{equation}
  - i \rho {{\xi}}_i^* v^j \nabla_j {\xi}^i =
  \rho \Omega \left[ m \left| \hat{\xi} \right|^2 +
  i ({\hat{\xi}}^* \times \hat{\xi})_z \right].
\end{equation}
But
\begin{equation}
  \left| ({\hat{\xi}}^* \times \hat{\xi})_z \right| \le
  \left| \hat{\xi} \right|^2
\end{equation}
and hence we must have (for uniform rotation)
\begin{equation}
  \sigma_\mathrm{p} - \Omega \left( 1 + \frac{1}{m} \right) \le
  \frac{J_\mathrm{c}/m^2}{\left< \hat{\xi}, \rho\hat{\xi} \right>} \le
  \sigma_\mathrm{p} - \Omega \left( 1 - \frac{1}{m} \right).
  \label{ineq1}
\end{equation}

Eq.~(\ref{ineq1}) forms a key part of the proof that rotating perfect fluid
stars are generically unstable in the presence of
radiation \citep{friedman78:_secul_instab}. The argument goes as
follows: Consider
modes with finite frequency in the $\Omega \to 0$ limit. Then
Eq.~(\ref{ineq1}) implies that co-rotating modes (with $\sigma_\mathrm{p}>0$) must
have $J_\mathrm{c}>0$, while counter-rotating modes (for which $\sigma_\mathrm{p} < 0$) will
have $J_\mathrm{c}<0$. In both cases $E_\mathrm{c}>0$, which means that both classes of
modes are stable. Now consider a small region near a point where
$\sigma_\mathrm{p}=0$ (at a finite rotation rate). Typically, this corresponds to
a point where the initially counter-rotating mode becomes co-rotating. In
this region $J_\mathrm{c}<0$. However, $E_\mathrm{c}$ will change sign at the point where
$\sigma_\mathrm{p}$ (or, equivalently, the frequency $\omega$) vanishes. Since the
mode was stable in the non-rotating limit this change of sign indicates the
onset of instability at a critical rate of rotation. The situation for the fundamental f-mode of a rotating star is illustrated in figure~\ref{fmode}.

\begin{figure}[htb]
    \centerline{\includegraphics[width=0.6\textwidth]{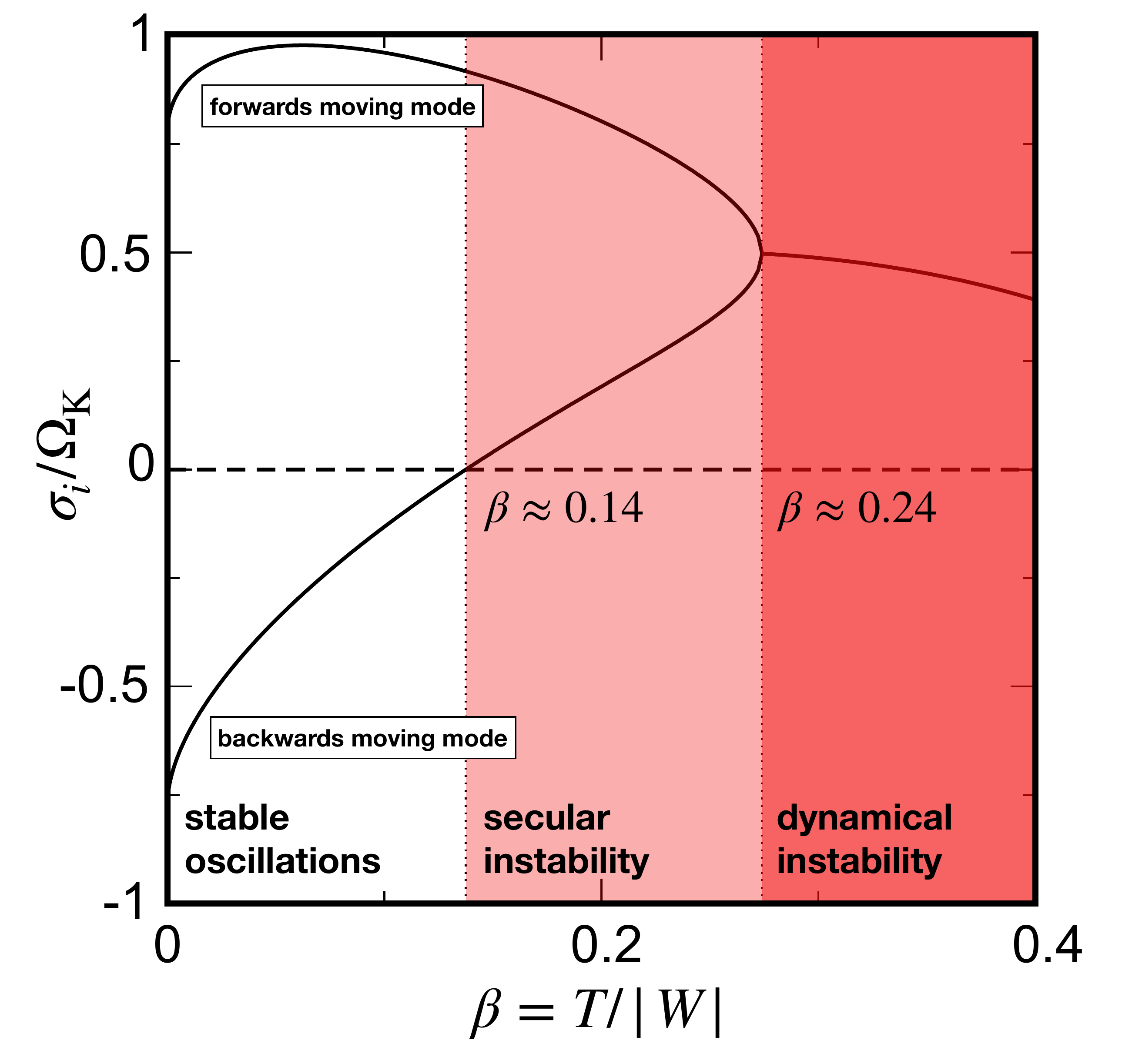}}
    \caption{An illustration of the instabilities affecting the fundamental f-mode of a rotating neutron star. The horizontal axis represents the rotation, expressed in terms of the ratio between the kinetic energy and the gravitational potential energy ($\beta = T/|W|$). The angular velocity is not a (particularly) useful parameter as values beyond (something like) $\beta\approx 0.11$ requires some degree of differential rotation. That is, rigidly rotating bodies never reach the dynamically unstable regime (at least not in Newtonian gravity). The vertical axis gives the pattern speed of the mode, with waves that appear to move forwards (according to a distance observer) having positive values, while backwards moving modes lead to negative values. The originally backwards moving f-mode becomes secularly unstable at $\beta\approx 0.14$, at the point where the mode first appears to move forwards (because of the rotation of star). The mode becomes dynamically unstable (this is the so-called bar-mode instability) when the two modes merge at  $\beta\approx 0.24$. (Adapted from \citealt{narev}). }
    \label{fmode}
\end{figure}

In order to further demonstrate the usefulness of the canonical energy,
let us prove the instability of the inertial r-modes (these are oscillation modes that owe their existence to the rotation of the star, and which are predominantly associated with the Coriolis force). For a general inertial
mode we have (cf.\ \citealt{lockitch99:_r-modes} for a discussion of the
single fluid problem using notation which closely resembles the one we adopt
here)
\begin{equation}
  v^i \sim \delta v^i \sim \dot{\xi}^i \sim \Omega
  \qquad \mathrm{and} \qquad
  \delta \Phi \sim \delta n \sim \Omega^2.
\end{equation}
In particular, modes like the r-modes are dominated by convective currents, so we have $\delta v_r
\sim \Omega^2$ and the continuity equation leads to
\begin{equation}
  \nabla_i \delta v^i \sim \Omega^3
  \qquad \Longrightarrow \qquad
  \nabla_i \xi^i \sim \Omega^2.
\end{equation}

Under these assumptions we find that $E_\mathrm{c}$ becomes (to order
$\Omega^2$)
\begin{equation}
  E_\mathrm{c} \approx \frac{1}{2} \int \rho \left[
  \left|\partial_t {\xi} \right|^2 - \left| v^i \nabla_i{\xi} \right|^2 +
  \xi^{i*} \xi^{j} \nabla_i \nabla_j
  \left( \Phi + \tilde{\mu} \right) \right] \mathrm{d} V.
  \label{ec1}
\end{equation}
We can rewrite the last term using the equation governing the axisymmetric
equilibrium. Keeping only terms of order $\Omega^2$ we have
\begin{equation}
  \xi^{i*} \xi^{j} \nabla_i\nabla_j
  \left( \Phi + \tilde{\mu} \right) \approx
  \frac{1}{2} \Omega^2  \xi^{i*} \xi^{j} \nabla_i \nabla_j (r^2 \sin^2 \theta).
\end{equation}
A bit more work then leads to
\begin{equation}
  \frac{1}{2} \Omega^2  \xi^{i*} \xi^{j} \nabla_i \nabla_j
  (r^2 \sin^2 \theta) = \Omega^2 r^2
  \left[ \cos^2 \theta \left|\xi^\theta \right|^2 +
  \sin^2\theta \left| \xi^\varphi \right|^2  \right] \ , 
\end{equation}
and
\begin{multline}
  \left| v^i \nabla_i \xi_j \right|^2 =
  \Omega^2 \left\{ m^2 \left| \xi \right|^2 -
  2imr^2 \sin \theta \cos \theta
  \left[ \xi^\theta \xi^{\varphi *} - \xi^\varphi \xi^{\theta *} \right] \right. \\
  + \left.
  r^2 \left[ \cos^2 \theta \left|\xi^\theta\right|^2 +
  \sin^2\theta \left| \xi^\varphi\right|^2 \right] \right\},
\end{multline}
which means that the canonical energy can be written in the form
\begin{multline}
  E_\mathrm{c} \approx - \frac{1}{2} \int \rho
  \left\{ (m \Omega - \omega)(m \Omega + \omega) |\xi|^2 \right. \\
  \left. -
  2 i m \Omega^2 r^2 \sin \theta \cos \theta
  \left[ \xi^\theta \xi^{\varphi *} - \xi^\varphi \xi^{\theta *}
  \right] \right\} \mathrm{d} V \ ,
\end{multline}
for an axial-led mode.

Introducing the axial stream function $U$ we have
\begin{eqnarray}
  \xi^\theta &=&
  - \frac{iU}{r^2 \sin \theta} \partial_\varphi Y_l^m e^{i \omega t},
  \\
  \xi^\varphi &=&
  \frac{iU}{r^2 \sin \theta} \partial_\theta Y_l^m e^{i\omega t},
\end{eqnarray}%
where $Y_l^m=Y_l^m(\theta,\varphi)$ are the spherical harmonics.
This now leads to
\begin{equation}
  |\xi|^2 = \frac{|U|^2}{r^2} \left[ \frac{1}{\sin^2 \theta}
  |\partial_\varphi Y_l^m|^2 + |\partial_\theta Y_l^m|^2 \right] \ , 
\end{equation}
and
\begin{multline}
  ir^2 \sin \theta \cos \theta
  \left[ \xi^\theta \xi^{\varphi *} - \xi^\varphi \xi^{\theta *} \right] \\
  =
  \frac{1}{r^2} \frac{ \cos \theta}{\sin \theta} m |U|^2
  \left[ Y_l^m \partial_\theta Y_l^{m*} +  Y_l^{m *} \partial_\theta Y_l^{m}\right].
\end{multline}

After performing the angular integrals, we find that
\begin{equation}
  E_\mathrm{c} = - \frac{ l(l+1) }{2}
  \left\{ (m \Omega - \omega)(m \Omega + \omega) -
  \frac{2 m^2 \Omega^2}{l(l+1)} \right\} \int \rho |U|^2 \, \mathrm{d} r.
\end{equation}
Combining this with the r-mode frequency \citep{lockitch99:_r-modes}
\begin{equation}
  \omega = m \Omega \left[ 1 - \frac{2}{l(l+1)} \right] \ , 
\end{equation}
we see that $E_\mathrm{c} < 0$ for all $l>1$ r-modes, i.e., they are all unstable.
The $l=m=1$ r-mode is a special case, as it leads to $E_\mathrm{c}=0$.

\subsection{The relativistic problem}

The theoretical framework for studying stellar stability in General Relativity
was mainly developed during the 1970s, with key contributions from
\cite{cf72a, cf72b} and \cite{bfs72a, bfs72b}.
Their work extends the Newtonian analysis discussed above. There are
basically two reasons why a relativistic analysis is more complicated than the
Newtonian one. First of all, the problem is algebraically more complex because
one must solve the Einstein field equations in addition to the fluid equations
of motion.

This is apparent from the perturbation relations we have written down already. 
For any given equation of state---represented by $\Lambda(n)$---we can express the perturbed equations of 
motion in terms of the displacement vector $\xi^a$ and the Eulerian variation of the metric, 
$\delta g_{ab}$. In doing this it is worth noting that the usual approach to relativistic stellar 
perturbations is to work with this combination of variables (see, e.g., 
\citealt{1992PhRvD..46.4289K}). Essentially, we need the Eulerian perturbation of the Einstein field 
equations and the Lagrangian variation of the momentum equation \eqref{force}. The description 
of the perturbed Einstein equations is standard (see, e.g., \citealt{nabook}), so we focus on the fluid aspects here.

The perturbations of \eqref{euler1f} are easy to work out once we note that the Lagrangian 
variation commutes with the exterior derivative.  We immediately get
\beq
(\Delta n^a) \nabla_{[a}\mu_{b]} + n^a \nabla_{[a}\Delta \mu_{b]} = 0 \ .
\label{pmom1}
\eeq
This simplifies further if we use \eqref{dna0} and assume that the background is such that 
\eqref{euler1f} is satisfied. The first term then vanishes, and we are left with
\beq
 n^a \nabla_{[a}\Delta \mu_{b]} = 0 \ .
\label{euler1}
\eeq
To complete this expression, we need to work out $\Delta \mu_a$. This is a straightforward task 
given the above results, and we find
\beq
\Delta \mu_a = \left( \mathcal{B} + n {d \mathcal{B} \over dn} \right) g_{ab} \Delta n^b + \left( \mu^b \delta_a^d - {d \mathcal{B} \over d n^2} n_a n^b n^d \right) \Delta g_{b d} \ .
\eeq

An additional complication is associated with the fact that one must account for gravitational waves, leading to the system being dissipative. The work culminated in a
series of papers \citep{fs0, friedman78:_lagran, friedman78:_secul_instab, jf}
in which the role that gravitational radiation plays in these problems was
explained, and a foundation for subsequent research in this area was
established. The main result was that gravitational radiation acts in the
same way in the full theory as in a post-Newtonian analysis of the problem.
If we consider a sequence of equilibrium models, a mode becomes secularly
unstable at the point where its frequency vanishes (in the inertial frame).
Most importantly, the proof does not require the completeness of the modes
of the system.



\section{A step towards multi-fluids}

Returning to the relativistic setting, let us consider what happens if one tries to extend the off-the-shelf 
analysis from Sect.~\ref{shelve} to the case of two components. Take, for example, the case of 
a single particle species at finite temperature; a case where we have to account for the 
presence of entropy. In general, one would have to allow for the heat to
(i.e.~entropy) flow relative to the matter (see Sect.~\ref{sec:heat}), but we will assume that this is not the case here. If the 
entropy is carried along with the matter flow, we are dealing with a single-fluid problem and we should 
be able to make progress with the tools we have at hand.  The equation of state is, however, no longer barotropic since we 
have $\varepsilon=\varepsilon(n,s)$, with $n$ the matter number density and $s$ the entropy density (as before). Nevertheless, the stress-energy tensor can still be expressed in terms of 
the pressure $p$ and the energy density $\varepsilon$, as in Sect.~\ref{shelve}. The 
fluid equations obtained from its divergence will take the same form as in the barotropic case. 
The difference becomes apparent only when we try to close the system of equations. Now the 
energy variation takes the form 
\be
d\varepsilon = \mu dn + T ds , 
\ee
where the temperature is identified as the chemical potential of the entropy:
\begin{equation}
T = \left( {\partial \varepsilon \over \partial s} \right)_n \ .
\end{equation}
This means that we have 
\begin{equation}
    T^{ab} = (n\mu + sT) u^a u^b + p g^{ab} 
\end{equation}
and, if we note that 
\be
dp = n d\mu + s dT \quad \Longrightarrow \quad \nabla_a p = n \nabla_a \mu + s \nabla_a T \ ,
\ee
it follows that 
energy conservation leads to
\begin{equation}
    \mu \nabla_a n^a + T \nabla_a s^a = 0 \ ,
\end{equation}
or
\be
\mu \left( \dot n + n \nabla_a u^a \right) + T \left( \dot s + s \nabla_a u^a \right) = 0 \ , 
\label{comb}\ee
\begin{equation}
\dot n = {dn \over d\tau} = u^a \nabla_a n \ .
\end{equation}
At this point we need to make additional assumptions. If, for example, the motion is adiabatic then the entropy is conserved and the second term on the left-hand side vanishes. It then follows that the first bracket must vanish as well, so the matter flux is also conserved. If the flow is not adiabatic, the
situation is different. Suppose there are no sources or sinks for the matter. Then the matter flux should still be conserved, but now
the entropy is not. So the first term in \eqref{comb} still vanishes, but the second can not. We obviously have a problem, unless we relax the 
assumption that the entropy flows with the matter. Introducing a heat flux relative to the matter, we avoid the issue.
However, by doing so, we introduce extra degrees of freedom that need to be accounted for and understood. We will consider this 
problem in detail once we have extended the variational formalism to deal with additional flows. We could also consider the implication the other way; in order for a single particle flow to be adiabatic, the entropy must be carried
along with the matter.

Moving on to the momentum equations arising from $\nabla_a T^{a b}=0$, replicating the 
analysis from Sect.~\ref{shelve},
recalling the definition $\mu_a = \mu u_a$ and introducing the analogous quantity 
$\theta_a = T u_a$, we can write \eqref{euler1f} as
\be
2 n^a \nabla_{[a}\mu_{b]}+ 2 s^a \nabla_{[a}\theta_{b]}=0 
\ee
That is, we arrive at a ``force balance'' equation with two vorticity terms instead of the single one  we had before. The implication is that, 
even in the absence of external agents we have to consider possible interactions between the two components. By extending the variational approach we gain insight that helps address this issue (also in more complicated situations). 

It is also worth highlighting that, by using notation that highlights the entropy component we have 
made the problem look less ``symmetric'' than it really is. In many situations it is practical to 
introduce constituent indices (labels telling us which component the quantity belongs to), 
e.g., use $n_\n^a$ and $n_\s^a$ instead of $n^a$ and $s^a$. Noting also that the temperature is 
the chemical potential associated with the entropy, i.e. $\theta_a = \mu^\s_a$, we can write the 
above result as
\be
\sum_{\x=\n,\s} f_a^\x = \sum_{\x=\n,\s} 2 n_\x^b \nabla_{[b}\mu^\x_{a]} 
                                    = \sum_{\x=\n,\s} 2 n_\x^b \omega^\x_{b a} = 0 \ .
\ee
The generalisation of this result to situations where additional components are carried along by 
the same four velocity is now obvious. The problem with distinct four velocities, which we turn to 
in Sect.~\ref{sec:twofluids}, requires additional thinking.

\subsection{The two-constituent, single fluid}
\label{pbonemc} 

Before we move on to the general problem, let us consider how the problem discussed in the 
previous Sect.~\ref{ldyn} would be described in the variational approach. Generally speaking, the total 
energy density $\varepsilon$ can be a function of independent parameters other than the particle 
number density $n_\n$, like the entropy density $s=n_\s$ in the case we just considered, 
assuming that the system scales in the manner discussed in Sect.~\ref{sec:thermo} so that only 
densities need enter the equation of state. 

\vspace*{0.1cm}
\begin{tcolorbox}
\textbf{Comment:} There is an an important transition happening at this point. In the following we will, almost exclusively, work with the constituent 
indices $\X,~\Y$, etc., which range over the individual components of the system (here $\{\n,\s\}$) and which do not satisfy any 
kind of summation convention.
\end{tcolorbox}
\vspace*{0.1cm}

  As we have already suggested, if there is no heat 
flow (say) then this is a single fluid problem, meaning that there is still just one 
flow velocity $u^a$. This is what we mean by a two-constituent, single fluid. We assume that the 
particle number and entropy are both conserved along the flow. Associated which each parameter there is then a conserved current  
flux, i.e.~$n^a_\n = n_\n u^a$ for the particles and $n_\s^a = n_\s u^a$ for the entropy. 
Note that the ratio $x_\s = n_\s/n_\n$ (the specific entropy) is co-moving in the sense that 
\begin{equation} 
     u^a \nabla_a x_\s = \dot{x}_\s= 0 \ . \label{comoving} 
\end{equation}
This is, of course, the relation \eqref{comb} from before.

Making use of the constituent indices, the associated first law can be written in the form 
\begin{equation} 
     { d} \varepsilon = \sum_{\x = \n,\s} \mu^\x {d} n_\x 
             =  - \sum_{\x = \n,\s} \mu^\x_a {d} n^a_\x \ , 
\end{equation} 
since $\varepsilon = \varepsilon(n_{\n},n_{\s})$, where 
\begin{equation} 
     n^a_\x = n_\x u^a \quad , \quad 
     n^2_{\x} = - g_{a b} n^a_\x n^b_\x \ , 
\end{equation} 
and 
\begin{equation} 
     \mu^\x_a = g_{a b} \B^{\x} n^b_\x \quad , \quad 
     \B^{\x} \equiv 2 \frac{\partial \varepsilon}{\partial n^2_{\x}} \ . 
\label{bxdef}
\end{equation} 

Given that we only have one four-velocity, the system will still just have one fluid element per 
spacetime point. But unlike before, there is an additional conserved number, $N_\s$, that can 
be attached to each worldline, like the particle number $N_\n$ of Fig.~\ref{pullback}. In order to 
describe the worldlines we can use the same three scalars $X^A(x^a)$ as before. But how do 
we get a construction that allows for the additional conserved number? \ Recall that the 
intersections of the worldlines with some hypersurface, say $t = 0$, is uniquely specified 
by the three $X^A(0,x^i)$ scalars. Each worldline will also have the conserved numbers $N_\n$ 
and $N_\s$ assigned to them. Thus, the values of these numbers can be expressed as 
functions of the $X^A(0,x^i)$. But most importantly, the fact that each $N_\x$ is conserved, 
means that this specification must hold for all of spacetime, so that the ratio $x_\s$ is 
of the form $x_\s(x^a) = x_\s(X^A(x^a))$. Consequently, we now have a construction where 
this ratio identically satisfies Eq.~(\ref{comoving}), and the action principle remains a variational 
problem in terms of the three $X^A$ scalars. 

The variation of the action follows just like before, except now a constituent index $\x$ must be 
attached to the particle number density current and three-form: 
\begin{equation} 
     n^\x_{a b c} = \epsilon_{dabc} n^d_\x 
                               \ . 
\end{equation} 
Once again it is convenient to introduce the momentum form, now defined as 
\begin{equation} 
     \mu^{a b c}_\x = \epsilon^{ d a b c } 
                                \mu^\x_d \ . \label{momformx} 
\end{equation} 
Since the $X^A$ are the same for each $n^\x_{a b c}$, the above discussion indicates that the 
pull-back construction  is now to be based on 
\begin{equation} 
     n^\x_{a b c} = \psi^A_a \psi^B_b \psi^C_c N^\x_{A B C}  , 
\end{equation} 
where $N^\x_{A B C}$ is completely antisymmetric and a function only of the $X^A$. After a little 
thought, it should be obvious that the only thing required here (in addition to the 
single-component arguments) is to attach an $\x$ index to $n^a$ and $n$ in 
Equations ( (\ref{delnvec0}) and (\ref{dens_perb}), respectively. 

If we now define the Lagrangian to be 
\begin{equation} 
    \Lambda = - \varepsilon 
\end{equation}
and the generalized pressure $\Psi$ as 
\begin{equation} 
     \Psi = \Lambda - \sum_{\x = \n,\s} \mu^\x_a n^a_\x 
          = \Lambda + \sum_{\x = \n,\s} \mu^\x n_\x \ , 
\end{equation}
then the first-order variation of $\Lambda$ is  (ignoring a surface term, as usual)
\begin{multline}
     \delta \left(\sqrt{- g} \Lambda\right) = \frac{1}{2} \sqrt{- g} 
     \left[\Psi g^{a b} + \left(\Psi - \Lambda\right) u^a u^b 
     \right] \delta g_{a b} \\  - \sqrt{- g} \left(\sum_{\x = \n,\s} 
     f^\X_a\right) \xi^a
      + \nabla_a \left(\frac{1}{2} \sqrt{-g} \sum_{\X = \n,\s} 
     \mu^{a b c}_\x n^\x_{b c d} \xi^d\right) 
     \ , 
\end{multline}  
where 
\begin{equation} 
     f^\x_a = 2 n^b_\x \omega^\x_{b a} \ , \label{forcex} 
\end{equation} 
and 
\begin{equation} 
     \omega^\x_{a b} = \nabla_{[a} \mu^\x_{b]} \ . 
\end{equation} 
At the end of the day, the equations of motion are 
\begin{equation} 
   \sum_{\x = \n,\s} f^\X_a = 0 \ , \label{eueqn1} 
\end{equation} 
and 
\begin{equation} 
   \nabla_a n^a_\x = 0 \ , 
\end{equation} 
while the stress-energy tensor takes the form  
\begin{equation} 
    T^{a b} = \Psi g^{a b} + (\Psi - \Lambda) u^a u^b \ . 
\end{equation} 
Not surprisingly, these results accord with the expectations from the previous analysis.

\subsection{Speed of sound (again)}
\label{sndspeed2}

We have already considered the problem of wave propagation in the case of a single 
component (barotropic) fluid, see Sect.~\ref{ldyn}. Now we are equipped to revisit this 
problem in the more complex case of a two-constituent single-fluid---a fluid that is ``stratified'' either by thermal or composition gradients. As before, the analysis is 
local---assuming that the speed of sound is a locally defined quantity---and performed using local 
inertial frame (Minkowski) coordinates $x^a = (t,x^i)$. The purpose of the analysis is twofold: 
The main aim is to  illuminate how the presence of various constituents impacts on the local 
dynamics, but we also want to illustrate how the problem works out if we take the variational 
equations of motion as our starting point. An additional motivation is to develop notation that is 
flexible enough that we can deal with problems of increasing complexity, ideally without losing 
sight of the underlying physics.

Focussing on a small spacetime region, we can make the same argument as in 
Sect.~\ref{ldyn} that the configuration of the matter with no waves present is locally 
isotropic, homogeneous, and static.  Thus, for the background $n^a_\x = [n_\x,0,0,0]$ and the 
vorticity $\omega^\x_{a b}$ vanishes. The general form of the (Eulerian) variation of the force 
density $f^\x_a$ for each constituent is then
\begin{equation} 
    \delta f_a^\x = 2 n^b_\X \partial_{[b} \delta \mu^\x_{a]} \ . 
\label{df}
\end{equation} 
Similarly, the conservation of the flux $n^a_\x$ gives 
\begin{equation} 
    \partial_a \delta n^a_\x = 0 \ . \label{consvar} 
\end{equation} 
We are now taking the view that the $n^a_\x$ are the fundamental fluid fields and thus plane-wave propagation 
means that we have (the covariant analogue fo \eqref{ppwave})
\begin{equation} 
    \delta n^a_\X = A^a_\x e^{i k_b x^b} \ , \label{pwave} 
\end{equation} 
where the amplitudes $A^a_\x$ and the wave vector $k_a$ are constant. Combining 
Eqs.~(\ref{consvar}) and (\ref{pwave}) we see that
\begin{equation} 
    k_a \delta n^a_\x = 0 \ , \label{transverse} 
\end{equation} 
i.e.~the waves are ``transverse'' in the spacetime sense. It is worth pointing out that this requirement is 
not in contradiction with the fact that sound waves are longitudinal (in the spatial sense), as 
established in Sect.~\ref{ldyn}. It is easy to see that \eqref{transverse} is exactly what 
we should expect, if we note that $\delta n_\x^a = \delta n_\X u^a + n_\x \delta v^a$ and 
identify $k_0 = - k \sigma$ where, recall, $\sigma$ is the mode speed and $k$ is the spatial part 
magnitude obtained from $k^2 = k_j k^j$ ($k^i = g^{i j} k_j$).

Moving on to the equations of motion, as given by \eqref{df}, we need the perturbed momentum 
$\delta \mu^\x_a$. For future reference, we will work out its general form, and only afterwards 
assume a static, homogeneous, and isotropic background. However, in order to establish the 
strategy, it is useful to start by revisiting the barotropic case. Suppose there is only one 
constituent, with index $\x = \n$.\ The Lagrangian $\Lambda$ then depends only on $n^2_{\n}$, 
and the variation in the chemical potential due to a small disturbance $\delta n^a_\n$ is 
\begin{equation} 
     \delta \mu^\n_a = \B^{\n}_{a b} \delta n^b_\n \ , 
\end{equation} 
where 
\begin{equation} 
    \B^{\n}_{a b} = \B^{\n} g_{a b} - 2 
    \frac{\partial \B^{\n}}{\partial n^2_{\n}} n^\n_a n^\n_b \ .
    \label{knn} 
\end{equation} 
There are two terms, simply because we need to perturb both $\B^\n$ and $n_\n^a$ in 
\eqref{bxdef}.

The single-component equation of motion is $\delta f^\n_a = 0$. It is not difficult to 
show, by using the condition of transverse wave propagation,  
Eq.~(\ref{transverse}), and contracting with the spatial part of the wave 
vector $k^i$ (the time part is trivial because \eqref{df} is orthogonal to $n_\n^a$ which 
in turn is aligned with $u^a$), that the equation of motion reduces to 
\begin{equation} 
    \left(\B^{\n } + \B^{\n}_{0 0} \frac{k_j k^j}{k^2_0}\right) k_i 
    \delta n^i_\n = 0 \ . 
\end{equation} 
From this we see that the  dispersion relation  takes the form
\be
\sigma^2 = {k_0^2 \over k_jk^j} = - {\B^{\n}_{0 0}\over \B^\n} = 1 + 2 { n_\n^2 \over \B^\n}  
    \frac{d \B^{\n}}{d n^2_{\n}} = 1 + \frac{d \ln \B^{\n }}{d \ln n_\n} \ . 
\label{ssp}
\ee
We have used the fact that we are working in a locally flat spacetime, so that 
$g_{a b} = \eta_{a b}$. If we have done this right, then we should recover the expression for the 
speed of sound $C_s^2$ from before, cf.~Eq.~\eqref{csdmudn}. To see that this is the case, 
recall that  $\mu_\n = n_\n \B^\n$ and work out the required derivative.
That is
\begin{equation}
    C_s^2 = \sigma^2 = {n\over \mu} {d \mu \over dn} = {dp \over d\varepsilon}\ .
    \end{equation}

In order to ensure that the behaviour of the system is ``physical'', we need to consider two conditions:
\begin{enumerate}
\item absolute stability, $\sigma^2 \geq 0$ \ , and
\item causality, $C^2_s \leq 1$ \ .
\end{enumerate}
These conditions provide constraints which can be imposed on, say, parameters in equation of 
state models, the net effect being absolute limits on the possible forms for the master function 
$\Lambda$. As an example, take the result from Eq.~\eqref{ssp} and impose the two constraints
to find that
\beq
       0 \leq 1 + \frac{d \ln \B^{\n }}{d \ln n_\n} \leq 1 
       \quad \implies \quad 
       - 1 \leq \frac{d \ln \B^{\n }}{d \ln n_\n} \leq 0 \ . \label{abstab_caus}
\eeq
From the definition of $\Bn$, cf.~Eq.\eqref{bxdef}, we have two bounds on 
$\Lambda$.

Even with the aid of the constraint from Eq.~\eqref{abstab_caus}, the mode frequency 
solution in Eq.~\eqref{ssp} is obviously less transparent than the simple statement of the 
speed of sound as the variation of the pressure with changing density. However, as we will 
establish, the formalism we are developing readily deals with much more complex situations 
(such as multiple sound speeds and so-called ``two-stream'' instabilities). The main reason  
is that the fluxes enter the formalism on equal footing as four-vectors, whereas starting with 
energy density typically requires the introduction of an ad-hoc reference frame (e.g., the $U^a$ from 
Sect.~\ref{tab1}), in order to define what the energy density is, and any independent fluid motion 
(like heat flow) is then defined as a three-velocity with respect to this frame.

As a further example, let us consider the case when there are the two constituents with densities 
$n_\n$ and $n_\s$, two conserved density currents $n^a_\n$ and $n^a_\s$, two chemical 
potential covectors $\mu^\n_a$ and $\mu^\s_a$, but still only one four-velocity $u^a$. (We are 
primarily thinking about matter and entropy, as before, but it could be any two individually 
conserved components which move together.) The matter Lagrangian $\Lambda$ may now 
depend on both $n^2_{\n}$ and $n^2_{\s}$ meaning that 
\begin{equation} 
    \delta \mu^\x_a = \B^{\x}_{a b} \delta n^b_\x 
                        + {\cal X}^{\x \y}_{a b} \delta n^b_\y \ , \quad \y \neq \x \ ,
\end{equation} 
where we recall that summation is not implied for repeated constituent indices, and we have 
defined
\beq
    \Xcalab = - \Ccc \sqrt{\BX \BY} u^\x_a u^\x_b \ , \label{crosscoup}
\eeq
(with $u^\x_a = u^\y_a = u_a$ in this specific example)
where
\beq
    \Ccc^2 \equiv \frac{1}{\BX \BY} \left(2 n_\x n_\y \frac{\partial \BX}{\partial \ny}\right)^2 \ .
\eeq
The $\B^{\n}_{a b}$ coefficient is defined as before and  $\B^{\s}_{a b}$ is given by the same expression (Eq.~\eqref{knn}) with each $\n$ replaced by $\s$. The  
$\Ccc$ coefficient represents a true multi-constituent effect, which depends on the 
composition (e.g., the entropy per baryon $x_\s = n_\s/n_\n$ used in the discussion surrounding 
Eq.~\eqref{comoving}). 

The fact that $n^a_\s$ is parallel to $n^a_\n$ implies that it is only the magnitude of the entropy 
density current that is independent.  One can show that the condition of transverse propagation, 
as applied to both currents, implies 
\begin{equation}
    \delta n^a_\s = x_\s \delta n^a_\n \ . 
\label{nrel}
\end{equation}
It is worth taking a closer look at this condition. First of all, the time component leads to
\be
\delta n_\s = x_\s \delta n_\n = \frac{n_\s}{n_\n} \delta n_\n  
\qquad \Longrightarrow \qquad 
\delta x_\s =  0 \ .
\ee 
That is, the entropy per particle is constant---the perturbations are adiabatic.  Meanwhile, it is easy to show that the spatial part of \eqref{nrel} is trivial, since the 
two components move together.

Now, we proceed as in the previous example. Noting that the equation of motion is 
\begin{equation}
    \delta f^\n_a + \delta f^\s_a = 0\ ,
\end{equation}
we find
\begin{equation} 
    \left[\left(\B^{\n} + x^2_\s \B^{\s}\right) \sigma^2 - \left(\B^{\n} c^2_\n + x^2_\s \B^{\s} 
    c^2_\s - 2 x_{\s} {\cal X}^{\n \s}_{0 0} \right)\right] k_i 
    \delta n^i_\n = 0 \ , 
\end{equation} 
where, inspired by the result for the speed of sound in the single component case 
[cf.~Eq.~\eqref{ssp}], we have defined
\begin{equation} 
    c^2_\x \equiv 1 + \frac{\partial \ln \B^{\x}}{\partial \ln n_\x} 
                  \ . \label{cs1}
\end{equation} 
We find that the  speed of sound is given by  
\begin{equation} 
    C_s^2 = \sigma^2 = \frac{\B^{\n} c^2_\n + x^2_\s \B^{\s} c^2_\s - 2 x_\s  
              {\cal X}^{\n \s}_{0 0}}{\B^{\n} + x^2_\s \B^{\s}} \ . 
\end{equation}
As this result looks quite complicated,  let us  see if we can manipulate it to make it more 
intuitive.
The obvious starting point is to replace the abstract coefficients we have introduced with the 
underlying thermodynamical quantities, i.e. use $\mu_n = n_\n \B^\n = \mu$ and 
$\mu_\s = n_\s \B^\s = T$ leading to
\be
c_\n^2 = {n\over \mu} \left( {\partial \mu \over \partial n} \right)_s 
              \qquad \mbox{ and } \qquad 
c_\s^2 = {s\over T} \left( {\partial T \over \partial s} \right)_n \ .
\ee
We also see that 
\be
 {\cal X}^{\n \s}_{0 0} = -  \left( {\partial \mu \over \partial s} \right)_n = -  \left( {\partial T \over \partial n} \right)_s \ , 
\ee
where the identity follows since we have mixed partial derivatives (both $\mu$ and $T$ arise as 
derivatives of $\varepsilon$). Given these results, we find that
\be
C_s^2 = {1 \over p+\varepsilon} \left[ n^2  \left( {\partial \mu \over \partial n} \right)_s + 2sn \left( {\partial T \over \partial n} \right)_s
+ s^2 \left( {\partial T \over \partial s} \right)_n \right] \ ,
\ee
which already looks a little bit more transparent. However, we can also use the fact that $dp = nd\mu+sdT$ to rewrite this as
\be
C_s^2 = {1 \over p+\varepsilon} \left[  n  \left( {\partial p \over \partial n} \right)_s + s  \left( {\partial p \over \partial s} \right)_n \right] \ .
\label{sou2}
\ee
Finally, let us ask what happens if we work with $x_\s$ instead of $s$. 

To do this, we need 
\begin{multline}
dp  =  \left( {\partial p \over \partial n} \right)_{x_\s} dn + 
 \left( {\partial p \over \partial x_\s} \right)_n d x_\s 
 \\
 = \left[   \left( {\partial p \over \partial n} \right)_{x_\s} - {s \over n^2}   \left( {\partial p \over \partial x_\s} \right)_{n} \right] dn +  {1 \over n}   \left( {\partial p \over \partial s} \right)_n ds \ .
\end{multline}
From this we see that 
\be
  \left( {\partial p \over \partial n} \right)_{x_\s} = \left( {\partial p \over \partial n} \right)_s  + {s \over n}  \left( {\partial p \over \partial s} \right)_n 
\ee
and once we combine with the fact that, when $x_\s$ is kept constant we have
\be
d\varepsilon = {p+\varepsilon \over n} dn \ ,
\ee
we get the expected result for the adiabatic sound speed:
\be
C_s^2 = \left( {\partial p \over \partial \varepsilon} \right)_{x_\s} \ .
\ee

\subsection{Multi-component cosmology}

The modern description of cosmology draws on ideas from fluid dynamics. In the simplest picture---after averaging up to a suitably large scale---planets, stars and galaxies are treated as collisionless ``dust'', represented by the simple stress-energy tensor
\begin{equation}
    T^{ab} = \varepsilon u^a u^b \ .
\end{equation}
This introduces a natural flow of cosmological time---associated with the proper time linked to $u^a$---and the associated fibration of spacetime \citep{tsagas}. The focus on the ``fluid observer'' worldlines means that the model is closely related to our description of fluid dynamics, and it is fairly straightforward to build more complex (read:realistic) models by, for example, adding the cosmological constant to the Einstein equations (or viewing it as a ``dark energy'' contribution with negative pressure, $p=-\varepsilon$) or accounting for more complicated description of the matter content in the Universe. The matter description relies on ideas we have already introduced. In particular, the cosmological principle states that the Universe is homogeneous and
isotropic, suggesting that the relevant matter Lagrangian should be built from scalars. 
Given the increased quality of cosmological observations, this 
fundamental principle is now becoming testable, and (perhaps) questionable. 

The most pressing issues that arise in cosmology relate to the simple fact that  we do not 
have a good handle on the nature of dark components that appear to dominate the ``standard model'' \citep{PPJPU}.  A number of alternative 
models---including alternatives to Einstein's relativistic gravity---have been suggested, but few of these are compelling. The treatment of the different matter components, in 
particular, tends to remain based on the notion of coupled perfect fluids or scalar 
fields. If we are to understand the bigger picture, we may need to review this aspect, especially if we want to be able to consider issues like heat flow 
\citep{modak,pavon,AL11}, dissipative mechanisms 
\citep{Weinberg71,PK91,Velten:2011bg}, Bose--Einstein condensation 
of dark matter \citep{SY09,Harko11} and possibly many others. Many issues are similar to ones that arise in more realistic models of neutron star astrophysics. 

A particularly interesting aspect, given the focus of this review, may be the suggestion that there could have been 
phases during which the Universe would have effectively been anisotropic (see \citealt{TCM08} for a  useful review), with 
different components evolving ``independently'' \citep{2012PhLB..715..289C,2012PhRvD..85j3006C}. For the most part, models 
considered in the current literature, including initially anisotropic geometries, 
describe the matter content in terms of either effectively many component 
single fluid models \citep{Gromov:2002ek}, or a single component 
\citep{EGcCP07,PPU08,KM10}; although an evolution towards isotropy is expected in such 
settings, as required to end up with a realistic (read: in agreement with 
 observational data) model \citep{DLH09}. Having said that, interesting new consequences 
may be inferred by enhancing an initially vanishingly small 
non-Gaussian signal \citep{DP11}.

Within this context, it is relevant to ask how distinct fluid flows may 
lead to anisotropy, with the spacetime metric taking the form of a Bianchi~I 
solution of the Einstein equations. In this case there is a 
spacelike privileged vector, associated with the relative flow between  
two matter components. As we will soon establish, such a feature is natural in the multi-fluid context, but it can never arise in 
the usual multi-constituent single fluid. This point has been considered in some detail in \cite{2012PhLB..715..289C,2012PhRvD..85j3006C}.
It has been 
suggested \citep{BT07,ABDR11,ACM11} that, since Bianchi universes---seen as 
averaged inhomogeneous and anisotropic spacetimes---can have effective
strong energy condition violating stress-energy tensors, they could
be part of a backreaction driven acceleration model. 

Yet another reason for studying such cosmological models stem,
perhaps surprisingly, from the observations: Large angle anomalies in the Cosmic 
Microwave Background (CMB) have been observed and discussed for 
quite some time \citep{SSHC04,CHSS10,Perivolaropoulos11,MEC11} and may be related with underlying Bianchi models \citep{PC07,Pontzen09}.



\section{The ``pull-back'' formalism for two fluids}
\label{sec:twofluids}

Having discussed the single fluid model, and how one accounts for
stratification (either thermal or composition gradients), it is time to move on to the problem of modeling
multi-fluid systems. We will experience for the first time novel
effects due to a relative flow between two
interpenetrating fluids, and the fact that there is no longer a
single, preferred rest-frame. This kind of formalism is necessary, for
example, for the simplest model of a neutron star, since it is generally
accepted that the inner crust is permeated by an independent neutron
superfluid, and the outer core is thought to contain superfluid
neutrons, superconducting protons, and a highly degenerate gas of
electrons. Still unknown is the number of independent fluids required for neutron stars that have deconfined
quark matter in the deep core \citep{alford00:_cfl}. The model can also
be used to describe superfluid Helium and heat-conducting fluids, problems which relate to the incorporation of dissipation (see
Sect.~\ref{sec:viscosity}). We will focus on this example here, as a natural extension of the case considered in the previous section. It
should be noted that, even though the particular system we concentrate on
consists of only two fluids, it illustrates all new features of a general
multi-fluid system. Conceptually, the greatest step is to go from one to
two fluids. A generalization to a system with further degrees of freedom
is straightforward.

In keeping with the previous section, we will rely on use of  constituent indices, 
which throughout this section will range over $\x,\y = \n,\s$. In the example we consider the two fluids represent the particles ($\n$) and the entropy ($\s$). Once again, 
the number density four-currents, to be denoted $n^a_\x$, are taken to be separately conserved, 
meaning that
\begin{equation}
  \nabla_a n^a_\x = 0 \ .
\end{equation}
As before, we use the dual formulation, i.e., introduce the three-forms
\begin{equation}
  n^\x_{a b c} =
  \epsilon_{d a b c } n^d_\x \ ,
  \qquad
  n^a_\x = \frac{1}{3!}
  \epsilon^{b c d a} n^\x_{b c d}.
\end{equation}
Also like before, the conservation rules are equivalent to the individual three-forms being closed (the arguments proceeds in exactly the same way);
i.e.
\begin{equation}
  \nabla_{[a} n^\x_{b c d]} = 0.
  \label{multiclosed}
\end{equation}
However, we need a formulation whereby such conservation obtains
automatically, at least in principle.

We make this happen by introducing the three-dimensional matter space, the
difference being that we now need two such spaces. These will be labelled
by coordinates $X^A_\X$, and we recall that $A,B,C,\mathrm{etc.} = 1,2,3$.
The idea is illustrated in Fig.~\ref{pullback2}, which indicates the important facts
that (i) a given point in space can be intersected by each fluid's worldline
and (ii) the individual worldlines are not necessarily parallel at the
intersection, i.e., the independent fluids are interpenetrating \emph{and} can
exhibit a relative flow with respect to each other. Although we have not
indicated this in Fig.~\ref{pullback2} (in order to keep the figure as
uncluttered as possible) attached to each worldline of a given constituent
will be a fixed number of particles $N^\X_1$, $N^\X_2$,
etc.\ (cf.\ Fig.~\ref{pullback}). For the same reason, we have also not
labelled (as in Fig.~\ref{pullback}) the ``pull-backs'' (represented by the
arrows) from the matter spaces to spacetime.

\begin{figure}[htb]
    \centerline{\includegraphics[scale = 0.5]{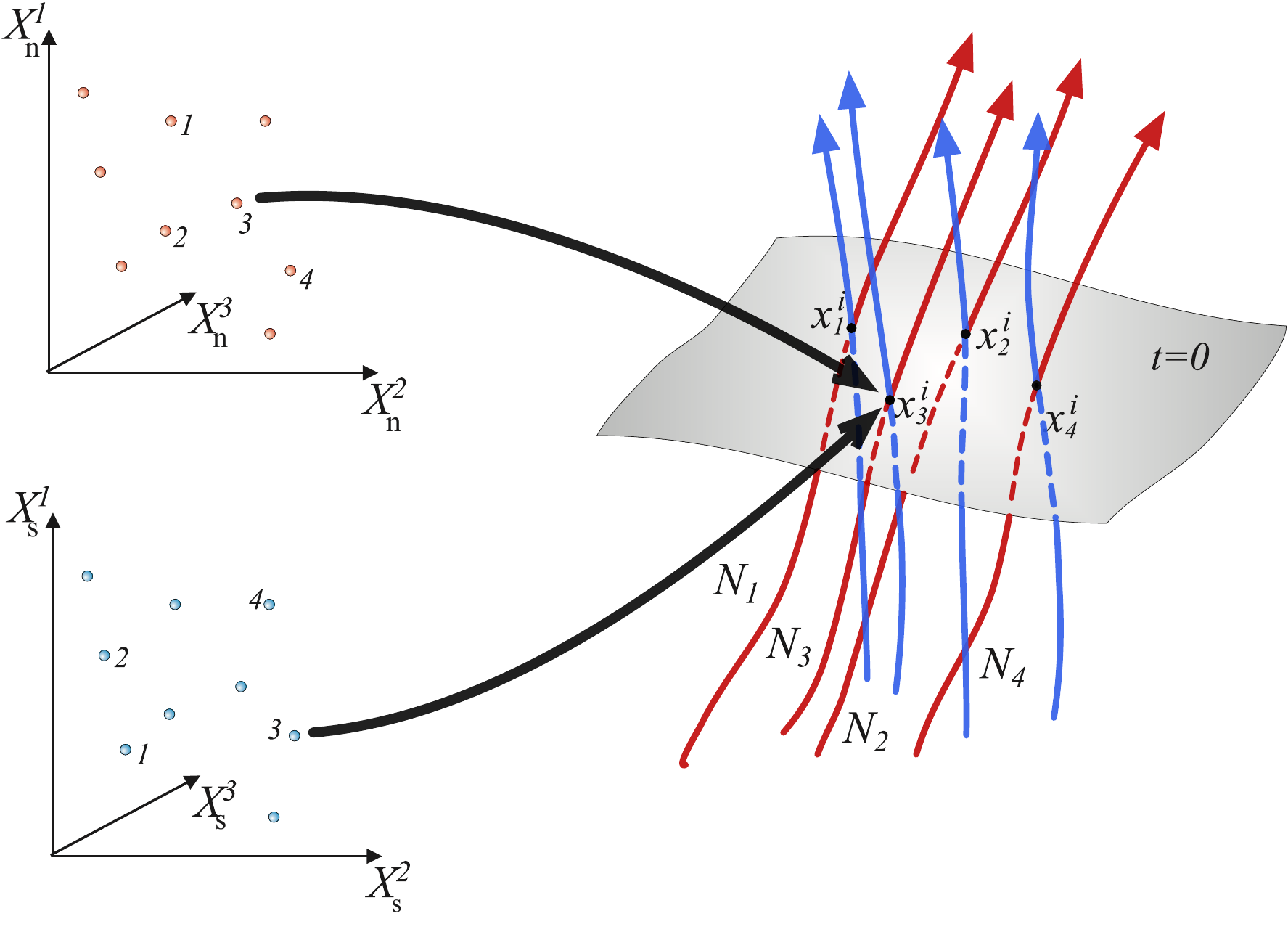}}
    \caption{The pull-back from a point in the
    $\x^{\mathit{th}}$-constituent's three-dimensional matter space
    (on the left) to the corresponding ``fluid-particle'' worldline in
    spacetime (on the right). The points in matter space are labelled
    by the coordinates $\{X^1_\x,X^2_\x,X^3_\x\}$, and the constituent
    index $\x = \n,\s$. There exist as many matter spaces as there are
    dynamically independent fluids, which for this case means two.}
    \label{pullback2}
\end{figure}

By ``pushing forward'' each constituent's three-form onto its respective
matter space we can once again construct three-forms that are
automatically closed on spacetime, i.e., let
\begin{equation}
  n^\x_{a b c} = \psi_{\x a}^A \psi_{\x b}^A \psi_{\x c}^C 
  N^\x_{A B C}  \ ,
\end{equation}
where 
\begin{equation}
\psi_{\x a}^A    = {\partial X_\x^A \over \partial x^a } \ , 
\end{equation}
and
$N^\x_{A B C}$ is completely antisymmetric in its indices and is a
function only of the $X^A_\x$. Using the same reasoning as in the single fluid
case, the construction produces three-forms that are automatically
closed, i.e., they satisfy Eq.~(\ref{multiclosed}) identically. If we let
the scalar fields $X^A_\x$ (as functions on spacetime) be the fundamental
variables, they yield a representation for each particle number density
current that is automatically conserved. The variations of the three-forms
can now be derived by varying them with respect to the $X^A_\x$.

The Lagrangian displacements on spacetime for each fluid, to be denoted
$\xi^a_\x$,  are related to the
variations $\delta X^A_\x$ via
\begin{equation}
  \Delta_\x X^A = \delta X^A_\x +\xi^a_\x \partial_a X^A_\x=  \delta X^A_\x +\xi^a_\x \psi^A_{\x a}= 0 \ .
\end{equation}
In general, the various single-fluid equations we have considered are easily extended to the
two-fluid case, except that each
displacement and four-current will now be associated with a constituent index,
using the decomposition
\begin{equation}
  n^a_\x = n_\x u^a_\x \ ,
  \qquad
  u^\x_a u^a_\x = - 1 \ .
\end{equation}

Associated with each constituent's Lagrangian displacement is its own
Lagrangian variation. As above, these are naturally defined to be
\begin{equation}
  \Delta_\x \equiv \delta + {\cal L}_{\xi_\x},
\end{equation}
so that it follows that
\begin{equation}
  \Delta_\X n^\x_{a b c} = 0,
\end{equation}
as expected for the pull-back construction. Likewise, two-fluid analogues of 
Eqs.~(\ref{lagradelu}, \ref{lagradeleps}, \ref{lagradeln}) exist which take the same form 
except that the constituent index is attached. However, in contrast to the ordinary fluid case, 
there are more options to consider. For instance, we could also look at the Lagrangian 
variation of the first constituent with respect to the second constituent's flow, i.e., 
$\Delta_\s n_\n $, or the other way around, i.e., $\Delta_\n n_\s$. The Newtonian analogues of 
these Lagrangian displacements were essential to an analysis of instabilities in rotating 
superfluid neutron stars \citep{andersson04:_canon_energy}.

We are now in a position to construct an action principle that yields
the equations of motion and the stress-energy tensor. Again, the central
quantity is the matter Lagrangian $\Lambda$, which is now a function of
all the different scalars that can be formed from the $n^a_\x$,
i.e., the scalars $n_\x$ together with
\begin{equation}
  n^2_{\x \y} = n^2_{\y \x} = - g_{a b} n^a_\X n^b_\Y.
\end{equation}
In the limit where all the currents are parallel, i.e., the fluids are
comoving, $- \Lambda$ corresponds (As before) to the local thermodynamic
energy density. In the action principle, $\Lambda$ is the Lagrangian
density for the fluids. 

\vspace*{0.1cm}
\begin{tcolorbox}
\textbf{Comment:} It should be noted that our choice to use only
the fluid currents to form scalars implies that the system is ``locally
isotropic'' in the sense that there are no a priori preferred directions---the fluids are equally free to move in any direction. Structures
like the  crust close to the surface of  a neutron star
generally could be locally anisotropic, e.g., with sound waves 
moving in a preferred direction associated with the lattice or the local magnetic field.
\end{tcolorbox}
\vspace*{0.1cm}

An unconstrained variation of $\Lambda$ with respect to the independent
vectors $n^a_\x$ and the metric $g_{a b}$ takes the form
\begin{equation}
  \delta \Lambda = \!\!\!\!\!\sum_{\x = \{\n,\s\}}\!\!\!
  \mu^\x_a \, \delta n^a_\x + \frac{1}{2} 
  \left( \sum_{\x = \{\n,\s\}}\!\!\! n^a_\x \mu^b_\x \right)
  \delta g_{a b},
\end{equation}
where
\begin{eqnarray}
  \mu^\x_a &=& \B^{\x } n_a^\x + \A^{\x \y} n_a^\y \ ,
  \label{mudef2}
  \\
  \A^{\x \y} &=& \A^{\y \x} =
  - \frac{\partial \Lambda}{\partial n^2_{\x \y}},
  \qquad
  \mathrm{for\ } \x \neq \y \ .
  \label{coef12}
\end{eqnarray}%
The momentum covectors $\mu^\x_a$ are each dynamically, and
thermodynamically, conjugate to their respective number density currents
$n^a_\x$, and their magnitudes are the chemical potentials. Here we note something new: the 
$\A^{\X \Y}$ coefficient represents the fact that each fluid momentum  $\mu^\x_a$
may, in general, be given by a linear combination of the individual
currents $n^a_\x$. That is, the current and momentum for a particular
fluid do not have to be parallel. This is known as the entrainment
effect. We have chosen to represent it by the letter $\A$ for
historical reasons. When Carter first developed his formalism he opted
for this notation, referring to the ``anomaly'' of having misaligned
currents and momenta. It has since been realized that the entrainment
is a key feature of most multi-fluid systems and it would, in fact, be
anomalous to leave it out!

In the general case, the momentum of one constituent carries along some
mass current of the other constituents. The entrainment only vanishes
in the special case where $\Lambda$ is independent of $n^2_{\x \y}$
($\x \neq \y$) because then we obviously have $\A^{\x \y} = 0$.
Entrainment is an observable effect in laboratory
superfluids \citep{putterman74:_sfhydro, tilley90:_super} (e.g., via
flow modifications in superfluid ${}^4{\mathrm{He}}$ and mixtures of
superfluid ${}^3{\mathrm{He}}$ and ${}^4{\mathrm{He}}$). In the case of neutron
stars, entrainment---in this case related to the mobility of the superfluid neutrons that  permeate the neutron star crust---plays a key role in the discussion of  pulsar glitches glitches \citep{radha69:_glitches,
  reichly69:_glitches}. As we will see later (in Sect.~\ref{sec:heat}), these ``anomalous'' terms are necessary for causally
well-behaved heat conduction in relativistic fluids, and by extension
necessary for building well-behaved relativistic equations that
incorporate dissipation (see also \citealt{2010RSPSA.466.1373A,andersson11:_ent_ent}).

In terms of the constrained Lagrangian displacements, a variation of
$\Lambda$ now yields 
\begin{multline}
  \delta \left( \sqrt{- g} \Lambda \right) =
  \frac{1}{2} \sqrt{- g} \left(\! \Psi g^{a b} +
  \!\!\!\!\!\sum_{\x = \{\n,\s\}}\!\!\! n^a_\x \mu^b_\x  \!\right)
 \delta g_{a b} -
  \sqrt{- g} \!\!\!\sum_{\x = \{\n,\s\}}\!\!\! f^\x_a \xi^a_\x \\ +
  \nabla_a \left(\! \frac{1}{2} \sqrt{-g} \!\!\!\sum_{\x = \{\n,\s\}}\!\!\!
  \mu^{a b c}_\x n^\x_{b c d} \xi^d_\x \!\right) \ ,
\end{multline}
where $f^\x_a$ is as defined in Eq.~(\ref{forcex}) except that the
individual velocities are no longer parallel. The generalized pressure
$\Psi$ is now
\begin{equation}
  \Psi = \Lambda -
  \!\!\!\!\!\sum_{\x = \{\n,\s\}}\!\!\! n^a_\x \mu^\x_a\ .
\end{equation}
At this point we  return to the view that $n^a_\n$ and
$n^a_\s$ are the fundamental variables. Because
the $\xi^a_\x$ are independent variations, the equations of motion
consist of the two original conservation conditions from Eq.~(\ref{consv2}),
plus two Euler-type equations
\begin{equation}
  f^\x_a = n_\x^b \omega^\x_{ba} = 0 \ ,
  \label{eueqn2}
\end{equation}
and of course the Einstein equations (obtained exactly as before by adding
in the Einstein--Hilbert term, see Sect.~\ref{sec:efevar}). We also find that the stress-energy
tensor is
\begin{equation}
  T^a{}_b = \Psi \delta^a{}_b +
  \!\!\!\!\!\sum_{\x = \{\n,\s\}}\!\!\! n^a_\x \mu^\x_b.
  \label{seten2}
\end{equation}
When the complete set of field equations is satisfied then it is
automatically true that $\nabla_b T^b{}_a = 0$. One can
also verify that $T_{a b}$ is symmetric. The momentum form
$\mu^{a b c}_\x$ entering the boundary term is the natural
extension of Eq.~(\ref{momformx}) to this two-fluid case.

It must be noted that Eq.~(\ref{eueqn2}) is significantly different from
the multi-constituent version from Eq.~(\ref{eueqn1}). This is true even
if one is solving for a static and spherically symmetric configuration,
where the fluid four-velocities would all necessarily be parallel.
Simply put, Eq.~(\ref{eueqn2})  represents two independent
equations. If one takes entropy as an independent fluid, then the
static and spherically symmetric solutions will exhibit thermal
equilibrium \citep{comer99:_quasimodes_sf}. This explains, for instance,
why one must specify an extra condition (e.g., convective
stability; \citealt{weinberg72:_book}) to solve for a double-constituent
star with only one four-velocity.



\section{Waves in multi-fluid systems}
\label{sec:soundspeed}

Crucial to the understanding of black holes is the effect of spacetime curvature on the light-cone 
structures, that is, the totality of null vectors that emanate from each spacetime point. Crucial to 
the propagation of massless fields (and gravitational waves!) is the light-cone structure. In the case of fluids, it is both the 
speed of light and the speed (and/or speeds) of sound that dictate how waves propagate through 
the matter. We have already used  a local analysis of plane-wave propagation to derive the 
speed of sound for both the single-fluid case (in Sect.~\ref{ldyn}) and the two-constituent single-fluid case (in Sect.~\ref{sndspeed2}). We will now repeat the analysis for a general two-fluid system, 
using the same assumptions as before (see~\citealt{carter89:_covar_theor_conduc} for a more
rigorous derivation). However, we will provide an important extension by allowing a relative flow 
between the two fluids in the background/equilibrium state. While this extension is  
straight-forward, we will see that the final results are quite astonishing---demonstrating the existence of a 
two-stream instability.


\subsection{Two-fluid case}

As a reminder, we first note that the analysis is, in principle, performed in a small region (where the meaning of
``small'' is dictated by the particular system being studied) and we assume that the configuration 
of the matter with no waves present is locally isotropic, homogeneous, and static. Thus, for the 
background, $n^a_\x = [n_\x,0,0,0]$ and the vorticity $\omega^\x_{a b}$ vanishes.  The 
linearized fluxes take the plane-wave form given in Eq.~\eqref{pwave}.

The two-fluid problem is qualitatively different from the previous cases, since there are now 
two independent currents. This  impacts on the analysis in two crucial ways: (i) The 
Lagrangian $\Lambda$ depends on $n^2_{\n}$, $n^2_{\s}$, \emph{and} 
$n^2_{\n \s} = n^2_{\s \n}$ (i.e.~entrainment is present), and (ii) the equations of motion, after
taking into account the transverse flow condition of Eq.~\ref{transverse} for both fluids, are 
doubled to $\delta f^\n_a = 0 = \delta f^\s_a$. The key point is that there can be 
\emph{two} simultaneous wave propagations, with each distinct mode having its own sound 
speed.

Another ramification of having two fluids, is that the variation $\delta \mu^\x_a$ has more 
terms than in the previous, single-fluid analysis. There are individual fluid bulk effects, 
cross-constituent effects due to coupling between the fluids, and entrainment. We can isolate 
these various effects by writing $\delta \mu^\x_a$ in the form
\beq
    \delta \mu_a^\x = \left(\BXab + \Acalxab\right) \delta n^b_\x + 
                      \left(\Xcalab + \Acalxyab\right) \delta n^b_\y \ . 
                      \label{muvarx}
\eeq
The bulk effects are contained in 
\beq
    \BXab = \BX \left(\perp^\x_{a b} - c^2_\x u^\x_a u^\x_b\right) 
            \ , \label{bulk}
\eeq
which is just the two-fluid extension of Eq.~\eqref{knn} [with $\n$ replaced by $\x$ and using
Eq.~\eqref{cs1}]. The cross-constituent coupling enters via $\Xcalab$ [defined 
already in Eq.~\eqref{crosscoup}]. Finally, entrainment enters through the coefficients 
$\Acalxab$ and $\Acalxyab$ given by, respectively,
\begin{equation}
    \Acalxab = - \left[\BX_{, \x \y} \left(u^\x_a u^\y_b + u^\x_b u^\y_a
                 \right) + \frac{n_\y}{n_\x} \AXY_{, \x \y} u^\y_a u^\y_b
                 \right] \ , \label{entraincoup1}
\end{equation}
\begin{multline}
    \Acalxyab = \AXY \perp^\x_{a b} \\
    - \left[\left(\AXY + \frac{n_\x}{n_\y} 
                  \BX_{, \x \y}\right) u^\x_a u^\x_b + \frac{n_\y}{n_\x} 
                  \BY_{, \x \y} u^\y_a u^\y_b + \AXY_{, \x \y} u^\y_a 
                  u^\x_b\right] \ , \label{entraincoup2}
\end{multline}
where we have introduced the notation 
\be
    \BX_{, \x \y} \equiv n_\x n_\y \frac{\partial \BX}{\partial \nxy}
                    \ , 
\ee
and
\be
    \AXY_{, \x \y} \equiv n_\x n_\y \frac{\partial \AXY}{\partial 
                   \nxy} \ . \label{twoderivs}
\ee

The same procedure as in the previous two examples---the single fluid with one and then two 
constituents---leads to the dispersion relation
\begin{multline}
  \left( \B^{\n} \sigma^2 - \left[ \B^{\n}_{0 0} + \A^{\n \n}_{0 0} \right]\right) \left( \B^{\s} \sigma^2 -
  \left[ \B^{\s}_{0 0} + \A^{\s \s}_{0 0} \right]\right) \\
  - \left( \A^{\n \s} \sigma^2 -
  \left[ {\cal X}^{\n \s}_{0 0} + \A^{\n \s}_{0 0} \right]\right)^2 = 0 \ , \label{2fldisrel}
\end{multline}
recalling from Eq.~\eqref{ssp} that $\sigma^2 = k^2_0/k_i k^i$.  This is  a quadratic in $\sigma^2$, 
meaning that there are two sound speeds. This is a natural result of the doubling of fluid degrees of
freedom. 

To finish this discussion of local mode solutions in the two-fluid problem, it is useful to consider 
what constraints the simplest solutions of zero interaction imposes on the equation of state. The 
dispersion relation becomes simply
\begin{equation}
\label{dr.f}
(\sigma^2 - c_\n^2)(\sigma^2 - c_\s^2) = 0 \ ,
\end{equation}
so the mode speed solutions $\sigma_\n$ and $\sigma_\s$ are
\beq
     \sigma^2_\n = c^2_\n = 1+ \frac{\partial \log \Bn}{\partial \log n_\n} \ , \quad
     \sigma^2_\s = c^2_\s = 1+ \frac{\partial \log \Bs}{\partial \log n_\s} \ .     
\eeq
The constraints of absolute stability and causality implies that $\Lambda$ must 
be such that
\begin{equation}
- 1 \le \frac{\partial \log \Bn}{\partial \log n} \le 0 \ , \quad 
- 1 \le \frac{\partial \log \Bs}{\partial \log s} \le 0 \ .
\end{equation}
A general analysis which keeps in entrainment and cross-constituent coupling has been 
performed by \cite{samuelsson10:_rel2st}.

While the sound speed analysis is local, the doubling of the fluid degrees of freedom naturally 
carries over to the global scale relevant for the analysis of modes of oscillation of a fluid body. 

\vspace*{0.1cm}
\begin{tcolorbox}
\textbf{Comment:} For a neutron star, the full spectrum of modes is quite impressive (see \citealt{mcdermott88:_modes}): polar (or spheroidal) f-, p-, and g-modes, and the axial (or 
toroidal) r-modes. \cite{epstein88:_acoust_proper_ns} was the first to suggest that there 
should be even more modes in superfluid neutron stars because the superfluidity allows the 
neutrons to move independently of the protons. \cite{mendell91:_superflnodiss}
developed this idea further by using an analogy with coupled pendulums. He argued that the new 
modes should feature a counter-motion between the neutrons and protons, i.e., as the neutrons
move out radially, say, the protons will move in. This is in contrast to ordinary fluid motion that 
would have the neutrons and protons move in more or less ``lock-step''. Analytical and numerical
studies \citep{lee95:_nonrad_osc_superfl_ns,
  lindblom95:_does_gravit_radiat_limit_angul, comer99:_quasimodes_sf,
  andersson01:_dyn_superfl_ns,2015PhRvD..92f3009K} have confirmed this basic picture and the new modes of 
  oscillation are commonly known as superfluid modes.
\end{tcolorbox}
\vspace*{0.1cm}

\subsection{The two-stream instability}

Consider a system having two components between which there can be a relative flow, such as 
ions and electrons in a plasma, entropy and matter in a superfluid, or even the rotation of a 
neutron star as viewed from asymptotically flat infinity. If the relative flow reaches a speed where 
a mode in one of the components looks like it is going one direction with respect to that 
component, but the opposite direction with respect to the other component, then the mode will 
have a negative energy and become dynamically unstable. This kind of  ``two-stream'' 
instability has a long history of investigation in the area of plasma physics (see 
\citealt{farley63:_2stream,buneman63:_2stream}).  The Chandrasekhar--Friedman--Schutz (CFS) instability 
\citep{chandrasekhar70:_grav_instab, friedman78:_lagran,friedman78:_secul_instab} 
(already discussed in Sect.~\ref{sec:cfs}) develops when a mode in a rotating star appears to be 
retrograde with respect to the star itself, and yet prograde with respect to an observer at infinity. The possible link between two-stream instability in the superfluid in the inner crust and pulsar glitches  is more recent \citep{acp03:_twostream_prl,andersson04:_twostream}. Another relevant
discussion considers a cosmological model consisting of a relative flow between matter and 
blackbody radiation \citep{Comer12:_cosmo_two_stream}. Two-stream instability between two 
relativistic fluids in the linear regime has been examined in general by  
\cite{samuelsson10:_rel2st}, and extended to the non-linear regime by 
\cite{Hawke13:_non_lin_2str}. Finally, a discussion on the relationship between energetic 
and dynamical instabilities, starting from a Lagrangian for two complex scalar fields, was given 
by \cite{haber2016:_2scal_field_2strm}. 

Repeating the key steps from \cite{samuelsson10:_rel2st}, we  start with a system having plane-wave propagation (as before, in a locally flat region of 
spacetime) on backgrounds such that $\omega^\x_{a b} = 0$. The various background quantities 
are considered constant, and there is a relative flow between the fluids. As in the previous 
sound-speed analyses, we  let $u^a_\x$ represent the background four-velocity of the 
$\x$-fluid. Its total particle flux then takes the form
\beq
       n^a_\x = n_\x u^a_\x + A^a_\x \exp^{i k_b x^b} \ ,
\eeq
Because $\omega^\x_{a b} = 0$ for the background and there is flux conservation, the analysis 
still leads to the linearized equations;
\beq
    \nabla_a \delta n^a_\x = 0 
         \quad , \quad 
    n_\X^a \nabla_{[a}\delta \mu^\x_{b]} = 0 \ . \label{pereqsx}
\eeq
The variation $\delta \mu^\x_a$ is the same as in Eq.~\eqref{muvarx}. 

However, the system flow is now such that $u^a_\x$ does {\em not} equal $u^a_\y$, the $\y$-fluid 
four-velocity. There is a non-zero relative velocity of, say, the $\y$-fluid with respect to 
the $\x$-fluid given by
\beq
    \gamma_{\x \y} v^a_{\x \y} = \perp^{\x a}_b u^b_\y \ , \label{vxy} 
\eeq
where $v_{\x \y} = v_{\y \x}$ represents the magnitude of the relative flow,
\beq
    \perp^{\x b}_a = \delta_a{}^b + u^\x_a u^b_\x 
         \quad , \quad 
    \perp^{\x b}_a u^a_\x = 0 \ , \label{perp}
\eeq
and 
\beq
    \gamma_{\x \y} = \gamma_{\y \x} = - u^c_\x u^\y_c = 
                     \frac{1}{\sqrt{1 - v^2_{\x \y}}} \ .
\eeq
This leads to (adapting \eqref{varel} to the present context)
\beq
    u^a_\y = \gamma_{\x \y} \left(u^a_\x + v^a_{\x \y}\right) \ . 
\eeq

For convenience, we will work in the material frame associated with the fluid, meaning that $k_a$ 
and $A^a_\x$ will be decomposed into timelike and spatial pieces as defined locally by $u^a_\x$. 
For $k_a$ we write
\beq
    k_a = k_\x \left(\sigma_\x u^\x_a + \hat{k}^\x_a\right) \ , \label{kdef}
\eeq
where $\sigma_\x$, $k_\x$, and the unit wave vector $\hat{k}^\x_a$ are obtained from $k_a$ via
\beq
    k_\x \sigma_\x = - k_a u^a_\x \ , \quad 
    k^a k_a = - k^2_\x \left(1 - \sigma^2_\x\right) \ , \quad 
   \hat{k}^\x_a = \frac{1}{k_\x} \perp^b_{\x a} k_b \equiv \hat{k}^\x_a \ . 
\eeq
Similarly, the wave amplitude $A^a_\x$ becomes
\beq
    A^a_\x = A^\x_{||} u^a_\x + A_{\x \perp}^a \ ,
\eeq 
where
\beq
    A^\x_{||} = - u^\x_a A^a_\x \quad , \quad A_{\x \perp}^a = \perp^a_{\x b} 
                A^b_\x \ . 
\eeq

It is necessary to point out that the three quantities $\sigma_\x$, $k^\x_a$, and $v^a_{\x \y}$ are 
determined by an observer moving along with the $\x$-fluid. Of course, we could choose the 
frame attached to the other fluid. Fortunately, there are well-defined transformations between the 
two frames, which we determine as follows: The relative flow $v^a_{\y \x}$ of the 
$\x^{\rm th}$-fluid with respect to the $\y^{\rm th}$-fluid frame is related to $v^a_{\x \y}$ via
\beq
    v^a_{\y \x} = - \gamma_{\x \y} \left(v^2_{\x \y} u^a_\x + 
                  v^a_{\x \y}\right) \ , 
\eeq
using the fact that $v_{\y \x} = v_{\x \y}$. Since $k_a$ is a tensor, we must have
\beq
    k_a = k_\y \left(\sigma_\y u^\y_a + \hat{k}^\y_a\right) = k_\x 
         \left(\sigma_\x u^\x_a + \hat{k}^\x_a\right) \ . \label{kequal}
\eeq
Noting that
\be
    u^a_\x= - v^{- 2}_{\x \y} \left(v^a_{\x \y} + \gamma^{- 1}_{\x \y} 
               v^a_{\y \x}\right) \ , 
\ee
\be
    u^a_\y =
- v^{- 2}_{\x \y} \left(v^a_{\y \x} + \gamma^{- 1}_{\x \y} 
               v^a_{\x \y}\right) \ , \label{423vel}
\ee
and contracting each with the wave-vector $k_a$, we obtain the matrix equation 
\beq
    \left[\begin{array}{cc}
    v_{\x \y} \sigma_\x - \cos \theta_{\x \y} &  - \gamma^{- 1}_{\x \y} 
    \cos \theta_{\y \x} \\
    - \gamma^{- 1}_{\x \y} \cos \theta_{\x \y} & 
    v_{\x \y} \sigma_\y - \cos \theta_{\y \x}
    \end{array}\right] 
    \left[\begin{array}{c}
    k_\x \\
    k_\y
    \end{array}\right] = 
    \left[\begin{array}{c} 0 \\ 0 \end{array}\right]
    \ . 
\eeq
The non-trivial solution requires that the determinant of the $2 \times 2$ matrix vanishes; 
therefore, 
\beq
    \sigma_\y = \cos \theta_{\y \x} \frac{\sigma_\x - v_{\x \y} \cos 
    \theta_{\x \y}}{v_{\x \y} \sigma_\x - \cos \theta_{\x \y}} \ . 
    \label{sigtrans}
\eeq
It is not difficult to show that if $\sigma^2_\x \le 1$ then $\sigma^2_\y \le 1$, and clearly if 
$\sigma_\x$ is real then so is $\sigma_\y$. 

The equation of flux conservation is the same as \eqref{transverse} (except $\x$ 
ranges over two values). Here, it implies for each mode that
\beq
     - \sigma_\x A^\x_{||} + \hat{k}^\x_a A_{\x \perp}^a = 0 \ .
\eeq
The two-fluid Euler equations become 
\bea
    0 &=& K^\x_{a b} A^b_\x + K^{\x \y}_{a b} A^b_\y \ , \label{pereqns1} \\
    0 &=& K^\y_{a b} A^b_\y + K^{\y \x}_{a b} A^b_\x \ , \label{pereqns2}
\eea
where the ``dispersion'' tensors are
\bea
    K^\x_{a b} &=& n^c_\x \left(k_{[c} \BX_{a]b} + k_{[c} \mathcal{A}^\x_{a]b}
                   \right) \ , \cr
    K^{\x \y}_{a b} &=& n^c_\x \left(k_{[c} \Xcal^{\x \y}_{a]b} + 
                        k_{[c}\mathcal{A}^{\x \y}_{a]b}\right) \ . 
                        \label{disptens}
\eea
Note that $K^\y_{a b}$ and $K^{\y \x}_{a b}$ are obtained via the interchange of 
$\x \leftrightarrow \y$ in \eqref{disptens}. 

The general solution to \eqref{pereqns2} requires, say, using Eq.~\eqref{pereqns2} to determine 
$A^a_\y$, and then substitute that into Eq.~\eqref{pereqns1}. This means we need the four 
inverses
\beq
    \tilde{K}^{a c}_\x K^\x_{c b} = \delta^a{}_c
         \quad , \quad 
    \tilde{K}^{a c}_{\y \x}  K^{\x \y}_{c b} = \delta^a{}_c \ .
\eeq
With these in hand, we can write 
\beq
    0 = \left(\tilde{K}^{a c}_\y K^{\y \x}_{c b} - \tilde{K}^{a c}_{\y \x} 
        K^\x_{c b}\right) A^b_\x \equiv {\cal M}^a{}_b A^b_\x \ .
\eeq
Having a non-trivial solution requires that $k_a$ be such that $\det \mathcal M^a_{\ b} = 0 $. 
However, the examples which follow will be kept simple enough that the general procedure will not be required. For example, we will focus on the case of aligned flows. 

\cite{samuelsson10:_rel2st} have shown that the relative flow between the 
two fluids enters through the inner product $\hat{v}^a_{\x \y} \hat{k}^\x_a$ (where 
$\hat{v}^a_{\x \y} = v^a_{\x \y}/v_{\x \y}$), and so it is natural to introduce the angle 
$\theta_{\x \y}$ between the two vectors. This means that, the inner product becomes
\beq
    \hat{v}^a_{\x \y} \hat{k}^\x_a = \cos \theta_{\x \y} \ .
\eeq 
Having an aligned flow means, say, setting $\theta_{\x \y} = 0$ and $\theta_{\y \x} = \pi$. The 
wave vector takes the form
\beq
      k^a = \frac{1}{\gamma_{\x \y} v_{\x \y}} \left(k_\x u^a_\y - k_\y u^a_\x\right) \ , \label{alignk}
\eeq
and the flux conservation becomes
\beq
      k_\x u_a^\y A^a_\x = k_\y u_a^\x A^a_\x \ . \label{fluxconred}
\eeq
This, in turn, implies that the problem is reduced from four equations with four unknowns to a much simpler $2\times2$ system. Finally, we note that Eqs.~\eqref{kequal} and \eqref{sigtrans} imply, respectively,
\beq
       \frac{k_\y}{k_\x} = \sqrt{\frac{1- \sigma^2_\x}{1 - \sigma^2_\y}} 
\eeq
and 
\beq
      \sigma_\y = \frac{\sigma_\x - v_{\x \y}}{1 - v_{\x \y} \sigma_\x} \ .
\eeq
It will prove useful later to note that this last result implies
\beq
      1 - \sigma^2_\y = \frac{1}{\gamma^2_{\x \y}} \frac{1 - \sigma^2_\x}{\left(1 - v_{\x \y}
      \sigma_\x\right)^2} 
\eeq
and therefore
\beq
       \frac{k_\y}{k_\x} = \gamma_{\x \y} \sqrt{\left(1- v_{\x \y} \sigma_\x\right)^2} \ .
\eeq

Another place where we will simplify the analysis is the choice of equation of state; namely, 
to consider forms with just enough complexity in the $\BX_{a b}$, $\mathcal{A}^\x_{a b}$, 
$\Xcal^{\x \y}_{a b}$, and $\mathcal{A}^{\x \y}_{a b}$ coefficients to establish the main feature we are interested in: the two-stream 
instability. Obviously, any fluid must have non-zero bulk properties; the other two properties of 
entrainment and cross-constituent coupling depend on the particular features of the fluid system 
incorporated into the equation of state. We will first consider the case where only bulk features 
are present and then follow this up by incorporating entrainment. 


Let us first set both the entrainment and cross-constituent coupling to zero. This implies 
$K^{\x \y}_{a b} = 0$ and the mode equations are
\bea
    0 &=& K^\x_{a b} A^b_\x = - \frac{1}{2} \BX n_\x k_\x \left(\sigma_\x \perp^\x_{a b} + c^2_\x 
    \hat{k}^\x_a u^\x_b\right) A^b_\x \ , \cr
    0 &=& K^\y_{a b} A^b_\y = - \frac{1}{2} \BY n_\y k_\y \left(\sigma_\y \perp^\y_{a b} + c^2_\y 
    \hat{k}^\y_a u^\y_b\right) A^b_\y \ .
\eea
We contract each mode equation with $k_a$ to find
\beq
    0 = \left(\sigma^2_\x - c^2_\x\right) A^\x_{||} \ , \quad 
    0 = \left(\sigma^2_\y - c^2_\y\right) A^\y_{||} \ ,
\eeq
and the solution reduces to the $2\times2$ matrix problem
\beq
    \left[\begin{array}{cc}
    \left(\sigma^2_\x - c^2_\x\right) &  0 \\
    0 & \left(\sigma^2_\y - c^2_\y\right)
    \end{array}\right] 
    \left[\begin{array}{c}
    A^\x_{||} \\
    A^\y_{||}
    \end{array}\right] = 
    \left[\begin{array}{c} 0 \\ 0 \end{array}\right]
    \ ,
\eeq
and it is easy to see that the resulting dispersion relation is  
\beq
    \left(\sigma^2_\x - c^2_\x\right) \left(\sigma^2_\y - c^2_\y\right) = 0 
    \ . \label{freedisp}
\eeq
The modes of this system are the ``bare'' sound waves with speeds $c_{\x}$ or $c_\y$, as one would have expected. There are no interactions between the two fluids and so there is no sense in 
which they ``see'' each other. Generally, we conclude that the existence of a two-stream 
instability requires more than just a background relative flow. Some coupling agent is required.

With this in mind, we include coupling via entrainment. As we are ignoring a cross-constituent coupling term we 
still have $\Xcalab = 0$. The simplest inclusion of entrainment is to set $\BX_{, \x \y} = 0$ and 
$\AXY_{, \x \y} = 0$. This means $\Acalxab = 0$, $\Acalxyab = \AXY g_{a b}$, and therefore
\bea
   K^\x_{a b} &=& - \frac{1}{2} \BX n_\x k_\x \left(\sigma_\x \perp^\x_{a b} + c^2_\x 
    \hat{k}^\x_a u^\x_b\right) \ , \cr
   K^{\x \y}_{a b} &=& - \frac{1}{2} \AXY n_\x k_\x \left(\sigma_\x \perp^\x_{a b} +  
    \hat{k}^\x_a u^\x_b\right) \ . 
\eea
The mode equations then become
\bea
    0 &=& \BX \left(\sigma_\x \perp^\x_{a b} + c^2_\x \hat{k}^\x_a u^\x_b\right) A^b_\x 
              + \AXY \left(\sigma_\x \perp^\x_{a b} + \hat{k}^\x_a u^\x_b\right) A^b_\y \ , \cr
    0 &=& \BY \left(\sigma_\y \perp^\y_{a b} + c^2_\y \hat{k}^\y_a u^\y_b\right) A^b_\y 
              + \AXY \left(\sigma_\y \perp^\y_{a b} + \hat{k}^\y_a u^\y_b\right) A^b_\x \ .
\eea
By contracting each with $k_a$, using Eqs.~\eqref{alignk} and \eqref{fluxconred}, we get
\begin{multline}
    0 = \frac{1}{k_\x}\left\{\BX \left(\sigma_\x \perp^\x_{a b} k^a + c^2_\x k_\x u^\x_b\right) 
              A^b_\x \right. \\
              \left. + \AXY \left[\sigma_\x k_a + k_\x \left(1 - \sigma^2_\x\right) u^\x_a\right] 
              A^a_\y\right\} \\
      = \BX \left[\frac{\sigma_\x}{\gamma_{\x \y} v_{\x \y}} \left(\frac{k_\y}{k_\x} - \gamma_{\x \y}
              \right) + c^2_\x\right] u^\x_a A^a_\x + \AXY \left(1 - \sigma^2_\x\right) \frac{k_\x}{k_\y} 
              u^\y_a A^a_\y \ , 
\end{multline}
\be
   0 = \BY \left[\frac{\sigma_\y}{\gamma_{\x \y} v_{\x \y}} \left(\frac{k_\x}{k_\y} - \gamma_{\x \y}
              \right) + c^2_\y\right] u^\y_a A^a_\y + \AXY \left(1 - \sigma^2_\y\right) \frac{k_\y}{k_\x} 
              u^\x_a A^a_\x \ .
\ee
The dispersion relation now becomes 
\beq
    0 = \left(\sigma^2_\x - c^2_\x\right) \left(\sigma^2_\y - c^2_\y\right) - \left(
          \frac{\AXY}{\sqrt{\BX \BY}}\right)^2 \left(1 - \sigma^2_\x\right) \left(1 - \sigma^2_\y\right)
           \ .
\eeq
This can be rewritten in a form more useful for numerical solutions; namely,
\beq
    0 = \left(x^2 - b^2\right) \left[\left(x - y\right)^2 - \left(1 - c_\y^2 y x\right)^2\right] -  
          a^2 \frac{\left(1 - c^2_\y x^2\right)^2}{\gamma^2_{\x \y}} \ ,
\eeq
where $x = \sigma_\x/c_\y$, $y = v_{\x \y}/c_\y$, $b = c_\x/c_\y$. and
\beq
       a^2 = \left(\frac{\AXY}{c^2_\y \sqrt{\BX \BY}}\right)^2 \ .
\eeq

The immediate thing to note is that the relative speed changes the equation from a 
quadratic in $\sigma^2_\x$ to being fully quartic in $\sigma_\x$; thus, it is inevitable that complex 
solutions will result. The question is if the imaginary contributions can be realized for physical 
parameters. Recall that this means the system must exhibit absolute stability and causality. 
\cite{samuelsson10:_rel2st} have shown that these are guaranteed when
\beq
      0 \leq \left(\frac{\AXY}{\sqrt{\BX \BY}}\right)^2 \leq c^2_\x c^2_\y 
               \quad \Longrightarrow \quad 
       a^2 \leq b^2 \ . \label{abcons}
\eeq 

In the Newtonian limit the dispersion relation takes the same mathematical form for entrainment 
as it does for non-zero cross-constituent coupling; namely,
\beq
    \frac{\left(x^2 - b^2\right)}{a^2} \left[\left(x - y\right)^2 - 1\right] = 1 \ .
\eeq
As this is quartic in $x$, the exact solutions are known. However, they are quite tedious and 
their main use is to serve as the basis for numerical evaluations of the modes. A basic algorithm would be to fix $a$ and 
$b$, subject to the constraint in Eq.~\eqref{abcons}, and then evaluate the real and imaginary 
parts of $\sigma_\x$ as functions of $y$. The end result of this process is to reveal that the 
instability exists in a ``window'' of $y$-values
\citep{acp03:_twostream_prl,andersson04:_twostream,samuelsson10:_rel2st}. As an illustration we may consider the example from \cite{andersson04:_twostream}, illustrated in Fig.~\ref{2stream}.
A more recent study \citep{2019arXiv190804275A}, in the framework of relativity, highlights the fact that the system will be prone to an energy instability (closely related to the CFS instability from Sect.~\ref{sec:cfs}, as it sets in at the point where originally backwards moving modes are dragged forwards by the background flow). As indicated by the left panel of Fig.~\ref{2stream} this energy instability tends to set in before the system suffers the (dynamical) two-stream instability.

\begin{figure}[htb]
    \centerline{\includegraphics[width=\textwidth]{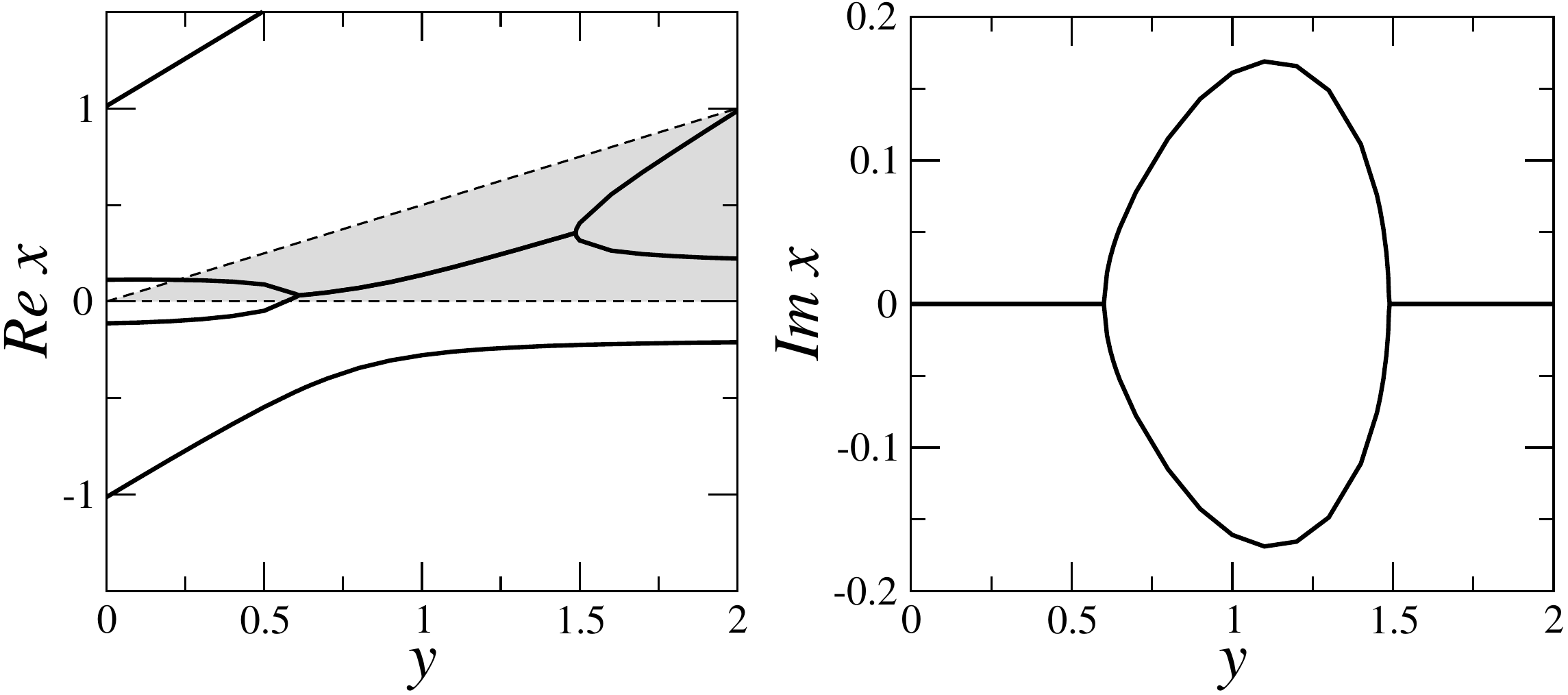}}
    \caption{An illustration of the two-stream instability, showing the real (left panel) and imaginary (right panel) parts of the four roots of the 
dispersion relation for  the model parameters
($a^2=0.0249$ and $b^2=0.0379$) used in \cite{andersson04:_twostream}. For these parameters the quartic dispersion relation 
has four real roots for both
$y=0$ and $y=2$, while it has two real roots and a complex
conjugate pair for $y$ in the range $0.6<y<1.5$. In this range, the 
two-stream instability is active. (Reproduced from \citealt{andersson04:_twostream}.)}
    \label{2stream}
\end{figure}

Finally, let us take the opportunity to note that the relativistic two-stream instability has also been analyzed in the non-linear regime \citep{Hawke13:_non_lin_2str}. This  first nonlinear numerical 
simulation of the effect in relativistic multi-species hydrodynamical systems shows that the 
onset and initial growth of the instability match closely  the results of linear perturbation 
theory. But, in the later stages of the evolution, the linear and nonlinear description have only 
qualitative overlaps. The main conclusion is that the instability does not saturate in the nonlinear 
regime by purely ideal hydrodynamic effects.


\section{Numerical simulations: fluid dynamics in a live spacetime}
\label{sec:numsim}

Many  astrophysical phenomena involve violent nonlinear matter dynamics. Such systems cannot (meaningfully) be described within perturbation theory. Instead, the modelling requires fully nonlinear---and multi-dimensional, given the lack of symmetry of (say) turbulent flows---simulations, taking into account the live spacetime of General Relativity. The last decades have seen considerable progress in the development of the relevant computational tools, especially for gravitational-wave sources like supernova core collapse \citep{SNreview} and neutron star mergers \citep{2017RPPh...80i6901B}. The state-of-the-art technology includes the consideration of fairly sophisticated matter models. In the case of supernova modelling, neutrinos are expected to play an important role in triggering the explosion \citep{Janka2012} and the role of magnetic fields may also be significant \citep{moestanature}. Meanwhile, for neutron star mergers, finite temperature effects are central as shock heating ramps up the temperature of the merged object to levels beyond that expected even during core collapse (see, e.g., \citealt{Bauswein2010} or \citealt{Kastaun2015}). Magnetic fields are expected to have decisive impact on the post-merger dynamics are likely to leave an observational signature, e.g., in terms of short gamma-ray bursts (e.g., \citealt{Kumar2015}).

\subsection{Spacetime foliation}

 We have already explored some aspects of the problem (like the thermodynamics and the matter equation of state, see Sect.~\ref{sec:thermo})  and we have considered features that arise in models of increasing complexity (in particular when we need to account for the relative flow of distinct fluid components). So far, the discussion has assumed a fibration of spacetime  associated with a family of fluid observers. This approach is natural if one is mainly interested in the local fluid dynamics (e.g., wave propagation) and it also leads to the 1+3 formulation often used in cosmology (where ``clocks'' associated with the fluid observers define the notion of cosmic time), see \cite{tsagas} for a relevant discussion. The strategy is, however, not natural for  numerical  simulations with a live spacetime. Instead, most such work makes use of a 3+1 spacetime foliation (see \citealt{baum} for a relevant discussion), where progression towards the ``future'' is associated with a set of Eulerian observers. Hence, we need to understand how we extend the multifluid model from fibration to foliation.

The standard approach to numerical simulations takes as its starting point a ``foliation''of spacetime into a family of spacelike hypersurfaces, $\Sigma_t$, which arise as level surfaces of a scalar time $t$ (see, e.g., \citealt{Alcubierre:2008}). Given the normal to this surface
\be
N_a = - \alpha \nabla_ a t \ , 
\label{normal}
\ee
where the function $\alpha$ is known as the lapse,
we have
\be
N_a = (-\alpha,0,0,0) \ ,
\ee
and the normalisation $N_a N^a=-1$ (we are thinking of the normal as associated with an observer moving through spacetime in the usual way) leads to $\alpha^2 = -1/g^{tt}$.
The sign in \eqref{normal}  ensures that time flows into the future.  The dual to $\nabla_a t$ leads to a time vector
\be
t^a = \alpha N^a + \beta^a \ ,
\ee
where the so-called shift vector $\beta^a$ is spatial, in the sense that $N_a \beta^a = 0$. It follows that
\be
N^a = \alpha^{-1} ( 1,-\beta^i) \ ,
\ee
and the spacetime can be written in the  Arnowitt--Deser--Misner (ADM) form \citep{2008GReGr..40.1997A,1979sgrr.work...83Y}:
\be
ds^2 = - \alpha^2 dt^2 + \gamma_{ij} \left( dx^i + \beta^i dt \right)   \left( dx^j + \beta^j dt \right) \ ,
\label{adm}
\ee
where the (induced) metric on the spacelike hypersurface is
\be
\gamma_{ab} = g_{ab} + N_a N_b \ .
\ee
Note that $\gamma^a_b$ represents the projection orthogonal to $N_a$ and that $\gamma_{ab}$ and its inverse can be used to raise and lower indices of purely spatial tensors. For example, we have $\beta_i = \gamma_{ij} \beta^j$.

In essence, the lapse $\alpha$ determines the rate at which proper time advances from one time slice to the next, along the normal  $N_a$, while the vector $\beta^i$ determines how the coordinates  shift from one spatial slice to the next. This is illustrated in Fig.~\ref{foliate}. The two functions encode the coordinate freedom of General Relativity.

\begin{figure}[htb]
    \centerline{\includegraphics[width=0.7\textwidth]{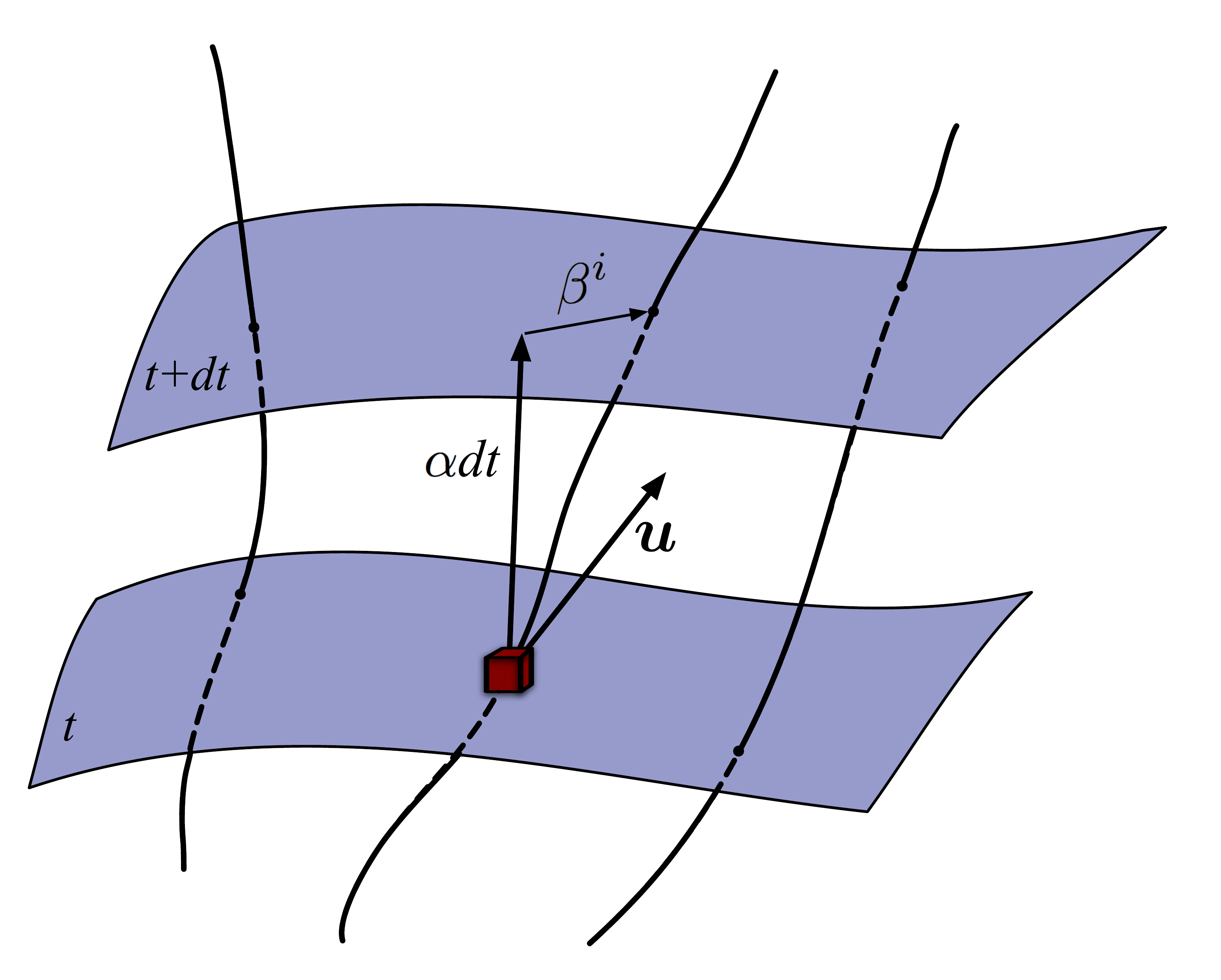}}
    \caption{An illustration of the two formulations for the relativistic fluid problem. The fibration approach, which focuses on the worldline associated with a given fluid element (and a four velocity $\boldsymbol u$ with components $u^a$), provides a natural description of the microphysics and issues relating to thermodynamics. Meanwhile, a spacetime foliation, based on the use of spatial slices and normal observers (with the coordinate freedom encoded in the lapse $\alpha$ and the shift vector $\beta^i$), is typically used in  numerical simulations. In order to ensure that the local physics is appropriately implemented in simulations, we need to understand the translation between the two descriptions.}
    \label{foliate}
\end{figure}

Reading off the metric from the line element, we have
\be
g_{ab} = \left(\begin{array}{cc} -\alpha^2 + \beta_i \beta^i & \beta_i \\ \beta_i & \gamma_{ij} \end{array} \right) \ ,
\ee
with inverse 
\be
g^{ab} = \left(\begin{array}{cc} -1/\alpha^2 & \beta^i/\alpha^2 \\ \beta^i/\alpha^2 & \gamma^{ij}-\beta^i \beta^j /\alpha^2 \end{array} \right) \ .
\ee

Having specified the spacetime foliation, we can decompose any tensor  into time and space components (adapting the logic from the discussion of the stress-energy tensor in Sect.~\ref{tab1}). Suppose, for example, that we have a fluid associated with a four velocity $u^a$. Then we can introduce the decomposition\footnote{In order to make the distinction clear, we are using the convention that all velocities measured by the Eulerian observer have hats, while velocities relative to the fluid frame do not.}
\be
u^a = W (N^a + \hat v^a)  = {W\over \alpha} \left( t^a - \beta^a + \alpha\hat  v^a \right) \ , 
\label{fluidobs}
\ee
where $N_a \hat v^a =0$ and the Lorentz factor is given by
\be
W= - N_a u^a =  \alpha u^t = (1-\hat v^2)^{-1/2} \ ,
\ee
where $\hat v^2 = \gamma_{ij} \hat v^i \hat v^j$ and
the last equality follows from $u^a u_a=-1$, as usual.
From this relation it is easy to see that
\be
\hat v^t = 0 \ , \qquad \hat v^i = {u^i\over W} - N^i = {1\over \alpha} \left( {u^i \over u^t} + \beta^i\right) \ ,
\ee
and it then follows that
\be
\hat v_t = g_{ta} \hat v^a = \beta_i \hat v^i \ , \qquad \hat v_i = \gamma_{ia} \hat v^a = {\gamma_{ij} \over \alpha} \left( {u^j\over u^t} + \beta^j \right) \ .
\ee

We also need to consider derivatives. First of all, we introduce a derivative associated with the hypersurface. Thus, we use the (totally) projected derivative
\be
D_a = \gamma_a^b \nabla_b \ ,
\ee
where all free indices should be projected into the surface.
This derivative is compatible with the spatial metric (see  Sect.~\ref{sec:gr}) in the sense that
\be
D_a\gamma_{bc} = \gamma_a^d \gamma_b^e \gamma_c^f\nabla_d \gamma_{ef} = 0 \ ,
\ee
which means that it acts as a covariant derivative in the surface orthogonal to $N^a$. The upshot of this is that we can construct a tensor algebra for the three-dimensional spatial slices.
In particular, we can introduce a three-dimensional Riemann tensor. This projected Riemann tensor does not contain all the information from its four-dimensional cousin; the missing information is encoded in the extrinsic curvature, $K_{ab}$. This is a symmetric spatial tensor, such that $N^a K_{ab}=0$. The extrinsic curvature provides a measure of how the $\Sigma_t$ surfaces curve relative to spacetime. In practice, we measure how the normal $N_a$ changes as it is parallel transported along the hypersurface. That is,  we define\footnote{Note that it follows from the definition of $N_a$ in terms of the lapse (and the projections) that $K_{ac}$ is symmetric. The symmetry is also evident from \eqref{lieK}.}
\be
K_{ac} = -D_a N_c = - \gamma_a^b \gamma_c^d \nabla_b N_d = - \nabla_a N_c - N_a (N^b\nabla_b N_c) \ ,
\label{Kdef}
\ee
where the second term is analogous to the fluid four-acceleration. We also have
\be
K= K^a_a = g^{ab}K_{ab} = - \gamma^{ab} D_a N_b = - \nabla_a N^a \ .
\ee
Alternatively, we can use the properties of the Lie derivative to show that
\be
K_{ij} = - {1 \over 2}\mathcal L_N \gamma_{ij} \ ,
\label{lieK}
\ee
but since
\be
\mathcal L_N = {1\over \alpha} ( \mathcal L_t - \mathcal L_\beta) = {1\over \alpha } ( \partial_t - \mathcal L_\beta) \ ,
\label{tder}
\ee
we have
\be
\partial_t \gamma_{ij} = - 2\alpha K_{ij} + \mathcal L_\beta \gamma_{ij} \ .
\ee
From the trace of this expression we get
\be
\alpha K = - \partial_t \ln \gamma^{1/2} + D_i \beta^i \ ,
\ee
where $\gamma=g^{ab}\gamma_{ab}$ and $\gamma^{ij} \partial_t \gamma_{ij} = \partial_t \ln \gamma$.

\subsection{Perfect fluids}

The spacetime foliation  provides us with  the tools we need to formulate relativistic fluid dynamics in a way suitable for numerical simulations (compatible with the solution of the Einstein field equations for the spacetime metric, which needs to be carried out in parallel; \citealt{Alcubierre:2008,baumgarte2010numerical}). However, our immediate focus is on the equations of fluid dynamics (see \citealt{Fontlivrev} for more details).

 Let us start with the simple case of baryon number conservation. That is, we assume the flux $n u^a$ is conserved, where $n$ is the baryon number density according to an observer moving along with the fluid. Thus, we have
\be
\nabla_a (n u^a) = \nabla_a [ Wn (N^a + \hat v^a) ]= 0 \ .
\ee
First we note that the particle number density measured by the Eulerian observer is
\be
\hat n =-N_a n u^a = nW \ ,
\ee
so we have
\be
N^a \nabla_a \hat n + \nabla_i (\hat n \hat v^i) = - \hat n \nabla_a N^a =  \hat n K \ ,
\ee
(since $\hat v^i$ is spatial). Making use of the Lie derivative and \eqref{tder} this can be written
\be
N^a \nabla_a \hat n = \mathcal L_N \hat n = {1\over \alpha} ( \partial_t - \mathcal L_\beta) \hat n = - \nabla_i (\hat n \hat v^i) + \hat n K
\ ,
\ee
or
\be
\partial_t \hat n + (\alpha \hat v^i - \beta^i )\nabla_i \hat n + \alpha \hat n \nabla_i \hat v^i = \alpha  \hat n K \ .
\ee
Finally, since $\hat v^i$ and $\beta^i$ are already spatial, we have
\be
\partial_t \hat n + (\alpha \hat v^i - \beta^i )D_i \hat n + \alpha \hat n D_i \hat v^i = \alpha  \hat n K =- \hat n   \partial_t \ln \gamma^{1/2} + \hat n D_i \beta^i \ ,
\ee
or
\be
\partial_t \left( \gamma^{1/2} \hat n\right) +  D_i \left[ \gamma^{1/2}\hat n (\alpha \hat v^i - \beta^i )\right] = 0 \ ,
\label{baryons}
\ee
This simply represents the advection of the baryons along the flow, as seen by  an Eulerian observer. In arriving at this result, we have used the fact that
\be
\left( -g\right)^{1/2} = \alpha \gamma^{1/2} \ ,
\ee
so
\be
 \nabla_a (-g)^{1/2} = \nabla_a ( \alpha \gamma^{1/2}) = 0 \ .
\ee
For future reference, it is also worth noting that   
\be
D_i \gamma^{1/2} = \partial_i \gamma^{1/2} - \Gamma^j_{ji} \gamma^{1/2} = 0 \ ,
\ee
where the Christoffel symbol is the one associated with the covariant derivative in the hypersurface.

\vspace*{0.1cm}
\begin{tcolorbox}
\textbf{Comment:} As a slight aside, we have expressed \eqref{baryons} in the usual flux-conservative form. However, in some situations it may be useful to pay closer attention to the local physics experienced by a family of observers that ride along with the fluid (e.g., when we consider the microphysics). Then we have (at least) two alternatives. We can choose to describe the physics in a local fluid frame associated with the four velocity $u^a$ (as we have done) or we can try to make the equations look ``similar''  to the more familiar flat space (Newtonian) ones. In this latter approach [see for example Thorne+Macdonald] one would introduce a global time (associate with $t^a$) and use a spatial tetrad (relative to this time coordinate) to describe the fluid. In essence,  the fluid  then has four velocity 
$$
u^a = {\gamma \over \alpha} \left( t^a +  V^a \right)  \ .
$$
Comparing to \eqref{fluidobs} we have $\gamma=W$ and 
\be
V^i = \alpha \hat v^i - \beta^i  \ .
\label{fluidframe}
\ee
Making use of this result, we can rewrite \eqref{baryons} as
$$
\left( \partial_t + \mathcal L_V \right) \left( \gamma^{1/2} \hat n \right) + \gamma^{1/2} \hat n D_i V^i = 0 \ , 
$$
or, if we define $\bar n = \gamma^{1/2} \hat n$, 
$$
\partial_t \bar n + D_i \left( \bar n V^i \right) = 0 \ .
$$
\end{tcolorbox}
\vspace*{0.1cm}

Moving on, the fluid equations of motion follow from $\nabla_a T^{ab}=0$, where we recall that a perfect fluid is described by the stress-energy tensor
\be
T^{ab} = (p+\varepsilon) u^a u^b + p g^{ab} \ .
\ee
Here $p$ and $\varepsilon$ are the pressure and the energy density, respectively. As discussed in Sect.~\ref{sec:thermo} these quantities are related by the equation of state, which encodes the relevant microphysics. In order to make contact with this discussion, a numerical simulation must allow us to extract these quantities from the evolved variables.

However, a numerical simulation is naturally carried out using quantities measured by the Eulerian observer. That is, we decompose the stress-energy tensor into normal and spatial parts as (again, see the discussion in Sect.~\ref{tab1})
\be
T^{ab} = \rho N^a N^b + 2 N^{(a} S^{b)} + S^{ab} \ ,
\label{TM}
\ee
with (noting the conflict in notation from the discussion in Sect.~\ref{sec:numsim}, where $\rho$ represented the mass density)
\be
 \rho = N_a N_b T^{ab} =  \varepsilon W^2 - p \left( 1 - W^2\right) \ ,
\ee 
\be
S^i = - \gamma^i_c N_d T^{cd} =  \left(p+\varepsilon \right) W^2 \hat v^i \ ,
\ee
and
\be
S^{ij} = \gamma^i_c \gamma^j_d T^{cd} = p \gamma^{ij}  + \left( p +\varepsilon \right) W^2  \hat v^i \hat v^j \ .
\ee

A projection of the equations of motion along $N_a$ then leads to the energy equation. From
\be
N^a \nabla_a   \rho +   \rho \nabla_a N^a + \nabla_ a S^a - N_b N^a \nabla_a S^b - N_b \nabla_a S^{ab} =   0 \ ,
\ee
we get
\be
N^a \nabla_a  \rho + \nabla_a S^a=  \rho K-S^b N^a\nabla_a N_b - S^{ab}\nabla_a N_b \ ,
\ee
 where we have used
\be
N^a\nabla_a N_b = D_b \ln \alpha
\ee
We also have
\be
{1\over \alpha} \left( \partial_t - \mathcal L_\beta\right)   \rho + \nabla_a S^a=  \rho K-S^b D_b \ln \alpha  + S^{ab}K_{ab} \ ,
\ee
leading to
\be
\partial_t  \left(\gamma^{1/2}  \rho\right) + D_i \left[ \gamma^{1/2} \left( \alpha S^i -\rho \beta^i\right) \right]  =  \gamma^{1/2} \left( \alpha S^{ij}K_{ij} -S^i D_i  \alpha 
\right) \ .
\label{energyq}
\ee

\vspace*{0.1cm}
\begin{tcolorbox}
\textbf{Comment:} It is common to evolve $\tau = \rho - m_0 \hat n$ (where $m_0$ is the  mean baryon rest mass density) rather than $\rho$. This is done to avoid numerical issues arising from the fact that \eqref{energyq}
matches (to leading order in velocity) the evolution equation for the conserved proper rest-mass density [$m_0$ times \eqref{baryons}]. This  has no impact on the formal discussion here, but it is nevertheless an important point.
Note also that, one may opt to evolve the entropy instead of the energy. Indeed, in the multifluid formalism it is natural to focus on the entropy and  it is easy to show that the energy equation leads directly to an advection equation for the entropy. However, the energy equation is typically preferred in numerical simulations as its balance law form is compatible with conservative evolution schemes and ensures suitable behaviour when shocks appear \citep{lrr_font}. 
\end{tcolorbox}
\vspace*{0.1cm}

Turning to the momentum equation, which is obtained by a projection orthogonal to $N_a$, we have
\be
 \rho N^a \nabla_a N^c + \gamma^c_{\ b}N^a \nabla_a S^b + S^c \nabla_a N^a + S^a \nabla_a N^c + \gamma^c_{\ b} \nabla_a S^{ab}
=0 \ ,
\ee
which leads to
\be
\left( \partial_t - \mathcal L_\beta\right) S_i - S^j \left( \partial_t - \mathcal L_\beta\right)\gamma_{ij} - \alpha K S_i + \rho D_i \alpha
+ \alpha \gamma_{ij} D_k S^{kj} = 0 \ ,
\ee
where we have used
\be
N^a \nabla_a S^c = \mathcal L_N S^c + S^a \nabla_a N^c = \mathcal L_N S^c - S^a K_a^c \ .
\ee
This leads to the final result
\be
\partial_t (\gamma^{1/2} S_i) + D_j \left[ \gamma^{1/2} \left( \alpha S_i^j -S_i \beta^j \right) \right] = \gamma^{1/2} \left( S_j D_i \beta^j - \rho D_i \alpha \right) \ .
\label{momentum}
\ee
This completes the set of equations we need in order to carry out a perfect fluid simulation. The extension to more general setting follows, at least formally, the same steps. 

\subsection{Conservative to primitive}

We have written down the set of evolution equations we need for a single-component problem. This leaves us with one important issue to resolve. How do we connect the evolution to the underlying microphysics and the equation of state? In order to do this, we have to consider the inversion from the variables used in the evolution to the ``primitive'' fluid variables associated with the equation of state. 

Let us, in the interest of conceptual clarity, focus on the case of a cold barotropic fluid, such that the equation of state provides the energy as a function of the baryon number density $\varepsilon = \varepsilon(n)$ (see Sect.~\ref{sec:thermo}). This then leads to the chemical potential
\be
\mu = {d\varepsilon \over dn} \ ,
\ee
and the pressure $p$ follows from the thermodynamical relation:
\be
p = n \mu - \varepsilon \ .
\label{pdef}
\ee
We see that, in order to connect with the thermodynamics we need the evolved number density. We also need to decide which observer measures  equation of state quantities. In the single-fluid case this question is relatively easy to answer; we need to express the equation of state in the fluid frame (use the fibration associated with $u^a$). 

In the simple case we consider here the  evolved system, \eqref{baryons} and \eqref{momentum}, provides (assuming that $\gamma^{1/2}$ is known from the evolution of the Einstein equations)
\be
\hat n = nW = n (1-\hat v^2)^{-1/2} \ , 
\label{one}
\ee
and
\be
S^i = (p+\varepsilon) W^2 \hat v^i \ .
\label{twop}
\ee
We need to invert these two relations to extract the primitive variables, $n$ and $\hat v^i$. This can be formulated as a one-dimensional root-finding problem. For example, we may start by guessing a value for $n=\bar n$. This then allows us to work out $\varepsilon$ from the equation of state and $p$ from \eqref{pdef}. With these variables in hand we can solve
\be
{S^2 \over (p+ \varepsilon)^2} =  W^4 \hat v^2 \ ,  \quad \mbox{with} \quad S^2 = \gamma_{ij}S^i S^j \ ,
\ee
for $\hat v^2$.  This, in turn, allows us to work out the Lorentz factor $W$ and then $\hat v^i$ follows from \eqref{twop}. Finally, we get
$n=\hat n/W$ from \eqref{one}. The result  can be compared to our initial guess $\bar n$. Iterating the procedure  gives a  solution consistent with the conserved quantities, and hence all primitive quantities.

Unfortunately, the numerical implementation of this strategy may not be as straightforward as it sounds. For example, the result may be sensitive to the initial guess and the algorithm may not converge. This is particularly true for more complex situations (e.g., multi-parameter equations of state or problems involving magnetic fields; \citealt{lrr_font,2013PhRvD..88d4020D}). However, our aim here is not to resolve the possible numerical issues. We are only outlining the logic of the approach.

\subsection{The state of the art}

Without attempting an exhaustive survey of the relevant literature, it is useful to provide comments on the current state of the art along with suggestions for further reading. The area of numerical simulations of general relativistic fluids is developing rapidly, stimulated by the breakthrough discoveries in gravitational-wave astronomy---in particular, the astonishing GW170817 neutron star binary merger event \citep{2017ApJ...848L..12A,2017ApJ...848L..13A}, observations of which engaged a large fraction of the global astronomy community.

Focus on nonlinear simulations with a live spacetime, one may identify (at least) four (more or less) separate bodies of work:
\begin{itemize}
    \item First of all, numerical simulations have been used to explore the problem of instabilities in rotating stars and disks. This is a classic problem in applied mathematics/fluid dynamics, where perturbative studies may be used to establish the existence of an instability (for simpler models) but where numerical simulations are required for a higher level of realism and also to investigate the nonlinear evolution of an unstable system (to what extent the nonlinear coupling of different oscillation models leads to an instability saturating at some level, etcetera). The archetypal problems---basically because they involve instabilities that grow sufficiently rapidly that they can be tracked by (expensive) multi-dimensional simulations---are the bar-mode instability of (rapidly and differentially) rotating stars \citep{1985ApJ...298..220T,1987ApJ...315..594W,2000PhRvD..62f4019N,2000ApJ...542..453S,2007PhRvD..75d4023B} and the run-away instability of (thick) accretion disks \citep{2003MNRAS.341..832Z}.
    \item A second setting that has been explored since the early days of numerical relativity \citep{1985PhRvL..55..891S,1986NYASA.470..247P} involve the gravitational collapse to form a black hole \citep{2005PhRvD..71b4035B,2007CQGra..24S.139O,2011PhRvL.106p1103O}. The typical collapse time-scale is short enough that these simulations can be carried out without extortionate cost, but the problem involves a number of complicating issues relating to the formation of the black-hole horizon.  The typical set-up involves initial data representing a stable fluid body from which pressure support is artificially removed to trigger the collapse. The main conclusion drawn from this body of work may be that the gravitational-wave signal from collapse and black-hole formation tends to be dominated by quasinormal mode ringing. 
    \item Realistic modelling of the core-collapse of star that reaches the endpoint of its main-sequence life is exceedingly complicated \citep{2007PhR...442...38J,2018ApJ...861...10M}. The problem involves complex physics and a vast range of scales that need to be accurately tracked in a simulation. In spite of the challenges, there has been huge progress on understanding the problem in the last two decades. From the fluid dynamics point of view, the  main developments  involve the implementation of a (more) realistic matter description (based on nuclear physics and accounting for thermal effect; \citealt{2017PhRvD..95f3019R}) and developments towards an accurate implementation of neutrinos \citep{2016ApJ...831...98R,2017MNRAS.468.2032A,2019ApJ...873...45G,2020EPJA...56...15E}. The latter is crucial, as the  neutrinos  are thought to be necessary to trigger the supernova explosion. 
    \item The final problem setting---attracting a lot of interest at the present time \citep{2017RPPh...80i6901B,2020arXiv200406419B}---involves the inspiral and merger of binary neutron stars. Many of the challenges, regarding the physics, are the same as in the case of core-collapse simulations. The problem involves a vast range of scale, not so much involved with an explosion as the outflow of matter that is unbound during the merger, undergoes rapid nuclear reactions and give rise to a kilonova signal \citep{2011ApJ...738L..32G,2012PhRvD..86f3001B,2015MNRAS.450.1777K, 2018ApJ...869..130R,2019ApJ...880L..15M}. At the same time the hot merger remnant oscillates wildly \citep{2011MNRAS.418..427S,2015PhRvL.115i1101B,2016PhRvD..93l4051R} until it loses enough angular momentum (or cools enough) that it (most likely) collapses to form a black hole. An important additional complication involves the presence of magnetic fields \citep{pale}, hugely relevant as neutron star mergers are expected to be the source of observed short gamma-ray bursts \cite{2011ApJ...732L...6R,2015ApJ...806L..14P}. This connection was observationally confirmed by the GW170817 event, but numerical simulations have not yet reached the stage where the detailed engine of of these events can be explored \citep{2020arXiv200307572C}.
    
\end{itemize}



\section{Relativistic elasticity}
\label{sec:relastic}

Shortly after a neutron star is born, the outer layers freeze to form an 
elastic crust and the temperature of the high-density core drops below the level where superfluid 
and superconducting components are expected to be present. The different phases of matter 
impact on the observations in a number of ways. The crust is important as
\begin{itemize}
\item it 
anchors the star's magnetic field (and provides dissipative channels leading to the gradual field evolution; 
\cite{2013MNRAS.434..123V}), 

\item there is an immediate connection between observed quasi-periodic oscillations in the tails of magnetar flares \citep{2005ApJ...632L.111S} and the dynamics of 
the elastic nuclear lattice. An understanding of the properties of the crust is essential for 
 efforts to match the theory to observed seismology features 
\citep{2007MNRAS.374..256S,2009CQGra..26o5016S},

\item the ability of the crust to sustain elastic 
strain is key to the formation of asymmetries which may lead to detectable gravitational waves 
from a mature spinning neutron star. Continuous gravitational-wave searches with the LIGO-Virgo network of interferometers is beginning to set interesting upper limits for such signals for a 
number of known pulsars \citep{2017ApJ...839...12A}, in some instances reaching significantly 
below the expected maximum ``mountain'' size estimated from state of the art molecular 
dynamics simulations of the crustal breaking strain 
\citep{2009PhRvL.102s1102H,2013PhRvD..88d4004J}. 
\end{itemize}
In essence,  the elastic properties of the crust 
are crucial for an understanding of neutron-star phenomenology. In order for such models to reach the required level of realism we must consider the problem in the context of General Relativity. Interestingly, relativistic elasticity turns out to represent a (more or less) natural extension of the variational framework, with the key step involving the structure of matter space. 

\subsection{The matter space metric}
\label{relelas}

The modern view of elasticity \citep{1972RSPSA.331...57C,1975AnPhy..95...74C,1975ApJ...202..511C,1992JGP.....9..207K,1997RpMP...39...99K,2003CQGra..20..889B,2003RSPSA.459..109B} relies on comparing the actual matter configuration to an 
unstrained/relaxed reference shape. In order to keep track of the reference state relative to which the 
strain is measured, we introduce a positive definite and symmetric tensor field, $k_{a b}$ 
\citep{Karlovini03:_elas_ns_1}. The geometric meaning of this object is quite intuitive; it encodes the (three-)geometry of the solid (as 
seen by the solid itself). 

From the point of view of the variational framework, the tensor $k_{a b}$ is similar to $n_{a b c}$ in the sense that it is flow-line orthogonal \citep{1972RSPSA.331...57C}
\be
u^a k_{a b} = 0 \ .
\ee
The main properties of $k_{a b}$ are established by introducing the corresponding matter 
space object, $k_{A B} (= k_{B A})$, via the usual map: 
\beq
      k_{a b} = \psi^A_a \psi^B_b k_{A B} \ . \label{kAB}
\eeq
The tensor $k_{A B}$ is ``fixed'' on matter space, in 
the same sense as $n_{A B C}$, because it is (assumed to be) a function of its own matter space 
coordinates $X^A$ only. The associated volume form is $n_{A B C}$ (see the Appendix of \cite{supercrust} for a detailed discussion). 
If we introduce
\beq
      g^{A B} = \psi^A_a \psi^B_b g^{a b} = \psi^A_a \psi^B_b \perp^{a b} \ ,
\eeq
as before, 
and use Eqs.~\eqref{ndual} and \eqref{pb3form}, then we can show that
\beq
     n^2 = - g_{a b} n^a n^b = \frac{1}{3!} \det{\left(k_{A B}\right)} \det{\left(g^{A B}\right)} \ . 
     \label{n2eqkg}
\eeq
Moreover, using the relations~\eqref{DelX} and \eqref{kAB}, we can easily establish that the Lagrangian variation of 
$k_{a b}$ vanishes. That is, we have
\beq
      \delta k_{a b} = - {\mathcal L}_\xi k_{a b} \quad \Longrightarrow \quad \Delta k_{a b} = 0 \ .
      \label{Delkab}
\eeq
Finally, since $u^a \psi^A_a = 0$, and $k_{A B}$ is a function of $X^A$, we have
\beq
      {\mathcal L}_u k_{A B} = u^a \psi^C_a \frac{\partial k_{A B}}{\partial X^C} = 0 \ ,
\eeq
and it follows that
\begin{multline}
       {\mathcal L}_u k_{a b} = k_{A B} {\mathcal L}_u \left( \psi^A_a \psi^B_b \right)  
      \\
      = k_{A B} \left[u^c \frac{\partial}{\partial x^c} \left(\psi^A_a \psi^B_b\right) + \psi^A_c \psi^B_b \frac{\partial u^c}{\partial x^a} + \psi^A_a \psi^B_c \frac{\partial u^c}{\partial x^b}\right] \\
       = k_{A B} u^c \left[\frac{\partial^2 X^A}{\partial x^c \partial x^a} \psi^B_b + \psi^A_a \frac{\partial^2 X^B}{\partial x^c \partial x^b} - \frac{\partial^2 X^A}{\partial x^a \partial x^c} \psi^B_b - \psi^A_a \frac{\partial^2 X^B}{\partial x^b \partial x^c}\right] = 0 \ .
\end{multline}

Following \cite{Karlovini03:_elas_ns_1} we now introduce the matter space tensor 
$\eta_{A B}$ to quantify the 
{\em unsheared} state.  Its defining characteristic is that it is the inverse to $g^{A B}$ but only for 
the relaxed configuration (when the energy density $\varepsilon = \check{\varepsilon}$, using a check to indicate the reference shape from now on):
\beq
       g^{A C} \eta_{C B} = \delta^A_B \quad , \quad \varepsilon = \check{\varepsilon} \ . \label{etainv}
\eeq 
If we introduce
\beq
      \epsilon^{A B C} = \psi^A_a \psi^B_b \psi^C_c u_d \epsilon^{d a b c} \ ,
\eeq
then it follows from \eqref{pb3form} that
\beq
      n_{A B C} = n \epsilon_{A B C} \ .
\eeq
In other words, 
\beq
      \epsilon_{A B C} = \sqrt{\det{\left(\eta_{A B}\right)}} \left[A \ B \ C\right] \ .
\eeq
The tensor $\eta_{AB}$  is useful because it provides us with a straightforward way to model conformal elastic 
deformations. Specifically, if $f$ is the conformal factor, we let
\beq
       k_{A B} = f \eta_{A B} \quad \Longrightarrow \quad  \det{\left(k_{A B}\right)} = f^3  \det{\left(\eta_{A B}\right)} \ .
\eeq
But,
\beq
      n_{A B C} = \sqrt{ \det{\left(k_{A B}\right)}} \left[A \ B \ C\right] = n \epsilon_{A B C} = n \sqrt{\det{\left(\eta_{A B}\right)}} \left[A \ B \ C\right] \ ,
\eeq
which shows that $f = n^{2/3}$. This demonstrates that  $k$ (the determinant of $k_{a b}$) is such 
that $k = n^2$ \citep{Karlovini03:_elas_ns_1}, even though $k_{a b}$ does not itself depend on the 
number density. 

\vspace*{0.1cm}
\begin{tcolorbox}
\textbf{Comment}: It is possible to develop a framework for elasticity such that the map $\psi^A_a$ is elevated to a dynamical variable. This is, indeed, the strategy of one of the few ventures into numerical simulations of elastic materials in relativity \citep{gundhawk}. It is an interesting approach, but we will not go into the details here.
\end{tcolorbox}
\vspace*{0.1cm}

\subsection{Elastic variations}

Let us now consider the variational derivation of the equations of motion for an elastic system. 
First of all, the fact that the Lagrangian variation of $k_{a b}$ vanishes means 
that $k_{a b}$, in addition to being a natural quantity for describing the elastic configuration, is 
 useful in the development of Lagrangian perturbation theory. 
 
Letting the Lagrangian $\Lambda$ depend also on the new tensor (in essence, incorporating the 
energy associated with elastic strain) we have
\beq
    \delta \left(\sqrt{- g} \Lambda\right) = \sqrt{- g} \left[
    \mu_a \delta n^a + \left( \frac{1}{2}\Lambda g^{a b} +  
    {\partial \Lambda \over \partial g_{ab}} \right) \delta g_{a b} + {\partial \Lambda \over \partial k_{ab} }\delta k_{ab} \right] \ . \label{dlamb2}
\eeq
We proceed as in Sect.~\ref{sec:pullback} and replace $\delta n^a$ with the Lagrangian displacement $\xi^a$. In 
addition, it follows from \eqref{Delkab} that
\begin{equation}
       \delta k_{ab} = - \xi^d \nabla_d k_{ab} - k_{d b} \nabla_a \xi^d - k_{a d} \nabla_b \xi^d \ .
\end{equation}
Again ignoring surface terms, we have (as $k_{ab}$ is symmetric)
\begin{equation}
{\partial \Lambda \over \partial k_{ab} }\delta k_{ab} =  \xi^a \left[ 2 \nabla_b \left( {\partial \Lambda \over \partial k_{bd} } k_{a d}\right) - {\partial \Lambda \over \partial k_{bd} }\nabla_a k_{b d}  
\right] \ .
\end{equation}
Making use of this result, we arrive at
\beq
    \delta \left(\sqrt{- g} \Lambda\right) = \sqrt{- g} \left\{ \left[ \frac{1}{2}\left( \Lambda - n^d \mu_d\right) g^{a b} +  
    {\partial \Lambda \over \partial g_{ab}} \right] \delta g_{a b} + \tilde f_a \xi^a \right\} \ , \label{dlamb3}
\eeq
where
\begin{equation}
\tilde f_a = 2 n^b \nabla_{[a}\mu_{b]} + 2 \nabla_b \left( {\partial \Lambda \over \partial k_{bd} } k_{a d}
\right) - {\partial \Lambda \over \partial k_{bd} }\nabla_a k_{b d} = 0  \ .
\label{tforce}
\end{equation}
As in the fluid case, this result provides the equations of motion for the system. However, we  
need to do a bit of work in order to get the result into a more user-friendly form. To start with, 
we read off the stress-energy tensor from \eqref{dlamb3}:
\begin{equation}
T^{ab} =  \left( \Lambda - n^d \mu_d\right) g^{a b} + 2 {\partial \Lambda \over \partial g_{ab}} \ .
\label{stressen}
\end{equation}

The next step involves  giving physical meaning to $k_{ab}$. This involves quantifying 
the deviation of a given state from the relaxed configuration. This is where the additional matter space tensor $\eta_{A B}$ comes into play \citep{Karlovini03:_elas_ns_1}. This object depends on $n$, and relates 
directly to the relaxed state, see~\eqref{etainv}.  Its spacetime counterpart is
\beq
      \eta_{a b} = \psi^A_a \psi^B_b \eta_{A B} \ .
\eeq 
and we have already seen that
\begin{equation}
\eta_{a b} = n^{- 2/3} k_{a b} \ . \label{etadef}
\end{equation}
This relation is important, as we have already established that $k_{ab}$ is a fixed matter space 
tensor.

Let us now imagine that the system evolves away from the relaxed state. This means that 
\eqref{etainv} no longer holds:  $\eta_{AB}$ retains the value set by the initial state, but 
$g^{AB}$ evolves along with the spacetime.  This leads to the build up of elastic strain, simply 
quantified in terms of the strain tensor
\begin{equation}
s_{a b} = {1\over 2} ( \perp_{a b} - \eta_{a b}) =  {1\over 2} \left( \perp_{a b} - n^{-2/3} k_{a b} \right)\ .
\label{sab}
\end{equation}
In the relaxed configuration, we have $\eta_{ab} = \perp_{ab}$ by construction so it is obvious that 
$s_{ab}$ vanishes. 

This model is fairly intuitive, but in practice it is more natural to work with scalars formed 
from $\eta_{ab}$  (which can be viewed as ``invariant''). This helps make the model less abstract. 
Hence, we introduce  the  strain  scalar $s^2$ (not to be confused with the entropy density from before) as a suitable combination of the 
invariants of $\eta_{ab}$:
\be
I_1 = \eta^a_{\ a} = g^{A B} \eta_{A B} 
         \ , 
\ee
\be
I_2 = \eta^a_{\ b} \eta^b_{\ a} = g^{A D} g^{B E} \eta_{E A} \eta_{D B} 
          \ , 
\ee
\be
I_3 = \eta^a_{\ b} \eta^b_{\ d} \eta^d_{\ a} = g^{A E} g^{B F} g^{D G} \eta_{E B} \eta_{F D} 
              \eta_{G A} 
\ . \label{invs}
\ee
However,  the number 
density $n$ also can be seen to be a combination of invariants, since
\begin{equation}
     k = n^2 = {1\over 3!} \left( I_1^3 - 3 I_1I_2+2I_3 \right) \ . 
\end{equation}
Given this,  it makes sense to replace one of the $I_N$ ($N=1, 2, 3$) with $n$, which now becomes one of 
the required invariants. Then we define $s^2$ to be a function of two of the other invariants. We 
can choose different combinations, but we must ensure that $s^2$ vanishes for the relaxed state. 
For example, \cite{Karlovini03:_elas_ns_1} work with
\begin{equation}
s^2 = {1\over 36} \left( I_1^3- I_3-24 \right) \ .
\label{Lars}
\end{equation}
In the limit $\eta_{ab} \to \perp_{ab}$ we have $I_1 , I_3 \to 3$ and we see that the combination for 
$s^2$ in Eq.~\eqref{Lars} vanishes.

Next, we assume that the Lagrangian of the system depends on $s^2$, rather than the tensor 
$k_{ab}$. In doing this, we need to keep in mind that Eqs.~\eqref{etadef} and \eqref{invs} show 
that the invariants $I_N$  depend on $n$ (and hence both $n^a$ and $g_{ab}$) as well as 
$k_{ab}$.

So far, the description is nonlinear, but in most situations of astrophysical interest it should be 
sufficient to consider a slightly deformed configuration\footnote{Note that this assumption is distinct from that of linear perturbations describing the dynamics.}. In effect, we may focus on a Hookean model, 
such that
\begin{equation}
\Lambda = - \check \varepsilon(n) - \check \mu(n) s^2 = - \varepsilon \ ,
\label{hooke}
\end{equation}
where $\check\mu$ is the shear modulus (not to be confused with the chemical potential). 
As mentioned earlier, the checks indicate that quantities are calculated for the 
unstrained state, with the specific understanding that 
$s^2=0$, and it should be apparent from \eqref{hooke} that we have an expansion in (a 
supposedly small) $s^2$. 

Since the strain scalar is given in terms of invariants, as in \eqref{Lars}, 
it might be tempting to suggest a change of variables such that $s^2=s^2(I_1,I_3)$. Our final 
equations of motion will, indeed, reflect this, but it would be premature to make the change at this 
point. 
Instead we note that the momentum is now given by
\begin{multline}
\mu_a = {\partial \Lambda \over \partial n^a} = {\partial n^2 \over \partial n^a} {\partial \Lambda\over  \partial n^2} \\
= - {1 \over n} {\partial \Lambda\over  \partial n} g_{ab}n^b
=  {1\over n} \left( {d \check \varepsilon \over dn} + {d\check \mu\over dn} s^2 + \check \mu {\partial s^2 \over \partial n} \right) g_{ab}n^b \ ,
\end{multline}
while
\begin{equation}
{\partial \Lambda \over \partial g_{ab} }=-  \left( {d\check \varepsilon \over dn} + {d\check \mu\over dn} s^2 + \check \mu  {\partial s^2 \over \partial n} \right) {\partial n \over \partial g_{ab}} - \check \mu 
 {\partial s^2 \over \partial g_{ab}} \ .
 \label{Lambder}
\end{equation}
Here we need (note that $n^a$ is held fixed in the partial derivative)
\begin{equation}
 {\partial n \over \partial g_{ab}} = - {1\over 2n} n^a n^b  \ ,
\end{equation}
and it is useful to note that 
\begin{equation}
 {\partial s^2 \over \partial g_{ab}} = - g^{ad} g^{be} {\partial s^2 \over \partial g^{de}} \ .
\end{equation}
Also, when working out this derivative, we need to hold $n$ fixed [as is clear from \eqref{Lambder}].  
At the end of the day, we have for the stress-energy tensor
\begin{multline}
T^{ab} = \left[ \Lambda + n \left( {d\ec \over dn} + {d\check \mu\over dn} s^2 + \check \mu {\partial s^2 \over \partial n} \right)\right] g^{a b}   
    \\
    + {1\over n} \left( {d\ec \over dn} + {d\check \mu\over dn} s^2 + \check \mu  {\partial s^2 \over \partial n} \right) n^a n^b  +2  \check \mu g^{ad} g^{be} {\partial s^2 \over \partial g^{de}}
    \\
    =  \Lambda g^{ab} + n \left( {d\ec \over dn} + {d\check \mu\over dn} s^2 + \check \mu {\partial s^2 \over \partial n} \right) h^{ab}  +  2 \check \mu g^{ad} g^{be} {\partial s^2 \over \partial g^{de}} \ .
    \label{stress2}
\end{multline}

Let us now make the change of variables we hinted at previously. In order to establish the procedure, let us consider a 
situation where $s^2$ depends only on $I_1$. Then we need
\begin{equation} 
I_1 = \eta^a_{\ a} = n^{-2/3} g^{ab} k_{ab} \ ,
\end{equation}
\begin{equation}
\left({\partial s^2 \over \partial n}\right)_1= - {2I_1 \over 3n}  { \partial s^2 \over \partial I_1 } \ ,
\end{equation} 
\begin{equation}
\left({\partial \Lambda \over \partial k_{ab}}\right)_1 = - \check \mu 
{\partial s^2 \over \partial k_{ab} } = - \check \mu n^{-2/3} g^{ab} { \partial s^2 \over \partial I_1 } 
\  ,
\end{equation}
(recall the comment on the partial derivative from before) and
\begin{equation}
\left({\partial s^2 \over \partial g^{de}}\right)_1 =  { \partial s^2 \over \partial I_1 }\eta_{de} \ .
\end{equation}
Making use of these results, we readily find
\begin{multline}
T^{ab} =  -\varepsilon g^{ab} + n \left( {d\ec \over dn} + {d\check \mu\over dn} s^2 \right) \perp^{ab} + 2  \check \mu { \partial s^2 \over \partial I_1 }  \left( \eta^{ab} - {1\over 3} I_1 \perp^{ab} \right)   \\
=  -\varepsilon g^{ab} + n \left( {d\ec \over dn} + {d\check \mu\over dn} s^2 \right) \perp^{ab} + 2  \check \mu { \partial s^2 \over \partial I_1 }  \eta^{\langle ab \rangle} \ , \label{stress3}
\end{multline}
where the $\langle \ldots \rangle$ brackets indicate the symmetric, trace-free part of a tensor with 
two free indices. In our case, we have
\beq
\eta_{\langle ab \rangle} = \eta_{(ab)} - { 1 \over 3} \eta^d_{\ d} \perp_{ab} \ . \label{angdef}
\eeq

Comparing this result to the standard decomposition of the stress-energy tensor, 
\beq
T^{ab} = \varepsilon u^a u^b + \bar p \perp^{ab} + \pi^{ab}\ , \qquad \mbox{where} \qquad \pi^a_{\ a} = 0 \ ,
\label{stress4}
\eeq
and $\bar p$ is the isotropic pressure (which differs from the fluid pressure, $p$, as it accounts 
for the elastic contribution). We see that  elasticity introduces an anisotropic 
contribution
\begin{equation}
\pi^1_{ab} = 2 \check \mu {\partial s^2 \over \partial I_1}  \eta_{\langle ab \rangle}  \ .
\end{equation}

Following the same steps  for the other two invariants (see \citealt{supercrust} for details), $I_2$ and $I_3$, we find that 
\begin{equation}
\pi^2_{ab} = 4 \check \mu {\partial s^2 \over \partial I_2}   \eta_{d \langle a} \eta_{b \rangle}^{\ d} \ ,
\end{equation}
and
\begin{equation}
\pi^3_{ab}  = 6 \check \mu {\partial s^2 \over \partial I_3}  \eta^{d e} \eta_{d \langle a} \eta_{b \rangle e} \ , 
\end{equation}
respectively.
Combining these results with \eqref{Lars}, we have
\begin{equation}
\pi_{ab} = \sum_N \pi^N_{ab} =  {\check\mu\over 6}  \left[ \left(\eta^d_{\ d}\right)^2 \eta_{\langle ab\rangle}-  \eta^{d e} \eta_{d \langle a} \eta_{b\rangle e}\right] \ ,
\label{piab}
\end{equation}
which agrees with equation (128) from \cite{Karlovini03:_elas_ns_1}. 

Now consider the final stress-energy tensor. Note first of all that, if we consider $n$ and $s^2$ as 
the independent variables of the energy functional, then the isotropic pressure should follow from
\beq
\bar p = n \left( {\partial \varepsilon \over \partial n} \right)_{s^2} - \varepsilon = \check p + \left( \frac{n}{\check \mu} {d\check \mu\over dn} -1 \right) {\check \mu} s^2 \ ,
\eeq
where
\beq
\check p = n {d\ec \over dn}  - \ec  \ , 
\eeq
 is identical to the fluid pressure from before. However, we may also introduce a corresponding momentum, such that
\beq
\bar \mu_a = - \left( {\partial \Lambda \over \partial n^a} \right)_{s^2} = \left( {d\ec \over dn} + {d\check \mu\over dn} s^2 \right) n_a\ , \label{elmom}
\eeq
which  leads to 
\beq
\bar p = \Lambda - n^a \bar \mu_a =  \pc + \left( {n \over \muc} {d \muc \over dn} - 1 \right) \muc s^2 \ .
\eeq

Finally, in order to obtain the equations of motion for the system we can either take the 
divergence of \eqref{stress4} or return to \eqref{tforce} and make use of our various definitions. 
The results are the same (as they have to be). After a little bit of work we find that \eqref{tforce} 
leads to
\beq
2n^b\nabla_{[b}\bar \mu_{a]} + \perp_a^d \left( \nabla^b \pi_{b d} - \muc \nabla_d s^2\right) = 0 \ ,
\label{finalmom}
\eeq
where it is worth noting that the combination in the parentheses is automatically flow line 
orthogonal. 

\subsection{Lagrangian perturbations of an unstrained medium}

Many applications of astrophysical interest---ranging from neutron star oscillations to tidal deformations in binary systems and mountains on spinning neutron stars---are adequately modelling within perturbation theory. As should be clear from the development of the elastic model, this requires the use of a Lagrangian framework. Luckily, we have already done most of the work needed to consider this problem. In 
particular, we know that 
\beq
  \Delta k_{ab} = 0 \ .
\label{delkab}\eeq
We now make maximal use of this fact. 

If we assume that the background configuration is relaxed, i.e. that $s^2=0$ vanishes for the 
configuration we are perturbing with respect to, then the fluid results from Sect.~\ref{sec:pullback}
together with \eqref{delkab} make the elastic perturbation problem straightforward (although it 
still involves some algebra). 

Consider, first of all, the strain scalar. A few simple steps lead to
\beq
\Delta s^2 = 0 \ .
\label{dels2}\eeq
To see this, recall that $s^2$ is a function of the invariants, $I_N$. Express these in terms of the 
number density $n$, the spacetime metric and  $k_{ab}$. Once this is done, make use of  
\eqref{delkab} and the fact that the background is unstrained, i.e. $\eta_{ab} = \perp_{ab}$, to see 
that $\Delta I_N=0$, which makes intuitive sense. Since the strain scalar is quadratic,  
linear perturbations away from a relaxed configuration should vanish. An important implication of 
this result is that the last term in \eqref{finalmom} does not contribute to the perturbed equations 
of motion.  

This leads to
\beq
\Delta \eta_{ab} = {1\over 3} \eta_{ab} \perp^{de} \Delta g_{de} \ , 
\eeq
and 
\beq
\Delta \eta^{ab} = \left[ - 2 g^{a(e} \eta^{d)b} + {1\over 3} \eta^{ab} \perp^{de} \right] \Delta g_{de} \ .
\eeq
It then follows from \eqref{sab} and \eqref{piab}, that
\beq
\Delta \pi_{ab} = - 2 \muc \Delta s_{ab} \ ,  
\eeq
where 
\beq
 2 \Delta s_{ab} = \left( \perp^e_{\ a} \perp^d_{\ b}  -  \frac13 \perp_{ab} \perp^{de}  \right) \Delta g_{de}  \ .
\eeq

It is worth noting that the final result for an isotropic material agrees with, for example, \cite{1983MNRAS.203..457S} where the relevant strain term is simply added to the stress-energy tensor (without detailed justification). 

Next, let us consider the perturbed equations of motion. In the case of an unstrained background, 
it is easy to see that the argument that led to \eqref{euler1} still holds. This gives us the 
perturbation of the first term in \eqref{finalmom} (after replacing $\mu_a\to \bar \mu_a$). 
Similarly, since $\pi_{ab}$ vanishes in the background, the Lagrangian variation commutes with 
the covariant derivative in the second term. Thus, we end up with a perturbation equation of form
\beq
 2n^a\nabla_{[a} \Delta \bar \mu_{b]} + \nabla^a \Delta \pi_{ab} = 0  \ .
\eeq
This is the final result, but
in order to arrive at an explicit expression for the perturbed momentum, it is useful to note that
\beq
\Delta \mu_a = - {1 \over 2n} \betac u_a \perp^{{b} d} \Delta g_{{b} d} + \mu \left( \delta_a^{{b}} u^d
+ { 1 \over 2} u_a u^{{b}} u^d \right) \Delta g_{{b} d} \ , 
\eeq
where we have defined the bulk modulus $\betac$ as
\beq
\betac = n {d\pc \over dn} = (\pc + \check\varepsilon) {d\pc \over d \check\varepsilon} =  (\pc + \check \varepsilon) \check{C}^2_s \ ,
\eeq
$\check{C}^2_s$ is the sound speed in the elastic medium and we have used the fundamental 
relation $\pc + \check \varepsilon = n \mu$.  It also follows that 
\beq
\Delta p = - {\check \beta \over 2} \perp^{ab} \Delta g_{ab} \ .
\eeq

When we consider perturbations of an elastic medium we need to pay careful attention to the 
magnitude of the deviation away from the relaxed state. If the perturbation is too large,  the 
material will yield \citep{2009PhRvL.102s1102H}.  It may fracture or behave in some other fashion 
that is not appropriately described by the equations of perfect elasticity. We need to quantify the 
associated breaking strain. In applications involving neutron stars, this is important if we want to 
consider star quakes in a spinning down pulsar, establish to what extent crust quakes in a 
magnetar lead to the observed flares \citep{2016RvMP...88b1001W} and whether the crust breaks 
due to the tidal interaction in an inspiralling binary 
\citep{2005ApJ...632L.111S,2012ApJ...749L..36P,2012PhRvL.108a1102T}.
A  commonly used criterion to discuss elastic yield strains in engineering involves the von Mises 
stress, defined as
\beq\label{vMdef}
  \Theta_{\mathrm{vM}} = \sqrt{\frac32 s_{ab}s^{ab}}
\eeq
When this scalar exceeds some critical value 
$\Theta_{\mathrm{vM}} > \Theta^{\mathrm{crit}}_{\mathrm{vM}}$, say, the material no longer 
behaves elastically. In order to 
work out the dominant contribution to the von Mises stress in general we need to (at least 
formally) consider second order perturbation theory \citep{supercrust}, but in the simple case of an unstrained background we have  
\beq\label{vM1}
  \Theta_{\mathrm{vM}} = \sqrt{\frac32 \Delta s_{ab} \Delta s^{ab}} 
     = \sqrt{\frac38 \perp^{a\langle c}\perp^{d\rangle b}\Delta g_{ab}\Delta g_{cd}}
\eeq
This allows us to quantify when a strained crust reaches the point of failure. This allows us to work out the maximal deformation, but unfortunately it is difficult to model what happens beyond this point. The same is true for terrestrial materials.



\section{Superfluidity}
\label{sec:superfluids}

Low temperature physics continues to be a vibrant area of research, providing
interesting and exciting challenges, many of which are associated
with the properties of superfluids/superconductors. Basically, matter appears
to have two options when the temperature decreases towards absolute zero.
According to classical physics one would expect the atoms in a liquid to slow
down and come to rest, forming a crystalline structure. It is, however,
possible that quantum effects become relevant before the liquid solidifies,
leading to the formation of a superfluid condensate (a quantum liquid). This
will only happen if the interaction between the atoms is attractive and
relatively weak. The
archetypal superfluid system is Helium. It is well established that $^4$He
exhibits superfluidity below $T=2.17$~K. Above this temperature liquid Helium
is accurately described by the Navier-Stokes equations. Below the critical
temperature the modelling of superfluid $^4$He requires a ``two-fluid''
description. Two fluid degrees
of freedom are required to explain, for example, ``clamped'' flow through narrow capillaries and the presence of a second
sound (associated with heat flow).

Many other low temperature systems are known to exhibit superfluid properties.
The different phases of $^3$He have been well studied, both theoretically and
experimentally, and there is considerable current interest in
atomic Bose--Einstein condensates. The relevance of
superfluid dynamics reaches beyond systems that are accessible in the
laboratory. It is generally expected that neutron stars will contain a number of superfluid phases. This expectation
is natural given the extreme core density (reaching several times the nuclear
saturation density) and low temperature (compared to the nuclear scale of the
Fermi temperatures of the different constituents, about $10^{12}$~K) of these stars. 

The rapid spin-up and subsequent relaxation associated with radio pulsar
glitches provides strong, albeit indirect, evidence for neutron-star
superfluidity \citep{2018ASSL..457..401H}. The standard model for these events is based on, in the first
instance, the pinning of superfluid vortices (e.g., to the crust lattice) which
allows a rotational lag to build up between the superfluid and the part of the
star that spins down electromagnetically, and secondly the sudden unpinning
which transfers angular momentum from one component to the other, leading to
the observed spin-change.
Recent observations of the youngest known neutron star in the galaxy, the compact object in the 
Cassiopeia A supernova remnant, with an estimated age of around 330 years, are also 
relevant in this context. The cooling of this objects seems to accord with 
our understanding of neutron stars with a superfluid component in the core \citep{casa1,casa2}.
The data can be used to infer the pairing gap for neutron superfluidity in the core, 
which helps constrain current theory.  Similarly, the slow thermal relaxation observed in neutron stars that enter quiescence at the end of an accretion phase requires a superfluid component to be present in the neutron star crust \citep{2017JApA...38...49W}.

Basically,
neutron star astrophysics provides ample motivation for us to  develop
a relativistic description of superfluid systems. At one level this turns out to be  
straightforward, given the general variational multi-fluid model. However, when we consider the fine print we uncover a number of 
 hard physics questions. In particular, we need to make contact with 
microphysics calculations that determine the various parameters of the relevant 
multi-fluid systems. We also need to understand how to incorporate quantized vortices \citep{2001LNP...571.....B}, and the
associated mutual friction, in the relativistic context. In order to establish the proper context for the discussion, it makes sense to 
first discuss the multi-fluid approach to Newtonian superfluids. We do this for the particular case of Helium, the
archetypal laboratory two-fluid system.

\subsection{Bose--Einstein condensates}
\label{sec:bec}

In order to understand the key aspects of the connection between the fluid model and the underlying quantum system, it is natural to consider the problem of a single component Bose--Einstein condensate. 
In recent years there has been a virtual explosion of interest in such systems. A key reason for this is that 
atomic condensates lend themselves to precision experiments, allowing researchers to probe the nature of the associated macroscopic
quantum behaviour \citep{2008bcdg.book.....P} In addition, from the relativity point of view, the description of Bose--Einstein condensates is relevant as it connects with issues that may play a role  in cosmology \citep{SY09,Harko11}.

On a sufficiently large scale,  atomic condensates are accurately represented by a fluid model,  similar to that used for superfluid Helium (described below). Consider as an example a uniform Bose gas, in a volume $V$, with an effective (long-range) interaction energy $U_0$. The relevant interaction
arises in the Born approximation, and is related to the s-wave scattering length $a$ through
\be
U_0 = {4\pi \hbar^2 a\over m}\ , 
\ee
where $m$ is the atomic mass. This effectively means that the model is appropriate only for dilute gases, where short-range corrections 
to the interaction can be ignored. In essence, we are focussing on the long-wavelength behaviour. Given the interaction, the energy of a state with $N$ bosons (recalling that we need to multiply by the number of ways that these can be arranged in pairs) is
\be
E = {N(N-1)\over 2} {U_0\over V} \approx {N^2\over 2} {U_0\over V} = {1\over 2} n^2 V U_0 \ , 
\ee
where we have defined the number density $n=N/V$. From this we see that the chemical potential is
\be
\mu = {dE\over dN } = {N\over V} U_0 = n U_0  \ .
\ee
Alternatively, we may work with the energy density 
\be
\varepsilon = {E \over V} \qquad \Longrightarrow \qquad \mu = {d\varepsilon \over dn} \ , 
\ee
as in Sect.~\ref{sec:thermo}. From the usual thermodynamical relation we see that the pressure of the system 
follows from
\be
dp = n d\mu \ .
\ee

The main theoretical tool for studying the dynamics of atomic Bose--Einstein 
condensates is the Gross--Pitaevskii equation. This equation, which takes the form
\be
- {\hbar^2 \over 2m} \nabla^2 \Psi +V_\mathrm{ext} \Psi + U_0 |\Psi|^2\Psi = i \hbar \partial_t\Psi \ , 
\label{GP}
\ee
encodes the dependence of the order parameter $\Psi$ (note that this is not the many-body quantum wave-function) on the interaction $U_0$ and an external potential $V_\mathrm{ext}$. In laboratory 
systems the external potential usually represents an optical trap. In an astrophysical setting it can be taken as a proxy for the coupling to the 
gravitational field.

At low temperatures (such that we can ignore thermal excitations) the order parameter is normalized in such a way that the density of the condensate 
equals the density of the gas
\begin{equation}
|\Psi|^2 = n\ .
\end{equation}
With this identification, we may consider the 
simplest problem;  the stationary solution to \eqref{GP}, representing the ground state of the system. Letting the time dependence 
be of form $\Psi=\Psi_0 \exp(-i\mu t/\hbar) $ we see that a uniform, stationary solution corresponds to
\be
\mu = n U_0+V_\mathrm{ext} \ .
\ee

Moving on to the time-dependent dynamics, we note that \eqref{GP} describes a complex-valued function $\Psi$. In effect, there are 
two degrees of freedom to consider. Given the connection to $n$ it is useful to consider the magnitude of $\Psi$. Multiplying 
\eqref{GP} with $\Psi^*$ (where the asterisk represents complex conjugation) and subtracting the result from its own complex conjugate, we readily arrive at
\be
\partial_t |\Psi|^2 + {\hbar \over 2mi} \nabla_i \left( \Psi^* \nabla^i \Psi - \Psi \nabla^i \Psi^* \right) = 0 \ . 
\ee
Comparing this result with the continuity equation, we see that the two take the same form provided that we identify (in analogy with the momentum operator in quantum mechanics)
the velocity
\be
v^i = {p^i \over mi} =  {\hbar \over 2mi} {1\over |\Psi|^2} \left( \Psi^* \nabla^i \Psi - \Psi \nabla^i \Psi^* \right) \ .
\label{viden}
\ee
In other words, we have
\be
\partial_t n + \nabla_i \left(n v^i \right) = 0 \ .
\label{contBE}
\ee

Having already made use of the magnitude, it makes sense to let the second degree of freedom in the problem be represented by the phase of 
$\Psi$. Letting $\Psi = \sqrt{n} \exp(iS)$ we can write the real part of \eqref{GP} as
\be
-\hbar \partial_t S =   \mu + V_\mathrm{ext}  + {mv^2 \over 2}- {\hbar^2 \over 2m} {1\over \sqrt{n}} \nabla^2 \sqrt{n} \ .
\label{pheq}\ee
Here we have identified the chemical potential as before. We have also used
\be
{\hbar^2 \over 2m} (\nabla_i S) (\nabla^i S) =  {mv^2 \over 2} \ , 
\ee
which follows from \eqref{viden}. Finally, we take the gradient of \eqref{pheq} to get
\begin{multline}
m \partial_t v_i  + \nabla_i \left[   \mu + V_\mathrm{ext}  + {mv^2 \over 2}- {\hbar^2 \over 2m} {1\over \sqrt{n}} \nabla^2 \sqrt{n} \right]  \\
= m \left( \partial_t +v^j \nabla_j \right)v_i +  \nabla_i \left(  \mu + V_\mathrm{ext} \right) \\
+m \epsilon_{ijk} v^j \left( \epsilon^{klm} \nabla_l v_m\right) 
- \nabla_i \left({\hbar^2 \over 2m} {1\over \sqrt{n}} \nabla^2 \sqrt{n}  \right) = 0 \ .
\end{multline}
By definition, the flow is potential and hence irrotational (at least as long as we ignore quantum vortices, which we consider later), so 
\be
 m \left( \partial_t +v^j \nabla_j \right)v_i +  \nabla_i \left(  \mu + V_\mathrm{ext} \right)
- \nabla_i \left({\hbar^2 \over 2m} {1\over \sqrt{n}} \nabla^2 \sqrt{n}  \right) = 0  \ . 
\label{eulBE}\ee
Comparing to the standard fluid result, we see that only the last term differs. Notably, it is also the only term that (explicitly) retains the 
quantum origins of the model (Planck's constant!).

So far, we have not made any simplifications. The two equations \eqref{contBE} and \eqref{eulBE} contain the same information as 
the Gross--Pitaevskii equation \eqref{GP}. The  equations differ from those for irrotational fluid flow only by the presence of the 
final term in \eqref{eulBE}. This term, which represents a ``quantum pressure'' is, however, irrelevant as long as we focus on the large-scale dynamics.
To see this, assume that the order parameter varies on some length-scale $L$. It then follows that 
\be
\nabla \mu \sim {nU_0 \over L} \qquad \mbox{ and } \qquad \nabla \left({\hbar^2 \over 2m} {1\over \sqrt{n}} \nabla^2 \sqrt{n}  \right)  \sim {\hbar^2 \over mL^3} \ .
\ee
In other words, the quantum pressure can be neglected as long as
\be
{\hbar^2 \over mn L^2 U_0} \ll 1 \ .
\ee
In order to give this relation a clearer meaning, we introduce the coherence length $\xi$, roughly the length-scale on which the kinetic energy balances the pressure. This leads to
\be
{\hbar^2 \over 2m \xi^2 } \approx n U_0 \ , 
\ee
and we can neglect the quantum pressure as long as
\be
\left( {\xi \over L} \right)^2 \ll1 \ .
\ee
As long as this condition is satisfied,  a low temperature Bose--Einstein condensate is faithfully represented by a fluid model. In the atomic condensate literature this regime is sometimes referred to as the Thomas--Fermi limit. It is worth noting that, even though the above condition implies that the 
fluid model is appropriate on larger scales, it is fundamentally not the same averaging argument that leads to the notion of a fluid element in the
usual discussion. In the case of quantum condensates, the fluid model may in fact be appropriate at much shorter scales since it tends to be the case that the coherence length is vastly smaller than the mean-free path of the various particles that make up a normal ``fluid''. This scale enters the 
quantum problem once we consider finite temperature excitations, being relevant for the second component that then comes into play.

\vspace*{0.1cm}
\begin{tcolorbox}
\textbf{Comment:} The example we have considered provides a direct connection between a quantum system and fluid dynamics. Similar arguments apply for general systems that exhibit macroscopic quantum behaviour, like superfluids and superconductors. In particular, the coherence length replaces the mean-free path argument, typically leading to fluid behaviour being expected on much smaller scales. The discussion also provides a direct example of the notion that fluid behaviour arises in the long-wavelength limit of a quantum field theory. This is an important aspect, as it motivates the use of a derivative expansion (systematically representing shorter wavelength corrections) to account for dissipative effects, see Sect.~\ref{sec:viscosity}.
\end{tcolorbox}
\vspace*{0.1cm}

\subsection{Helium: The original two-fluid model}

Phenomenologically, the  behaviour of superfluid Helium is ``easy'' to
understand if one first considers a system at absolute zero temperature. Then
the dynamics is entirely due to the quantum condensate (as in the previous example). There exists a single
quantum wavefunction, and the momentum of the flow follows directly from the
gradient of its phase. This immediately implies that the flow is irrotational.
At finite temperatures, one must also account for thermal excitations (like phonons)---not all atoms remain in the ground state. A second dynamical degree of
freedom arises since the excitation gas may drift relative to the atoms. In
the standard two-fluid model, one makes a distinction between a ``normal''
fluid component\footnote{The model obviously assumes that the excitations can be modelled as a ``fluid'', e.g., that 
the mean-free path of the phonons is sufficiently short. This may not be the case at very low temperatures.} and a superfluid part. The identification of the associated densities is to a
large extent ``statistical'' as one cannot physically separate the ``normal'' component
from the ``superfluid'' one. It is important to keep this in mind.

We take as our starting point the Newtonian version of the multi-fluid framework. We consider the simplest conducting system
corresponding to a single particle species exhibiting superfluidity. Such systems tyically have two degrees of freedom, c.f. 
$\mathrm{He}^4$ \citep{putterman74:_sfhydro, tilley90:_super} where
the entropy can flow independently of the superfluid Helium
atoms. Superfluid $\mathrm{He}^3$ can also be included in the mixture,
in which case there will be a relative flow of the $\mathrm{He}^3$
isotope with respect to $\mathrm{He}^4$, and relative flows of each
with respect to the entropy \citep{voll_book}. The model we advocate here distinguishes the  atoms  from the massless ``entropy''---the former will be identified by
a constituent index $\n$, while the latter is represented by $\s$. As this
description is different (in spirit) from the standard two-fluid model for
Helium, it is relevant to explain how the two descriptions are
related. 

First of all, we need to allow for a difference in the two
three-velocities
\begin{equation}
  w_i^{\Y \X} =  {v}_{i}^{\Y} - {v}_{i}^{\X} \ , \quad \y\neq \x \ .
\end{equation}
Letting the square of this difference be given by
$w^2$, the equation of state then takes the form ${\cal E} =
{\cal E}(n_{\mathrm{n}},n_{\mathrm{s}},w^2)$. Hence, we have
\begin{equation}
{d} {\cal E} = \mu^{\mathrm{n}} \, {d} n_{\mathrm{n}} +
  \mu^{\mathrm{s}} \, {d} n_{\mathrm{s}} + \alpha \, {d} w^2,
  \label{flaw}
\end{equation}
where
\begin{equation}
  \mu^{\mathrm{n}} =
  \left. \frac{\partial \mathcal{E}}{\partial n_{\mathrm n}}
  \right|_{n_{\mathrm{s}},w^2}\!\!\!,
  \qquad
  \mu^{\mathrm{s}} =
  \left. \frac{\partial \mathcal{E}}{\partial n_{\mathrm s}}
  \right|_{n_{\mathrm{s}},w^2}\!\!\!,
  \qquad
  \alpha =
  \left. \frac{\partial \mathcal{E}}{\partial w^2}
  \right|_{n_{\mathrm{s}},n_{\mathrm{s}}}\!\!\!.
\end{equation}
The $\alpha$ coefficient reflects the effect of entrainment on the equation of
state. Similarly, entrainment causes the fluid momenta to be modified
to
\begin{equation}
  \frac{p^\X_i}{m^\X} =
  v^i_\X + 2 \frac{\alpha}{\rho_\X} w^i_{\Y \X}.
\end{equation}

The number density of each fluid obeys a continuity equation:
\begin{equation}
  \frac{\partial n_{\X}}{\partial t} + \nabla_{j} (n_{\X} v_{\X}^{j}) = 0.
  \label{eq:Cont}
\end{equation}
Each fluid also satisfies an Euler-type equation, which ensures the
conservation of total momentum. This equation can be written
\begin{equation}
  \left( \frac{\partial}{\partial t} + {v}^{j}_{\X}\nabla_{j} \right)
  \left[ {v}_{i}^{\X} + \varepsilon_{\X} w_i^{\Y\X} \right] +
  \nabla_{i} (\Phi + \tilde{\mu}_{\X}) +
  \varepsilon_{\X} w_j^{\Y \X} \nabla_{i} v^{j}_{\X} = 0\ ,
  \label{eq:MomEqu}
\end{equation}
where
\begin{equation}
  \tilde{\mu}_\X = \frac{\mu^\X}{m^\X} \ ,
\end{equation}
and
the entrainment is now included via the coefficients
\begin{equation}
  \varepsilon_\X = 2 \rho_\X \alpha.
\end{equation}
For a detailed discussion of these equations,
see \cite{prix04:_multi_fluid, andersson05:_flux_con}.

We have already seen that the entrainment means that 
each momentum does not have to be parallel to the associated flux. In the case
of a two-component system, with a single species of particle flowing with
$n^\n_i = n v^\n_i$ and a massless entropy with flux $n^\s_i = sv^\s_i$ (i.e., letting $n_\n=n$ and $n_s=s$, where $n$ is the particle number density and $s$ represents 
the entropy per unit volume),  the
momentum densities are
\be
   \pi_i^\n = n p_i^\n = mn v_i^\n - 2 \alpha w_i^{\n\s} \ ,
\ee
and
\be
   \pi^\s_i = s p^\s_i = 2 \alpha w_i^{\n\s} \ .
\ee

\vspace*{0.1cm}
\begin{tcolorbox}
\textbf{Comment:} At this point it is worth stressing the association between the entropy entrainment and inertia of heat. The entropy may be massless, but this does not mean that the corresponding momentum vanishes. This may seem somewhat novel, as there may (at first) 
seem to be no reason to consider ``entrainment'' between particles and entropy.
However, the effect arises naturally in the variational model, and if we consider the entrainment as altering the effective mass 
of a constituent, then it would be very natural for this mechanism to affect also the entropy. 
This interpretation is  conceptually elegant, and turns out to be practically useful as well. This will become particularly apparent when we consider the problem of heat flux in Sect.~\ref{sec:heat}.
\end{tcolorbox}
\vspace*{0.1cm}

In order to understand the physical relevance of the entrainment better, let us compare the two-fluid model to the 
orthodox model used to describe laboratory superfluids. This  also clarifies the dynamical role of the 
thermal excitations in the system. 

Expressed in terms of the momentum densities, the two momentum equations can be
written, cf. \eqref{eq:MomEqu},
\be
    \partial_t \pi_i^\n + \nabla_j \left(v_\n^j \pi_i^\n \right)+ n \nabla_i \left( \mu_\n - \frac{1}{2} m v_\n^2
   \right) + \pi_j^\n \nabla_i v_\n^j  = 0 \ , \label{eulern}
\ee
and
\be
 \partial_t \pi_i^\s + \nabla_j \left(v_\s^j \pi_i^\s\right) 
+ s \nabla_i T + \pi_j^\s \nabla_i v_\s^j = 0 \ , \label{eulers}
\ee
where we have used the fact that the temperature follows from $\mu_\s = T$. 
Let us now
assume that we are considering a superfluid system. For low temperatures and
velocities the fluid described by (\ref{eulern}) should be irrotational.
In order to impose this constraint we need to appreciate that it is the momentum that
is quantized in a rotating superfluid, not the velocity.
This means that we require
\be
\epsilon^{klm} \nabla_l p^\n_m = 0 \ .
\label{quantify}\ee
To see how this affects the equations of motion, we rewrite (\ref{eulern}) as
\be
n \partial_t p_i^\n + n \nabla_i \left[ \mu_\n - \frac{m}{2} v_\n^2 + v_\n^j
p_j^\n \right] - n \epsilon_{ijk} v_\n^j (\epsilon^{klm} \nabla_l p^\n_m) =
0
\label{eulvort}\ee
Using \eqref{quantify} we have
\be
\partial_t p_i^\n + \nabla_i \left[ \mu_\n - \frac{m}{2} v_\n^2 + v_\n^j
p_j^\n \right] = 0
\ . \label{euler_sf}
\ee
We now have all the expressions we need to make a direct comparison with the standard two-fluid model 
for Helium.

It is natural to begin by identifying the drift velocity of the quasiparticle
excitations in the two models. After all, this is the variable that leads to
the ``two-fluid'' dynamics. Moreover, since it distinguishes the part of the flow that is
affected by friction it has a natural physical interpretation. In the standard
two-fluid model this velocity, $v_\N^i$, is associated with the ``normal
fluid'' component. In the variational framework, the excitations are directly associated
with the entropy of the system, which flows with $v_\s^i$. These two
quantities should be the same, and hence we have
\be
v_\N^i = v_\s^i \ .
\ee

The second fluid component, the ``superfluid'', is usually associated with a
``velocity'' $v_\S^i$. This quantity is directly linked to the gradient of
the phase of the superfluid condensate wave function. This means that it is,
in fact, a rescaled momentum. This means that we should identify
\be
v_\S^i = \frac{\pi^i_\n}{\rho_\n} = \frac{p_\n^i}{m} \ .
\ee
These identifications lead to
\be
\rho v_\S^i = \rho \left[ \left(1 - \varepsilon\right) v_\n^i + \varepsilon
              v_\N^i \right] \ ,
\ee
where $\varepsilon = 2\alpha/\rho$ and $\rho$ is the total mass density. We see that the total mass current is
\be
\rho v_\n^i = \frac{\rho}{1 - \varepsilon} v_\S^i -
\frac{\varepsilon \rho}{1 - \varepsilon} v_\N^i \ .
\ee
If we introduce the superfluid and normal fluid densities,
\be
\rho_\S = \frac{\rho}{1 - \varepsilon} \ , \qquad \mbox{ and } \qquad
\rho_\N =  - \frac{\varepsilon \rho}{1 - \varepsilon} \ ,
\ee
we arrive at the usual result \citep{khalatnikov65:_introd,putterman74:_sfhydro}
\be
\rho v_\n^i = \rho_\S v_\S^i  + \rho_\N v_\N^i \ .
\ee
Obviously, it is the case that $\rho = \rho_\S + \rho_\N$. This completes the
translation between the two formalisms. Comparing the two descriptions, it is
clear that the variational approach has identified the natural physical
variables---the average drift velocity of the excitations and the total
momentum flux. Since the system can be ``weighed'' the total density $\rho$
also has a clear interpretation. Moreover, the variational derivation 
identifies the truly conserved fluxes.  
In contrast, the standard model uses
quantities that only have a statistical meaning. The density
$\rho_\N$ is inferred from the mean drift momentum of the excitations. That
is, there is no ``group'' of excitations that can be identified with this
density. Since the superfluid density $\rho_\S$ is inferred from $\rho_\S =
\rho-\rho_\N$, it is a statistical concept, as well.
Furthermore, the two velocities, $v_\N^i$ and $v_\S^i$, are not individually
associated with a conservation law. From a practical point of view, 
this is not a problem. The various quantities can be calculated from
microscopic theory and the results are known to compare well to experiments.
At the end of the day, the two descriptions are (as far as applications are concerned)
identical and the preference of one over the other is very much a matter of taste
(or convention). 

The above results show that the entropy entrainment coefficient follows from
the ``normal fluid'' density according to
\be
\alpha = - \frac{\rho_\N}{2} \left( 1 - \frac{\rho_\N}{\rho} \right)^{-1} \ .
\ee
This shows that the entrainment coefficient diverges as the temperature
increases towards the superfluid transition and $\rho_\N \to \rho$. At first sight,
this may seem an unpleasant feature of the model. However, it is simply a
manifestation of the fact that the two fluids must lock together as one passes
through the phase transition. The model remains non-singular as long as
$v_i^\n$ approaches $v_i^\s$ sufficiently fast as the critical temperature is
approached. More detailed discussions of entrainment in finite temperature superfluids can be found in \cite{2013CQGra..30w5025A,2006MNRAS.372.1776G,2011PhRvD..83j3008K,2009PhRvC..80a5803G,2005NuPhA.761..333G}.

Having related the main variables, let us consider the form of the equations
of motion. We  start with the inviscid problem. It is common to work with the
total momentum. Thus, we combine (\ref{eulern}) and (\ref{eulers})
to get
\begin{multline}
\partial_t \left(\pi_i^\n + \pi_i^\s\right) +
      \nabla_l \left(v_\n^l \pi^\n_i + v_\s^l \pi_i^\s\right) + n \nabla_i
      \mu_\n + s \nabla_i T \\
       - n \nabla_i \left(\frac{1}{2} m v_\n^2 \right) + \pi_l^\n \nabla_i
      v_\n^l + \pi_l^\s \nabla_i v_\s^l = 0 \ .
\end{multline}
Here we have
\be
\pi_i^\n + \pi_i^\s = \rho v_i^\n \equiv j_i
\ee
which defines the total momentum density. From the continuity equations \eqref{eq:Cont} we see that
\be
\partial_t \rho + \nabla_i j^i  = 0 \ .
\ee
The pressure $\Psi$ follows from
\be
\nabla_i \Psi = n \nabla_i \mu_\n + s \nabla_i T - \alpha \nabla_i w_{\n\s}^2
     \ ,
\ee
and we also need the relation
\be
v_n^l \pi_i^\n + v_\s^l \pi_i^\s = v^\S_i j^l + v_\N^l j^0_i \ , 
\ee
where we have defined
\be
j^0_i = \rho_\N(v_i^\N - v_i^\S) = \pi_i^\s \ ,
\ee
and
\be
 \pi_l^\n \nabla_i v_\n^l + \pi_l^\s \nabla_i v_\s^l =  n \nabla_i \left(
\frac{1}{2} m v_\n^2 \right) - 2 \alpha w_l^{\n\s} \nabla_i w^l _{\n\s} \ .
\ee
Putting all the pieces together we have
\be
\partial_t j_i + \nabla_l \left(v_i^\S j^l + v_\N^l j^0_i\right) + \nabla_i
   \Psi = 0 \ . \label{mom1}
\ee

The second equation of motion follows directly from \eqref{euler_sf};
\be
\partial_t v_i^\S + \nabla_i \left( \tilde{\mu}_\S + \frac{1}{2} v_\S^2
\right) = 0 \ , 
\ee
where we have defined
\be
 \tilde{\mu}_\S = \frac{1}{m} \mu_\n - \frac{1}{2} \left(v_\n^i -
v_\S^i\right)^2 \ .
\label{mom2}
\ee

The above relations show that our inviscid equations of motion are identical
to the standard ones \citep{khalatnikov65:_introd,putterman74:_sfhydro}. The identified
relations between the different variables also provide a direct way to
translate the quantities in the two descriptions. For example, we can write down a generalized first law, starting from \eqref{flaw}. The key point is that we have
demonstrated how the ``normal fluid density'' corresponds to the entropy
entrainment in the variational model. This clarifies
the  role of the entropy entrainment; a quantity that arises in a natural way 
within the variational framework. 

\subsection{Relativistic models}

Neutron star physics provides ample motivation for the need to develop
a relativistic description of superfluid systems. As the typical core
temperatures (below $10^8 \mathrm{\ K}$) are far below the Fermi temperature of the
various constituents (of the order of $10^{12} \mathrm{\ K}$ for baryons) mature neutron stars
are extremely cold on the nuclear temperature scale. This means that---just
like ordinary matter at near absolute zero temperature---the matter in the
star will most likely freeze to a solid or become superfluid. While the
outer parts of the star, the so-called crust, form an elastic lattice, the
inner parts of the star are expected to be superfluid. In practice, this
means that we \emph{must} be able to model mixtures of superfluid
neutrons and superconducting protons. It is also \emph{likely} that
we need to understand superfluid hyperons and colour superconducting
quarks. There are many hard physics questions that need to be
considered if we are to make progress in this area. In particular, we
need to make contact with microphysics calculations that determine  parameters of such multi-fluid systems. 

One of the key features of a pure superfluid is that it is irrotational.
On a larger scale, bulk rotation is mimicked by the formation of vortices, slim ``tornadoes''
representing regions where the superfluid degeneracy is broken \citep{2001LNP...571.....B}. In
practice, this means that one would often, e.g., when modelling global neutron
star oscillations, consider a macroscopic model based on ``averaging'' over
a large number of vortices. The resulting model closely resembles the
standard fluid model. Of course, it is important to remember that the
vortices are present on the microscopic scale and that they may affect the parameters in the problem. There are also unique effects
that are due to the vortices, e.g., the mutual friction that is thought to
be the key agent that counteracts relative rotation between the neutrons and
protons in a superfluid neutron star core \citep{mendII}.

For the present discussion, let us focus on the case of
superfluid $\mathrm{He}^4$. We then have two fluids, the superfluid Helium atoms
with particle number density $n_\n$ and the entropy with particle number
density $n_\s$, as before. From the derivation in Sect.~\ref{sec:twofluids} we
know that the equations of motion can be written
\begin{equation}
  \nabla_a n_\x^a = 0 \ , 
\end{equation}
and
\begin{equation}
  n_\x^b \nabla_{[b} \mu^\x_{a]} = 0\ .
\end{equation}
To make contact with other discussions of the superfluid
problem \citep{carter92:_momen_vortic_helic,
  carter94:_canon_formul_newton_superfl, carter95:_equat_state,
  carter98:_relat_supercond_superfl}, we will use the notation $s^a=
n_\s^a$ and $\Theta_a = \mu_a^\s$. Then the equations that
govern the motion of the entropy  become
\begin{equation}
  \nabla_a s^a = 0
  \qquad \mathrm{and} \qquad
  s^b \nabla_{[b} \Theta_{a]} = 0 \ .
  \label{ent_eom}
\end{equation}
Now, since the superfluid constituent is irrotational we also have
\begin{equation}
  \nabla_{[a} \mu^\n_{b]} = 0 \ .
\end{equation}
The particle conservation law for the matter component is,
of course, unaffected by this constraint. This  shows how easy it is to restrict the multi-fluid
equations to the case where one (or several)
components are irrotational. It is worth emphasizing that it is the
momentum that is quantized, not the velocity. This is an important distinction in situations where entrainment plays a role.

It is instructive to contrast this description with other models, like the potential formulation
due to \cite{khal1, khal2}. We arrive at this
alternative formulation in the following
way \citep{carter94:_canon_formul_newton_superfl}. First of all, we
know that the irrotationality condition implies that the particle
momentum can be written as a gradient of a scalar potential, $\varphi$
(say). That is, we have
\begin{equation}
  V_a = - \frac{\mu^\n_a}{m} = - \nabla_a \varphi.
  \label{kh_eq1}
\end{equation}
Here $m$ is the mass of the Helium atom and $V_a$ is traditionally
(and somewhat confusedly, see the previous Section) referred to as the ``superfluid velocity''. It really is a rescaled momentum. Next assume that the momentum of the
remaining fluid (in this case, the entropy) is written
\begin{equation}
  \mu^\s_a = \Theta_a = \kappa_a + \nabla_a \phi \ .
\end{equation}
Here $\kappa_a$ is Lie transported along the entropy flow provided that
$s^a \kappa_a = 0$ (assuming that the equation of
motion~(\ref{ent_eom}) is satisfied). This leads to
\begin{equation}
  s^a \nabla_a \phi = s^a \Theta_a \ .
\end{equation}
There is now no loss of generality in introducing further scalar potentials
$\beta$ and $\gamma$ such that $\kappa_a = \beta \nabla_a \gamma$,
where the potentials are constant along the flow-lines as long as
\begin{equation}
  s^a \nabla_a \beta = s^a \nabla_a \gamma = 0.
\end{equation}
Given this, we have
\begin{equation}
  \Theta_a = \nabla_a \phi + \beta \nabla_a \gamma \ .
\end{equation}
Finally, comparing to Khalatnikov's formulation \citep{khal1, khal2} we
define $\Theta_a = - \kappa w_a$ and let $\phi \to \kappa \zeta$ and
$\beta \to \kappa \beta$. Then we arrive at the final equation of
motion
\begin{equation}
  - \frac{\Theta_a}{\kappa} =
  w_a = - \nabla_a \zeta - \beta \nabla_a \gamma \ .
  \label{kh_eq2}
\end{equation}
Equations~(\ref{kh_eq1}) and (\ref{kh_eq2}), together with the standard particle
conservation laws, are the key equations of the potential formulation. The content of this description is (obviously) identical to that of the
 variational picture, and we
have now seen how the various quantities can be related.

This example shows how easy it is to specify the 
equations that we derived earlier to the case when one (or several) 
components are irrotational/superfluid. 

Another alternative approach, related to the field theory inspired discussion in Sect.~\ref{sec:ftheory}, is based on the notion of broken symmetries. At a very basic level, a model 
with a broken $U(1)$ symmetry  corresponds to the superfluid model 
described above. In essence, the superfluid flow introduces a preferred direction which break the assumption that the model is isotropic. At first sight our equations differ from those 
used in, for example, \cite{son,pujol,zhang}, but it is easy to demonstrate that we can 
reformulate our equations to get those written down for a system with a 
broken $U(1)$ symmetry. The exercise is  of interest 
since it connects with models that have been used to describe other 
superfluid systems.

Take as starting point the general two-fluid system. From the 
discussion in Sect.~\ref{sec:twofluids}, we know that the momenta are in 
general related to the fluxes via
\begin{equation}
    \mu^\x_a = {\cal B}^\x n_a^\x + {\cal A}^{\x \y} n_a^\y \ .  
\end{equation}
Suppose that, instead of using the fluxes as our key variables, we 
consider a ``hybrid'' formulation based on a mixture of fluxes and momenta. 
 In the case of the particle-entropy system, we may use
\begin{equation}
    n_a^\n = \frac{1}{{\cal B}^\n} \mu_a^\n - 
    \frac{{\cal A}^{\n\s}}{{\cal B}^\n} n_a^\s \ . 
\end{equation}
Let us impose irrotationality on the fluid by representing the momentum as 
the gradient of a scalar potential $\varphi$. With $\mu_a^\n = 
\nabla_a \varphi$  we get
\begin{equation}
    n_a^\n = \frac{1}{{\cal B}^\n} \nabla_a \varphi - 
                \frac{{\cal A}^{\n \s}}{{\cal B}^\n} n_a^\s \ . 
\end{equation}
Now take the preferred frame to be that associated with the entropy flow, 
i.e.~introduce the unit four velocity $u^a$ such that $n_\s^a = n_\s 
u^a = s u^a$. Then we have
\begin{equation}
    n_a^\n = n u_a - V^2 \nabla_a \varphi 
\end{equation}
where we have defined
\begin{equation}
    n \equiv - \frac{s {\cal A}^{\n\s}}{{\cal B}^\n} \qquad \mbox{and} 
    \qquad V^2 = - \frac{1}{{\cal B}^\n} \ . 
\end{equation}
With these definitions, the particle conservation law becomes
\begin{equation}
    \nabla_a n_\n^a = \nabla_a \left( n u^a - V^2 \nabla^a 
                            \varphi \right) = 0 \ . 
\end{equation}
Meanwhile, 
the chemical potential in the entropy frame follows from
\begin{equation} 
    \mu = - u^a \mu^\n_a = - u^a \nabla_a \varphi \ . 
\end{equation}  
One can also show that the stress-energy tensor becomes
\begin{equation}
    T^a{}_b  = \Psi \delta^a{}_b + (\Psi+ \rho) u^a u_b - V^2 
                   \nabla^a \varphi \nabla_b \varphi \ ,
\end{equation}
where the generalized pressure is given by $\Psi$ as usual, and we have 
introduced
\begin{equation}
    \Psi + \rho = {\cal B}^\s s^2 + {\cal A}^{\s \n} s n \ .  
\end{equation}
The equations of motion can now be obtained from $\nabla_b T^b{}_a = 
0$. (Keeping in mind that the equation of motion for $\X=\n$ is automatically 
satisfied once we impose irrotationality, as before.) This 
essentially completes the set of equations written down by, for example, 
\cite{son} (see also \citealt{2006MNRAS.372.1776G,2011PhRvD..83j3008K}). The argument in favour of this formulation is that it is close 
to the microphysics calculations, which means that the parameters may be 
relatively straightforward to obtain. Against the description is the fact 
that it is a---not very elegant---hybrid where the inherent symmetry amongst 
the different constituents is lost, and there is also a risk of confusion 
since one is treating a momentum as if it were a velocity.

In the case when the superfluid rotates, the  two-fluid equations apply as long as the rotation is sufficiently fast that one can meaningfully
 average over the vortex array. In effect, we assume that we can ``ignore'' the smaller scales associated with, for example, the vortex cores. This may not be possible in all situations, and even if it is, the ``effective'' parameters on the averaged scale may depend on the more local physics. For example, averaging may be appropriate to describe rotating superfluid neutron stars, but it is easy to construct laboratory systems where  
 averaging is not appropriate. One may also envisage cosmological settings, e.g., involving dark matter condensates \citep{Harko11}, where averaging
 is not possible. In such situations we have to pay more careful attention to the forces acting on the vortices and the ensuing motion.
 
\subsection{Vortices and mutual friction}

Due to the fundamental quantum nature of superfluid (and for that matter, superconducting) condensates, the neutron component in a neutron star core
will be quantized into localized vortices that each carry a single quantum of momentum circulation. For simplicity, we will 
assume that the vortices are locally arranged in a rectilinear array, 
directed along a unit vector $\hat{\kappa}^i$, with surface density
$\mathcal{N} $. At the hydrodynamics level, after averaging and in the Newtonian gravity framework, we then have
\be
\mathcal{W}^i_\n = \frac{1}{m} \epsilon^{ijk} \nabla_j p^\n_k = \mathcal{N}  \kappa^i \ , 
\label{L1}
\ee
where we have used $\kappa^i = \kappa \hat{\kappa}^i $ with $\kappa = h / 2m$ the quantum of circulation (the factor of 2 arises from the underlying Cooper pairing, relevant for superfluid neutrons).
It is important to note that the quantized ``vorticities'' refer to the circulation of the canonical momentum
$p^i_\n$ rather than the circulation of velocity. It is the canonical
momentum which is related to the gradient of each condensate's wavefunction phase $ \varphi$,
leading to the Onsager-Feynman quantization condition
\be
\oint   p^i_\n dl_i= (\hbar/2) \oint  (\nabla^i \varphi)  dl_i = h/2 \ .
\ee

The variational analysis has already provided us with a two-fluid model that allows for vorticity (obviously).
However, if we want to understand the role of the vortices it is useful to consider the problem from a more
intuitive (albeit less general) point of view. To do this we generalize an approach  that was originally developed
in the context of two-fluid hydrodynamics for superfluid Helium \citep{1956RSPSA.238..215H}. This provides a conceptually
different derivation of the Euler equations, based on the kinematics of a conserved number of vortices. It also requires the input of the forces that determine the motion of a single isolated vortex. 
Thus, consistency between the two derivations allows us to identify the
total conservative force exerted on a single vortex, \emph{without} any need to study the detailed mesoscopic
vortex-fluid interaction. This will be useful when we  consider the  vortex mediated friction later.

The starting point of the derivation is the Onsager-Feynman condition (\ref{L1}).
We also need to use the fact that the vortex number density is conserved, i.e. $\mathcal{N} $ obeys a continuity equation of the form
\be
\partial_t \mathcal{N} + \nabla_j \left ( \mathcal{N}  v_\mathrm{v}^j\right ) =0 \ , 
\ee
where $v_\mathrm{v}^i$ is the collective vortex velocity within a typical fluid element---in a sense, this relation
defines this averaged vortex velocity.
Taking the time derivative of (\ref{L1}) we have
\be
\partial_t \mathcal{W}^i_\n = -\kappa^i \nabla_j ( \mathcal{N}   v_\mathrm{v}^j ) + \mathcal{N}  \partial_t \kappa^i \ .
\ee
Reshuffling terms and using the identity $\nabla_i \mathcal{W}^i_\n =0 $ we obtain
\be
\partial_t \mathcal{W}^i_\n = \nabla_j \left ( \mathcal{W}^j_\n v_\mathrm{v}^i \right ) -  \nabla_j \left ( \mathcal{W}^i_\n v_\mathrm{v}^j \right )
+ \mathcal{N}  \left ( \partial_t \kappa^i + v_\mathrm{v}^j \nabla_j \kappa^i -\kappa^j \nabla_j v_\mathrm{v}^i \right ) \ .
\label{Hall0}
\ee
The motion of a single vortex can be expressed as the Lie-dragging of the vector $\kappa^i $ (which designates the local vortex direction) by the $v_\mathrm{v}^i $ flow, leading to
\be
\partial_t \kappa^i + {\cal L}_{v_\mathrm{v}} \kappa^i=0 \ .
\label{Lie_kappa}
\ee
Then (\ref{Hall0}) reduces to 
\be
\partial_t \mathcal{W}^i_\n + \epsilon^{ijk} \nabla_j \left ( \epsilon_{klm} \mathcal{W}^l_\n v_\mathrm{v}^m\right ) =0 \ .
\label{vort1}
\ee
which states that the canonical vorticity $ \mathcal{W}^i_\n $ is locally conserved and advected by the $v_\mathrm{v}^i $ flow.
Rewriting the result in terms of the momentum, we have
\be
\partial_t p^i_\n -\epsilon^{ijk} v_{\mathrm{v}_j} \epsilon_{klm} \nabla^l p^m_\n = \nabla^i \Psi \ ,
\label{Halleqn}
\ee
where $\Psi$ is a (so far unspecified) scalar potential. 

Making use of the relative velocity, $ w^i_{\rm nv} = v^i_\n - v_\mathrm{v}^i$, we subsequently write (\ref{Halleqn}) as
\be
n_\n \partial_t p^i_\n - \epsilon^{ijk} n_j^\n \epsilon_{klm} \nabla^l p^m_\n -n_\n \nabla^i \Psi_\n
= \mathcal{N}  \rho_\n \epsilon^{ijk}  \kappa_j w_k^{\rm nv}\ .
\label{Halleqn1}
\ee
The left-hand-side of this equation coincides with the \emph{vortex-free} Euler equations of motion
(\ref{eulvort}) after a suitable identification of the potential $\Psi$. The right-hand side 
appears only in the presence of vortices. We can trace the origin of this contribution back to the Magnus force
exerted on a vortex (per unit length) by the associated fluid given by
\be
f^i_{\rm M} = -\rho_\n \epsilon^{ijk} \kappa_j w^{\rm nv}_k \ .
\ee
Thus,  we identify $-\mathcal{N}  f^i_{\rm M}$, the right-hand side of \eqref{Halleqn1}, as the \emph{averaged} reaction force
exerted on a fluid element by the vortex array. In the absence of balancing forces, like dissipative scattering off thermal excitations, the equation 
of motion for a single vortex leads to $f^i_{\rm M} =0$, implying that the vortices must move along with $v_\n^i$ flow. In this case, 
we retain (\ref{eulvort}) as the appropriate equation of motion.

This situation is, of course, somewhat artificial. In order for the argument to make sense, something must prevent the vortices from moving with the bulk flow. Of course, in order to describe a real superfluid, either at finite temperatures or co-existing with some other component (as in a neutron star core) we need (at least) two components. The interaction between the vortices and this second component effects the relative vortex flow. This interaction tends to be dissipative. The standard example of this is the so-called \emph{mutual friction} which assumes that the Magnus force acting on each vortex is balanced by resistivity with respect to the second component in the system (e.g., the thermal excitations in Helium, represented by $\x=\p$ here). That is we have  \citep{1956RSPSA.238..215H,1991ApJ...380..530M,2006MNRAS.368..162A} 
\begin{equation}
    f^i_{\rm M} = -\rho_\n \epsilon^{ijk} \kappa_j w^{\rm nv}_k  = - \mathcal R w^i_{\rm vp}
    \label{sfmag}
\end{equation}
which leads to---after repeated cross products to isolate the vortex velocity;
\be
f^\n_i =   \rho_\n \mathcal N  \kappa\left( \mathcal{B}' \epsilon_{ijk} \hat \kappa^j w_{\n\p}^k
+  \mathcal{B} \epsilon_{ijk}\hat{\kappa}^j \epsilon^{klm}\hat  \kappa_l w_m^{\n\p} \right) \ 
\label{mf}\ee
with
\be
\mathcal{B}' = \mathcal{R} \mathcal{B} = { \mathcal{R}^2 \over 1 + \mathcal{R}^2 } \ .
\label{bvsr}\ee
The mutual friction has decisive impact on superfluid dynamics. In particular, it provides one of the main mechanisms for damping (or even preventing) the CFS instability in rotating superfluid neutron stars \citep{1995ApJ...444..804L}.

\subsection{The Kalb--Ramond variation}

Moving on to the relativistic description of the quantized vortex problem, we have two options. We could ``simply'' generalize the steps from the Newtonian case. This is helpful, as it assists the intuition. However, it may be more instructive to take an alternative route. Opting for this strategy---with the view that it will allow us to introduce additional aspects---we now set out to derive the fluid results from a different perspective. The ultimate aim is to arrive at an alternative description of  the (suitably averaged) dynamics of a collection of quantized  vortices. 

The new strategy builds on efforts to relate  string dynamics to the forces acting on a superfluid vortex \citep{lund,kalb,ps1,ps2}. 
We start by recalling that the superfluid  velocity (technically; the momentum) can be linked  the gradient of a scalar potential $\varphi$. We  identify this velocity as the dual\footnote{In this section we use tildes to indicate duals, rather than the $\star$ notation. This is simply to avoid cluttering up expressions that already have both sub- and superscripts.} 
 \beq
 \tilde H_a = \eta \partial_a \varphi = {1\over 3!} \epsilon_{abcd} H^{bcd} \ , 
 \eeq
and introduce the so-called Kalb--Ramond field \citep{kalb}, such that
 \beq
 H^{abc} =  \partial^{[a} B^{bc]} \ .
 \eeq
 It is now easy to see that the scalar wave equation
 \beq
\Box \varphi = 0 \ , 
 \eeq 
 is automatically satisfied, as long as
 \beq
 \nabla_a \left( \nabla^a B^{bc} + \nabla^c B^{ab} + \nabla^b B^{ca} \right) = 0 \ .
 \eeq
 In effect, we can shift the focus from $\varphi$ to $B^{ab}$, treating this object as an independent variable. The relevant dynamical equations are then automatically solved by expressing this field in terms of a scalar potential. The two descriptions are complementary, as they have to be \citep{ps1}. However, as we will soon demonstrate, the  Kalb--Ramond representation makes the introduction of topological defects (vortices/strings) intuitive.

First, let us return to the fluid problem but shift the attention from the matter flux to the vorticity. 
Following \cite{kr1,kr2,kr3}, we do this by noting that we can ensure that the conservation law \eqref{consv2} is automatically satisfied by introducing a two-form $B_{ab}$  (the Kalb-Ramond field) such that
\beq
n_{abc} = 3 \nabla_{[a}B_{bc]}
\label{Bdef}\eeq
That is, we have
\beq
n^a = {1\over 2} \epsilon^{abcd} \nabla_b B_{cd}
\eeq
and the flux conservation \eqref{consv2}  follows as an identity---we no longer need to introduce the three-dimensional matter space. 

Second, in order to find an action that reproduces the perfect fluid results,  we elevate 
 the vorticity $\omega_{ab}$ to an additional variable. A Legendre transformation---designed in such a way that the stress-energy tensor remains unchanged \citep{kr2}---leads to the Lagrangian
\beq
\bar \Lambda = \Lambda - {1\over 4} \epsilon^{abcd} B_{ab} \omega_{cd}  = \Lambda - {1\over 2} \tilde \omega^{ab} B_{ab} \ , 
\label{newlambda}
\eeq
where we have used the dual
\beq
\tilde \omega^{ab} = {1\over 2}  \epsilon^{abcd} \omega_{cd} \ .
\eeq

Assuming that $\Lambda =\Lambda(n)$ we get (ignoring the perturbed metric for clarity)
\beq
\delta \bar \Lambda = -{1\over 3!} \mu^{abc} \delta n_{abc} - {1\over 2} B_{ab} \delta \tilde \omega^{ab} - {1\over 2} \tilde \omega^{ab} \delta B_{ab} \ , 
\eeq
where we note that, cf. Sect.~\ref{sec:pullback},
\beq
 {\partial \Lambda \over \partial n_{abc} } = - {1\over 3!} \mu^{abc} \ .
\eeq
However, we now have
\beq
\delta n_{abc} = 3 \nabla_{[a} \delta B_{bc]} \ , 
\eeq
which means that
\beq
\delta \bar \Lambda  ={1\over 2}  \left(  \nabla_{a}  \mu^{abc} -  \tilde \omega^{bc}  \right) \delta B_{bc}- {1\over 2} B_{ab} \delta \tilde \omega^{ab} -{1\over 2} \nabla_a \left( \mu^{abc} \delta B_{bc}\right) \ .
\label{krvar}\eeq
Ignoring the surface term (as usual), we see that a variation with respect to $B_{ab}$ requires
\beq
\tilde \omega^{bc} =  \nabla_{a}  \mu^{abc} \ , 
\label{vortdef}\eeq
which leads back to \eqref{omdef}.
However, with a free variation we would also have $B_{ab}=0$.  That is, we need to constrain the variation of $\tilde\omega^{ab}$ (or rather $\omega_{ab}$). Fortunately, the matter space argument comes to the rescue, providing us with the strategy for doing this. The only difference is that we now make use of a two-dimensional space with coordinates $\chi^I$ (here, and in the following $I,J,\ldots$ represent two-dimensional coordinates). We obtain this two-dimensional space either via a map from the original matter space
\begin{equation}
 \hat \psi^I_A = {\partial \chi^I \over \partial X^A} \ ,    
\end{equation}
or directly from spacetime, using
\begin{equation}
  \bar \psi^I_a = {\partial \chi^I \over \partial x^a}\ .   
\end{equation}
The two descriptions are  consistent since
\begin{equation}
\bar \psi^I_a =\hat \psi^I_A  \psi^A_a  = {\partial \chi^I \over \partial X^A}  {\partial X^A \over \partial x^a} = {\partial \chi^I \over \partial x^a}  \ .
\end{equation}
The different coordinates and the maps are illustrated in Fig.~\ref{maps}.

\begin{figure}
\centering
\includegraphics[width=0.8\textwidth]{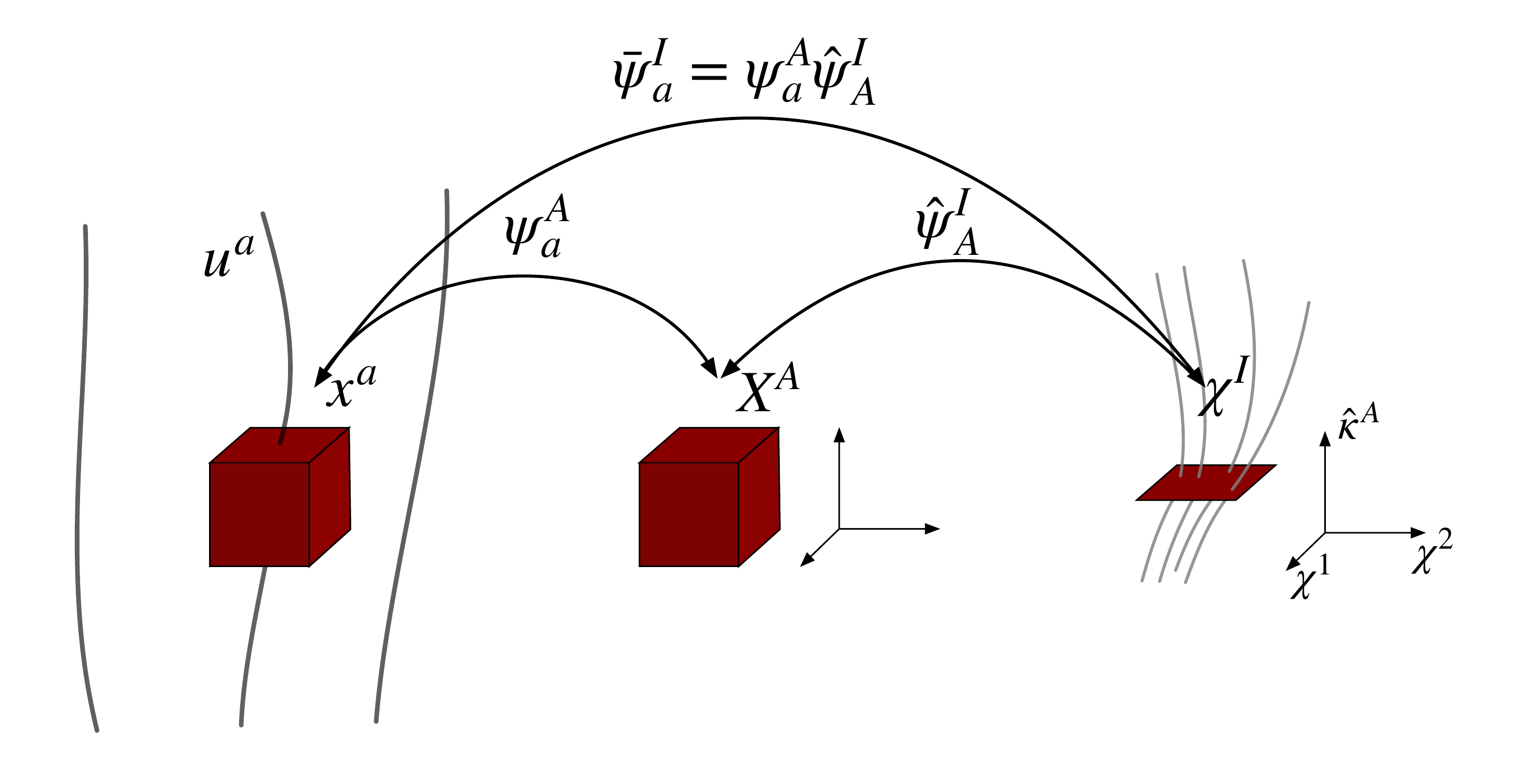}
\caption{An illustration of the matter space maps and  the coordinates used in the analysis of vortex dynamics and elasticity.}
\label{maps}
\end{figure}

The third step involves introducing  the four velocity $u^a$ associated with the motion of the vortices in spacetime, which may be different from the motion of the ``fluid'' (in turn related to $n^a$). In order for the vorticity to be a purely spatial object---orthogonal to the flow--we must have
\beq
u^a\omega_{ab} = 0  \ .
\eeq
In addition, we want it to be ``fixed'' in the (new) matter space, in the sense that
\beq
\mathcal L_u \omega_{ab} = 0 \ .
\eeq
Since $\omega_{ab}$ is anti-symmetric, this leads to 
\beq
u^c \nabla_{[a}\omega_{bc]} = 0 \ ,
\eeq
which will be satisfied if
\beq
\nabla_{[a}\omega_{bc]} = \partial_{[a}\omega_{bc]} = 0 \ .
\label{twop1}\eeq
Adapting the logic that led to the conserved matter flux in Sect.~\ref{sec:pullback}, we  introduce the matter space tensor $\omega_{IJ}$, such that 
\beq
\omega_{ab} = \psi^A_a \psi^B_b \omega_{AB} = \bar \psi^I_a \bar \psi^J_b \omega_{IJ} \ .
\eeq
Noting that \eqref{twop1} becomes
\beq
 \partial_{[a}\omega_{bc]} = \bar \psi^I_a \bar \psi^J_b \bar \psi^K_c \partial_{[I} \omega_{JK]} = 0  \ , 
 \label{vortder}
\eeq
it follows that the  condition holds as long as  $\omega_{IJ}$ only depends on the $\chi^I$ coordinates. It should (by now) be a familiar argument.

Next, we introduce Lagrangian perturbations such that
\beq
\Delta \chi^I = 0 \longrightarrow \delta \chi^I = - \mathcal L_\xi \chi^I \ , 
\eeq
and we  have 
\beq
\Delta \omega_{ab} = 0 \ .
\eeq
Again leaving out the metric variations, we have
\beq
\delta \tilde \omega^{ab} = {1\over 2} \epsilon^{abcd} \delta \omega_{cd}
= - \xi^c \nabla_c \tilde \omega^{ab} - \epsilon^{abcd} \omega_{ed} \nabla_c \xi^e \ , 
\eeq
and, after a little  bit of work,  the middle term in \eqref{krvar} becomes
\beq
- {1\over 2} B_{ab} \delta \tilde \omega^{ab} 
 = {3\over 2} \xi^c \tilde \omega^{ab} \nabla_{[c} B_{ab]} + \nabla_c \left(\omega^{ab} B_{ab} \xi^c\right) \ .
\eeq
We have have noted that, \eqref{vortdef} implies that 
\beq
\nabla_a \tilde \omega^{ab} = 0 \ .
\eeq
Finally, we see that a variation with respect to $\xi^a$ leads to 
\beq
{3\over 2}  \tilde \omega^{ab} \nabla_{[c} B_{ab]} = {1\over 4}  \epsilon^{abde} \omega_{de}   n_{cab}  =   n^d \omega_{dc} = 0 \ , 
\eeq
and we recover the usual  fluid equations.
This completes the initial argument. The introduction of the Kalb-Ramond field shifts the focus onto the vorticity,  which is associated with a two-dimensional subspace (replacing the usual three-dimensional matter space). The key point is that we arrive at fluid equations without explicitly associating the fluid flux $n^a$ with the four-velocity $u^a$.  

\subsection{String fluids}

In order to form a complete picture---including connections with related problems---and develop the tools we need to make progress, it is useful to take a slight detour in the direction of string theory. The key point is that, a one-dimensional string moving through spacetime traces out a two-dimensional world sheet. This world sheet is spanned by two vectors, one timelike (here taken to be the four velocity of the string, $u^a$) and one spacelike (intuitively, the tangent vector to the string, represented by $\hat \kappa^a$). These vectors are  associated with two-dimensional coordinates\footnote{These coordinates are not the same as the $\chi^I$ from the Kalb-Ramond action. When combined, the two sets of coordinates provide us with the means to completely represent spacetime.} such that  $x^a = x^a (\phi^I)$, leading to
 the tangent surface element
 \beq
S^{ab} = \epsilon^{IJ} {\partial x^a \over \partial \phi^I} {\phi x^b \over \partial \phi^J} \ , 
\label{surfel}
 \eeq
 with $\epsilon^{IJ}$ the (normalised) two-dimensional Levi-Civita tensor (density). 

Associated with this world sheet we have a bivector (read: an anti-symmetric tensor of rank 2), to be denoted $\Sigma^{ab}$. This object can be expressed in terms of the linearly independent vectors that span the surface; as the bivector spans a surface, it is natural to think of it as a contravariant object. Noting that   a simple timelike bivector can be written as the alternating product of a timelike and a spacelike vector \citep{stachel} (such that its dual will be a simple spacelike bivector) and  assuming the normalisation 
\beq
 \Sigma_{ab} \Sigma^{ab} = -2 \ , 
 \label{norm}
 \eeq
we may use
\beq
 \Sigma^{ab} = u^a  \hat\kappa^b  - u^b \hat \kappa^a \ , 
 \label{biv}
\eeq
such that
\beq
\hat \kappa^a = \Sigma^{ab} u_b \ .
\eeq
The projection into the two-dimensional space spanned by $u^a$ and $\hat \kappa^a$ is then given by
 \beq
 \Sigma^{ac} \Sigma_{cb} =  \hat \kappa^a \hat \kappa_b - u^a u_b \ . 
 \label{simple1}
 \eeq
 
 Introducing the dual
\beq
\tilde \Sigma_{ab} = {1\over 2} \epsilon_{abcd} \Sigma^{cd}  =   \epsilon_{abcd} u^c \hat\kappa^d \ , 
\eeq
we also have the  orthogonal projection 
 \beq
 \tilde \perp^a_{\ b} = \tilde \Sigma^{ac} \tilde \Sigma_ {cb} = \delta^a_b + u^a u_b - \hat \kappa^a \hat \kappa_b \ ,
 \eeq
 and we see that
\beq
\tilde \Sigma_{ab} \Sigma^{bc} = 0  \ .
\eeq
In fact, this result follows immediately from the condition that the bivector is simple:
\beq
\Sigma^{[ab} \Sigma^{c]d} = 0 \quad \Leftrightarrow \quad \Sigma^{ab} \Sigma^{cd} \epsilon_{abce}  = 0  \ .
\label{simple}
\eeq
Finally, the bivector is surface forming, as long as \citep{stachel}
 \beq
\tilde \Sigma_{ab} \nabla_c \Sigma^{bc} =  \tilde \Sigma_{ab} \partial_c \Sigma^{bc} = 0 \ .
 \label{surf}
 \eeq
  
With this set up, we may take the bivector to be proportional to the surface element. Letting
\beq
\Sigma^{ab} = \alpha^{-1/2} S^{ab} \ , 
\eeq
we have
\beq
\Sigma^{IJ} = \alpha^{-1/2} S^{IJ} = \alpha^{-1/2}  \epsilon^{IJ} \ .
\eeq
Making use of the induced metric (which we also use to raise and lower indices in the two-dimensional subspace)
 \beq
 \gamma_{IJ} = g_{ab} {\partial x^a \over \partial \phi^I} {\partial x^b \over \partial \phi^J} \ , 
 \eeq
 we have
 \beq
  \gamma_{IK} \gamma_{JL}  \epsilon^{IJ}  \epsilon^{KL} = - 2 \alpha \ , 
 \eeq
 and hence we identify
 \beq
 \alpha = - \gamma = - \mathrm{det}\ \gamma_{AB} \ .
 \eeq
 That is, we arrive at
 \beq
 \Sigma^{ab} = \sqrt{-\gamma} {\partial x^a \over \partial \phi^I} {\partial x^b \over \partial \phi^J}  \epsilon^{IJ} \ .
 \eeq
 Geometrically, the dual of $\Sigma^{ab}$ is a two-form that represents (when integrated) the flux carried by vortices (string) across a surface in spacetime. The variable $\gamma$ is a measure of this flux.

 Let us now assume that the Lagrangian of the system depends on $\gamma$, with 
 \beq
 \gamma =  {1\over 2} \Sigma^{ab} \Sigma_{ab} = {1\over 2} \Sigma^{IJ} \Sigma_{IJ} = -1 \ .
 \eeq
 Moreover, as we want to compare to a model based on averaging over a set of vortices---treated as a fluid described by a small number of fields (density, velocity, tension etcetera)---it is natural to consider the analogous example of a coarse-grained ``string fluid'' \citep{shub1,shub2,shub3}. In effect, we take 
 $\sqrt{-g} \Lambda(\gamma)$ to be the matter contribution to the action.  Further, if we let $\Lambda = M \sqrt{-\gamma}$ this leads to the coarse-grained version of the standard Nambu--Goto string action \citep{letel,vile}, with $M$ the string tension.

For the stress-energy tensor we now need
\begin{multline}
\delta \Lambda = {d\Lambda \over d\gamma} \left( {\partial \gamma \over \partial \Sigma^{ab} } \delta \Sigma^{ab} +  {\partial \gamma \over \partial g_{ab}} \delta g_{ab} \right)
\\
= {d\Lambda \over d\gamma} \left( \Sigma_{ab} \delta \Sigma^{ab} + \Sigma_c^{\ a} \Sigma^{cb} \delta g_{ab} \right) \ , 
\end{multline}
which leads to
\beq
T^{ab}  =  \Lambda g^{ab} + 2 {\delta \Lambda \over \delta g_{ab}} = \Lambda g^{ab} + 2 {d\Lambda \over d\gamma} \Sigma_c^{\ a} \Sigma^{cb}  \ .
\eeq
From this it follows that the equations of motion are
\beq
\nabla_a T^{ab} = g^{ab} \nabla_a \Lambda + 2  \Sigma_c^{\ a} \Sigma^{cb} \nabla_a \left( {d\Lambda \over d\gamma} \right) +  2 {d\Lambda \over d\gamma} \nabla_a   \left( \Sigma^a_{\ c}  \Sigma^{cb}\right) = 0 \ .
\eeq
 However, we have
 \beq
  \nabla_a \Lambda = {d\Lambda \over d\gamma} \nabla_a \gamma = 0  \ , 
 \eeq
 and
 \beq
 \nabla_a \left( {d\Lambda \over d\gamma} \right) = \left( {d^2\Lambda \over d\gamma^2} \right) \nabla_a \gamma = 0 \ , 
 \eeq
 since $\gamma=-1$. This means that we have
 \begin{multline}
 \nabla_a   \left( \Sigma^a_{\ c}  \Sigma^{cb}\right) 
 =  \Sigma^{cb} \nabla_a \Sigma^a_{\ c}  + {1 \over 2} \Sigma_{ca} \left(  \nabla^a  \Sigma^{cb} +\nabla^c \Sigma^{ba} + \nabla^b \Sigma^{ca} \right) 
 \\
 =  \Sigma^{cb} \nabla_a \Sigma^a_{\ c}  + 3 \Sigma_{ca}   \nabla^{[a}  \Sigma^{cb]}  =0 \ , 
  \end{multline}
  where we have used \eqref{norm}.
Following \cite{stachel}, we contract with $\Sigma_{db}$ to get 
  \beq
 \Sigma_{db}  \Sigma^{cb} \nabla_a \Sigma^a_{\ c}  + 3 \Sigma_{[ac} \Sigma_{b]d}   \nabla^{[a}  \Sigma^{cb]}  =0 \ , 
  \eeq
  where the second term vanishes since the bivector is simple, cf. \eqref{simple}. Noting also that 
  \beq
   \Sigma_{db}  \Sigma^{cb} \nabla_a \Sigma^a_{\ c} = 0 \Longrightarrow  \Sigma_{dc} \nabla_a \Sigma^{ac} =0 \ .
  \eeq
  and considering \eqref{surf}, we infer the conservation law \cite{stachel,shub3} 
  \beq
  \nabla_a \Sigma^{ab} = 0 \ .
  \label{conse}
  \eeq
Basically, if the contractions of a vector with both the bivector and the dual vanish then the vector must itself be zero.  Returning to the equations of motion, we are left with 
 \beq
  \Sigma^a_{\ c}  \nabla_a  \Sigma^{cb} = 0\ , 
  \eeq
  or
  \beq
   \perp^c_b \left( \hat \kappa^a \nabla_a \hat \kappa^b - u^a \nabla_a u^b \right) = 0 \ .
  \eeq
This is the simplest version of the model and it is all we need for now. Still, it is interesting to note extensions like the dissipative case considered in \cite{shub3} and the discussion of charged cosmic strings in \cite{cart}.

 Before we move on, let us establish two useful results. First of all, we have
  \beq
\hat \kappa^a = \Sigma^{ab} u_b \Longrightarrow  \nabla_a \hat \kappa^a  + u^a u_b  \nabla_a \hat \kappa^b = u_b \nabla_a \Sigma^{ab} = 0 \ , 
\label{divk}
 \eeq
 by virtue of \eqref{conse}. 
 Similarly 
 \beq
  u^a = \Sigma^{ab} \hat \kappa_b   \Longrightarrow
  \nabla_a u^a  - \hat \kappa^a \hat \kappa_b  \nabla_a u^b = \hat \kappa_b \nabla_a \Sigma^{ab} = 0 \ .
  \label{divu}
 \eeq
 These will be required later.

\subsection{Vortex dynamics}

A natural extension to the fluid model allows $\Lambda$ to depend on both $n_{abc}$ and $\omega_{ab}$ from the outset. Starting from $\Lambda = \Lambda (n_{abc}, \omega_{ab}, g^{ab})$ we immediately have 
\beq
\delta \Lambda = -{1\over 3!} \mu^{abc} \delta n_{abc} - {1\over 2} \lambda^{ab} \delta \omega_{ab} + {\delta \Lambda \over \delta g^{ab}} \delta g^{ab} \ , 
\eeq
where
\beq
\lambda^{ab} = - 2 {\partial \Lambda \over \partial \omega_{ab} } \ .
\eeq
From \eqref{newlambda} it then follows that (ignoring the metric variation and the surface term, as before)
\beq
\delta \tilde \Lambda 
= {1\over 2} \left( \nabla_{c}  \mu^{cab}- \tilde \omega^{ab} \right) \delta B_{ab}  - {1\over 2} \left(  \lambda^{cd} + {1\over 2}  \epsilon^{abcd} B_{ab} \right) \delta \omega_{cd} \ , 
\eeq
which leads us back to \eqref{krvar} and  \eqref{vortdef}. However,  we now have an additional term involving $\delta \omega_{ab}$. 
Making use of \eqref{twop1}, this new term can be written
\beq
-{1\over 2}   \lambda^{cd}  \delta \omega_{cd} = {1\over 2}  \lambda^{cd}\left( \xi^a \nabla_a \omega_{cd} + 2 \omega_{ad} \nabla_c \xi^a \right) \\
=  - \xi^a  \omega_{ad} \nabla_c \lambda^{cd} \ .
\eeq
Combining this  with the result from the previous section, we see that a variation with respect to the displacement leads to (see \citealt{kr1,kr2,kr3})
\beq
n^a \omega_{ab} =  \omega_{ab} \nabla_c \lambda^{ca} = - 2  \omega_{ab} \nabla_c \left({\partial \Lambda \over \partial \omega_{ca} } \right) \ .
\label{neweuler}
\eeq
The explicit dependence on the vorticity has led to amended equations of motion. In order to interpret the term on the right-hand side of \eqref{neweuler} we, first of all, note that we may write \eqref{neweuler} as
\beq
\left[ n^a + 2  \nabla_c \left({\partial \Lambda \over \partial \omega_{ca} } \right) \right] \omega_{ab}  \equiv \bar n^a \omega_{ab} = 0 \ , 
\eeq
with 
\beq
\bar n^a = n^a + 2  \nabla_c \left({\partial \Lambda \over \partial \omega_{ca} } \right) \ .
\eeq
This makes the result appear more ``familiar'', but it does not really help us understand the contributions to \eqref{neweuler}. 

Let us dig deeper. Consider the implications of the two-dimensional matter space we introduced for the vorticity, see Fig.~\ref{maps}. Intuitively, the idea makes sense for a collection of (locally) aligned quantized vortices as one can always introduce a two-dimensional surface orthogonal to the vortex array. Points in this surface are described by the $\chi^I$ coordinates. 
Not surprisingly, we can adapt the logic from the usual matter-space construction to this new setting---although  in doing so we focus on the map from the original three-dimensional space to the two-dimensional one. As is evident from \eqref{vortder}, we  also need the map from spacetime to either low-dimensional space. The original fluid derivation involved
\beq
\psi^A_b \psi^a_A = \perp^a_{\ b} \ ,
\eeq
while the corresponding map to the two-dimensional stage takes the form 
\beq
\hat \psi^I_B \hat \psi^A_I = \delta^A_B - \hat \kappa^A \hat \kappa_B \ ,
\eeq
with a suitable spatial unit vector $\hat \kappa^a$, automatically orthogonal to the four velocity $u^a$ since
\beq
u^a \hat \kappa_a = (u^a \psi^A_a) \hat \kappa^A = 0 \ .
\eeq
We will take the new vector $\hat \kappa^a$ to be normal to the area spanned by the $\chi^I$ coordinates (and identify it with the spacelike coordinate used to describe the string world sheet). That is, we have
\beq
\hat \kappa^A \hat \psi^I_A = 0 \ .
\eeq
In essence, $\hat \kappa^A$ is aligned with the quantized vortices. 
It also follows that 
\begin{multline}
\bar \psi^I_a \bar \psi^b_I = ( \psi^A_a \hat \psi^I_A)  ( \psi^b_B \hat \psi^B_I)  = \psi^A_a \psi^b_B (\delta_A^B - \hat \kappa_A \hat \kappa^B ) 
\\
=  \delta_a^b+u_a u^b - \hat \kappa_a \hat \kappa^b \equiv  \tilde \perp^b_a \ .
\end{multline}

Turning to the vorticity, it is natural to introduce a vector
\beq
W^A = {1\over 2} \epsilon^{ABC} \omega_{BC} \longrightarrow \omega_{AB} = \epsilon_{ABC} W^C \ .
\eeq
In spacetime, we then have the vorticity  vector 
\beq
W^a =  {1\over 2} \psi^a_A \epsilon^{ABC} \omega_{BC} =  {1\over 2} \psi^a_A \psi^b_B \psi^c_C \epsilon^{ABC} \omega_{bc} = {1\over 2} u_d \epsilon^{dabc} \omega_{bc} \ , 
\eeq
which is simply related to the dual:
\beq
W^a = u_d \tilde \omega^{da} \ .
\eeq

We may also work in the two-dimensional space, where it makes sense to let
\beq
\omega_{IJ} = \mathcal N \kappa \epsilon_{IJ} \longrightarrow \omega_{AB} = \mathcal N  \kappa  \epsilon_{AB} \ , 
\eeq
with 
\beq
\epsilon_{IJ} \epsilon^{JK} = \delta_I^K \ , 
\eeq
\beq
\epsilon_{IJ} \epsilon^{IJ} = 2 \ , 
\eeq
and
\beq
\epsilon_{AB} = \hat \kappa^C \epsilon_{CAB} \ .
\eeq	

Letting $\kappa^A = \kappa \hat \kappa^A$, we now have
\beq
 \omega_{AB} = \mathcal N   \kappa^C \epsilon_{CAB}  \ , 
\eeq
so
\beq
\kappa^A \omega_{AB} = 0 \ .
\eeq
In fact, we have
\beq
W^A = \mathcal N  \kappa^A \ .
\eeq
The interpretation of this is intuitive---we have a collection of vortices, each associated  with a quantum $\kappa$ of circulation---with number density (per unit area) $\mathcal N $. It is also worth noting the close resemblance to the various relations for $n_{ABC}$ from Sect.~\ref{sec:pullback}.
We also have
\begin{multline}
W^2 = (\mathcal N  \kappa)^2 =   {1\over 2} \omega_{IJ} \omega^{IJ} = {1\over 2} \omega_{AB} \omega^{AB} \\
=  {1\over 2} \omega_{ab} \omega^{ab} = {1\over 2} g^{ac} g^{bd} \omega_{ab} \omega_{cd}
\end{multline}

Finally, the spacetime vorticity takes the (expected)  form 
\beq
\omega_{ab} = \mathcal N  u^c \kappa^d \epsilon_{cdab} \ .
\label{omnew}
\eeq
We also have
\begin{multline}
\mathcal L_u \kappa_a = \mathcal L_u \left( \psi_a^A \kappa_A \right) = \psi_a^A  \mathcal L_u \kappa_A  = \psi^A_a u^c \partial_c \kappa_A
\\
=  \psi^A_a (u^c  \psi^B_c) {\partial \kappa_A\over \partial X^B } = 0 \ , 
\end{multline}
\beq
u^b \nabla_b  \mathcal N  = (u^b \tilde \psi^I_b) {\partial \mathcal N  \over \partial \chi^I} = 0 \ ,
\label{uNperp}
\eeq 
as well as
\begin{multline}
\kappa^a \nabla_a \mathcal N  = \kappa^a \tilde \psi^I_a  {\partial \mathcal N  \over \partial \chi^I}  = \psi_A^a \kappa^A \psi^B_a \hat \psi^I_B   {\partial \mathcal N  \over \partial \chi^I}  \\
= \kappa^A \delta^A_B \hat \psi^I_B  {\partial \mathcal N  \over \partial \chi^I}  = \kappa^A \hat \psi^I_A  {\partial \mathcal N  \over \partial \chi^I}  = 0 \ .
\label{kNperp}
\end{multline}	
These results are quite intuitive. It is worth noting that 
\beq
    (u^a u_b - \hat \kappa^a \hat \kappa_b )  \nabla_a \mathcal N   = 0 \ , 
    \label{Nperp2}
 \eeq
 and we also need to recall \eqref{divk} and \eqref{divu}. 
 
Let us now return  to the equations of motion \eqref{neweuler}. If we consider an explicit model where $\Lambda = \Lambda(n^2 , \mathcal N ^2)$, we have
\beq
{\partial \Lambda \over \partial \omega_{ab} } =  {\partial \Lambda \over \partial \mathcal N ^2}  {\partial \mathcal N ^2 \over\partial \omega_{ab} }=  {\partial \Lambda \over \partial \mathcal N ^2} \omega^{ab}  = -{1\over 2} \lambda^{ab} \ , 
\eeq
and we arrive at 
\beq
n^a\omega_{ab} =  - {2 \over \kappa^2} \omega_{ab} \nabla_c \left(  {\partial \Lambda \over \partial \mathcal N ^2} \omega^{ca}\right) = - {1 \over \kappa} \omega_{ab} \nabla_c \left(  {\partial \Lambda \over \partial \mathcal N } {1\over \mathcal N  \kappa} \omega^{ca}\right) \ .
\label{final}\eeq

Making use of \eqref{omnew} we then have
\begin{multline}
 {1 \over \kappa} \omega_{ab} \nabla_c \left(  {\partial \Lambda \over \partial \mathcal N } {1\over \mathcal N  \kappa} \omega^{ca}\right)   \\
 = -  \mathcal N  \perp^a_b \left[  \nabla_a \left( {\partial \Lambda \over \partial \mathcal N } \right) -  {\partial \Lambda \over \partial \mathcal N }  \left( \hat \kappa^c \nabla_c \hat \kappa_a  -  u^c \nabla_c u_a \right)\right] \ .
 \label{fvort}
\end{multline}
Here it is worth noting that $-\partial \Lambda /\partial \mathcal N $ is naturally interpreted as the energy per vortex (assuming that all vortices carry the same circulation and that the averaged energy is simply proportional to the vortex density.  It is straightforward to make a connection  with the ``thin vortex'' limit  considered by \cite{kr3} but we will not do so here.

Suppose that we also introduce a four-velocity associated with the matter flux, i.e. let
\beq
n^a = n u_\n^a \ ,
\eeq
such that (as usual)
\beq
u_\n^a = \gamma ( u^a + v^a) \ , \quad u^a v_a = 0 \ , \quad \gamma = (1-v^2)^{-1/2} \ , 
\eeq
We then have 
\beq
n^a \omega_{ab} =n  \gamma \mathcal N  v^a \kappa^d \epsilon_{dab}   =  n \gamma \mathcal N  \epsilon_{bac} \kappa^a v^c \ , 
\eeq
which represents the  Magnus force  that acts on a set of  vortices moving relative to a superfluid condensate (represented by $n^a$), cf. Eq.~\eqref{sfmag}. Also recognizing the surface tension associated with vortex world sheet, we have the final equations of motion
\beq 
 \underbrace{n \gamma  \epsilon_{bac} \kappa^a v^c}_\mathrm{Magnus\ force}
  =    \perp^a_b \left[  \nabla_a \left( {\partial \Lambda \over \partial \mathcal N} \right) - \underbrace{ {\partial \Lambda \over \partial \mathcal N}  \hat \kappa^c \nabla_c \hat \kappa_a +  {\partial \Lambda \over \partial \mathcal N}  u^c \nabla_c u_a }_\mathrm{surface\ tension} \right] \ .
  \label{vomo}
\eeq

For completeness, we should also work out the stress-energy tensor for this model. This is fairly straightforward.  With $\Lambda = \Lambda(n^2, \mathcal N^2) = \Lambda(n_{abc}, \omega_{ab}, g^{ab})$ we need 
\beq
{\partial \Lambda \over \partial \mathcal N^2 } \delta \mathcal N^2  ={1\over 2 \mathcal N \kappa^2 } {\partial \Lambda \over \partial \mathcal N}   \left( g^{cd} \omega_{ca}   \omega_{db} \delta g^{ab}  +   \omega^{bd} \delta \omega_{bd} \right) \ , 
\eeq
leading to a contribution  (using \eqref{omnew})
\beq
{\partial \Lambda \over \partial \mathcal N^2 } {\delta \mathcal N^2 \over \delta g^{ab}} =  {1\over 2}   \mathcal N   {\partial \Lambda \over \partial \mathcal N} \perp_{ab} \ .
\eeq
Combining this with the previous (fluid) result, we have
\beq
T_{ab} = \left( \Lambda - n^c \mu_c \right) g_{ab} + n_a \mu_b -  \mathcal N   {\partial \Lambda \over \partial \mathcal N} \perp_{ab} \ .
\label{stressvort}
\eeq
A direct calculation verifies that the divergence of this expression leads us back to \eqref{vomo}. 

We can extend the vortex model---following the steps from the Newtonian case---to account for mutual friction \citep{2016CQGra..33x5010A}. We may also consider the implications of the long-range nature of the vortex-vortex interaction, which implies that the vortex lattice has elastic properties \citep{1983JLTP...50...57B,1986JLTP...62..119C,Andersson_2020}. In principle, this means that the vortex lattice supports a set of elastic oscillation modes known as Tkachenko modes \citep{2014JETPL..98..758S}. These were first proposed in the 1960s \citep{1966JETP...22.1282T,1966JETP...23.1049T}, and have been discussed for superfluid helium, superfluid atomic condensates \citep{2002cond.mat.10063A,2009RvMP...81..647F} and neutron stars \citep{1970Natur.225..619R,2008PhRvD..77b3008N,2011PhRvD..83d3006H}. The experimental verification of the idea is, however, quite recent \citep{2003PhRvL..91j0402C}.


\section{Perspectives on electromagnetism}

Magnetic fields are ubiquitous in the Universe---electricity and magnetism are of obvious importance to our every day existence, and electromagnetism also plays a crucial role in astrophysics. In the  context of general relativistic fluid dynamics, we are particularly interested in situations where 
strong gravity couples to charged flows.
 A typical example of such a problem would be two magnetized neutron stars crashing together at the end of a slow  inspiral driven by the emission of gravitational radiation \citep{2017RPPh...80i6901B}. Another interesting problem concerns ultra-relativistic jets 
associated with active galactic nuclei (and some stellar mass objects, as well), thought to be generated by the spin of the central object (via the so-called Blandford--Znajek mechanism; \citealt{1977MNRAS.179..433B,1982MNRAS.198..345M}).  Neutrons stars come into focus as the strongest known magnetic fields (above $10^{14}$~G) are found in a subclass aptly referred to as magnetars \citep{dynamo,magnetar}, systems that also form the largest (and hottest!) known superconductors \citep{casa1,casa2}. Magnetic fields are equally relevant on the vastly larger scale of entire galaxies, and are likely to have played a role in the early Universe as well \citep{ellis,ellis2,tsagas}.  These are just a few---fairly obvious---examples that illustrate why we need to develop an understanding of the interaction between charged fluids (generating and maintaining the electromagnetic field) and relativistic gravity.

\subsection{The Lorentz force}

We laid the foundation for the covariant description of electromagnetism in Sect.~\ref{sec:emvar} (see also \citealt{efsth}). Starting from a suitable Lagrangian
that couples the vector potential $A_a$ (in the form of the Faraday tensor $F_{a b}$) to the four-current $j^a$, we established that the electromagnetic field is governed by
\be
\nabla_b F^{a b} = \mu_0 j^a \ .
\label{max1}\ee
Moreover, since $F_{a b}$ is anti-symmetric, it will automatically satisfy 
\be
\nabla_{[c} F_{a b]} = 0 \ . 
\label{max2}\ee
However, up to this point we had to take the claim that these equations describe electromagnetism on faith. In order for the model to make more intuitive sense, we need to make contact with the standard description in terms of the electric and magnetic fields and Maxwell's equations. 

This exercise is, in principle, straightforward, but at the same time one must tread carefully. In order to be consistent, we need to be mindful of the units of the various quantities involved. Unfortunately, the issue of units is somewhat thorny in electromagnetism. The underlying reason for this is that the theory involves two 
``coupling constants", which we will call $\mu_0$ and $\epsilon_0$. We have already seen the first of these, and we know that it represents the strength
of the coupling between the field and the current. As we will soon see, the second of the two coefficients represents the coupling to the charge density. 
The two coefficients combine in such a way that $\mu_0 \epsilon_0 = 1/c^2$, defining the speed of light\footnote{Note that we generally use geometric units, so $c^2=1$.}. However, splitting this ``constraint'' involves an element of choice, which leads to different (perfectly consistent) sets of units. In fact, in his celebrated textbook \cite{jackson} makes the point that the two constants \emph{must} be 
chosen arbitrarily.  In the following, we will opt to work in (what is essentially) SI units, occasionally providing the ``translation'' to the Gauss units that are common in astrophysics.

Another issue that makes the problem non-trivial arises from the fundamental principle of electromagnetism; varying electric fields generate magnetic fields and vice versa. This implies that the decomposition into electric and magnetic fields must be observer dependent. If two observers move in different ways then they will observe different charge currents and therefore different fields.

According to an observer moving with four-velocity\footnote{Adapting the convention from Sect.~\ref{tab1} that $U^a$ is associated with a general observer, in order to distinguish between the two specific choices considered later.} $U^a$, the Faraday tensor takes the form\footnote{Our discussion differs from alternatives like \cite{ellis} in a few subtle ways.  First of all the sign of the magnetic field $B^a$ is different, but this is later compensated for by a difference in the definition of $\epsilon_{abc}$. These differences mean that any comparison with the literature must be carried out with care.}
\be
F_{ab} =  2 U_{[a} E_{b]}  + \epsilon_{abcd}U^c B^d \ .
\label{Faraday}\ee
This defines the electric and magnetic fields as
\be
E_a = -  U^b F_{ba} \ , 
\ee
and
\be
B_a = - U^b \left( {1 \over 2} \epsilon_{abcd}F^{cd}\right) \ .
\ee
The physical fields are both orthogonal to $U^a$, so 
each has three components, just as in non-relativistic physics. 

In the presence of a medium, we also need an expression for the charge current, and it is natural 
to decompose this in a similar way; namely,
\be
j^a = \sigma U^a + J^a \ , \qquad \mbox{where} \qquad J^a U_a = 0  \ .
\label{ecurr}
\ee

Intuitively, the electromagnetic field couples to the moving fluids through the Lorentz force. It is easy to see how this notion comes about. 
The overall stress-energy tensor for the system combines a ``matter'' part with the relevant electromagnetic contribution. The overall divergence has to vanish, as usual. This means that we can define the magnetic force $f^a_\mathrm{L}$ as
\be
\nabla_b T^{b a}_\mathrm{fluid} = - \nabla_b T^{b a}_\mathrm{EM} \equiv f^a_\mathrm{L} \ .
\ee
Making use of the explicit  stress-tensor for the electromagnetic field from Sect.~\ref{sec:variational}; 
\be
T_{ab}^\mathrm{EM} =  {1\over \mu_0} \left[g^{cd}F_{ac}F_{bd}-{1\over 4}g_{ab}\left( F_{cd}F^{cd}\right) \right] \ .
\label{TEM}\ee
 we find that
\be
f^a_\mathrm{L} =  j_b F^{ab} \ . 
\ee
Alternatively, making use of the decomposition into the electric and magnetic fields, we have 
\be
f^a_\mathrm{L} = \sigma E^a + \epsilon^{a b c d} J_b U_c B_d + U^a \left( J_b E^b\right) \ .
\ee

This exercise prompts a  fundamental question. What exactly is the current $j^a$? Intuitively, we know 
the answer. A net current results from different charged components flowing relative to one another. However, the single-fluid 
picture that we have considered so far (with a single observer) does not consider this aspect. It only provides the final result, which is the charge current that is required to source the electromagnetic field. In order to understand the physics, we need to consider a system of coupled charged fluids.
It is natural to do this by extending the variational approach to account for charged flows. Fortunately, this is straightforward and we will do this shortly. However, before going in this direction, let us convince ourselves that we have (indeed) a formulation that leads back to Maxwell's equation.

\subsection{Maxwell in the fluid frame}

As a step towards making  contact with applications, it is useful to consider the form of Maxwell's equations in the fluid frame. That is, we introduce a fibration of spacetime associated with the fluid four velocity $u^a$ (again, as in the discussion of the stress-energy tensor in Sect.~\ref{sec:pullback}). This leads to the formulation that is commonly used to discuss electromagnetism, especially in cosmology \citep{ellis,ellis2,tsagas}. 

In order to write down Maxwell's equation it is useful to introduce the general decomposition
\be
\nabla_a u_b  = \sigma_{ab} + \varpi_{ab} - u_a \dot u_b + {1 \over 3} \theta \perp_{ab}  \ ,
\label{decomp}\ee
where the co-moving time derivative leads to the four acceleration 
\begin{equation}
\dot u^a = u^b \nabla_b u^a \ ,
\label{fourac}
\end{equation}
(and similarly for other variables in the following).
We also have
the expansion scalar
\be
\theta = \nabla_a u^a \ ,
\ee
the shear
\be
 \sigma_{ab} = \bar D_{\langle a}u_{b\rangle} \ , 
\ee
where the angle brackets indicate symmetrization and trace removal (as in \eqref{angdef}), and
\be
\bar D_a u_b = \perp_a^{\ c} \perp_{b}^{\ d} \nabla_c u_d \ , 
\ee
is the fibration equivalent of the totally projected derivative we already introduced for spacetime foliations. 
The merit of using this (totally projected) derivative is that the individual terms in \eqref{decomp} are perpendicular to $u^a$.
 We have also defined the vorticity
 \be
 \varpi_{ab} = \bar D_{ [a}u_{b]} \ .
 \ee
 
 Making use of these quantities, we find that \eqref{max1} and \eqref{ecurr} (with $U^a\to u^a$)
lead to
\be
\perp^{ab} \nabla_b e_a = \nabla_a e^a - u_a \dot e^a = \mu_0 \sigma + \bar \epsilon^{abc} \varpi_{ab} b_c  = \mu_0 \sigma + 2W^a b_a\ ,
\ee
where we use $e^a$ and $b^a$ for the electric and magnetic field in the fluid frame, respectively, in order to avoid confusion later.
We have also defined the vector
\be
W^{a} = {1 \over 2} \bar \epsilon^{abc} \varpi_{bc} \ , \quad \mbox{so that} \quad \varpi_{ab} = \bar \epsilon_{abc}W^c \ , \quad \mbox{and} 
\quad u^a W_a = 0 \ , 
\ee
where
\begin{equation}
    \bar \epsilon_{abc} = u^d \epsilon_{dabc} \ .
\end{equation}

 \vspace*{0.1cm}
 \begin{tcolorbox}
 \textbf{Comment:} At this point it is useful to make a few remarks. First of all, we  add bars to the projected derivative $\bar D_a$ and the $\bar \epsilon_{abc}$ in order to avoid confusion with the corresponding quantities for foliations. As comparisons are only made in this section, we only use this notation here. Note also that we define the vorticity tensor to have the opposite sign compared to \cite{ellis}. This is obviously just convention, but it is important to keep it in mind if one wants to compare the various  relations. Note also that $\varpi_{ab}$ is distinct from the vorticity two-form $\omega_{ab}$ used in the variational fluid model.
 \end{tcolorbox}
 \vspace*{0.1cm}
 
Next we get
\be
\perp_{ab}\dot{e}^b - \bar \epsilon_{abc} \bar D^b b^c + \mu_0 J_a =  \left( \sigma_{ab} -\varpi_{ab} - {2\over 3} \theta \perp_{ab}\right) e^b + \bar \epsilon_{abc}\dot{u}^b b^c \ .
\label{maxe}
\ee

The second set of equations follow from 
\be
\nabla_{[a}F_{bc]} = 0 \ ,
\ee
which leads to
\be
\perp^{ab}\nabla_b b_a = \bar D_a b^a = - 2W^a e_a\ ,
\ee
and
\be
\perp_{ab}\dot{b}^b +  \bar \epsilon_{abc}\bar D^b e^c   = - \bar \epsilon_{abc}\dot{u}^b e^c +  \left( \sigma_{ab} -\varpi_{ab} - {2\over 3} \theta \perp_{ab}\right) b^b \ .
\ee
It is easy to see that, if we consider an inertial observer (simply ignoring all derivatives of the four velocity),  these results reduce to the  text-book form of Maxwell's equations. The complete  expressions given here are, however, useful as they highlight the coupling between the electromagnetic field and a given fluid flow (with shear, vorticity and expansion). This also makes the coupling to spacetime apparent (through the presence of the covariant derivative).


In the context of astrophysics, most models involve some version of magnetohydrodynamics. In effect, this involves assuming that the local electric field vanishes, or at least that the electric field contribution to \eqref{maxe} can be ignored, e.g., via a low velocity argument involving the characteristic length- and time-scales. In the non-relativistic setting this argument is not particularly controversial, although one may take the view that  magnetohydrodynamics is more an assumption than an approximation \citep{schnack}. 

Effectively, we assume $e^a\approx 0$ which then implies that $\sigma \approx 0$ and \eqref{maxe} reduces to
\beq
 \mu_0 J_a  \approx \bar\epsilon_{abc}\bar D^b b^c \ .
\eeq
Once we have a handle on the magnetic field and the fluid flow, we can work out the charge current. This leads to \emph{ideal} magnetohydrodynamics. An alternative route to (basically) the same conclusions would be to start from a resistive model. The vanishing of the electric field then follows if the medium is assumed to be a perfect conductor, i.e. when the resistivity vanishes (or equivalently, the conductivity becomes infinite). However, this approach requires some version of Ohm's law, so we will return to this later.

\subsection{Variational approach for coupled charged fluids}

The description of electromagnetism is, of course, not complete until we consider the coupling to the fluid medium. This is the point where the variational model comes to the fore.
As we will now demonstrate; the advantage of having a well-grounded action principle for coupled fluids and an
identification of the true momenta is that it is relatively easy to incorporate
electromagnetism into the system.  To do this, we extend the standard 
procedure of introducing a (minimal) gauge coupling between the matter and the Faraday 
field, already discussed in Sect.~\ref{sec:emvar}. The only difference  is that we  now consider multiple charge carriers with identifiable fluxes, $n_\x^a$, and 
individual charges, $q_\x$. The charge current (density) associated with each flow is 
\beq
    j^a_\x = q^\x n^a_\x \ ,
\eeq
and the total current, that sources the electromagnetic field, is simply the sum
\be
j^a = \sum_\x j_\x^a \ .
\ee
It is worth noting that the variational derivation in Sect.~\ref{sec:emvar} requires that the current is conserved. This constraint 
is automatically satisfied if each individual current is conserved, as assumed in the variational derivation. Hence, we simply change the  
electromagnetic Lagrangian to 
\beq
  L_\mathrm{EM} = - \frac{1}{4\mu_0} F_{a b} F^{a b} + 
                           A_a \sum_\x j^a_\x \ , 
\eeq
and the equations that govern the electromagnetic field remain exactly as before. In addition,
the gauge coupling leads to a modified fluid momentum 
\beq
    \bar{\mu}^\x_a = \mu^\x_a + q^\x A_a \ , \label{guagecoup}
\eeq
which satisfies the equations of motion\footnote{As a slight aside, it is worth noting that \eqref{mageul} provides a useful starting point for a discussion of conservation laws  \citep{gour1,gour2}.}
\beq
    n^b_\x \bar{\omega}^\x_{b a} = 0 \ ,
\label{mageul}\eeq
where
\beq
    \bar{\omega}^\x_{a b} = 2 \nabla_{[a} \bar{\mu}^\x_{b]} \ . 
\eeq
 Finally, the total stress-energy tensor takes the form
\beq
    T^a{}_b = \Psi \delta^a{}_b + \sum_\x n^a_\x \mu^\x_b 
               - {1\over\mu_0} \left[ F^{c a}F_{c b} - {1\over 4} \delta^a{}_b \left( F_{c d}F^{c d}\right)  \right] \ ,
\eeq
simply representing the sum of the fluid and the electromagnetic contributions.

As an alternative, we may consider writing the momentum equation \eqref{mageul} as a force-balance relation.
Moving the electromagnetic contribution to the right-hand side, we get 
\be
   n^b_\x \omega^\x_{b a} = n_\x^b q^\x F_{a b} = j_\x^b F_{a b} \equiv f^\x_a \ .
\ee
Making contact with the previous section, we have
\be
f_\mathrm{L}^a = \sum_\x f_\x^a \ .
\ee

It is also worth considering the four-current in more detail. Let us consider the current and charge density inferred by the fluid observer from above, moving with 
four-velocity $u^a$. We can then express the various fluxes as 
\be
n_\x^a = n_\x \gamma_\x \left( u^a + v_\x^a\right) ,
\ee
where
\be
\gamma_\x = \left( 1 - v_\x^2 \right)^{-1/2} \ ,
\quad \mbox{and } \quad v^\x_a u^a = 0 \ .
\ee
It follows that the charge density $\sigma$ used in the previous section takes the form;
\be
\sigma =  \sum_\x n_\x q^\x \gamma_\x \approx   \sum_\x n_\x q^\x
\ee
in the low-velocity limit. Meanwhile, the spatial components of the current are given by
\be
j^i = \sum_\x j_\x^i =  \sum_\x n_\x q^\x \gamma_\x v_\x^i \approx   \sum_\x n_\x q^\x  v_\x^i =  J^i \ .
\ee

For two-fluid systems, our analysis readily reproduces the results for electron-positron plasmas \citep{koide08,koide09,kandus}. Moreover, the charged multi-fluid system can be extended to account for ``non-ideal'' effects like resistivity and particle reactions (i.e. non-conserved flows). In essence, if we want to  account for resistivity, we need to add a phenomenological ``force'' term to \eqref{mageul}.
This additional term should describe the dissipative interaction between the two components, and the standard intuition \citep{schnack,bellan} tells us that  it should be linear in the relative velocity between the two components. We then see from \eqref{mageul} that the 
required force must be orthogonal to each respective flux \citep{2012PhRvD..86d3002A} (note that this condition must be relaxed if we want to allow for particle creation/destruction).

Developments in this direction are (particularly) important for realistic neutron-star modelling. The most advanced step in this direction \citep{2017CQGra..34l5002A} considers  a four-component system composed of neutrons (n), protons (p), electrons (e) and entropy (s). The relative flow of the protons and electrons  leads to the charge current that couples the material motion to electromagnetism. The entropy flow is key if we want to account for the redistribution of heat, which we need to track if we want to consider (say) the cooling of a young neutron star. Finally, the neutrons need to be singled out, not just  because they make up the bulk of the star but, as the star matures  they  become superfluid and (at least partially) decouple from the other components. In order to explore the evolution and dynamics of maturing neutron stars, one has to allow for the relative flows of these four components. 

\subsection{The foliation equations}

We have seen how---once we introduce a fluid observer---the relativistic formulation for electromagnetism leads back to the, familiar looking, set of Maxwell's equations. Let us now connect the description with the foliation approach from Sect.~\ref{sec:numsim}, as required if we want to carry out nonlinear simulations. For clarity,  let us assume that we work with the electric and magnetic fields\footnote{Noting that there are good reasons for considering a mixed formulation using, for example, the electric field and the vector potential $A^a$ [Baumgarte].} $E^a$ and $B^a$, now measured by an Eulerian observer (defined by the spacetime foliation, as usual). We then have
\be
F_{ab} =  2 N_{[a} E_{b]}  + \epsilon_{abcd}N^c B^d = 2 N_{[a} E_{b]}  +\epsilon_{abd} B^d \ ,
\label{Faraday_appendix}\ee
where we have introduced\footnote{Note that, in the discussion of the 3+1 results we define $\epsilon_{abc}$ to be with respect to the Eulerian observer moving with $N^a$, not the fluid flow and $u^a$.}
\be
\epsilon_{abd} = \epsilon_{cabd}N^c \ .
\ee
That is, the electric and magnetic fields measured in the Eulerian frame are
\be
E_a = -  N^b F_{ba} \ ,
\ee
and
\be
B_a = - N^b \left( {1 \over 2} \epsilon_{abcd}F^{cd}\right) = {1\over 2} \epsilon_{acd} F^{cd}\ .
\ee
Both fields  are manifestly orthogonal to $N^a$ so
each  has three components, as expected.

It is instructive to relate the fields to those associated with the fluid frame. We then need to first of all recall that
\be
u^a = W(N^a + \hat v^a) \ , 
\ee
(where it is worth noting that we  use hats to indicate fluid quantities observed in the frame associated with $N^a$, as in Sect.~\ref{sec:numsim}), with $W$ the relevant Lorentz factor. 
This means that we have
\begin{multline}
e_a = - u^b F_{ba} = - W(N^b + \hat v^b) F_{ba} \\
= W\left[ E_a +N_a (\hat v^b E_b)\right] - W \hat v^b \epsilon_{bad}  B^d \\
= W \left[E_a + N_a (\hat v^b E_b) + \epsilon_{abc}\hat v^b B^c \right] \ ,
\end{multline}
and
\begin{multline}
b_a = - u^b \left( {1 \over 2} \epsilon_{abcd} F^{cd}\right) = - W(N^b +\hat v^b) \left( {1 \over 2} \epsilon_{abcd} F^{cd}\right) \\
=  W \left[ B_a +N_a ( \hat v^b B_b) - \epsilon_{abc} \hat v^b E^c  \right] \ .
\end{multline}
It is evident from this expression that, in general, the electric field inferred by the local observer has a component parallel to $N^a$
\be
e^\parallel = - e^a N_a =  W \left(\hat v^b E_b\right) \ ,
\ee 
as well as an orthogonal piece
\be
e_a^\perp = W  \left( E_a + \epsilon_{abc} \hat v^b B^c \right) \ .
\label{mhd1}
\ee
This is important. Let us assume that the observer can be chosen in such a way that the perpendicular component vanishes---the  assumption that leads to ideal magnetohydrodynamics. That is, let
\be
e_a^\perp = 0 \quad \Longrightarrow \quad  E_a + \epsilon_{abc} \hat v^b B^c= 0 
\label{EMHD}
\ee
It is easy to see that this also means that $e^\parallel=0$, so we actually have $e^a=0$; the electric field vanishes according to the ``fluid'' observer. 
We need to keep this result in mind later.

Turning to the matter equations, rather than working with the divergence of the total stress-energy tensor for the system we can isolate  the electromagnetic contribution. The right-hand side of the matter equations then have additional terms which follow from the Lorentz force
\be
f_\mathrm{L}^a = - j_a F^{ab} = N^b (\hat J^a E_a) + ( \hat \sigma E^b + \epsilon^{bac} \hat J_a B_c ) \ ,
\ee
where we have used the charge current 
\be
j^a = \hat \sigma N^a + \hat J^a \ .
\ee
From this we see that means that we need to add, first of all, a term
\be
\alpha \gamma^{1/2} ( \hat J^i E_i) \ , 
\ee
to the right-hand side of \eqref{energyq}, representing the electromagnetic contribution to the energy flow and including the Joule heating. Secondly, we need a term
\be
\alpha \gamma^{1/2}  ( \hat  \sigma E^i + \epsilon^{ijk} \hat  J_j B_k ) \ , 
\ee
on the right-hand side of \eqref{momentum}, representing the (spatial) Lorentz force. 

Finally,  we need to add the foliation version of Maxwell's equations to the evolution system. First of all, Eq.~\eqref{max1}
leads to
\be
\gamma^{ab} \nabla_b E_a = \mu_0 \hat \sigma + \epsilon^{abc}\left( \nabla_a N_b\right) B_c \ , 
\ee
or
\be
\gamma^{b}_a\nabla_b E^a - \mu_0 \hat \sigma = - \epsilon^{abc}K_{ab} B_c = 0 \ ,
\ee
since $K_{ab}$ is symmetric. That is, using the projected derivative $D_a$ from Sect.~\ref{sec:numsim} (not to be confused with $\bar D_a$ from above), we have
\be
D_i E^i = \mu_0 \hat \sigma \ .
\label{divE}
\ee

We also get
\begin{multline}
\gamma_{ab} N^c\nabla_c E^b - \epsilon_{abc}\nabla^b B^c + \mu_0 \hat J_a \\
=  E^b \nabla_b N_a - E_a \nabla_b N^b  + \epsilon_{abc}( N^d \nabla_d {N}^b) B^c \\
=  - E^b K_{ba} + E_a K  + \epsilon_{abc}( N^d \nabla_d {N}^b) B^c \ ,
\end{multline}
and we end up with
\be
\left( \partial_t - \mathcal L_\beta\right) E^i - \epsilon^{ijk} D_j (\alpha B_k) + \alpha \mu_0 J^i = \alpha K E^i \ .
\label{dtE}
\ee

The second pair of Maxwell equations follow from Eq.~\eqref{max2}, 
which leads to
\be
\gamma^{ab}\nabla_b B_a = -  \epsilon^{abc} E_a \nabla_b N_c \ ,
\ee
or
\be
\gamma^{b}_a\nabla_b B^a =  \epsilon^{abc} E_a K_{bc} = 0 \ ,
\ee
so we have
\be
D_i B^i = 0 \ .
\label{divB}
\ee

Finally,
\begin{multline}
\gamma_{ab}N^c\nabla_c {B}^b +  \epsilon_{abc}\nabla^b E^c   \\
= - \epsilon_{abc}(N^d \nabla_d N^b) E^c +   B^b \nabla_b N_a - B_a \nabla_b N^b \\
= - \epsilon_{abc}(N^d \nabla_d N^b) E^c -   B^b K_{ba} + B_a K \ ,
\end{multline}

leads to
\be
\left( \partial_t - \mathcal L_\beta\right) B^i + \epsilon^{ijk} D_j (\alpha B_k) = \alpha K B^i \ .
\label{dtB}
\ee

The four Maxwell equations can be written in different forms, depending on what is convenient. For example, in order to formulate a system suitable for numerical simulations it may be necessary to replace the covariant derivatives with partials, making the connections coefficients explicit \citep{2013PhRvD..88d4020D,2017CQGra..34l5003A}. However, such a reformulation does not add (much) to our understanding so we will settle for the equations in the present form.

\subsection{Electron dynamics and Ohm's law}

So far we have not explored the multi-fluid aspects of the problem. These inevitably enter if we try to add features like resistivity. Then we have to consider the ``friction'' between the separate flows. From the multi-fluid point of view,   we need to keep track of additional number densities. When these fluxes are conserved, we have 
\be
\nabla_a n_\x^a = 0 \  \Longrightarrow \ 
\left( \partial_t - \mathcal L_\beta\right) \left( \gamma^{1/2} \hat n_\x \right) + D_i \left[ \gamma^{1/2} \hat n_\x \left( \alpha \hat v_\x^i - \beta^i \right) \right] = 0\ . 
\label{necon} 
\ee

It is fairly straightforward (if  a bit messy) to write down the complete set of charged multi-fluid equations, representing a generic plasma setting. 
However, if we want to arrive at a set of equations representing ``magnetohydrodynamics'' we need to reduce the problem to (effectively) a single fluid degree of freedom. A natural step in this direction involves assuming that the relative flow between the different components in the system is modest enough that it can be represented as a linear drift. The idea is simple. 
Take the fluid frame (represented by $u^a$) to be associated with the baryons and let another component flow relative to it (with four velocity $u_\x^a$). In general, we then have
\be
u_\x^a = \gamma_\x \left( u^a + v_\x^a \right) \ , \qquad u_a v_\x^a = 0 \ , 
\ee
where (as usual)
\be
\gamma_\x = \left( 1 - v_\x^2 \right)^{-1/2} \ .
\ee
AT this level---for each component that exhibits a relative flow ($v_\x^a\neq0$)---we need to keep track of the individual Lorentz factor (relative to the chosen observer), $\gamma_\x$. To avoid this, we assume that the relative drift is  slow enough that we can linearize the relations. In effect, we assume that $\gamma_\x \approx 1$. This is an essential part of the ``single fluid reduction'' as we no longer need to keep track of the individual Lorentz factors. Moreover, it helps make contact with the thermodynamics and the equation of state. 

To illustrate this point, note that the fluid observer  measures each chemical potential as (introducing tildes to avoid confusion with the discussion in Sect.~\ref{sec:thermo})
\be
\tilde \mu_\x = - u^a \mu^\x_a \ .
\ee
If we ignore entrainment, then
\be
\mu^\x_a = \mu_\x u^\x_a  
\ee
so we need
\be
\tilde \mu_\x = - \mu_\x ( u^a u^\x_a ) \ .
\ee
Within the linear drift model, it is straightforward to show that $\tilde \mu_\x \approx \mu_\x$. Similarly, if we define the measured number density as
\be
\tilde n_\x = - u_\x n_\x^a \ , 
\ee
then we also have $\tilde n_\x \approx n_\x$. In essence, different fluid observers  agree on both number densities and  chemical potentials \citep{2017CQGra..34l5002A}. This is crucial as it means that there is no ambiguity in the concept of chemical equilibrium. For the outer core of neutron star (for example) we need to consider the Urca reactions, so chemical  equilibrium corresponds to
\be
\beta = \mu_\n - \mu_\p - \mu_\e = - u^a \left( \mu^\n_a - \mu^\p_a - \mu^\e_a \right) = 0 \ .
\label{beta}
\ee
As long as this condition is  satisfied, we can consistently ignore reactions and assume that the different particle species are conserved. The situation would be much less clear if we allowed for a nonlinear drift. Different observers would measure different number densities/chemical potentials and determining the frame with which one should associate the thermodynamics becomes an issue.  

Assuming that the linear drift argument holds on the evolution scale (as we have to in order to arrive at an effective one-fluid description)  and translating to the point of view of an Eulerian observer it makes sense to assume that the difference between the two (three-) velocities $\hat v_\x^a$ and $\hat v^a$ is small, as well. 
Linearizing in the Eulerian velocity difference, we then have
\be
W_\x = (1- \hat v_\x^2)^{-1/2}  \approx W \left[ 1 + W^2 \hat v_a (\hat v_\x^a - \hat v^a ) \right] \ .
\ee
Combining this with
\be
u_\x^a = W_\x \left(N^a +  \hat v_\x^a \right)  \approx   W \left( N^a +  \hat v^a \right) +v_\x^a  \ ,
\label{relate}
\ee
we find that
\be
v_\x^a \approx W\left[\delta^a_b + W^2 \hat v_b ( N^a +\hat v^a)\right] (\hat v_\x^b - \hat v^b ) \ ,
\label{fluidvel} 
 \ee
 This shows that the linearization argument is consistent.

In the present case, where the focus is on charged flows if electrons and protons (say), we now have
\begin{multline}
\hat \sigma = e (\hat n_\p - \hat n_\e ) = e ( W_\p n_\p - W_\e n_\e) \\
= e W \left[ \left(n_\p - n_\e\right) - W^2 n_\e \hat v_a (\hat v_\e^a - \hat v^a ) \right] \ ,
\end{multline}
and
\begin{multline}
\hat J^a =  e  ( \hat n_\p  \hat  v^a -  \hat n_\e  \hat v_\e^a) = e W \left[ n_\p \hat v^a - n_\e \hat v_\e^a -  W^2 n_\e \hat v_b (\hat v_\e^b - \hat v^b ) \hat v_\e^a \right] \\
= eW (n_\p - n_\e) \hat v^ a - eW n_\e (\hat v_\e^a-\hat v^a ) - eW^3 n_\e \hat v_b (\hat v_\e^b - \hat v^b) (\hat v_\e^a - \hat v^a + \hat v^a) \\
\approx \hat \sigma \hat v^ a - eW n_\e (\hat v_\e^a-\hat v^a ) \ .
\end{multline}
That is,
\begin{equation}
 \hat v_\e^a-\hat v^a \approx { 1\over e W n_\e} \left[ \hat \sigma \hat v^ a - \hat J^a \right] \ ,
\label{Jhat}
\end{equation}
 where we have used the fact that the linear drift assumption leads to
\be
\hat n_\e = n_\e W_\e \approx n_\e W \left[ 1 + W^2 \hat v_a \left( \hat v_\e^a-\hat v^a \right) \right]  \approx n_\e W \left[ 1 -  {\hat \sigma \over e n_\e }  \right] \ .
\ee

The momentum equation for a general component is\footnote{From here on we correct a number of typos---basically removing a term involving extrinsic curvature tracing back to Eqs.~(78)--(80) from \cite{2017CQGra..34l5003A}, and which propagate through to (129) in the paper.}
\begin{multline}
\left[ \partial_t + (\alpha \hat v_\x^j - \beta^j ) D_j \right]   S^\x_i  + S^\x_j D_i \left( \alpha \hat v_\x^j - \beta^j  \right) \\
+  D_i \left[ \alpha \left(  \hat \mu_\x - \hat v_\x^j S^\x_j\right) \right]   
= {\alpha \over \hat n_\x} \mathcal F^\x_i  \ , 
\label{momx}
\end{multline}
where
\be
 \mathcal F^\x_i = e_\x \hat n_\x \left( E_i + \epsilon_{ijk} \hat v_\x^j B^k \right) + \gamma^a_i R^\x_a \ ,
\ee
with the last term representing resistivity (implementing the model outlined by \citealt{2017CQGra..34l5001A}).

Noting that (in absence of entrainment) we have
\be
S_\x^i = \hat \mu_\x v_\x^i \ , 
\ee
and recalling \eqref{fluidframe}---that the fluid velocity is $V_\x^i = \alpha \hat v_\x^i - \beta^i$---we see that \eqref{momx} can be concisely written;
\be
\left( \partial_t + \mathcal L_{V_\x}\right) S^\x_i + D_i \left( {\alpha \hat \mu_\x \over W_\x^2} \right) = {\alpha \over \hat n_\x} \mathcal F^\x_i   \ ,
\label{momenew}
\ee
noting that the result relies on the linear drift assumption. In essence, we  keep only linear terms in velocity differences in a frame determined by the global time coordinate. This means that 
\be
V_\x^a = V^a + \alpha  ( \hat v_\x^a - \hat v^a ) \ .
\label{fluidx}
\ee
 
\vspace*{0.1cm}
\begin{tcolorbox}
As a slight aside, we may combine \eqref{fluidx} with  \eqref{necon},  making use of the global time argument and the expression for charge conservation, to show that
the electron fraction $x_\e = n_\e/n$ satisfies
$$
\left( \partial_t +\mathcal L_V\right) x_\e =0
$$
In essence, the electron fraction is advected by the fluid flow \citep{2013PhRvD..88f4009G}. Note that  no relativistic effects other than frame dragging enter the equation. 
\end{tcolorbox}

In the particular case of the electrons we then have
\begin{multline}
\left[ \partial_t + (\alpha \hat v_\e^j - \beta^j ) D_j \right]   S^\e_i  + S^\e_j D_i \left( \alpha \hat v_\e^j - \beta^j  \right) \\
+  D_i \left[ \alpha \left(  \hat \mu_\e - \hat v_\e^j S^\e_j\right) \right]   
= {\alpha \over \hat n_\e} \mathcal F^\e_i \ ,  
\label{finohm2}
\end{multline}
where
 \be
S_\e^i =  \hat \mu_\e  \hat v_\e^i =   \mu_\e W_\e \left[ \hat  v^i  + {1\over e n_\e W} \left(  \hat \sigma \hat v^i -  \hat J^i \right) \right]  \ .
\label{finohme}
\ee

Finally, we need an expression for the resistivity.  From \cite{2017CQGra..34l5001A,2017CQGra..34l5002A,2017CQGra..34l5003A} we have the general result (neglecting nuclear reactions, as we have assumed that the fluid remains in chemical equilibrium) 
\be
\gamma^a_c R^\x_a = \gamma^a_c \sum_{\y\neq\x} \mathcal R^{\x\y} \left(\delta^b_a + v_\x^b u_a \right) w^{\y\x}_b \ , 
\ee
where the velocities are with respect to the fluid. In the linear drift model, these are related to the Eulerian velocities through \eqref{fluidvel}. Thus, we arrive at
\be
\gamma^a_c R^\x_a = \sum_{\y\neq\x} \mathcal R^{\x\y} W \left(\delta^b_a + W^2 \hat v^b \hat v_a \right)  \left( \hat v^\y_b - \hat v^\x_b\right)\ .
\ee
In the two-component case we are considering, this reduces to the intuitive relation
\be
\gamma^a_c R^\e_a = \mathcal R W \left(\delta^b_a +W^2 \hat v^b \hat v_a \right)  \left( \hat v_b - \hat v^\e_b\right) \\
= { \mathcal R \over e n_\e} \hat J_a
\ee
It is worth noting that there are no $\hat \sigma$ terms in the final expression.

Resistivity is usually implemented at the level of some version of Ohm's law, typically viewed as a closure condition added to the magnetohydrodynamics relation \eqref{EMHD}.
In the multi-fluid model, the required relation follows from the electron momentum equation  \citep{2017CQGra..34l5003A}. As a first step, let us assume that we can ignore the electron inertia. Then it follows   from \eqref{momenew} that
\be
 \mathcal F^\e_i  \approx - e  n_\e W_\e \left( E_i + \epsilon_{ijk} \hat v_\e^j B^k \right) + { \mathcal R \over e n_\e} \hat J_a \approx { n_\e W_\e \over \alpha}  D_i \left( {\alpha  \mu_\e \over W_\e} \right)  
\ee
That is, we have
\be
 E_i + \epsilon_{ijk} \hat v_\e^j B^k  + { 1 \over \alpha}  D_i \left( {\alpha  \mu_\e \over W_\e} \right)  = { \mathcal R \over e n_\e^2 W_\e} \hat J_i \approx { \mathcal R \over e n_\e^2 W} \hat J_i
 \equiv \eta \hat J_i
 \label{genohm}
\ee
which defines the scalar resistivity coefficient $\eta$.
It is reassuring to note that \eqref{genohm} is consistent with the text-book result for non-relativistic two-fluid systems, e.g., Eq.~(2.75) in \cite{bellan} or \cite{1999stma.book.....M}, once we set $\alpha = W_\e = W \to 1$ at the same time as we assume that $\hat \sigma \to 0$. 

Ignoring the chemical gradient term, we have
\be
E_i + \epsilon_{ijk} \hat v^j B^k  + {1\over e n_\e W} \epsilon_{ijk} \left( \hat \sigma \hat v^j - \hat J^j\right)B^k  = \eta \hat J_i \ . 
\ee
Also neglecting (without particular justification at this point) the Hall term, we are left with
\be
E_i + \epsilon_{ijk} \hat v^j B^k   = \eta \hat J_i \ .
\label{finohms}
\ee
Through a hierarchy of approximations and simplifications  we have moved from a model that retains the properties of a charged two-component  plasma  to a simple expression for Ohm's law. 

The sequence of arguments leading to \eqref{finohms} provides insight into the applicability of ``ideal'' magnetohydrodynamics, which corresponds to the assumption that the local electric field vanishes
\be
 e^a\approx 0 \quad \longrightarrow \quad E_i + \epsilon_{ijk} \hat v^j B^k  =0 \ .
\ee
The usual argument for this is that the medium is a perfect conductor, i.e. $\R\to 0$ ($\eta \to 0$). However, this limit only affects the resistive term in \eqref{genohm}. We still have to argue that the remaining terms are unimportant. This is less straightforward. 

It is instructive to compare the final result to the standard argument from the literature  \citep{bekor,wata,pale,taka}, which starts from magnetohydrodynamics and arrives at Ohm's law by taking the current to be proportional to the Lorentz force acting on a particle in the fluid frame. Assuming
\be
 \perp_a^b j_b = \bar \eta F_{ab} u^b \ , 
 \label{ohmp}
\ee
and recalling that 
\be
u^a = W (N^a + \hat v^a) \ , 
\ee
we have 
\begin{multline}
 j_a = \hat \sigma N_a + \hat J_a  
= \bar \eta W (N^b +\hat v^b) \left(  N_{a} E_{b} - N_b E_a  + \epsilon_{abc} B^c \right) \\
= \bar \eta W  \left[  N_{a} (\hat v^b E_b) + E_a  + \epsilon_{abc} \hat v^b B^c \right] \ . 
\end{multline}
Project along $N^a$ to get
\be
 \hat \sigma  + W^2 (\hat v_i\hat J^i - \hat \sigma)     = \bar \eta W  (\hat v^i E_i) \ , 
 \label{barsigma}
\ee
while the orthogonal projection leads to
\be
\hat  J_a  - W^2 \hat v_a (\hat \sigma - \hat v_i\hat J^i)
=  \bar \eta W \left( E_a  + \epsilon_{abc} \hat v^b B^c \right) \ . 
\label{hatj}
\ee
It follows that
\be
\hat v^i \hat J_i - W^2 \hat v^2  (\hat \sigma - \hat v_i\hat J^i)= \bar \eta W  (\hat v^i E_i) \ , 
\ee
and we finally arrive at
\be
E_i  + \epsilon_{ijk} \hat v^j B^k = {1\over \bar \eta W} \left[ \hat J_i - W^2  (\hat \sigma - \hat v_l\hat J^l) \hat v_i \right] \ . 
\ee
This version of Ohm's law---notably identical to \eqref{finohms} once we identify $\eta = 1/\bar \eta W$---has been implemented in recent numerical simulations, see for example Eq.~(22) in \cite{pale}. 
The comparison provides a nice ``sanity check'' of the logic, but the multi-fluid derivation clearly provides a better understanding of the  physics. Moreover, it allows us to extend the model to account for additional aspects (should we want to do so). In fact, if we were to retain  the time variation of the charge current we would add in most of the relevant plasma features (the only restriction being that we assumed a linear drift fairly early on in the developments). 

\subsection{Tetrad formulation}
\label{sec:tetrad}

The general formalism we have outlined is fully nonlinear and includes the coupling to the dynamical spacetime. In essence, it is geared towards numerical simulations of violent phenomena in full General Relativity. However, there are relevant problems where the dynamical role of spacetime is less crucial (or, perhaps, not at all relevant).  A typical such problem would be the slow evolution of the magnetic field in a neutron star interior \citep{2013MNRAS.434..123V}. Assuming that we may take the spacetime as fixed, it can be useful to make the curved spacetime problem look ``as close to flat'' as possible. This typically involves using tetrads. As relevant parts of the literature draw on this strategy, it is useful to introduce the main ideas and steps here. We do this by adapting our magnetic field results to a fixed, slowly rotating spacetime. That is, we make contact with the Hartle-Thorne slow-rotation expansion \citep{1968ApJ...153..807H}, keeping only first order terms in the rotation, for simplicity. The metric is then given by
\be
ds^2 = - e^{2\nu} dt^2 - 2 \omega r^2 \sin^2 \theta d\phi dt + e^{2\lambda} dr^2 + r^2 d\theta^2 + r^2 \sin^2\theta d\phi^2 \ , 
\ee 
where the rotational frame-dragging $\omega$ is  a solution to 
\be
{1\over r^3} {d\over dr} \left[ r^4 e^{-(\nu+\lambda)} {d\bar\omega \over dr}\right] + 4 {d\over dr}\left[ e^{-(\nu+\lambda)}\right] \bar \omega = 0 \ , 
\ee
with
\be
\bar \omega = \Omega - \omega \ .
\ee
The solution external to a uniformly rotating body is
\be
\bar \omega_\mathrm{ext} = \Omega - {2J \over r^3} \ , 
\ee
where $\Omega$ is the rotation frequency of the star (as viewed by an asymptotic observer) and $J$ is the angular momentum. 

Comparing the slow-rotation line element to the 3+1 form from Eq.~\eqref{adm}
we identify the lapse
\be
\alpha = e^\nu \ , 
\ee
the shift vector
\be
\beta^i = -\omega \delta^i_\phi \ , 
\ee
and the spatial metric
\be
\gamma_{ij} = \left( \begin{array}{ccc} e^{2\lambda} & 0 & 0 \\ 0 & r^2 & 0 \\ 0 & 0 & r^2 \sin^2\theta \end{array}\right) \ .
\ee
The fact that $\gamma_{ij}$ is diagonal simplifies much of the following discussion. We also see that 
\be
\gamma^{1/2} = e^\lambda r^2 \sin \theta \ . 
\ee

Next, it is worth noting that
\be
\alpha K = - \partial_t \ln \gamma^{1/2} + D_i \beta^i = 0 \ , 
\ee
since the spacetime is stationary and axisymmetric. We also have
\be
\mathcal L_\beta \gamma^{1/2} = \partial_i \left( \gamma^{1/2} \beta^i\right) = 0  \ , 
\label{liegam}
\ee
since the spacetime is axisymmetric. This means that 
\be
\left( \partial_t -\mathcal L_\beta\right) \gamma^{1/2} = 0 \ ,
\ee
a  result which will be used in the following.

Up to this point, we have expressed all tensor relations in terms of components in a given coordinate basis.  However, when the focus is on measurements carried out by a given observer it may be helpful to work in a local inertial frame, using an orthonormal basis associated with a local tetrad \citep{1972ApJ...178..347B,1982MNRAS.198..339T}. This means that we (first of all) translate the equations into an orthonormal tetrad---changing the basis in such a  way that the metric appears flat. A simple way to do this is to rewrite the line element in terms of a new basis in such a way that
(using hats to denote quantities in the new orthonormal basis)
\be
ds^2 =  \eta_{\hat a \hat b}  dx^{\hat a}  dx^{\hat b} =  \eta_{\hat a \hat b}   \omega^{\hat a}_c   \omega^{\hat b}_d dx^c dx^d \ , 
\ee 
where $\eta_{\hat a \hat b} = \mathrm{diag}(-1,1,1,1)$.
Comparing to the slow-rotation metric, we see that we have
\begin{eqnarray}
\omega^{\hat 0}_a &=& e^{\nu} ( 1,0,0,0)\ ,  \\
 \omega ^{\hat 1}_a &=& e^{\lambda} ( 0,1,0,0) \ , \\
\omega^{\hat 2}_a &=& r( 0,0,1,0) \ , \\
\omega^{\hat 3}_a &=&  r\sin \theta ( -\omega,0,0,1) \ .
\end{eqnarray}
If we define the inverse through
\be
e _{\hat c}^a  \omega ^{\hat c}_b = \delta^a_b \ , 
\ee
it also follows that 
\begin{eqnarray}
 e _{\hat 0}^a &=& e^{-\nu} ( 1,0,0,\omega) \ , \\
 e _{\hat 1}^a &=& e^{-\lambda} ( 0,1,0,0) \ , \\
e _{\hat 2}^a &=& {1\over r} ( 0,0,1,0) \ , \\
e _{\hat 3}^a &=& {1\over r\sin \theta} ( 0,0,0,1) \ .
\end{eqnarray}
The $e_{\hat a}^b$ are usually referred to as the tetrad components. 

We now have  the tools we need to transform quantities from the coordinate basis to the orthonormal one. For instance;
\be
B_{\hat a} =  e_{\hat a}^b B_b \ , 
\label{cotran}
\ee
and
\be
B^{\hat a} =  \omega ^{\hat a}_b B^b \ .
\ee
An advantage of working in the orthonormal tetrad is that we can exhange co- and contravariant quantities without ``penalty'' (as the associated three-metric is flat). A disadvantage is that we have to be careful with derivatives. Before we consider this issue, let us provide an example of 
 why it is natural to work with the tetrad components of the various spatial objects. Let us take the Faraday tensor as example. First of all, according to an observer rotating with $\omega$ we have the coordinate basis result (see, for instance, \citealt{2001MNRAS.322..723R})
{\small
\be
F_{ab} = \left( \begin{array}{cccc} 0 & - e^\nu E_r - \omega e^\lambda r^2 \sin \theta B^\theta & - e^\nu E_\theta+ \omega e^\lambda r^2 \sin \theta B^r  & - e^\nu E_\phi \\
 e^\nu E_r + \omega e^\lambda r^2 \sin \theta B^\theta & 0 & e^\lambda r^2 \sin \theta B^\phi & - e^\lambda r^2 \sin \theta B^\theta \\
  e^\nu E_\theta- \omega e^\lambda r^2 \sin \theta B^r  &  - e^\lambda r^2 \sin \theta B^\phi & 0 & e^\lambda r^2 \sin \theta B^r
  \\
  e^\nu E_\phi & e^\lambda r^2 \sin \theta B^\theta & - e^\lambda r^2 \sin \theta B^r & 0
  \end{array}\right)  \ .
  \label{Far1}
\ee}
If we simply replace the field components with the corresponding quantities for the tetrad and
 project the tensor into the tetrad we get
\be
F_{\hat c \hat d} =  e_{\hat c}^a  e _{\hat d}^b F_{ab} = \left( \begin{array}{cccc} 0 & - E^{\hat r} & - E^{\hat \theta}  & -  E^{\hat \phi} \\
   E^{\hat r}   & 0 &   B^{\hat \phi} & -  B^{\hat \theta} \\
  E^{\hat \theta}   &  - B^{\hat \phi}& 0 &  B^{\hat r}
  \\
   E^{\hat \phi} &   B^{\hat \theta}& - B^{\hat r}
 & 0
  \end{array}\right) \ .
\ee
We recognize this as the usual flat-space form of the Faraday tensor, emphasizing that this is the natural description for a local observer.

As we move on to consider dynamics, we have to consider derivatives. For scalar quantities, this is relatively straightforward. For example, from \eqref{cotran} we
 see that
\be
\vec e_{\hat 0} =  \partial_\tau  =  e_{\hat 0}^a \vec e_a = e^{-\nu} \left( \vec e_t + \omega \vec e_\phi \right) = e^{-\nu} \left( \partial_t  + \omega  \partial_\phi \right) \ , 
\ee
allows us to introduce a natural time-derivative associated with the rotating frame. 
In fact, for a scalar $n$, we have 
\begin{multline}
(\partial_t - \mathcal L_\beta) (\gamma^{1/2} n) = \gamma^{1/2} (\partial_t - \mathcal L_\beta) n \\
=  \gamma^{1/2} (\partial_t - \beta^j \nabla_j ) n 
= \gamma^{1/2} (\partial_t - \beta^j \partial_j ) n \\
=  \gamma^{1/2} (\partial_t + \omega \partial_\phi ) n =  \gamma^{1/2} e^\nu \partial_\tau n =  \gamma^{1/2}  \partial_\tau \left( e^\nu n\right) \ .
\end{multline}
However, this is more of an aside because, for vector quantities this is not the appropriate time derivative. 
In order to understand the distinction, we need to reinstate the basis vectors (and forms). Using arrows to denote basis vectors (and tildes for basis one forms) we have the three-vector
\be
\boldsymbol B = B^b \vec e_b = e _{\hat c}^a  \omega ^{\hat c}_b B^b \vec e_a  = B^{\hat c} \vec e_{\hat c} \ .
\ee

If we want to make a connection with (more or less) text-book vector calculus, we need to understand derivatives of vectors in the ZAMO frame. First of all, we note that the (spatial) metric $\gamma_{ij}$ is diagonal (in fact, in 3D we can always find coordinates that lead to a diagonal metric) with scale factors $h_a$ given by (we are not summing over repeated indices for the rest of this section!);
\be
\vec e_{\hat a} = {1\over h_a } {\partial \over \partial x^a} = {1\over h_a } \vec e_a =  e_{\hat a}^a  \vec e_a \Longrightarrow  e_{\hat a}^a = {1\over h_a } \ .
\ee
Comparing (for later convenience) to the three-metric we see  that
\be
\gamma_{ac} = h_c^2 \delta_{ac}  \ .
\ee

Let us now define
\be
\boldsymbol \nabla =\sum_a  \tilde e^a D_a \ ,
\ee
such that 
the directional derivative is given by 
\be
D_a =  \vec e_{a} \cdot \boldsymbol \nabla \ ,
\ee
and we have
\begin{multline}
\boldsymbol \nabla \omega = \sum_a  \tilde e^a D_a \omega = \sum_a  \tilde e^a \partial_a \omega = \sum_{a,b,c}  e _{\hat c}^a  \omega ^{\hat c}_b \tilde e^b \partial_a \omega \\
= 
 \sum_{a,c} \tilde e^{\hat c} \left( e^a_{\hat c} \partial_a \omega \right) = \sum_a \tilde e^{\hat a} \left( {1\over h_a } \partial_a \omega \right)  \ .
\end{multline}
We see that we can express the components of the gradient in either frame, but in the orthonormal case we need to keep track of the scale factors. We obviously knew this already, but we can now make the connection explicit.

Turning to vectors, we have (the usual covariant derivative)
\begin{multline}
\boldsymbol \nabla \boldsymbol A =  \sum_{a,b}\tilde e^a D_a ( A^b \vec e_b) =   \sum_{a,b}\left[ (\partial_a A^b ) \tilde e^a \vec e_b + A^b \tilde e^a D_a \vec e_b \right]  \\
= \sum_{a,b,c} \left[  (\partial_a A^b ) \tilde e^a \vec e_b + A^b \tilde e^a \Gamma_{ba}^c \vec e_c  \right] \equiv  \sum_{a,b}  (D_a A^b ) \tilde e^a \vec e_b \ ,
\end{multline}
where $ \Gamma_{ba}^c$ is the connection associated with $\gamma_{ij}$.
We also have
\begin{multline}
\boldsymbol \nabla \boldsymbol A =  \sum_{a,b}\tilde e^a D_a ( A^{\hat b} \vec e_{\hat b}) =   \sum_{a,b}\left[ \partial_a \left( { A^{\hat b} \over h_b}  \right) \tilde e^a \vec e_b + {A^{\hat b} \over h_b} \tilde e^a D_a \vec e_b \right]  \\
=   \sum_{a,b}   D_a \left( {A^{\hat b} \over h_b}\right)  \tilde e^a \vec e_b=  \sum_{a,b}  {h_b \over h_a} D_a \left( {A^{\hat b} \over h_b}\right)  \tilde e^{\hat a} \vec e_{\hat b} \ ,
\end{multline}
and it follows that 
\be
D_{\hat a} A^{\hat c }=  {h_c \over h_a} D_a  \left({A^{\hat c} \over h_c} \right)
 = \sum_b   {h_c \over h_a} \left[ \partial_a \left({A^{\hat c} \over h_c} \right) + \Gamma_{ba}^c  \left({A^{\hat b} \over h_b} \right) \right] \ .
  \label{devcomp1}
\ee

Now, as $\gamma_{ij}$ is diagonal in the particular case we are considering (and likely in any problem one may be interested in), we have
\begin{multline}
\Gamma^c_{ab} = \sum_d \gamma^{cd} \left[  ( h_d \partial_b h_d)  \delta_{ad}+  ( h_d \partial_a h_d) \delta_{bd} -(h_b  \partial_d h_b) \delta_{ab}\right] \\
= {1\over h_c^2} \left[  ( h_c \partial_b h_c)  \delta^c_a+  ( h_c \partial_a h_c) \delta^c_b - \sum_d (h_b \partial_d h_b) \delta^{cd}  \delta_{ab}\right] \ .
\end{multline}
Using this in \eqref{devcomp1} we arrive at
\begin{multline}
D_{\hat a} A^{\hat c }= {h_c \over h_a}  {\partial \over \partial x^a} \left({ A^{\hat c} \over h_c} \right) \\
+ 
  \sum_{b,d} { 1  \over h_a h_c  } \left[  (h_c \partial_b h_c )\delta^c_{a}+  (h_c \partial_a  h_c) \delta^c_{b} - (h_b \partial_d h_b)\delta^{cd} \delta_{ab}\right]
 { A^{\hat b} \over h_b} \\
 = {1\over h_a} \partial_{a} A^{\hat c} - \sum_d \delta^{cd} {1  \over h_a h_c} (\partial_d h_a) A^{\hat a}+  \sum_{b} { 1  \over h_a h_b  }  ( \partial_b h_c) \delta_{a}^c A^{\hat b} \ .
\end{multline}

For the divergence we then need
\begin{multline}
\boldsymbol \nabla \cdot \boldsymbol B \equiv \sum_a D_{ a} B^{ a }  =  \sum_a D_{\hat a} B^{\hat a } \\
= \sum_a {1\over h_a} \partial_{a} B^{\hat a} - \sum_a  {1  \over h_a^2} \partial_a h_a B^{\hat a}+  \sum_{a,b} { 1  \over h_a h_b  }   \partial_b h_a B^{\hat b} \\
=  \sum_a {1\over h_a} \partial_{a} B^{\hat a} -  {1  \over h_1^2} \partial_1 h_1 B^{\hat 1} - {1  \over h_2^2} \partial_2 h_2 B^{\hat 2}  - {1  \over h_3^2} \partial_3 h_3 B^{\hat 3} \\
+  \sum_{a} \left[ { 1  \over h_a h_1  }   \partial_1 h_a B^{\hat 1} +  { 1  \over h_a h_2  }   \partial_2 h_a B^{\hat 2}+  { 1  \over h_a h_3  }   \partial_3 h_a B^{\hat 3} \right] \\
= {1\over h_1 h_2 h_3} \left[ {\partial \over \partial x^1} (h_2 h_3 B^{\hat 1}) +  {\partial \over \partial x^2} (h_1 h_3 B^{\hat 2})  +  {\partial \over \partial x^3} (h_1 h_2 B^{\hat 3}) \right] \ , 
\end{multline}
which is the textbook result. 

Similarly, it is straightforward to use \eqref{devcomp1} to show that we have the standard result for the curl: 
\begin{multline}
\boldsymbol \nabla \times \boldsymbol B = \sum_{a,b,c} \vec e_{\hat a}( \epsilon^{\hat a \hat b \hat c} \nabla_{\hat b} B_{\hat c}) \\
= \sum_{a,b,c}  \vec e_{\hat a} \omega^{\hat a}_b ( \epsilon^{bcd} \partial_c B_d )
= { 1 \over h_1 h_2 h_3} \left| \begin{array}{ccc} h_1 \vec e_{\hat 1}& h_2 \vec e_{\hat 2} & h_3 \vec e_{\hat 3} \\
\partial_r & \partial_\theta & \partial_\phi \\ h_1 B_{\hat 1} & h_2 B_{\hat 2} & h_3 B_{\hat 3} \end{array} \right| \ .
\end{multline}

Finally, we need time derivatives
\be
\sum_a \vec e_a \partial_t B^a = \partial_t \boldsymbol B \ , 
\ee
and
\begin{multline}
\sum_a \vec e_a ( \mathcal L_\beta B^a )=\sum_{a,b} \vec e_a \left(  \beta^b \partial_b B^a - B^b \partial_b \beta^a \right) 
\\
= 
\vec e_a \left(  \beta^b D_b B^a - B^b D _b \beta^a \right)  
= (\boldsymbol \beta \cdot \boldsymbol \nabla) \boldsymbol B - (\boldsymbol B \cdot \boldsymbol \nabla) \boldsymbol \beta \ , 
\end{multline}	
where
\be
\boldsymbol \beta = -  \omega \sum_a  \delta^a_\phi \vec e_a =  - \omega \sum_a  \delta^{\hat a}_\phi \vec e_{\hat a} = - \omega \boldsymbol n_\phi \ .
\ee
Thus, we see that
\be
\sum_a \vec e_a ( \partial_t B^a - \mathcal L _\beta B^a ) = \partial_t \boldsymbol B -  (\boldsymbol \beta \cdot \boldsymbol \nabla) \boldsymbol B + (\boldsymbol B \cdot \boldsymbol \nabla) \boldsymbol \beta
\ee

This is  all we need if we want to write various coordinate basis Maxwell equations in terms of three-vectors. As a start, consider \eqref{divB}. It is easy to see that, the scale factors associated with the spherical coordinates are $h_1 = e^\lambda$, $h_2 =r$ and $h_3=r \sin \theta$, and it follows immediately  that 
\be
\boldsymbol \nabla \cdot \boldsymbol B = 0 \ .
\label{divBtetrad}
\ee

Continuing in the spirit of making the equation look as close to the flat-space case as possible, we introduce
the charge density as $\hat \sigma = J^{\hat t}$. Then \eqref{divE} is
\be
\boldsymbol \nabla \cdot \boldsymbol E = 4 \pi \hat \sigma \ .
\ee

The time-dependent equations are a little bit messier, partly because the redshift factor $e^\nu$ needs to be accounted for (see \citealt{1982MNRAS.198..339T} for discussion).  Thus,  
we can write \eqref{dtB} as
\be
\partial_t \boldsymbol B -  (\boldsymbol \beta \cdot \boldsymbol \nabla) \boldsymbol B + (\boldsymbol B \cdot \boldsymbol \nabla) \boldsymbol \beta
+ \boldsymbol \nabla \times (e^\nu \boldsymbol E) = 0 \ .
\label{Bf}
\ee
Similarly, once we define
\be
\boldsymbol J = \sum_a J^{\hat a} \vec e _{\hat a} \ , 
\ee
Equation~\eqref{dtE} becomes
\be
\partial_t \boldsymbol E -  (\boldsymbol \beta \cdot \boldsymbol \nabla) \boldsymbol E + (\boldsymbol E \cdot \boldsymbol \nabla) \boldsymbol \beta
- \boldsymbol \nabla \times \left( e^\nu \boldsymbol B \right) = - 4\pi e^\nu \boldsymbol J \ .
\label{Ef}
\ee
The different relations  agree (as they have to) with Eqs.~(20)--(23) from \cite{1996A&A...307..665K}.

\subsection{A brief status report of magnetic field models}

Problems in astrophysics and cosmology involving magnetic fields are of obvious interest due to the (essentially) direct link to observation. Most objects of interest for astronomy tend to be endowed with magnetic fields and the large scale fields may have an impact on cosmology, as well.
Quite naturally, this means that the literature on the subject is vast and varied. We will not be able to give the different issues the attention they deserve, but it nevertheless makes sense to list some of the main issues that (may) require fully relativistic description of non-ideal magnetohydrodynamics.   Of most obvious relevance  are problems involving not only electromagnetism but the live spacetime of General Relativity. Key gravitational-wave sources immediately come to mind, like core-collapse supernovae \citep{kota1} and compact binary mergers 
\citep{merge1,merge2,2016ApJ...824L...6R,2019PhRvD..99h4032R,2020PhRvD.101f4042R}. Both cases involve strong gravity, a significant thermal component and magnetic fields. Going beyond ideal magnetohydrodynamics in these simulations is, however,  challenging both from a technical point of view and in view of the computational cost. This obviously does not mean that we should set our aim high---indeed, there have been several efforts in this direction \citep{wata,pale,taka,2013PhRvD..88d4020D}---but it is probably fair to say that this is work in progress. The step to a full plasma description and actual multi-fluid simulations \citep{zani} is also unlikely to be taken any time soon. 

The seemingly more innocuous problem of isolated compact stars also comes with unresolved issues. These range from the dynamics of the star's magnetosphere and the pulsar emission mechanism to the formation and evolution of the star's interior magnetic field.  In the case of the magnetosphere, the main focus has been on force-free models, but recent arguments \citep{spit} point to the need to account for resistivity. In the case of the formation and evolution of a compact star's 
global magnetic field, we need a better understanding of dynamo effects that may come into operation (see \citealt{dynamo} and also \citealt{brandenburg} for a recent review) and we also need
to understand the coupled evolution of the star's spin, temperature and magnetic field \citep{2013MNRAS.434..123V}. There are difficult issues to resolve, especially since it is becoming clear that the typical stationary and axisymmetric magnetic field models one would intuitively use as a starting point for the discussion tend to be unstable \citep{2012MNRAS.424..482L}.

In fact, it is clear that we need to develop the theory further. Typical issues that need to be addressed involve (i) the dynamics of the model, e.g.,  causality and stability of wave propagation and relation to issues like pulsar emission or the launch of outflows and jets, (ii) transitions between spatial regions where different simplifying assumptions are valid, such as a region in the magnetosphere where the fluid model applies and a low density region where the description breaks down and one would have to fall back on a kinetic theory description \citep{2003CQGra..20.1823M,2004ApJ...605..340M,1996PhRvL..76.3340G}, the transition from the  magnetosphere to the interior field at the star's surface or, indeed, accreting systems where an ion-electron plasma describes the inflowing matter while regions in the magnetosphere may still be appropriately modelled as a pair-plasma, (iii) the role of more complex physics, like the superconductor that is expected to be present in the star's core \citep{supercon} or regions where the assumption that the medium is electromagnetically ``passive'' does not apply, possibly in the pasta region near the crust-core transition \citep{2013NatPh...9..431P}. 

Another problem of key astrophysical interest concerns the launch of large-scale jet emission---either associated with core collapse or neutron star mergers---required to explain observed gamma-ray bursts \citep{2011ApJ...732L...6R}. The difficulties here remain technical and conceptual, with one of the main issues being the need to resolve the dynamics of the central engine (e.g., associated with the magnetorotational instability; \citealt{1991ApJ...376..214B,1991ApJ...376..223H,2018PhRvD..97l4039K}) while at the same time representing the large scale behaviour of the jet emission. One of the key challenges involves marrying the nonlinear dynamics of the strong-gravity central region with the evolution in the distant weak field region (where one may get away with treating spacetime as a fixed background, the typical assumption for jet simulations; \citealt{2007PhPl...14e6506U,2010LNP...794..265K,2018ApJ...863...58X}). 


\section{The problem with heat}
\label{sec:heat}

The fact that relativistic fluid dynamics is a mature field of study does not mean that there are no unresolved issues. In fact, there are quite a few. Some continue to be in focus and others are swept under the rug (perhaps to be rediscovered, cause confusion and then duly ignored again...) One of the main issues that continue to cause concern arises as soon as we consider dissipative systems. It is clear 
from the outset that we are facing a difficult problem. For example, the familiar
Fourier theory for heat conduction---which requires the introduction of thermal conductivity associated with the mobility of entropy carriers---leads to instantaneous propagation of
thermal signals (the heat equation is parabolic). The fact that this non-causality is built into the
description is unattractive already in the context of the classic Navier--Stokes
equations. Intuitively, one would expect heat to propagate at roughly the
mean molecular speed in the system. For a relativistic description
non-causal behavior would be totally unacceptable. Any acceptable formulation of the problem must circumvent this. In principle, we know what we have to do. There is a deep connection between
causality, stability, and hyperbolicity of a dissipative model \citep{hislin83}, so we need to make sure that we develop a fully hyperbolic formalism. The issue has been a main motivating factor behind the development of
extended irreversible thermodynamics \citep{joubook,mullerbook}, a model which introduces additional dynamical fields
in order to retain hyperbolicity and causality. 

From a formal point of view the debate has (at least to some extent) been settled since the late 1970s. 
The key contribution was the work of Israel and Stewart, who developed a model analogous to Grad's 
14-moment theory,  taking as its starting point  relativistic kinetic theory \citep{stew77,Israel79:_kintheo1,Israel79:_kintheo2}. This so-called ``second order'' theory, 
 extends the pioneering ``first order'' 
work of  \cite{eckart40:_rel_diss_fluid,landau59:_fluid_mech}, has been used in a number of different settings, including the
highly relativistic plasmas generated in colliders like RHIC at Brookhaven and the LHC at CERN \citep{2001PhLB..506..123E,Muronga2}. However, 
despite the obvious successes of the second-order  model, there are still
dissenting views in the literature, see for example \cite{2006JNET...31...11G,2009GReGr..41.1645G}. Particular objections concern the complexity of the formulation and the many additional  
``dissipation coefficients'' required to complete it. This is, however, a feature that is shared by all models within the 
extended thermodynamics framework \citep{joubook}. 

The simplest relevant problem involves heat flow, a problem with several interesting aspects and which also connects with fundamental physics questions, in particular in the context of nonlinear
phenomena,see for example
\cite{mr87,rms,jlmp,lrv} and \cite{llebot}. Non-linearities are  relevant for the development of both shocks and turbulence in real physical systems. 
However, at this point we aim to establish the viability of the
multi-fluids approach to the heat problem. For this purpose, a linear analysis should be adequate. If we dig deeper we uncover a  range of issues, including foundational problems like the nature of time (read: the role of the second law of thermodynamics) and the formation of structures at nonlinear deviations from thermal equilibrium.
Much recent work has been motivated by the modelling of complex  systems is astrophysics and
cosmology \citep{1996astro.ph..9119M}. The problem may date back to the origins of relativity theory \citep{1967Natur.214..903L}---is a moving body hot or cold?---but it remains an active challenge.

\subsection{The ``standard'' approach}
\label{section_14_1}

In order to illustrate the main principles, let us return to a situation we have considered several times already. Adding a thermal component to a single matter component, we envisage two distinct flows. The matter is represented by a flux $n^a$ which satisfies
\begin{equation}
  \nabla_a n^a = 0 \ ,
  \qquad \mathrm{where} \qquad
  n^a = n u^a \ .
  \label{ndiv}
\end{equation}
In the following (in order to be specific) we will work in the frame associated with the matter flow, $u^a$. Next we add the heat flux $q^a$ (which is spatial in the sense that $u^a q_a=0$) to the perfect fluid stress-energy tensor:
\begin{equation}
  T^{a b} =
  \varepsilon u^a u^b +  p \perp^{ab}+ 2 q^{(a} u^{b)}  \ .
  \label{Tdiss0}
\end{equation}
Finally, we need to incorporate the second law of thermodynamics. The requirement that the total
entropy must not decrease leads to the entropy flux $s^a$ having to be
such that
\begin{equation}
  \nabla_a s^a = \Gamma_\s \ge 0 \ .
  \label{sdiv}
\end{equation}
Assuming that the entropy flux is a combination of the available fluxes,
we have \citep{eckart40:_rel_diss_fluid} (we will connect this relation with the variational derivation later)
\begin{equation}
  s^a = s u^a + \beta q^a \ ,
\end{equation}
where $\beta$ is yet to be specified. It is easy to work out
the divergence of this, and we find (after introducing $x_\s = s/n$, as before,  and using \eqref{ndiv})
\begin{equation}
    n u^a \nabla_a x_\s + \beta \nabla_a q^a + q^a \nabla_a \beta = \Gamma_\s
\end{equation}
Next, we combine this result with 
\begin{equation}
  u_a \nabla_b T^{a b} = 0 \ ,
\end{equation}
and the thermodynamical relation\footnote{Note that this assumes thermodynamical equilibrium!} for an equation of state $\varepsilon=\varepsilon(n,s)$
\begin{equation}
    \nabla_a \varepsilon = \mu \nabla_a n + T \nabla_a s = {p+\varepsilon -sT \over n} \nabla_a n + T \nabla_a s \ , 
\end{equation}
to show that 
\begin{equation}
 T\Gamma_\s = \left( \beta T -1 \right) \nabla_a q^a + q^a \left( T \nabla_a \beta - u^b \nabla_b u_a \right) \ .
  \label{2ndlaw}
\end{equation}%
We want to ensure that the right-hand side of this equation is positive
definite (or indefinite). An easy way to achieve this is to make the
 identification
\begin{equation}
  \beta = 1/T\ ,
\end{equation}
and at the same time insist that the heat flux is such that
\begin{equation}
  q^a = - \kappa T \perp^{a b}
  \left( \frac{1}{T} \nabla_b T + u^c \nabla_c u_b \right) \ ,
  \label{hflux}
\end{equation}
with $\kappa \ge 0$ being the heat conductivity coefficient. 
This means that
\begin{equation}
    \Gamma_\s =
    \frac{q^a q_a}{\kappa T}  \ge 0 \ , 
\end{equation}
by construction, and the second law of thermodynamics is satisfied.

The energy equation now takes the form
\begin{equation}
    n T {d x_\s \over d\tau} + \nabla_a q^a + q^a \dot u_a = 0 
    \label{eneq}
\end{equation}
where $\dot u_a = u^b \nabla_b u_a$ is the four acceleration, as before. We also have the momentum equation
\begin{multline}
    \perp^c_b \nabla_a T^{ab} = 0 \\  \Longrightarrow  \quad
    (p+\varepsilon) \dot u^a + \perp^{ab}\left( \nabla_b p + \dot q_b \right) + q^b \nabla_b u^a + q^a \nabla_b u^b = 0 \ .
    \label{momeqs}
\end{multline}

This model seems quite generic. Unfortunately, it has some major problems. While it is built to pass the key test set by
the second law of thermodynamics, it fails at the next hurdle. A detailed analysis of perturbations away from
an equilibrium state \citep{hiscock85:_rel_diss_fluids} shows that small perturbations tend to be dominated
by rapidly growing instabilities (we will demonstrate this later), suggesting that the formulation may be practically useless.
From the mathematical point of view
it is also not acceptable since, being non-hyperbolic, it does not admit
a well-posed initial-value problem. We will discuss how we can fix these problems shortly. First we will take a slight detour towards an application.

\subsection{Case study: Neutron star cooling}

One situation where the model we have derived finds practical use is in the description of the thermal evolution of a maturing neutron star. This is (obviously) an interesting problem in itself, and from the present perspective it is worth clarifying the assumptions that lead to the equations commonly used in cooling simulations. The typical starting points tends to be the assumption that the configuration can be taken to be static, essentially meaning that we ignore the impact of the thermal pressure on the matter and the spacetime. Taking the spacetime to be spherically symmetric and static, we have the usual line element 
\begin{equation}
    ds^2 = - e^{2\nu} dt^2 + e^{2\lambda} dr^2 + r^2 d\theta^2 + r^2 \sin^2 \theta d\varphi^2 \ , 
\end{equation}
where $\nu$ and $\lambda$ are functions of $r$,
while the matter four velocity is take to be
\begin{equation}
    u^a = \left[ e^{-\nu}, 0, 0, 0 \right] \ .
\end{equation}
It is important to understand that this does not mean that $\dot u^a = 0$. We still get a contribution from the spacetime curvature. Ignoring the heat flux terms in \eqref{momeqs} we have (with primes denoting radial derivatives)
\begin{equation}
     (p+\varepsilon) \dot u^a + \perp^{ab} \nabla_b p = 0 \quad \Longrightarrow\quad p' = - (p+\varepsilon) \nu' 
     \label{TOV}
\end{equation}
It is worth taking a closer look at this (well-known) equation. Consider the case of a single fluid, for which we have (see Sect.~\ref{shelve})
\begin{equation}
    p+\varepsilon = n\mu \ , \qquad \mbox{and} \qquad \nabla_a p = n \nabla_a \mu
\end{equation}
and it follows that \eqref{TOV} simply represents the fact energies are affected by the gravitational redshift:
\begin{equation}
    {d\over dr} \left( \mu e^{\nu} \right) = 0 \ .
\end{equation}
In the situations where $q^a\neq0$, we are obviously ignoring the impact of the heat flux on the overall energy and the spacetime curvature. This is likely to be a good approximation in most situations of interest. 

Moving on to the equations that govern the thermal component, we first of all find that the radial component of \eqref{hflux} becomes
\begin{equation}
    q^r = - \kappa  e^{-2\lambda} \left( T' + {T} \nu' \right) = - \kappa e^{-2\lambda- \nu} \partial_r \left( T e^{\nu}\right) =  - \kappa e^{-2\lambda- \nu} \partial_r \left( T^\infty\right)
    \label{qf}
\end{equation}
where we have defined the temperature measured by an observer at infinity, $T^\infty$.
Finally, we need \eqref{eneq}. As we want to work with the temperature rather than the entropy, we use
\begin{equation}
    d\varepsilon = \mu dn + T ds = \left( {\partial \varepsilon \over \partial n} \right)_T dn + \left( {\partial \varepsilon \over \partial T} \right)_n dT \ . 
\end{equation}
We also note that, for a static configuration $\nabla_a u^a = 0 $ so \eqref{ndiv} means that 
\begin{equation}
    {dn \over d\tau} = 0  \ , 
\end{equation}
and we have 
\begin{equation}
    {ds \over d\tau} = {1\over T}\left( {\partial \varepsilon \over \partial T} \right)_n {dT \over d\tau} \ . 
\end{equation}
That is, we can write \eqref{eneq} as 
\begin{equation}
   \left( {\partial \varepsilon \over \partial T} \right)_n {dT \over d\tau} + \nabla_a q^a + q^b \dot u_b = 0 \ , 
\end{equation}
which (if we assume that the heat flux is radial) becomes
\begin{equation}
 \left( {\partial \varepsilon \over \partial T} \right)_n e^{-\nu} \partial_t T + {1\over r^2} e^{-(2\lambda+\nu)} \partial_r \left[ r^2 e^{(\lambda+\nu)} q^r \right] + {\nu' } q^r = 0 \ .
    \label{dT}
\end{equation}
In principle we now have the equations we need. We only need to massage them into a more intuitive form. The first step involves introducing the flux through a spherical surface with radius $r$:
\begin{equation}
    {L\over 4\pi r^2} = e^{\lambda} q^r = q^{\hat r} \ ,
\end{equation}
(based on using a tetrad description, see Sect.~\ref{sec:tetrad}). This means that \eqref{qf} becomes
\begin{equation}
    {L \over 4 \pi \kappa r^2} = -  e^{-(\lambda + \nu)} \partial_r \left( T e^{\nu}\right) \ ,
    \label{heat1}
\end{equation}
while \eqref{dT} can be written
\begin{equation}
    C_v e^{-\nu} \partial_t T + {1 \over 4\pi r^2} e^{-\lambda-2\nu} \partial_r \left( e^{2\nu} L \right)= 0  \ , 
\end{equation}
where we have identified the heat capacity at fixed volume 
\begin{equation}
    C_v =   \left( {\partial \varepsilon \over \partial T} \right)_n\ .
    \label{heatcap}
\end{equation}
Once we introduce the energy loss due to (say) the emission of neutrinos, we arrive at the equations discussed in the classic review by \cite{2004ARA&A..42..169Y}, which in turn originate from the classic work of \cite{1977ApJ...212..825T}.

\subsection{The multi-fluid view}

Let us now consider thermal dynamics from a multi-fluid perspective, with the view of comparing to the standard derivation. In order to do this we  assume  that the entropy component can be treated as a ``fluid" (analogous to the thermal excitations of a superfluid system, see Sect.~\ref{sec:superfluids}). In essence, 
this implies that  the mean free path of the phonons 
is taken to be small compared to the model scale. We then consider two fluxes, one corresponding to the matter flow
and one associated with the entropy. The latter is  treated as massless (zero rest-mass). The 
dynamics then follows from the usual two-fluid Lagrangian, which also depends on the relative flow of the two fluxes.
As we will see, the entropy entrainment
turns out to be a crucial feature of the model \citep{2010RSPSA.466.1373A,2011RSPSA.467..738L}.

As in the case of a general two-fluid system, the starting point is the definition of a relativistic invariant Lagrangian $\Lambda$.
Assuming that the system is isotropic, we take $\Lambda$ to be a function of the different scalars that can be formed
by the two fluxes. From the matter current $n^a$ and the entropy flux $s^a$ we can form three scalars (tweaking the multifluid notation to stay close to the previous derivation);
    \beq
    n^2  = -n_a n^a \ , \quad
    s^2  = -s_a s^a \ , \quad
    j^2  = -n_a s^a \ .
    \eeq
An unconstrained variation of $\Lambda$ then leads to
    \beq
    \label{var1}
    \delta \Lambda = \frac{\partial \Lambda}{\partial n}\delta n + \frac{\partial \Lambda}{\partial s} \delta s + \frac{\partial \Lambda}{\partial j} \delta j \ .
    \eeq
Replacing the passive density variations with dynamical variations of the worldlines (as  in Sect.~\ref{sec:pullback}) we find that
  \begin{multline}
    \label{var2}
    \delta \Lambda = \left[-2\frac{\partial \Lambda}{\partial n^2}n_a -\frac{\partial \Lambda}{\partial j^2}s_a \right]\delta n^a+
                \left[ -2\frac{\partial \Lambda}{\partial s^2}s_a -\frac{\partial \Lambda}{\partial j^2}n_a \right]\delta s^a  \\
           + \left[-\frac{\partial \Lambda}{\partial n^2}n^an^b - \frac{\partial \Lambda}{\partial s^2}s^a s^b - \frac{\partial \Lambda}{\partial j^2}n^a s^b\right]\delta g_{ab} \ .
 \end{multline}
From this  we can read off the conjugate momentum associated with each of the fluxes;
    \beq
    \mu_a=\frac{\partial \Lambda}{\partial n^a} =  g_{ab}(\Bn n^b + \Ans s^b) \ , \quad
    \theta_a=\frac{\partial \Lambda}{\partial s^a} =  g_{ab}(\Bs s^b + \Ans n^b) \ ,
    \eeq
where 
    \beq
    \label{var.coefs}
    \mathcal{B}^\n\equiv -2 \frac{\partial \Lambda}{\partial n^2}, \quad \mathcal{B}^\s\equiv -2 \frac{\partial \Lambda}{\partial s^2}, \quad \mathcal{A}^{\n\s}\equiv-\frac{\partial \Lambda}{\partial j^2} \ .
    \eeq

As usual, 
the stress-energy tensor is obtained by noting that the displacements of the conserved currents induce
a variation in the spacetime metric.  In this case,  we arrive at
    \beq
    \label{se-tensor}
    T_a^{\ b} = \mu_a n^b + \theta_a s^b + \Psi \delta_a^{\ b} \ ,
    \eeq
where we have defined the generalized pressure, $\Psi$, as
    \beq
    \label{psi}
    \Psi = \Lambda -\mu_a n^a - \theta_a s^a \ .
    \eeq
These results are completely analogous to the two-fluid model from Sect.~\ref{sec:twofluids}.

As  the divergence of the stress-energy tensor (\ref{se-tensor}) vanishes, we can  express the equations of motion as a force balance
    \beq
    \label{fbal}
    \nabla_b T_{a}^{\ b} = f^\n_a + f^\s_a = 0 \ ,
    \eeq
where the individual force densities are 
    \begin{align}
    \label{fn}
    f^\n_a &=2 n^b\nabla_{[b}\mu_{a]} + \mu_a \nabla_b n^b \ , \\
    \label{fs}
    f^\s_a &=2 s^b  \nabla_{[b}\theta_{a]}+ \theta_a \nabla_b s^b \ .
    \end{align}

Note that, in order to obtain the stress-energy tensor (\ref{se-tensor}), as in Sect.~\ref{sec:variational}, we needed to impose the conservation of the fluxes as constraints on the variation.
However, the equations of motion, (\ref{fn}) and (\ref{fs}), still allow for non-vanishing  production terms.
If we, for simplicity, consider a single particle species, the matter current is conserved (there can be no particle reactions) and we have
$\nabla_a n^a = 0$.
This removes the second term from the right-hand side of (\ref{fn}). In contrast, the entropy flux is  generally not conserved, but in accordance with the second law we must have
\beq
\label{divs}
\nabla_a s^a = \Gamma_\s \ge 0 \ .
\eeq

So far, the model is fairly general. 
To progress, we need to connect with  thermodynamics. In doing this it makes sense to consider a specific choice of frame.
In the context of a single (conserved) species of matter, we see that the force $f^\n_a$ is orthogonal to the matter flux, $n^a$, and therefore it has only three degrees of freedom. Furthermore, because of the
force balance \eqref{fbal}, we also have $n^a f^\s_a=0$. This suggests that it is natural to focus on observers associated with the matter frame. We therefore introduce the four-velocity $u^a$ such that $n^a = n u^a$,
where $u_a u^a = -1$ and $n$ is the number density measured in this frame. This is, of course, the same frame as in Sect.~\ref{sec:pullback}.

Having chosen to work in the matter frame, we can decompose the entropy current and its conjugate momentum into parallel  and orthogonal components. The entropy flux is then expressed as
    \beq
    s^a =  s^* (u^a + w^a) \ ,
    \label{entflux}
    \eeq
where $w^a$ is the relative velocity between the two fluid frames, and $u^a w_a=0$. Letting $s^a = s u^a_\s$ where $u_\s^a$ is the four-velocity associated with the
entropy flux, we see that $s^*=s\gamma$ where $\gamma$
is the redshift associated with the relative motion of the two frames~\footnote{In the following, we will use an asterisk to denote matter frame quantities.}.

Similarly, we can write the thermal momentum as
    \beq
    \theta_a = \left(\Bs s^* + \Ans n\right) u_a + \Bs s^* w_a \ .
    \eeq
This leads to a measure of the temperature
measured in the matter frame:
\beq
-u^a \theta_a = \theta^* =  \Bs s^* + \Ans n \ .
\label{tpar}\eeq
In essence, this quantity represents the effective mass of the entropy component.
Returning to the stress-energy tensor, and making use of the projection orthogonal to the matter flux, we find that
the heat flux (energy flow relative to the matter) is given by
    \beq
    \label{heat}
    q_a = -\perp_{ab}u_c T^{bc} = s^* \theta^* w_a \ . 
    \eeq
Defining the new  variables $\sigma^a = s^* w^a$ and $p_a = \Bs s^* w_a$, the energy density  measured in the matter frame can be obtained by a Legendre transform
on the Lagrangian. We have
    \beq
    \label{rhostar}
    \varepsilon^* = u_a u_b T^{ab} = - \Lambda + p_a \sigma^a \ .
    \eeq

The relevance of the new variables becomes apparent if we consider the fact that  the \emph{dynamical} temperature
in \eqref{tpar}  agrees with the  \emph{thermodynamical} temperature that an observer moving with the matter would measure. In other words, we have
\beq
\theta^* = \left. {\partial \varepsilon^* \over \partial s^*}\right|_{n,p} \ ,
\eeq
where $\varepsilon^* = \varepsilon^*(n,s^*,p)$. This is
the standard definition of temperature as the energy per degree of freedom of the system.
Formally, the temperature is obtained from the variation of the energy with respect to the entropy in the observer's frame (keeping the other thermodynamic variables fixed).

This result is not trivial. The requirement that the two temperature measures agree
determines the additional state parameter, $p$, to be held constant in the variation of $\varepsilon^*$.
The importance of the chosen state variables is emphasized further if we note that, when the system is out of equilibrium, the energy depends on the heat flux
(encoded in $\sigma^a$ and $p_a$).
This leads to an \emph{extended} Gibbs relation (similar to that postulated in many approaches to extended thermodynamics; \citealt{joubook});
  \beq
    d \varepsilon^* = \mu d n + \theta^* d s^* + \sigma d p \ .
    \eeq
This result arises naturally from the variational analysis. It is \emph{derived} rather than \emph{assumed}. 

Traditionally, thermodynamic properties like pressure and temperature are uniquely defined only in equilibrium.
Intuitively this makes sense since---in order to carry out a measurement---the measuring device must have time to
reach ``equilibrium'' with the system. A measurement is  only meaningful as long as the timescale required to obtain a result is
shorter than the evolution time for the system. However, this does not prevent a generalization of the various
thermodynamic concepts (as described above). The procedure may not be ``unique'', but one must at least require the generalized concepts to be internally consistent. 

The variational model encodes the finite propagation speed for heat, as required by causality.
To demonstrate this, we may use
the orthogonality of the entropy force density $f_\s^a$ with the
matter flux,  solve for the entropy production rate $\Gamma_\s$  and then impose the second law of thermodynamics.
It is natural to express the result in terms of the heat flux $q^a$, now given by
    \beq
    \label{sdecomp}
    s^a= s^*u^a + \frac{1}{\theta^*}q^a \ .
    \eeq
Meanwhile, the conjugate momentum takes the form
    \beq
    \label{thetadec}
    \theta_a = \theta^* u_a + \beta q_a \ ,
    \eeq
where 
    \beq
    \beta = \frac{1}{s^*} -  \frac{\Ans n }{s^* \theta^*}  \ .
    \eeq
With these definitions, we  impose the second law of thermodynamics by demanding that the entropy production is a quadratic in the sources, i.e.,
\beq
\Gamma_\s = {q^2 \over \kappa \theta_*^2}  \ge 0 \ ,
\label{gs1}\eeq
where   $\kappa>0$ is the thermal conductivity.
This means that the heat flux is governed by
    \beq
    \label{gato}
    \tau \left( \dot{q}^a + q_c \nabla^a u^c \right) + q^a = -\tilde\kappa \perp^{ab}\left( \nabla_b \theta^* + \theta^* \dot u_b\right) \ ,
    \eeq
    where $\dot{q}^a = u^b \nabla_b q^a$ and $\dot{u}^a$ is the four-acceleration (as before) and we have also
 introduced
    \beq
    \tilde\kappa \equiv \frac{\kappa}{1 + \kappa \dot \beta} \ ,
    \eeq
while the thermal relaxation time is given by
    \beq
    \tau = \frac{\kappa\beta}{1 + \kappa \dot \beta} \ .
    \eeq
The final result (\ref{gato}) is the relativistic version of the so-called Cattaneo equation \citep{cattaneo,andersson10:_caus_heat,2011RSPSA.467..738L}. It resolves the issue
of the instantaneous propagation of heat, see \cite{jcv} for a brief discussion. 
We also learn that the entropy entrainment, encoded in $\Ans$, plays a key role in determining the thermal relaxation time $\tau$.
This agrees with the implications of extended thermodynamics, as well as related results in the context of Newtonian gravity \citep{2010RSPSA.466.1373A}. Finally, as described by \cite{joubook}, the
 Cattaneo equation inspired
the development of the more general extended irreversible thermodynamics framework.

The heat problem (obviously) has two dynamical degrees of freedom, leading to the presence of a second sound in solids,
an effect that has been observed in  laboratory experiments on dielectric crystals
\citep{rms}. 
 So far, we focussed on the heat. In addition, we have
a momentum equation for the matter component. From \eqref{fn} it follows that this equation can be written
\beq
 \mu \dot{u}_a+\perp^b_a\nabla_b \mu +\alpha \dot{q}_a +\dot{\alpha} q_a +\alpha q^b \nabla_a u_b = {1 \over n} f^\n_a \ .
\label{mom}\eeq
Here we have represented the matter momentum by
\beq
\mu_a = \mu u_a + \alpha q_a \ ,
\eeq
where $\mu$ is the chemical potential (in the matter frame) and 
\beq
\alpha = {\Ans \over \theta^*} \ .
\label{alp}\eeq
That is, we have
\beq
\alpha = {1 - \beta s^* \over n} \ .
\eeq
Given these definitions, we have 
\beq
-f^\n_a = f^\s_a = - {1 \over \tilde{\kappa}}\left( s^* - {\beta q^2 \over \theta_*^2} \right) q_a  \ .
\eeq
It is useful to note that this implies that the force has a term that is linear in $q^a$. We will explore this fact in the following.

\vspace*{0.1cm}
\begin{tcolorbox}
Aiming to develop a simple model for heat conduction, \cite{carter_heat} suggested an ``off the peg'' model , similar to the model we have described, but with the entrainment between particles and entropy set to zero. However, as 
\cite{olshis90} have shown, this has disastrous consequences. The
 model violates causality in two simple model settings. As
discussed by \cite{priou91} and
\cite{carter92:_momen_vortic_helic}, this 
emphasizes the importance of the entrainment for this
problem. The problem is that, ignoring the entropy entrainment leaves us with no freedom to adjust the thermal relaxation timescale. Retaining this flexibility is important.
\end{tcolorbox}
\vspace*{0.1cm}

The two-fluid results can  be directly compared to the ``phonon hydrodynamics'' model developed by \cite{1966PhRv..148..778G}
(see \citealt{llebot}, and \citealt{cimmelli} for alternatives). This may be the most celebrated attempt to account for non-local
heat conduction effects, accounting for the interaction of phonons with each other and the conducting lattice.
The usefulness of this result is due to the fact that it can be used both in the
collision dominated and the ballistic phonon regime. In the former, the resistivity dominates, the nonlocal terms can be neglected and heat propagates as waves. In the opposite regime, the momentum conserving interactions
are dominant and we can neglect the thermal relaxation. In this regime, heat propagates by diffusion.
The transition between these two extremes has recently been discussed by \cite{vasquez}.

Interestingly, the non-local heat conduction model may also be useful for nano-size systems. If a system has characteristic size larger than the relevant mean-free path then one would not necessarily expect a fluid model to apply.
Nevertheless, \cite{alvarez09} have argued that the expected behaviour of the thermal conductivity
as the size of the system decreases (as discussed by \citealt{alvarez07}) can be reproduced provided that an appropriate
slip condition for $q^a$ is applied at the boundaries.  This is an interesting problem that deserves further  study.

Finally, it is worth commenting on dissenting perspectives. The main issue appears to stem from the presence of the term involving  the four-acceleration on the right-hand side of \eqref{gato}. We have already seen that this term encodes the impact of the gravitational redshift on the temperature, which obviously has no counterpart in the Newtonian problem. Dynamically, the effect results from the fact that the 
infinitesimal 3-spaces orthogonal to the matter world lines are not parallel, but ``tipped over'' because of the curvature of the  world line.
This leads to the interpretation of the four-acceleration contribution in terms of the effective inertia of heat \citep{ehlers}. This seems quite intuitive, but it has nevertheless been suggested \citep{2006JNET...31...11G,2009GReGr..41.1645G,2008PhLB..668..425T,2009PhyA..388.3765S} that this term causes instabilites and it should not be included. As this seems somewhat inconsistent, we will not analyse this suggestion in detail.

\subsection{A linear model and the second sound}

The variational model contains terms that enter as second order deviations from thermal equilibrium, e.g., pieces that are second order in the heat flux, $q^a$. In fact, it is clear that key effects (like the entropy entrainment) arise from the presence of 
such terms in the Lagrangian. Having said that, once we have written down the general model, we can opt to truncate the results at first order. Crucially, this does \emph{not} take us back to the original first-order model. The thermal relaxation 
remains, reflecting the simple fact that you need to know the energy of a system to quadratic order in order to develop the complete linear equations of motion. Noting this, it is interesting to consider the features of this new first-order model. After  all, this, much simpler, description may be adequate in many relevant situations.

We want to restrict our analysis to first order deviations from equilibrium.  
Thermal equilibrium corresponds to $q^a=0$, no heat flux, and $\dot{u}^a=0$, no matter acceleration (in essence, we are analyzing the problem at the local level, ignoring gravity). 
Moreover, in the simplest cases there should be no shear, divergence or vorticity associated with the flow, i.e., we  have $\nabla_a u^a = 0$ and 
$\nabla_b u^a=0$ as well. Treating all these quantities as first order, and noting that 
\beq
u_b \dot{q}^b = - q^b \dot{u}_b \ ,
\eeq
also contributes at second order, we arrive at  two momentum equations; from \eqref{mom} we have
\beq
\mu \dot{u}_a + \perp^b_{\ a} \nabla_b \mu + \alpha \dot{q}_a + \left( \dot{\alpha} - {s \over n\tilde{\kappa}} \right) q_a  =  0 \ ,
\label{umom}\eeq
while \eqref{gato} leads to 
\beq
\tau \dot{q}_a + q_a + \tilde{\kappa} \left(  \perp^b_{\ a} \nabla_b T + T \dot{u}_a \right)= 0  \ .
\label{qmom}\eeq
We also have the 
two conservation laws
\beq
\nabla_a n^a = 0 \ ,
\eeq
\beq
\nabla_a s^a = 0 \ , 
\eeq
noting that $\Gamma_\s$ is second order (by construction).
In these equations we have used the fact that $s^*$ and $\theta^*$ differ from the equilibrium values $s$ and $T$ only at second order.
To first order, the pressure $p$ is obtained from the standard equilibrium Gibbs relation
\beq
\nabla_a p = n \nabla_a \mu + s \nabla_a T \ .
\label{pres}\eeq
Finally, we have the fundamental relation
\beq
\varepsilon + p = \mu n + s T \ .
\label{fund1}\eeq
By comparing \eqref{umom} and \eqref{qmom} to the Eckart frame results it becomes apparent 
 to what extent the first-order model relies on its higher order origins.
Specifically, $\alpha$ and (therefore) $\tau$ depend on $\Ans$ and the entropy entrainment, c.f., \eqref{alp}. These effects rely on quadratic terms in the Lagrangian, and hence would not be present in a model that includes only first order terms from the start. 

In order to analyze the dynamics of the heat problem, we  consider perturbations (represented by $\delta$) away from a uniform
equilibrium state. First of all, recall that we have  $q_a=\dot{u}_a=0$ for a system in equilibrium. We can also ignore $\dot{\alpha}$ and $\dot{\beta}$, since the equilibrium configuration is uniform, which means that we can replace $\tilde{\kappa}$ by $\kappa$. 
This means that we are left with  two equations;
\beq
\mu \delta \dot{u}_a + \perp^b_{\ a} \nabla_b \delta \mu + \alpha \delta \dot{q}_a-{s \over n\kappa} \delta q_a= 0 \ ,
\label{eq1}\eeq
and
\beq
\tau \delta \dot{q}_a + \delta q_a + \kappa \perp^b_{\ a} \nabla_b \delta T + \kappa T \delta \dot{u}_a = 0  \ ,
\label{eq2}\eeq

We can combine these to get
\beq
\left( p+\varepsilon\right)\delta \dot{u}_a + \perp_a^b \nabla_b \delta p + \delta\dot{q}_a = 0  \ .
\label{eq3}\eeq
The last two equations [\eqref{eq2} and \eqref{eq3}] are, not surprisingly, identical to the first-order reduction of the Israel-Stewart model (see Sect.~\ref{sec:viscosity}), so the problem is relatively well explored. In particular, the conditions required for stability and causality were 
derived by \cite{hislin83,hislin87}, see also \cite{olshis90}. 

Working in the frame associated with the background flow, we note that \eqref{eq1} and
\eqref{eq2} only have spatial components. That is, we may erect a local Cartesian coordinate system associated with the matter frame and 
simply replace $a\rightarrow i$ where $i=1, 2, 3$.
Then taking the curl ($\epsilon^{jki} \nabla_k$) of the equations in the usual way, we arrive at
\beq
m_\star \dot{U}^i - {1 \over \tau}  \dot{Q}^i = 0 \ ,
\eeq
and
\beq
m_\star \dot{Q}^i +(p+\varepsilon) Q^i  = 0 \ ,
\eeq
where we have defined
\beq
U^i = \epsilon^{ijk} \nabla_j \delta u_k \ , \qquad \mbox{and} \qquad Q^i = \epsilon^{ijk} \nabla_j \delta q_k \  , 
\eeq
and
\beq
m_\star = n \left( \mu - { \alpha \kappa T \over \tau} \right) = p+ \varepsilon - {\kappa T \over \tau} \ .
\eeq

Assuming that the perturbations depend on time as $e^{i\omega t}$, where $t$ is the time-coordinate associated with the matter frame,
 we arrive at the dispersion relation
for transverse perturbations;
\beq
i\omega \left[  (p+\varepsilon)( 1 + i\omega \tau) - i \omega \kappa T \right]= 0 \ .
\eeq
Obviously $\omega = 0$ is a solution. The second root is
\beq
\omega =  {i(p+\varepsilon) \over m_\star\tau} \ .
\eeq
This result shows that the thermal relaxation time $\tau$ is essential in order for the system to be stable.
We need $m_\star>0$, i.e., the relaxation time must be such that
\beq
\tau > {\kappa T \over p+\varepsilon}  \ .
\label{tcon}\eeq
The analysis demonstrates why the Eckart model (for which $\tau=0$) is inherently unstable. 
Moreover, the constraint on the relaxation time agrees with one of the conditions obtained by \cite{olshis90} (cf.\ their Eq.~(41)), representing the inviscid limit of the exhaustive analysis of the Israel--Stewart model of \cite{hislin83}. We also note that the condition given in 
Eq.~(43) of \cite{olshis90} simply leads to the weaker requirement $\tau \ge 0$.

The problem of transverse oscillations is fairly simple since there are no corresponding restoring forces in  a pure fluid problem (these requires rotation, elasticity, the 
presence of a magnetic field etcetera). The  physical origin of the instability becomes clear once we note that $m_*$ plays the role of an ``effective'' inertial mass (density).
The importance of this quantity has been discussed in work by \cite{1997CQGra..14.2239H,1997MNRAS.287..161H,2002JMP....43.4889H}, especially in the context of gravitational collapse. Basically, the  instability of the Eckart formulation is due to the inertial mass of the fluid becoming negative. 
Once this happens the pressure gradient no longer provides a restoring force, rather it tends to push the system further away from equilibrium. 
This is a run-away process, associated with exponential growth of perturbations. Ultimately, the instability is due to the inertia of heat; an unavoidable consequence of the equivalence principle (heat carries energy, which means that it can be associated with an effective mass; \citealt{tolman34:_book}). The condition \eqref{tcon} may seem rather extreme (\citealt{hislin87} quote a timescale of $10^{-35}$~s for water at 300K), but it sets a sharp lower limit for the thermal relaxation in physical systems. A system with faster thermal relaxation can not settle down to equilibrium. However, it may still be reasonable to ask if a system may evolve in such a way that it enters the 
unstable regime (in the way discussed by \citealt{1997CQGra..14.2239H,1997MNRAS.287..161H}).

When we turn to the longitudinal case the situation changes. In a perfect fluid longitudinal perturbations propagate as sound waves, and when we add complexity to the model the dispersion relation soon gets  complicated.  The problem has been discussed in detail by \cite{2011RSPSA.467..738L}, so we will move straight to the results. The dispersion relation for the phase velocity, $\sigma = \omega/k$, is
\begin{multline}
m_\star\tau \sigma^4 - {i(p+\varepsilon) \over k} \sigma (\sigma^2 - C_s^2)
- \left[ (p+\varepsilon)\left( {\kappa\over n c_v} + C_s^2 \tau \right)-2\kappa T \alpha_s\right] \sigma^2  \\
+  \kappa \left[ {p+\varepsilon\over n} {C_s^2 \over c_{v}}  - T\alpha_s^2\right] = 0 \ , 
\label{quartz}
\end{multline}
where have introduced (i) the sound speed
\beq
C_s^2 = \left({\partial p \over \partial \varepsilon} \right)_{\bar{s}} = {n\over p+\varepsilon}  \left( {\partial p \over \partial n } \right)_{\bar{s}}  \ ,
\eeq
(ii) the specific heat at fixed volume
\beq
c_v = {C_v\over n} =  T \left( {\partial \bar{s} \over \partial T} \right)_{n} = {1\over n}  \left({\partial \varepsilon \over \partial T} \right)_{n} \ ,
\eeq
and (iii)
\beq
\alpha_s = {n\over T} \left( {\partial T \over \partial n} \right)_{\bar{s}} = {T\over n} \left( {\partial p \over \partial \bar{s}} \right)_{n} = T \left( {\partial p \over \partial s} \right)_{n} \ .
\eeq
For future reference, it is also useful to note the identity [cf.\ Eq.~(96) in \cite{hislin83}]
\beq
{1\over c_v} - {1\over c_p} = {n^3 \over T (p+\varepsilon) C_s^2 }  \left( {\partial T \over \partial n} \right)_{\bar{s}}^2 = {nT \over (p+\varepsilon ) C_s^2} \alpha_s^2 \  ,
\label{cvrel}\eeq
where $c_p$ is the specific heat  at fixed pressure.

The dispersion relation \eqref{quartz} is too complicated for us to be able to make definite statements about the solutions, but we can  
simplify the analysis by considering the long- and short-wavelength limits. The results we obtain in these limits illustrate the key features. At the same time, we should keep in mind that
both cases are somewhat ``artificial''. First of all, fluid dynamics is, fundamentally, an effective long-wavelength theory in the 
sense that it arises from an averaging over a large number of individual particles (constituting each fluid element).
In effect, the model only applies to phenomena on scales much larger than (say) the interparticle 
distance. However, the infinite wavelength limit represents a uniform system, which is  artificial since real physical systems tend to be finite.
Moreover, as we will not  account explicitly for gravity  we can only consider scales 
on which spacetime can be considered flat. While the plane-wave analysis holds on arbitrary scales in special relativity, a curved spacetime 
introduces a cut-off lengthscale beyond which the analysis is not valid (roughly, the size of a local inertial frame).

Let us first consider the long wavelength, $k\to0$, problem. This represents the true hydrodynamic limit, and it   easy to see that there are two sound-wave solutions and two modes that are predominantly
diffusive. The sound-wave solutions take the form
\beq
\sigma \approx \pm C_s \left[1 \pm i {\kappa T \over 2(p+\varepsilon) C_s^3} (C_s^2 - \alpha_s)^2 k \right]  \ .
\eeq
These solutions are clearly stable, since Im~$\sigma>0$. Using the Maxwell relations listed by \cite{hislin83}, we can show
that this results agrees with Eq.~(40) from \cite{hislin87}. Moreover, our result simplifies to [using \eqref{cvrel}]
\beq
\mathrm{Im}~\sigma \approx {\kappa \over 2n} \left( {1 \over c_v} - {1 \over c_p} \right) \ ,
\label{nonrel}\eeq
in the limit where $|\alpha_s|\gg C_s^2$, which is relevant since $C_s^2 \sim p/\rho$ becomes small in the non-relativistic limit. Indeed, we find that 
\eqref{nonrel}
agrees with the standard result for sound absorption in a heat-conducting medium \citep{1966RvMP...38..205M}.

In addition to the sound waves, we have a slowly damped solution
\beq
\sigma \approx i\kappa \left[ {1 \over n c_v} - {T \alpha_s^2 \over (p+\varepsilon) C_s^2 } \right] = {i\kappa \over n c_p}  \ .
\eeq
This is the classic result for thermal diffusion.
Finally, the system has a fast decaying solution;
\beq
\sigma \approx  {i (p+\varepsilon) \over m_\star  k \tau}  \ .
\eeq
Under most circumstances, this root decays too fast to be observable, so the model reproduces that standard 
``Rayleigh--Brillouin spectrum'' 
with two sound peaks symmetrically placed with respect to the broad diffusion peak at zero frequency \citep{1966RvMP...38..205M,2009PhRvE..79f6310G} 

The short wavelength limit probes different aspects of the problem.
Letting $k\to \infty$ we see that \eqref{quartz} reduces to a quadratic for $\sigma^2$. 
We have
\begin{equation}
    A\sigma^4 - B \sigma^2 + C = 0 \ , 
    \label{kvadrat}
\end{equation}
with 
\begin{equation}
    A = m_\star \tau > 0 \ , 
\end{equation}
(as required for stability)
\begin{equation}
    B= (p+\varepsilon) \left( {\kappa \over nc_v} + C_s^2 \tau\right) - 2\kappa T\alpha_s \ , 
\end{equation}
and
\begin{equation}
    C= \kappa \left( {p+\varepsilon \over n} {C_s^2 \over c_v}- T \alpha_s^2 \right) = \kappa {p+\varepsilon \over n} {C_s^2 \over c_p} > 0 \ .
\end{equation}
This allows us to write down the solutions in closed form and it is relatively straightforward
to establish the conditions required for the stability of the system in this limit. The analysis is a bit messy but at the same instructive as it demonstrates how the physics impacts on the mathematics. Moreover, the discussion allows us to make direct contact with many previous efforts to understand the problem.

In essence, we arrive at two conditions. First of all,  $\sigma^2$ is  real and positive as long as $B^2-4AC>0$, which leads to 
\begin{multline}
\left( C_s^2 \tau - {\kappa \over n c_v} -{2\kappa T \alpha_s \over p+\varepsilon} \right)^2 + {4\kappa T \alpha_s^2 \over p+\varepsilon} \left( \tau - {\kappa T \over p+\varepsilon} \right) \\
+ {4\kappa^2 T  \over (p+\varepsilon) n c_v} \left(C_s^2-2\alpha_s\right)> 0 \ .
\label{discriminate}\end{multline}
The first two terms are positive, as long as  \eqref{tcon} is satisfied. Hence, the condition is guaranteed to be satisfied as long as $C_s^2> 2\alpha_s$. In situations where this condition is not satisfied, \eqref{discriminate} provides a (complicated) constraint on the relaxation time.
We must also have $B> 0$, which leads to
\beq
\tau > {\kappa \over  C_s^2} \left[ {2T \over p+\varepsilon} \alpha_s - { 1 \over n c_v} \right] \ .
\label{ast}\eeq
This condition is identical to that given in Eq.~(146) of \cite{hislin83} (obtained in the limit where $\alpha_i\to0$ and $1/\beta_0$ and $1/\beta_2$ both also vanish, cf.\ \citealt{2006IJMPD..15.2197H,1996astro.ph..9119M}).

Let us move on to finite wavelengths. Letting $\sigma = \sigma_\pm+ \sigma_1/k$, where $\sigma_\pm$ solve \eqref{kvadrat}, 
and linearising in $1/k$, we find that
\beq
\sigma_1 =  {i(p+\varepsilon) \over 2}  \left( {\sigma_\pm^2 - C_s^2 \over 2A \sigma_\pm^2 - B}\right) \ .
\label{sig1}\eeq
Since all quantities in this expression are already constrained to be real,  we need $\mathrm{Im}\ \sigma_1 \ge 0$ (for real $k$)
in order for the system to be stable. From \eqref{kvadrat} we then have
that
\beq
2A \sigma_\pm^2 - B = \pm\left|B^2 - 4AC \right|^{1/2} \ ,
\eeq
which leads to the condition 
\beq
\sigma_-^2 \le C_s^2 \le \sigma_+^2 \ .
\label{fincon}\eeq
This is notably consistent with the notion that ``mode-mergers'' signal the onset of instability, see Sect.~\ref{sec:cfs}. 

As the waves in the system must remain causal, we must also insist that $\sigma^2< 1$. To ensure that this is the case, we adapt the strategy used by \cite{hislin83}. As \eqref{kvadrat} is a quadratic for $\sigma^2$ we can ensure that the roots are confined to the interval $0<\sigma^2 < 1$
(noting first of all that the roots are real since \eqref{discriminate} is satisfied). Given that $B$ and $C$ are both positive, the roots must be such that 
$\sigma^2>0$. Meanwhile, we can constrain the roots to $\sigma^2 <1$ by insisting that
\beq 
A-B+C > 0 \ , 
\eeq 
and
\beq
A-2B > 0 \ .
\label{two}\eeq
Combining these inequalities with the positive discriminant, we can show that $A> B/2> C$. The first of the two conditions can be written
\beq
(1-C_s^2) \left[ \tau - {\kappa \over n c_v} \right] > {\kappa T (1-\alpha_s)^2 \over p+\varepsilon }> 0 \ .
\label{hash}\eeq
Next, when combined with causality the condition \eqref{fincon} requires that $C_s^2 \le \sigma_+^2 < 1$. In other words, we must have 
$C_s^2< 1$, which means that \eqref{hash} implies that
\beq
\tau > {\kappa \over n c_v} \ .
\label{om6}\eeq
Comparing to the results of \cite{hislin83}, we recognize \eqref{hash} as their $\Omega_3>0$ condition (it is also Eq.~(4) of \citealt{1997MNRAS.287..161H}), while \eqref{om6} 
corresponds to $\Omega_6>0$.

Meanwhile, the condition \eqref{two} can be written
\beq
(2-C_s^2) \tau > {\kappa \over n c_v} + {2\kappa T \over p+\varepsilon} (1-\alpha_s) \ , 
\eeq
corresponding to eq. (148) of Hiscock and Lindblom.
Finally, $A>C$ leads to
\beq
\tau > {\kappa T \over p+\varepsilon} + {\kappa C_s^2 \over n c_p}  \ .
\eeq
This corresponds to Eq.~(3) in \cite{1997MNRAS.287..161H}, which derives from Eq.~(147) of \cite{hislin83}.
This completes the analysis of the stability and causality of the system. We have arrived at a set of conditions on the thermal relaxation
time (and  related them to the relevant literature). As long as these conditions are satisfied, the solutions to the problem should be well behaved. 

 To complete the analysis, let us briefly consider the nature of the solutions. Since the phase velocity $\sigma$ is obtained from a quartic,
we know that the problem has two (wave) degrees of freedom. This accords with the experience from superfluid systems
and experimental evidence for heat propagating as waves in low temperature solids. One of the solutions should be associated with 
the usual ``acoustic'' sound while the second degree of freedom will lead to a ``second sound'' for heat. 
It is instructive to demonstrate how these features emerge within our model. 

In order to explore the issue, it is natural to consider the large relaxation time limit. Taking the relaxation time 
$\tau$ to be long, the solutions to \eqref{kvadrat} take the form (up to, and including, order $1/\tau$ terms)
\beq
\sigma_+^2 \approx  c_s^2\left[ 1 +  {\kappa T \over (p+\varepsilon) \tau} \left( 1 + {\alpha_s^2 \over c_s^4} \right)\right]  \ ,
\eeq
which could be rewritten using \eqref{cvrel}, and
\beq
\sigma_-^2 \approx  {\kappa \over n \tau c_p} \ .
\eeq
The first of these solutions clearly represents the usual sound, while the other solution provides the second sound. In the latter case, the 
deduced speed is exactly what one would expect \citep{joubook}.
It is easy to see that the first root will satisfy \eqref{fincon}, and the associated roots will  be unstable in the long relaxation time limit. 
Moreover, the second solution leads to stable roots as long as
\beq
\tau \ge { \kappa \over n c_p c_s^2} \ .
\eeq
Basically, the finite wavelength condition implies that the second sound must propagate slower than the first sound. This is, indeed, what is measured in 
physical systems (like superfluid Helium). Moreover, it is easy to see that this condition must be satisfied in order for the long relaxation time approximation to be valid. The general behaviour is illustrated in
Fig.~\ref{heatplot}, which relates to degenerate matter. We see that the ordinary sound exists at all wavelengths. Meanwhile, at short long wavelengths (small $k$) the remaining two roots are exponentially damped, i.e. diffusive in character. One root has a relatively slow decay, corresponding to the expected thermal diffusion, while the other root decays so rapidly that it is unlikely to be observable by experiment. Below a critical lengthscale 
(corresponding to $k=10$ in Fig.~\ref{heatplot}) the second sound emerges as a result of the finite thermal relaxation time $\tau$. For very short lengthscales, heat signals will propagate as waves. However, as is evident, these solutions are always damped. In order to ``propagate'', the real part of the wave frequency must exceed the imaginary part (so that several cycles are executed before the motion is damped out). 
This conclusion is interesting if we consider systems that become superfluid. Suppose we consider a system which starts out in the diffusive regime (e.g., Helium above the superfluid transition temperature). When the system is cooled down through the relevant transition temperature, (non-momentum conserving) particle collisions  are suppressed. In effect, the critical value of $k$ decreases and the system may enter the regime where the second sound can propagate
on macroscopic scales. The second sound emerges in a natural way.

\begin{figure}[h]
\centering
\includegraphics[height=6cm,clip]{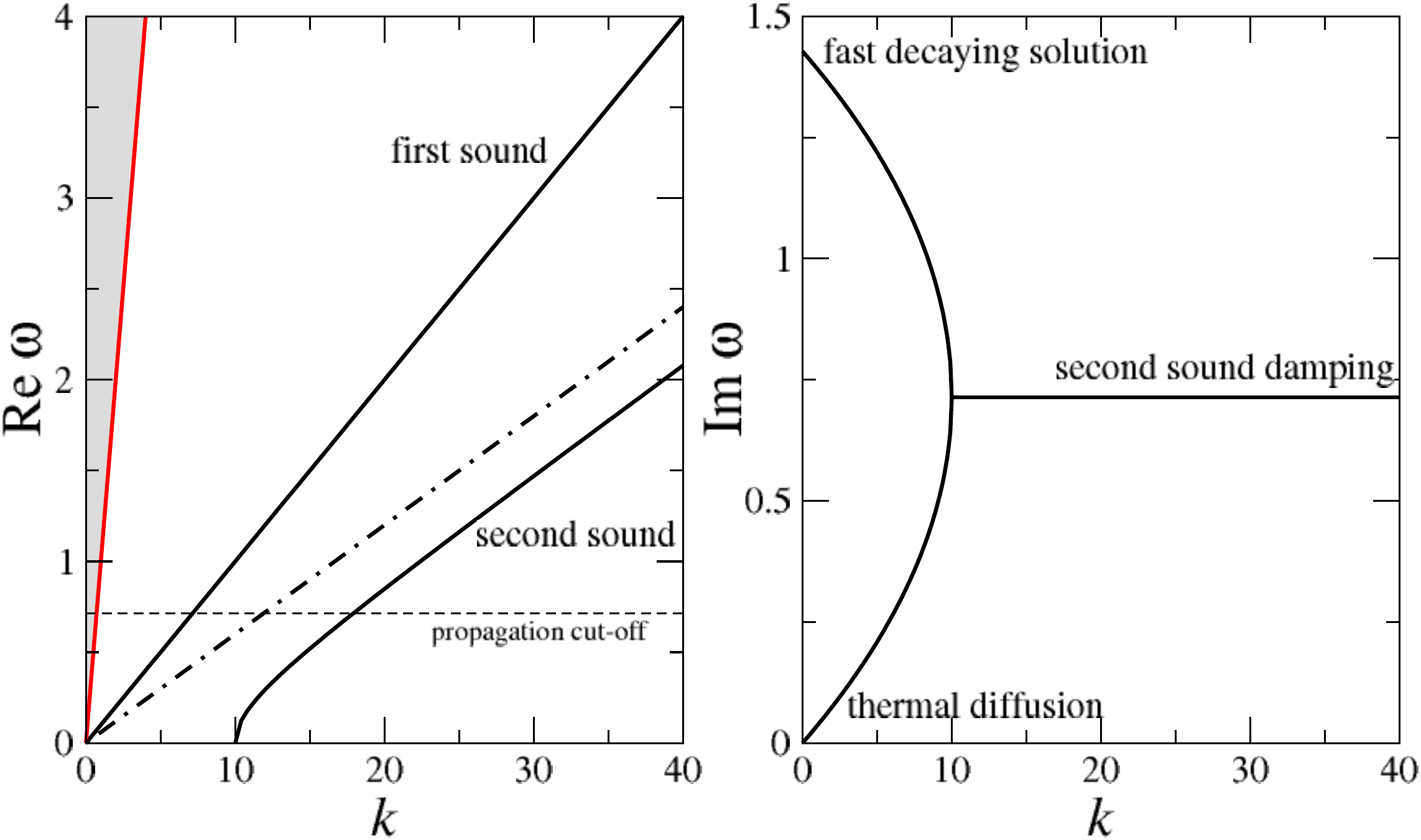}
\caption{An illustration of the qualitative nature of the behaviour of heat conducting degenerate matter, based on the first-order relativistic model. The parameters have been chosen in such a way that the speed of sound is 10\% of the speed of light, while the second sound (at short wavelengths, large $k$) propagate at $1/\sqrt{3}$ of this.  The phase velocity of the waves is $\sigma=\mathrm{Re}\ \omega/k$ (left panel).The thermal relaxation time $\tau$ has been chosen such that the critical wavenumber at which the second sound emerges is $k=10$. At lengthscales larger than this, the corresponding roots are diffusive (have purely imaginary frequency), and in the very long wavelength limit ($k\to0$) we retain the expected thermal diffusion. The damping time follows from $1/\mathrm{Im}\ \omega$ (right panel).We also indicate the noncausal region (grey area). 
The illustrated example is clearly both stable and causal. (Reproduced from \citealt{2011CQGra..28s5023A}.)}
\label{heatplot}
\end{figure}


\section{Modelling dissipation}
\label{sec:viscosity}

Although the inviscid model provides a natural starting point for any
investigation of the dynamics of a fluid system, the effects of
dissipation are often essential for the construction of a realistic
model. Consider, for example, the case of neutron star oscillations and
possible instabilities. While it is interesting from the conceptual point
of view to establish that an instability (such as the gravitational-wave
driven instability of the fundamental f-mode or the inertial r-mode
discussed in Sect.~\ref{sec:cfs}) may be present in an ideal fluid, it is crucial to
establish that the instability is able to grow on a
reasonably short timescale. To establish this, one must consider the most
important damping mechanisms and work out whether or not they will suppress the
instability. A discussion of these issues in the context of
the r-mode instability can be found in \cite{narev}.

As we have already seen for the particular case of heat flow, dissipation in a relativistic system raises difficult issues. 
According to the established consensus view, 
one must account for second-order deviations from thermal equilibrium in order to guarantee causality and stability.  This is 
certainly the lesson from the celebrated work of \cite{Israel79:_kintheo1,Israel79:_kintheo2}, see \cite{2010PhRvL.105p2501D,2011EPJWC..1307005B,2009JPhG...36f4029B} for more recent work on the problem. We have already introduced the main points in the context of heat conduction, taking a multi-fluid prescription based on  the variational formulation
as our starting point. This  approach has the flexibility required to account for the 
physics that we need to consider. A particularly appealing feature of the variational approach is that, once an ``equation of state'' for matter is provided, the theory provides
the relation between the various currents and their conjugate momenta. As we have seen, this leads to a model which has the key elements required for causality and stability, and clarifies the role of the inertia of heat (e.g., the effective mass associated with phonons). Moreover, as demonstrated by \cite{priou91} some time ago, the  variational model is formally equivalent to the Israel--Stewart construction. 
 At the end of the day, the theoretical framework becomes rather intuitive and the physics involved seems natural.

Does this mean that no issues remain in this problem area? Not really. First of all, it is clear that the need to introduce additional parameters (e.g., the relevant relaxation times) and keep track of higher order terms (fluxes of fluxes and so on) make  applications  complex. Secondly, we are not much closer to considering systems that deviate significantly from equilibrium, for which there is no natural ``small'' parameter to expand in. The variational model sheds some light on this regime by clarifying the role of the temperature in systems out of equilibrium, but there is some way to go before we understand issues associated with, for example, any ``principle of extremal entropy production'' and instabilities that lead to structure formation. Finally, despite the  successes of the extended thermodynamics framework \citep{joubook}, there is no universal agreement concerning the validity (and usefulness) of the results. To some extent this is natural given the interdisciplinary nature of the problem. To make progress we need to account for both 
thermodynamical principles and fundamental General Relativity. This leads to questions concerning, in 
particular, the  meaning of the variables involved in the different models (e.g., the entropy). The ultimate theory (if we imagine such a thing) should provide a clear link to statistical physics and even information theory. Our efforts are not yet at that level. 

In the following we will summarize the current thinking by describing the main models from the literature. We first consider the classic work of
\cite{eckart40:_rel_diss_fluid} and 
\cite{landau59:_fluid_mech}, which follow as a seemingly
natural extension of the inviscid equations. However, a detailed
analysis of \cite{hiscock85:_rel_diss_fluids,
  hislin87} has demonstrated that these descriptions have serious
flaws and must be considered unsuitable for practical use. Still,
 it is relatively ``easy'' to extend them in
the way proposed by \cite{stew77,
  Israel79:_kintheo2, Israel79:_kintheo1}. Their description, the
derivation of which was inspired by early work of \cite{grad} and
\cite{muller} and which results from relativistic kinetic theory,
provides a framework that is generally accepted as meeting the 
criteria for a relativistic model \citep{hislin83}. Next, we
describe Carter's more complete approach to the problem, which makes elegant use of the variational  argument. The
construction is also more general than that of, for example, Israel and Stewart. In
particular, it shows how one would account for several dynamically
independent interpenetrating fluid species. This extension is
important for, for example, the consideration of relativistic
superfluid systems. Finally, we consider recent progress on the development of an action principle for dissipative system, an approach that makes explicit use of the relevant matter space quantities.


\subsection{Eckart vs Landau/Lifschitz}

As in the heat problem (see Sect.~\ref{sec:heat}) we consider a single particle system, with a conserved matter flux $n^a$. However, we now allow for the possibility that we are not working in the matter frame. That is, we introduce a vector $\nu^a$ representing particle diffusion
\begin{equation}
  n^a = n u^a + \nu^a \ ,
\end{equation}
and assume that the diffusion satisfies the constraint $u_a \nu^a = 0$ (there is no particle production so $\nabla_a n^a = 0$).
This simply means that it is purely spatial according to an observer moving
with the particles in the inviscid limit, exactly what one would expect from
a diffusive process. Next we introduce the heat flux $q^a$ (as before) and the
viscous stress tensor, decomposed into a trace-part $\tau$ (not to be confused
with the proper time) and a trace-free piece $\tau^{a b}$, such that 
\begin{equation}
  T^{a b} = (p + \tau) \perp^{a b} +
  \varepsilon u^a u^b + 2 q^{(a} u^{b)} + \tau^{a b} \ ,
  \label{Tdiss1}
\end{equation}
subject to the constraints
\begin{eqnarray}
  u^a q_a = \tau^a{}_a &=& 0 \ ,
  \\
  u^b \tau_{b a} &=& 0 \ ,
  \\
  \tau_{a b} - \tau_{b a} &=& 0 \ .
\end{eqnarray}%
That is, both the heat flux and the trace-free part of the viscous stress
tensor are  spatial in the matter frame, and $\tau^{a b}$ is
symmetric. So far, the description is quite general (cf. the general decomposition of the stress-energy tensor discussed in Sect.~\ref{tab1}). The constraints
have simply been imposed to ensure that the problem has the anticipated
number of degrees of freedom.

The next step is to deduce the form for the additional fields
from the second law of thermodynamics. 
Assuming that the entropy flux is a combination of all the available vectors,
we have
\begin{equation}
  s^a = s u^a + \beta q^a - \lambda \nu^a \ ,
\end{equation}
where $\beta$ and $\lambda$ are yet to be specified (although we know already what $\beta$ will end up being from our previous discussion). It is easy to work out
the divergence of $s^a$. Then using the component of Eq.~(\ref{divT}) along
$u^a$, 
and the usual (equilibrium) thermodynamic relation for an equation of state $\varepsilon(n,s)$ (as in Sect.~\ref{sec:thermo}), we find that 
\begin{eqnarray}
  \nabla_a s^a &=& q^a
  \left( \nabla_a \beta - \frac{1}{T} u^b \nabla_b u_a \right) +
  \left( \beta - \frac{1}{T} \right) \nabla_a q^a
  \nonumber
  \\
  & & - \left( x_\s + \lambda - \frac{p + \varepsilon}{n T} \right)
  \nabla_a \nu^a - \nu^a \nabla_a \lambda -
  \frac{\tau}{T} \nabla_a u^a -
  \frac{\tau^{a b}}{T} \nabla_a u_b \ .
  \label{2ndlawb}
\end{eqnarray}%
We want to ensure that the right-hand side of this equation is positive
definite (or indefinite). An easy way to achieve this is to make the
following identifications:
\begin{equation}
  \beta = 1/T,
\end{equation}
and
\be
\lambda = {1\over nT} (p+\varepsilon - sT) = {\mu\over T}
\ee
We also identify 
\begin{equation}
  \nu^a = - \sigma T^2 \perp^{a b} \nabla_b \lambda \ ,
\end{equation}
where the ``diffusion coefficient'' $\sigma \ge 0$, and the projection is
needed in order for the constraint $u_a \nu^a = 0$ to be satisfied.
Furthermore, we find that the heat flux is given by the same expression as before (with $\beta = 1/T$) and we can use
\begin{equation}
  \tau = -\zeta \nabla_a u^a \ ,
\end{equation}
where $\zeta \ge 0$ is the coefficient of bulk viscosity. To complete the
description, we need to rewrite the final term in Eq.~(\ref{2ndlawb}). To
do this it is useful to note that the gradient of the four-velocity can
generally be written  (recall the discussion from Sect.~\ref{tab1})
\begin{equation}
  \nabla_a u_b =
  \sigma_{a b} + \frac{1}{3} \perp_{a b} \theta +
  \varpi_{a b} - a_b u_a \ ,
  \label{acce}
\end{equation}
with the usual four-acceleration, $a_b= u^a \nabla_a u_b$, 
the expansion  $\theta = \nabla_a u^a$, and the shear 
\begin{equation}
  \sigma_{a b} = \frac{1}{2}
  \left( \perp^c_b \nabla_c u_a +
  \perp^c_a \nabla_c u_b \right) -
  \frac{1}{3} \perp_{a b} \theta.
\end{equation}
Finally, the ``twist'' follows from\footnote{It is important to note
  the difference between the vorticity formed from the momentum and the corresponding
  quantity in terms of the four velocity. They differ because of the
  entrainment, and one can show that while the former is conserved along the
  flow, the latter is not. To avoid confusion we refer to $\varpi_{a b}$ as the
  ``twist'' here. This makes some sense because when we use it in
  Eq.~(\ref{acce}) we have not yet associated the four-velocity
  with the fluid flow.}
\begin{equation}
  \varpi_{a b} = \frac{1}{2}
  \left( \perp^c_b \nabla_c u_a -
  \perp^c_a \nabla_c u_b \right) \ .
\end{equation}
Since we want $\tau^{a b}$ to be symmetric, trace-free, and purely spatial
according to an observer moving along $u^a$, it is useful to introduce the
notation
\begin{equation}
  \left< A_{a b} \right> =
  \frac{1}{2} \perp^c_a \perp^d_b
  \left( A_{c d} + A_{d c} -
  \frac{2}{3} \perp_{c d} \perp^{e f}
  A_{e f} \right)
\end{equation}
for any $A_{a b}$. In the case of the gradient of
the four-velocity, it is easy to show that this leads to
\begin{equation}
  \left< \nabla_a u_b \right> = \sigma_{a b}
\end{equation}
and therefore it is natural to use
\begin{equation}
  \tau^{a b} = - \eta \sigma^{a b} \ ,
\end{equation}
where $\eta \ge 0$ is the shear viscosity coefficient. Given these
relations, we have
\begin{equation}
    T \, \nabla_a s^a =
    \frac{q^a q_a}{\kappa T} + \frac{\tau}{\zeta} +
    \frac{\nu^a \nu_a}{\sigma T^2 } +
    \frac{\tau^{a b} \tau_{a b}}{2 \eta} \ge 0 \ .
\end{equation}
By construction, the second law of thermodynamics is satisfied.

The model we have written down is quite general, especially since we did not yet specify the four-velocity $u^a$. By
doing this we can obtain both the formulation due to
\cite{eckart40:_rel_diss_fluid} and that of 
\cite{landau59:_fluid_mech}, see Sect.~\ref{tab1}. To arrive at the Eckart
description, we
associate $u^a$ with the flow of particles (as we did in the discussion of the heat problem). Thus we take $\nu^a = 0$
(or equivalently $\sigma =0$). This choice has the advantage of
being easy to implement. The Landau and Lifshitz model follows if we
instead choose the four-velocity to be a timelike eigenvector of the stress-energy
tensor. From Eq.~(\ref{Tdiss1}) it is easy to see that, by setting
$q^a = 0$, we get
\begin{equation}
  u_b T^{b a} = - \varepsilon u^a \ .
\end{equation}
This is equivalent to setting $\kappa = 0$. Unfortunately, these models,
which have been used in many applications to
date, are not that useful. While they pass the  test set by
the second law of thermodynamics, they fail  other requirements of a
relativistic description. A detailed analysis of perturbations away from
an equilibrium state \citep{hiscock85:_rel_diss_fluids} demonstrates serious
pathologies. The dynamics of small perturbations tends to be dominated
by rapidly growing instabilities. This suggests that these formulations may be practically useless. At the very least, they must be used with caution.

It has recently been argued that stability at linear order in a dissipative derivative expansion can be ensured by a judicious choice of frame \citep{2019JHEP...10..034K, 2019PhRvD.100j4020B} The argument is based on a general expansion, followed by a stability analysis to demonstrate that there exist constraints on the expansion parameters such that these models meet the stability and causality requirements. Intuitively, this argument seems somewhat at odds with the covariant nature of Einstein’s theory---the stability of a system should not depend on the chosen observer. \citet{2020arXiv200609843G} adds to the discussion by showing that the instability of the Landau-Lifschitz/Eckart models is due a failure to ensure maximum entropy at equilibrium. Meanwhile, the frame stabilised first-order models allow for violations of the second law.  As neither of these represent the anticipated physics, the issue of stability at linear order remains open.


\subsection{The Israel--Stewart approach}
\label{section_14_2}


From the above discussion we learn that the most obvious strategy for
extending relativistic hydrodynamics to include dissipation leads
to unsatisfactory results. Let us now explain how this problem can be solved.

The original strategy was based on describing the entropy current
$s^a$ as a linear combination of the fluxes in
the system, the four-velocity $u^a$, the heat-flux $q^a$ and the particle diffusion
$\nu^a$. In a series of now classic papers, \cite{stew77,
  Israel79:_kintheo2, Israel79:_kintheo1} contrasted this
``first-order'' theory with relativistic kinetic theory. Following
early work by \cite{muller} and connecting with Grad's
14-moment kinetic theory description \citep{grad}, they concluded that
a satisfactory model should be ``second order'' in the various
fields. If we, for simplicity, work in the Eckart frame (cf.\ 
\citealt{hislin83}) this means that we would use 
\begin{multline}
  s^a = s u^a + \frac{1}{T} q^a - \frac{1}{2 T}
  \left(\beta_0 \tau^2 + \beta_1 q_b q^b +
  \beta_2 \tau_{b c} \tau^{b c} \right) u^a \\
  +
  \frac{\alpha_0 \tau q^a}{T} +
  \frac{\alpha_1 \tau^a{}_b q^b}{T} \ .
  \label{2ndorder}
\end{multline}
This expression is arrived at by asking what the most general form of a
vector constructed from all the various fields in the problem may be. Of
course, we now have a number of new (so far unknown) parameters. The three
coefficients $\beta_0$, $\beta_1$, and $\beta_2$ have a thermodynamical
origin, while the two coefficients $\alpha_0$ and $\alpha_1$ represent the
coupling between viscosity and heat flow. From the above expression, we see
that in the frame moving with $u^a$  the effective entropy density is
given by
\begin{equation}
  - u_a s^a =  s - \frac{1}{2 T} \left( \beta_0 \tau^2 +
  \beta_1 q_a q^a + \beta_2 \tau_{a b}
  \tau^{a b} \right) \ .
\end{equation}
Since we want the entropy to be maximized in equilibrium, when the extra
fields vanish, we must have $[\beta_0, \beta_1, \beta_2]\ge0$. We also
see that the entropy flux
\begin{equation}
  \perp^a_b s^b = \frac{1}{T}
  \left[ (1 +  \alpha_0 \tau) q^a + \alpha_1 \tau^{a b} q_b \right]
\end{equation}
is affected only by the parameters $\alpha_0$ and $\alpha_1$.

Having made the assumption~(\ref{2ndorder}), the rest of the calculation
proceeds as in Sect.~\ref{sec:heat}. Working out the divergence of the entropy
current, and making use of the equations of motion, we arrive at
\begin{multline}
  \nabla_a s^a \\
  = - \frac{1}{T} \tau \left[ \nabla_a u^a +
  \beta_0 u^a \nabla_a \tau - \alpha_0 \nabla_a q^a -
  \gamma_0 T q^a \nabla_a \left( \frac{\alpha_0}{T} \right) +
  \frac{\tau T}{2} \nabla_a \left( \frac{\beta_0 u^a}{T} \right) \right]
  \\
 - \frac{1}{T} q^a \left[ \frac{1}{T} \nabla_a T +
  u^b \nabla_b u_a + \beta_1 u^b \nabla_b q_a -
  \alpha_0 \nabla_a \tau - \alpha_1 \nabla_b \tau^b{}_a \right.
  \\
   + \left. \frac{T}{2} q_a \nabla_b
  \left( \frac{\beta_1 u^b}{T} \right) - (1 - \gamma_0) \tau T
  \nabla_a \left( \frac{\alpha_0}{T} \right) -
  (1 - \gamma_1) T \tau^b{}_a \nabla_b
  \left( \frac{\alpha_1}{T} \right) \right]
  \\
 - \frac{1}{T} \tau^{a b} \left[ \nabla_a u_b +
  \beta_2 u^c \nabla_c \tau_{a b} -
  \alpha_1 \nabla_a q_b + \frac{T}{2} \tau_{a b} \nabla_c
  \left( \frac{\beta_2 u^c}{T} \right) -
  \gamma_1 T q_a \nabla_b \left( \frac{\alpha_1}{T} \right) \right].
  \qquad
\end{multline}%
In this expression  we have introduced (following
Lindblom and Hiscock) two further parameters, $\gamma_0$ and $\gamma_1$.
They are needed because, without additional assumptions, it is not clear how
the ``mixed'' quadratic term should be distributed. A natural way to fix
these parameters is to appeal to the Onsager symmetry
principle \citep{Israel79:_kintheo1}, which leads to the mixed terms
being distributed ``equally'' so  $\gamma_0 = \gamma_1 = 1/2$.

Denoting the comoving time derivative by a dot, i.e., using $u^a \nabla_a \tau
= \dot{\tau}$ (as before) we see that the second law of thermodynamics is
satisfied if we choose
\begin{multline}
  \tau = - \zeta \Bigg[ \nabla_a u^a + \beta_0 \dot{\tau} -
  \alpha_0 \nabla_a q^a \\
  - \gamma_0 T q^a \nabla_a
  \left( \frac{\alpha_0}{T} \right) + \frac{\tau T}{2} \nabla_a
  \left( \frac{\beta_0 u^a}{T} \right) \Bigg],
\end{multline}

\begin{multline}
  q^a  -\kappa T \perp^{a b} \Bigg[ \frac{1}{T} \nabla_b T +
  \dot{u}_b + \beta_1 \dot{q}_b - \alpha_0 \nabla_b \tau -
  \alpha_1 \nabla_c \tau^c{}_b + \frac{T}{2} q_b \nabla_c
  \left( \frac{\beta_1 u^c}{T} \right)
  \\
 \qquad \qquad \quad - (1 - \gamma_0) \tau T \nabla_b
  \left( \frac{\alpha_0}{T} \right) - (1 - \gamma_1) T \tau^c{}_b
  \nabla_c \left( \frac{\alpha_1}{T}\right) +
  \gamma_2 \nabla_{[b} u_{c]} q^c \Bigg],
 \end{multline}
 
 \begin{multline}
  \tau_{a b} = - 2 \eta \Bigg[ \beta_2 \dot{\tau}_{a b} +
  \frac{T}{2} \tau_{a b} \nabla_c
  \left( \frac{\beta_2 u^c}{T} \right) \\
  +
  \left< \nabla_a u_b - \alpha_1 \nabla_a q_b -
  \gamma_1 T q_a \nabla_b \left( \frac{\alpha_1}{T} \right) +
  \gamma_3 \nabla_{[a} u_{c]} \tau_b{}^c \right> \Bigg],
\end{multline}%
where the angular brackets denote symmetrization as before. In these
expressions we have added yet another two terms, representing the coupling to $\nabla_{[a} u_{b]}$. These bring two further ``free'' parameters, $\gamma_2$ and
$\gamma_3$. We are allowed to add these terms since
they do not affect the entropy production. In fact, a large number of
similar terms may, in principle, be considered (see note added in
proof in \citealt{hislin83}). The presence of coupling terms of the
particular form that we have introduced is suggested by kinetic
theory \citep{Israel79:_kintheo1}.

What is clear from these (very complicated) expressions is that we now have
evolution equations for the dissipative fields. Introducing characteristic
``relaxation'' times
\begin{equation}
  t_0 = \zeta \beta_0,
  \qquad
  t_1 = \kappa \beta_1,
  \qquad
  t_2 = 2 \eta \beta_2,
\end{equation}
the above equations can be written
\begin{eqnarray}
  t_0 \dot{\tau} + \tau &=& -\zeta [\dots] \ ,
  \\
  t_1 \perp^a_b \dot{q}^b + q^a &=& -\kappa T \perp^a_b [\dots] \ ,
  \\
  t_2 \dot{\tau}_{a b} + \tau_{a b} &=& - 2\eta [\dots] \ .
\end{eqnarray}

A detailed stability analysis by \cite{hislin83} shows
that the theory is causal for stable fluids. Then the
characteristic velocities are subluminal and the equations form a hyperbolic
system. An interesting aspect of the analysis concerns the 
stabilizing role of the extra parameters ($\beta_0,\dots,\alpha_0,\dots$).
Relevant discussions of the implications for the nuclear equation of state
and the maximum mass of neutron stars have been provided by 
\cite{olshis89, ols01}. A more detailed mathematical stability
analysis can be found in the work of \cite{kreiss}.

Although the Israel--Stewart model resolves the problems of the first-order
descriptions for near equilibrium situations,  issues remain to be
understood for nonlinear problems. This is highlighted in work by 
\cite{hislin88}, and \cite{olshis89b}. They
consider nonlinear heat conduction  and show that the Israel--Stewart
formulation becomes non-causal and unstable for sufficiently large deviations
from equilibrium. The problem appears to be more severe in the Eckart
frame \citep{hislin88} than in the frame advocated by 
\cite{olshis89b}. The fact that the formulation breaks down
in a nonlinear setting is not too surprising. After all, the basic
foundation is a ``Taylor expansion'' in the various fields. However,
it raises important questions. There are obvious physical situations
where a reliable nonlinear model may be crucial, e.g., heavy-ion
collisions and supernova core collapse. 

\subsection{Application: Heavy-ion collisions}
\label{heavyion}

Relativistic fluid dynamics has regularly been used as a tool to model heavy
ion collisions. The idea of using hydrodynamics to study the process of
multiparticle production in high-energy hadron collisions can be traced back
to work by, in particular, Landau in the early 1950s (see \citealt{bel_lan}).
In the early days these phenomena were observed in cosmic rays. The idea to
use hydrodynamics was resurrected as collider data became
available \citep{carrut} and early simulations were carried out at Los
Alamos \citep{amsden1, amsden2}. More recently, modeling has primarily
been focussed on reproducing data from RHIC at Brookhaven and the LHC at CERN. Useful
reviews of this active area of research can be found
in \cite{Clare_Strottman,roma2,2018ARNPS..68..339B,romatbook}. 

From the hydrodynamics perspective, a high-energy collision may be viewed in
the following way: In the centre-of-mass frame two Lorentz contracted nuclei
collide---at the typical energy of a nucleus-nucleus collision at RHIC (order 100 GeV per nucleon), each incoming nucleus is contracted by factor of about 100, making them  thin colliding
pancakes. After a complex microscopic process, a hot dense plasma is
formed. In the simplest description this matter is assumed to be in local
thermal equilibrium. The initial thermalization phase is out of
reach for hydrodynamics. In the model, the state of matter is simply
specified by the initial conditions, e.g., in terms of distributions of fluid
velocities and thermodynamical quantities. Then follows a hydrodynamical
expansion, which is described by the standard conservation equations for
energy, momentum, baryon number, and other conserved quantities, such as
strangeness, isotope spin, etc.\ (see \citealt{elze} for a variational
principle derivation of these equations). As the expansion proceeds, the
fluid cools and becomes increasingly rarefied. This stage may require a kinetic theory description. This eventually leads to the
decoupling of the constituent particles, which then do not interact until
they reach the detector.

Fluid dynamics provides a well defined framework for studying the stages
during which matter becomes highly excited and compressed and, later, expands
and cools down. In the final stage---when the nuclear matter is so dilute
that collisions are infrequent---hydrodynamics ceases to be
valid. At this point additional assumptions are necessary to predict the
number of particles, and their energies, which may be formed (to be compared
to  data obtained from the detector). These are often referred to as the
``freeze-out'' conditions. The problem is complicated by the fact that
the ``freeze-out'' typically occurs at a different time for each fluid cell.

Even though the application of hydrodynamics in this area has led to
useful results,  the theoretical foundation for this
description is not a trivial matter. Basically, the criteria required for
the equations of hydrodynamics to be valid are:
\begin{enumerate}
\item many degrees of freedom in the system,
\item a short mean free path,
\item a short mean stopping length,
\item a sufficient reaction time for thermal equilibration, and
\item a short de Broglie wavelength (so that quantum mechanics can be
  ignored).
\end{enumerate}
An interesting
aspect of the hydrodynamical description is that it makes use of concepts largely
outside traditional nuclear physics, e.g., thermodynamics, statistical
mechanics, fluid dynamics, and of course elementary particle physics. This
is natural since the very hot, highly excited matter has a large number of
degrees of freedom. But it is also a reflection of the basic lack of
knowledge. As the key dynamics is uncertain, it is comforting to resort to
familiar principles like the conservation of momentum and energy.

Another key reason why hydrodynamic models are favoured is the simplicity
of the input. Apart from  initial conditions that specify  masses
and velocities, one needs only an equation of state and an Ansatz for the
thermal degrees of freedom. If one includes dissipation one must also specify the form and magnitude of the viscosity and heat
conduction. The fundamental conservation laws are incorporated into the
Euler equations. In return for this relatively modest amount of input, one
obtains the differential cross sections of all the final particles, the
composition of clusters, etc. Of course, before one can confront the
experimental data, one must make additional assumptions about the freeze-out,
chemistry, and so on. A clear disadvantage of the hydrodynamics model is that
much of the microscopic dynamics is lost.

Let us discuss some specific aspects of the hydrodynamics that has been used
in this area. As we will recognize, the issues that need to be addressed
for heavy-ion collisions are very similar to those faced in studies of
relativistic dissipation theory and multi-fluid modeling. The one key
difference is that the problem only requires Special Relativity, so there is no
need to worry about the spacetime geometry. Of course, it is still
convenient to use a fully covariant description since one is then not tied
down to the use of a particular set of coordinates.

In many studies of heavy ions a particular frame of reference is chosen. As
we have already seen, this is an issue that must be approached with some
care. In the context of heavy-ion collisions it is common to choose $u^a$
as the velocity of either energy transport (the Landau--Lifshitz frame) or
particle transport (the Eckart frame). We have encountered both choices before. It is recognized that the Eckart
formulation is somewhat easier to use and that one can let $u^a$ be either
the velocity of nucleon or baryon number transport. On the other hand,
there are cases where the Landau--Lifshitz picture has been viewed as more
appropriate. For instance, when ultra-relativistic nuclei collide they
virtually pass through one another leaving the vacuum between them in a
highly excited state causing the creation of numerous particle-antiparticle
pairs. Since the net baryon number in this region vanishes, the Eckart
definition of the four-velocity cannot be easily employed. This discussion
is a reminder of the situation for viscosity in relativity, and the
resolution is likely the same. A true frame-independent description will
need to include several distinct fluid components.

Multi-fluid models have, in fact, been considered for heavy-ion
collisions. One can, for example, treat the target and projectile nuclei as
separate fluids to admit interpenetration, thus arriving at a two-fluid
model. One could also use a relativistic multi-fluid model to allow for
different species, e.g., nucleons, deltas, hyperons, pions, kaons, etc. Such
a model could account for the varying dynamics of the different species, as well as 
their mutual diffusion and chemical reactions. The derivation of such a
model would follow closely our discussion in Sect.~\ref{sec:twofluids}.
In the heavy-ion community, it has been common to confuse the issue somewhat
by insisting on choosing a particular local rest frame at each space-time
point. This is, of course, complicated since the different fluids move at
different speeds relative to any given frame. For the purpose of studying
heavy-ion collisions in baryon-rich regions of space, the standard option
seems to be to define the ``baryonic Lorentz frame''. This is the local
Lorentz frame in which the motion of the center-of-baryon number (analogous
to the center-of-mass) vanishes.

The main problem with the single-fluid hydrodynamics model is the requirement
of thermal equilibrium. In the fluid equations of motion it is
implicitly assumed that local thermal equilibrium is ``imposed'' via the
equation of state. In effect,  the relaxation timescale
and the mean-free path must be much smaller than both the hydrodynamical
timescale and the spatial size of the system. It seems reasonable to wonder
if these conditions can be met for hadron/nuclear collisions. On the
other hand, from the kinematical point of view (apart from the use of the
equation of state), the equations of hydrodynamics are nothing but
conservation laws of energy and momentum, together with other conserved
quantities such as charge. In this sense, for any process where the
dynamics of the flow is an important factor, a hydrodynamical framework is a
natural first step. The effects of a finite relaxation time and mean-free
path might be implemented later by using an effective equation
of state, incorporating viscosity and heat conductivity, or some simplified
transport equations. This does, of course, lead us back to the challenging
problem of designing a causal relativistic theory for dissipation. A discussion of numerical efforts can be found in \cite{roma2}. It is notable that very few
calculations have been performed using a fully three-dimensional,
relativistic theory with  dissipation. Considering  the
obvious importance of entropy, this may seem surprising
(although see \citealt{kapusta} for an exception). An interesting comparison of different
dissipative formulations is also provided in \cite{Muronga1, Muronga2}.


\subsection{The fluid-gravity correspondence}
\label{sec:flugrav}

The continued effort to explore the complex marriage between gravity and quantum theory has also led to (perhaps unexpected) developments in the modelling and understanding of relativistic fluids. The context for these developments is the AdS/CFT correspondence \citep{malda}, relating the dynamics
of a four-dimensional conformal field theory to (quantum) gravity in ten dimensions. The most commonly considered case---in essence the ``harmonic oscillator'' of the problem---relates to the duality between SU(N) $\mathcal N=4$ Super Yang-Mills theory and Type IIB string theory on AdS$_5\times$S$^5$. In general, these are both complicated theories, but the phenomenology simplifies in certain limits. The idea is attractive because it links a strongly coupled theory, for which perturbative calculations are not an option, to a weakly coupled system, for which one may be able to make progress.
This is the reason why AdS-CFT is referred to as a duality---the two descriptions
are valid in opposite regimes. However, this makes  the duality difficult to check.
In one regime we can calculate, but not in the other.

It is attractive to apply the idea to the 
state of matter explored in colliders---the
quark-gluon plasma. At the energies reached in experiments, the plasma is far from a weakly coupled gas of quarks and gluons. The system is well inside the non-perturbative
regime of QCD, where reliable tools are lacking. The AdS-CFT approach offers an avenue towards progress by reformulating the strongly coupled quantum systems as a dynamical problem in classical gravity. Perhaps the most important insight from this concerns the apparent universality of transport coefficients in gravity duals and the so-called entropy bound---the notion that for all thermal field theories (in the regime described by gravity duals) the ratio of shear viscosity 
to entropy density  is bounded by \citep{sostan}
\beq
{\eta \over s} \ge {1\over4\pi} \ .
\label{entb}
\eeq
If correct, this implies that a fluid with a  given volume density of entropy cannot be
arbitrarily close to being a perfect fluid (which would have zero viscosity).

The AdS-CFT correspondence is \emph{holographic} in the sense that the two dual
theories live in a different number of dimensions. Effectively, the  gauge theory lives ``on the boundary'' of AdS. The formalism provides a ``dictionary''
that translates  dynamical gauge theory questions
into the geometrical language associated with higher-dimensional General Relativity,  providing intriguing links between the two---traditionally separate---areas of research. 
Moreover, one can show that  long-wavelength solutions to the Einstein equations with a negative
cosmological constant (AdS)  are dual to solutions of the four-dimensional  fluid equations with a conformal symmetry. This has led to what is known as the fluid-gravity correspondence \citep{ranga}. The idea ties in with the fact that hydrodynamics may be viewed as an effective theory that governs the macroscopic behaviour of a system, on scales larger than some characteristic ``averaging'' scale (like the mean-free path).  

In practice, the fluid-gravity correspondence links a fluid system to the near-horizon dynamics of a higher dimensional black hole. This connection follows from the AdS-CFT correspondence, but at the same time it is somewhat separate from it. In fact, the connection between black holes and fluids/thermodynamics is not new at all---it dates back to the 1970s. Early work by, in particular, \cite{1} and \cite{2}, led to the appreciation that stationary black hole horizons have thermodynamic properties like temperature and entropy and the formulation of a generalized
second law of thermodynamics that treats black-hole entropy on a par with the usual matter entropy \citep{3}. This was followed by studies of analogue models of black holes \citep{4},  illustrating that fluids
can admit sonic horizons and even a version  of the Hawking temperature. Finally, through the membrane paradigm \citep{8,9} it was demonstrated that (for external observers) black holes behave like a fluid membrane, endowed
with physical properties such as viscosity and electrical conductivity  (see \citealt{gour} for a more recent discussion of this ``horizon fluid'').

The fluid-gravity correspondence takes the discussion to a different level, beyond the identification of 
holographic duals for given 
equilibrium field theory configurations, to a discussion of dynamics and dissipation. As it is instructive to understand how this comes about, let us consider a relatively simple example \citep{hube}.
Starting from an equilibrium black-hole solution we  can  generate a four-parameter family of solutions by scaling the radial coordinate $r$ and introducing a boost associated with a four-velocity $u^a$. Also introducing  ingoing Eddington--Finkelstein type coordinates we ensure that the metric is regular on the horizon. This leads to the planar Schwarzschild-AdS$_5$ black hole taking the form \citep{hube}
\beq
ds^2 = - 2 u_a dx^a dr + r^2 \left( \eta_{ab} + {\pi^4 T^4 \over r^4} u_a u_b\right) dx^a dx^b \ , 
\eeq
notably expressed in terms of the temperature $T$ and $u^a$. The boundary stress tensor induced by this (bulk) metric is (in suitable units) 
\beq
T^{ab} = \pi^4 T^4 (\eta^{ab} + 4 u^a u^b ) \ .
\eeq 
Effectively, we have 
a perfect fluid with energy $\varepsilon=3 \pi^4 T^4$ and pressure $p=\varepsilon/3$, moving with velocity $u^a$ on the flat four-
dimensional background, $\eta_{ab}$. The stress tensor is traceless, as expected for a 
conformal fluid. Also,  there is no dissipation in the system. This is natural since we still have an equilibrium solution. Let us now change this by perturbing the spacetime. This obviously leads to deviations from equilibrium, but we may execute the right to move the perturbed aspects of the metric to the other side of the equation and ``interpret'' them as contributions to the stress-energy tensor\footnote{This strategy is not too different from that used to defined the stress-energy tensor for gravitational waves.}. This leads to a time-dependent non-equilibrium fluid system, relaxing towards equilibrium as it evolves.  The relaxation/thermalization can be understood through an expansion in ``boundary derivatives'', leading to distinct dissipation channels (like shear viscosity).
The relevant transport coefficients may  be
extracted in this linearized regime, and one finds  that they can be associated with the quasinormal modes\footnote{The relevant quasinormal modes are different from those of (say) a Schwarzschild black hole [reference] in that they satisfy a vanishing Dirichlet condition at
the AdS boundary, $r=\infty$. 
This is also different from the boundary condition one uses to find the retarded
propagators in AdS/CFT, so the relation of the quasinormal modes to AdS/CFT correspondence
is not immediate.}
of the (planar AdS) black hole \citep{hube2,sostan}. This is conceptually interesting as it relates a problem in classical gravity to fluid behaviour.

Let us consider the implications of this argument. The holographic dictionary associates  low-energy phenomena  to the near
horizon dynamics. We  arrive at the usual argument describing fluid dynamics as an effective field theory for long wavelengths, albeit from an  unusual angle. Still, the logic is intuitive.
For a value to be assigned to the temperature $T$ at a given point, a fluid must have reached a local equilibrium. Basically, in order to insert a thermometer into the system to measure the temperature, the device must be able to reach some kind of equilibrium with the system. In order for this to work, we do not need a global equilibrium, but we must insist that any variations take place on a scale larger than that associated with the thermometer and the measurement. This naturally leads us to consider a long-wavelength expansion of the dynamics and a systematic expansion in derivatives (organised order by order to represent shorter scales), representing dissipative phenomena. Logically, this is close to writing down an effective field theory for a quantum system, at any given order 
taking into account all possible terms (derivatives) that may appear in the effective Lagrangian, consistent
with the underlying symmetry.

AdS-CFT and the fluid-gravity correspondence have led to progress in several interesting directions. In addition to efforts to explore issues relating to the entropy bound \eqref{entb}, 
work has been done to
construct the bulk duals of non-conformal fluids \citep{29},  charged fluids \citep{30,31}, 
superfluids \citep{32,33,34} and  anomalous fluids \citep{35}. The latter relate to the observation that some AdS black holes exhibit an instability that leads to the spontaneous formation of a scalar condensate below a critical temperature $T_c$, in analogy with the phase-transition seen in many low-temperature laboratory systems.
Not surprisingly, the more complicated the fluid system is, the more involved the gravity problem becomes. A typical example is the dissipative superfluid system considered by \cite{jyot}, which involves a map from locally hairy black brane solutions to the long wavelength solutions of  higher-dimensional Einstein-Maxwell
gravity and a phase where the global U(1) symmetry is spontaneously broken (as required to facilitate the superfluid flow). Similarly, a gravitational dual to a (type~II) superconductor can be obtained by 
coupling AdS gravity to a Maxwell field and a charged scalar \citep{gubs,hart1, hart2}. These developments are interesting given  that condensed matter physics involves a variety  of strongly coupled systems---often with unusual properties---that can
be engineered and explored in detail in laboratories \citep{hart3}.

\subsection{Completing the derivative expansion}

Taken at face value, the field theory approach to fluid dynamics prompts us to focus on the underlying symmetries (see Sect.~\ref{sec:ftheory}) and this has implications for a systematic derivative expansion aimed at representing dissipative effects. In practice, it means that---rather than introducing second order terms in order to fix the causality/stability issues of the first-order description---it is natural to ask what form second order terms may take, what the most general such model may be and how it is constrained by symmetries (e.g., of the dissipative stress-energy tensor) \citep{roma1,roma2}. Given the connection to AdS-CFT most efforts in this direction have focussed on conformal fluids, which (obviously) leaves out compressional degrees of freedom associated with bulk viscosity. Nevertheless, it is clear that the general dissipative second-order system must include a large set of parameters \citep{roma1}. It is interesting to note that, at second order the formal argument brings in coupling to the spacetime curvature. At first order, there can be no such terms since we require $\nabla_a g_{bc} = 0$, but second derivatives of the metric do not vanish so they could (perhaps should) be considered. In particular, we may have terms proportional to the Ricci scalar, $R$, and the contraction of the Ricci tensor with the fluid four-velocity, $u^a u^b R_{ab}$ \citep{baier}. The presence of such terms may come as a surprise, but they have been motivated by holographic arguments. At the same time, the situation seems a little bit confusing. By adding terms involving the Ricci tensor to the dissipative stress-energy tensor we introduce aspects that could equally well belong on the left-hand side of the Einstein equations. That is, we are modifying gravity into the general $f(R)$ class of theories (see for example \citealt{ricci}). This logic is supported by the observation that the specific terms are non-dissipative \citep{roma1}. This argument does not suggest that we should not account for these kinds of terms in a formal description, simply that we need to make more effort to understand why they should be present and what their role may be. In fact, this conclusion holds in a wider sense. General dissipative models include so many parameters---most of which we do not have any way of calculating from first principles---that  they are difficult to use in applications. Developments in this direction are important but it would perhaps make sense to shift the focus from generality to  specific questions concerning the manifestation of particular dissipation channels in  settings of practical interest.

\vspace*{0.1cm}
\begin{tcolorbox}
An important step in ``completing'' the fluid model involves mapping the formalism and the phenomenology onto the reality we want to describe. This inevitably brings in issues that can never be fully described at the averaged level; we need to consider the microphysics. There are many different ways to make this connection. In the case of neutron stars, the elusive matter equation of state has, for example, been modelled from first principle quantum calculations (often non-relativistic; \citealt{1998PhRvC..58.1804A}) and within chiral effective field theory \citep{2010PhRvC..82a4314H,2013PhRvC..88b5802K,2018ApJ...860..149T}. The latter provides an attractive strategy as it---at least in principle---provides ``error bars'' on the different parameters. These models allow us to model matter in equilibrium and study (using nonlinear simulations) the dynamics of dramatic events like neutron star mergers. However, the models do not provide us with much insight into non-equilibrium processes. This requires a more detailed understanding of transport properties. At the quantum level we need to account for stochastic fluctuations. Interesting progress in this direction---connecting with the variational strategy---aims to 
work out hydrodynamical correlation functions from an
effective action. This can be achieved by considering a classical effective
action with the characteristics of an effective
field theory suitable for an open system, formally building on the  Keldysh--Schwinger closed-time-path formalism \citep{2009AdPhy..58..197K,2018ScPP....5...53J,2018arXiv180509331G,2015JHEP...07..025H,2018JHEP...09..127J,2013arXiv1305.3670G}.  This approach is designed to describe non-equilibrium processes
at finite temperatures, at least for specific model problems. Real world applications require further developments.
\end{tcolorbox}


\subsection{Carter's canonical framework}

\cite{carter91} made a more formal attempt to construct a relativistic formalism for dissipative
fluids---taking the variational argument as its starting point. His
construction is quite general, which inevitably makes it more
complex. Of course, the generality could prove useful in more
complicated cases, e.g., for investigations of multi-fluid dynamics and/or
elastic media. Given the potential this formalism has for future
considerations, it is worth working through the details.

The overall aim is to extend the variational formulation in such a way that viscous ``stresses'' are accounted
for. Because the variational foundations are the same, the number currents
$n_\x^a$ play a central role. In
addition, we  introduce a number of viscosity tensors
$\tau^{a b}_\Sigma$, which we assume to be symmetric (even though it is
clear that such an assumption is not generally
correct, it is only to total stress-energy tensor that is required to be symmetric; \citealt{andersson05:_flux_con}). The index $\Sigma$ is
``analogous'' to the constituent index, although a bit more abstract as it represents different
viscosity contributions. It is introduced in recognition of the fact
that it may be advantageous to consider different kinds of viscosity,
e.g., bulk and shear viscosity, separately. As in the case of the
constituent index, a repeated index $\Sigma$ does not imply
summation in the following.

The key quantity in the variational framework remains the Lagrangian,
$\Lambda$. As it is a function of all the available fields, we  now have
$\Lambda(n_\x^a, \tau_\Sigma^{a b}, g_{a b})$, and a formal
variation leads to
\begin{equation}
  \delta \Lambda = \sum_\x \mu_a^\x \, \delta n_\x^a +
  \frac{1}{2} \sum_\Sigma \pi^\Sigma_{a b} \,
  \delta \tau_\Sigma^{a b} +
  \frac{\partial \Lambda}{\partial g^{a b}} \delta g^{a b} \ .
\end{equation}
Since the metric piece is treated in the same way as in the non-dissipative
problem we will leave it out from now on. In the
above expression we recognize the momenta $\mu^\x_a$ that are conjugate to
the fluxes. We also have a new set of ``strain'' variables (cf. the discussion of elasticity in Sect.~\ref{sec:relastic}) 
defined by
\begin{equation}
  \pi^\Sigma_{a b} = \pi^\Sigma_{(a b)} =
 \left. 2 \frac{\partial \Lambda}{\partial \tau^{a b}_\Sigma} \right|_{n_\x^a, g^{ab}} \ .
\end{equation}

As in the non-dissipative case, the variational framework suggests that the
equations of motion can be written as a force-balance equation,
\begin{equation}
  \nabla_b T^b{}_a =
  \sum_\x f_a^\x + \sum_\Sigma f^\Sigma_a = 0 \ ,
\end{equation}
where the generalized forces work out to be
\begin{equation}
  f_a^\x = \mu_a^\x \nabla_b n_\x^b +
  n_\x^b \nabla_{[b} \mu_{a]}^\x \ , 
\end{equation}
(as before), and
\begin{equation}
  f_a^\Sigma =
  \pi_{a b}^\Sigma \nabla_c \tau_\Sigma^{c b} +
  \tau_\Sigma^{c b} \left( \nabla_c \pi^\Sigma_{a b} -
  \frac{1}{2} \nabla_a \pi^\Sigma_{c b} \right) \ .
\end{equation}
Finally, the stress-energy tensor becomes
\begin{equation}
  T^a{}_b = \Psi \delta^a{}_b + \sum_\x \mu_b n_\x^a +
  \sum_\Sigma \tau_\Sigma^{a c} \pi^\Sigma_{c b} \ ,
\end{equation}
with the generalized pressure now given by
\begin{equation}
  \Psi = \Lambda - \sum_\x \mu_a^\x n^a_\x -
  \frac{1}{2} \sum_\Sigma \tau_\Sigma^{a b}
  \pi^\Sigma_{a b} \ .
\end{equation}

For reasons that will become clear shortly---basically, we want to be able to ensure that the different contributions to the entropy change are non-negative---it is useful to introduce a set
of ``convection vectors''. In the case of the currents, these are naturally taken as
proportional to the fluxes (as usual). This means that we introduce $\beta_\x^a$
such that
\begin{equation}
  h_\x \beta_\x^a = n_\x^a \ ,
  \qquad \mu^\x_a \beta_\x^a = -1
  \qquad \Longrightarrow \qquad
  h_\x = - \mu_a^\x n_\x^a \ ,
\end{equation}
and we see that, if we ignore entrainment then $h_\x$ is simply the chemical potential $\mu_\x$ measured by an observer riding along with the flow of the $\x$ component.
With this definition we can introduce a projection operator
\begin{equation}
  \perp_\x^{a b} = g^{a b} + \mu_\x^a \beta_\x^b
  \qquad \Longrightarrow \qquad
  \perp_{\x b}^a \beta_\x^b =
  \perp_\x^{a b} \mu^\x_b = 0 \ .
\end{equation}
From the definition of the force density $f_\x^a$ we can then show
that
\begin{equation}
  \nabla_a n_\x^a = - \beta_\x^a f^\x_a \ , 
\end{equation}
and
\begin{equation}
  h_\x {\cal L}_\x  \mu_a^\x = \perp^\x_{a b} f_\x^b \ ,
\end{equation}
where ${\cal L}_\x =
{\cal L}_{\beta_\x^a}$ represents the Lie-derivative along $\beta_\x^a$. We see that the component of the force 
parallel to the convection vector $\beta_\x^a$ is associated with particle
conservation. Meanwhile, the orthogonal component represents the change
in momentum along $\beta_\x^a$.

Next, we facilitate a similar decomposition for the viscous stresses by taking the conduction vector to be a unit null eigenvector (cf. \eqref{llframe})
associated with $\pi_\Sigma^{a b}$. That is, we introduce $ \beta_\Sigma^b$ such that
\begin{equation}
  \pi^\Sigma_{a b} \beta_\Sigma^b = 0 \ , 
\end{equation}
together with
\begin{equation}
  u_a^\Sigma = g_{a b} \beta_\Sigma^b
  \qquad \mathrm{and} \qquad
  u_a^\Sigma \beta_\Sigma^a = - 1 \ .
\end{equation}
Introducing the projection associated with this conduction vector,
\begin{equation}
  \perp^\Sigma_{a b} = g_{a b} + u^\Sigma_a u^\Sigma_b \ ,
\end{equation}
we (naturally) have
\begin{equation}
  \perp^\Sigma_{a b} \beta_\Sigma^b = 0 \ .
\end{equation}
Once we have introduced $\beta_\Sigma^a$, we can use it to reduce the degrees
of freedom of the viscosity tensors. So far, we have only required them to
be symmetric. However, in the standard case one would expect a viscous
tensor to have only six degrees of freedom. To ensure that this is the case
 we introduce the  degeneracy
condition 
%
\begin{equation}
  u_b^\Sigma \tau_\Sigma^{b a} = 0 \ .
\end{equation}
That is, we require the viscous tensor $\tau_\Sigma^{a b}$ to be purely
spatial according to an observer moving along $u^a_\Sigma$. With these
definitions one can show that
\begin{equation}
  \beta_\Sigma^a {\cal L}_\Sigma \pi^\Sigma_{a b} = 0 \ ,
\end{equation}
where ${\cal L}_\Sigma= {\cal L}_{\beta_\Sigma^a}$ is the Lie-derivative
along $\beta_\Sigma^a$, and
\begin{equation}
  \tau_\Sigma^{a b} {\cal L}_\Sigma \pi^\Sigma_{a b} =
  - 2 \beta_\Sigma^a f^\Sigma_a \ .
\end{equation}

Finally, let us suppose that we choose to work in a given observer frame, moving with
four-velocity $u^a$ (associated with the usual projection 
$\perp^a_b$). Then we can use the decompositions:
\begin{equation}
  \beta_\x^a = \beta_\x \left(u^a + v_\x^a\right)
  \qquad \mathrm{and} \qquad
  \beta_\Sigma^a = \beta_\Sigma \left(u^a + v_\Sigma^a\right) \ .
\end{equation}
As expected, $\mu^\x = 1/\beta_\x$ represents a chemical type potential for
species $\x$ with respect to the chosen frame. At the same time, we see that $\mu^\Sigma = 1/\beta_\Sigma$ is a Lorentz 
factor. Using the norm of $\beta^a_\Sigma$ we have
\begin{equation}
  \beta^a_\Sigma \beta^\Sigma_a =
  - \beta^2_\Sigma \left(1 - v_\Sigma^2 \right) = - 1 \ ,
\end{equation}
where $v_\Sigma^2 = v_\Sigma^a v^\Sigma_a$. Thus
\begin{equation}
  \mu^\Sigma = 1/\beta_\Sigma = \sqrt{ 1 - v_\Sigma^2},
\end{equation}
is analogous to the standard Lorentz factor.

So far the construction is quite formal. Let us now try to make it more intuitive by making contact
with the physics. First, we note that the above results allow us to
demonstrate that
\begin{multline}
  u^b \nabla_a T^a{}_b = -  \sum_\x
  \left( \mu^\x \nabla_a n_\x^a + v_\x^a f^\x_a \right)  
  \\
  -
  \sum_\Sigma \left( v_\Sigma^a f^\Sigma_a -
  \frac{1}{2} \mu^\Sigma \tau_\Sigma^{b a} {\cal L}_\Sigma
  \pi^\Sigma_{b a} \right) = 0 \ .
\end{multline}
Recall that similar results were central to expressing the second law of
thermodynamics in  Sect.~\ref{sec:heat}. To see how things work out
in the present case, and make contact with the previous discussion, let us single out the entropy fluid (with index
$\s$) by defining $s^a = n_\s^a$ and $T = \mu_\s$. To simplify the
final expressions it is also useful to assume that the remaining species
are governed by conservation laws of the form
\begin{equation}
  \nabla_a n_\x^a = \Gamma_\x \ ,
\end{equation}
subject to the constraint of total baryon number conservation; i.e.,
\begin{equation}
  \nabla_a n^a = \nabla_a \sum_{\x \neq \s} n_\x^a = \sum_{\x \neq \s} \Gamma_\x = 0.
\end{equation}
Given this, and the fact that the divergence of the stress-energy tensor
must vanish, we have
\begin{equation}
  T \nabla_a s^a = - \sum_{\x \neq\s} \mu^\x \Gamma_\x -
  \sum_\x v_\x^a f^\x_a - \sum_\Sigma
  \left(v_\Sigma^a f^\Sigma_a + \frac{1}{2} \mu^\Sigma
  \tau^{a b}_\Sigma {\cal L}_\Sigma \pi^\Sigma_{a b} \right) \ .
\end{equation}
Here we can bring the remaining two force contributions together by
introducing the linear combination 
\begin{equation}
  \sum_\x \zeta^\x_\Sigma v_\x^a = v_\Sigma^a \ ,
\quad \mathrm{with} \quad
  \sum_\x \zeta^\x_\Sigma = 1 \ .
\end{equation}
Then defining
\begin{equation}
  \tilde{f}^\x_a = f^\x_a + \sum_\Sigma \zeta^\x_\Sigma f^\Sigma_a \ ,
\end{equation}
we have
\begin{equation}
  T \nabla_a s^a = - \sum_{\x \neq \s} \mu^\x \Gamma_\x -
  \sum_\x v_\x^a \tilde{f}^\x_a -
  \frac{1}{2} \sum_\Sigma \mu^\Sigma \tau^{a b}_\Sigma
  {\cal L}_\Sigma \pi^\Sigma_{a b} \ge 0 \ .
  \label{2nd_final}
\end{equation}
The three terms in this expression represent, respectively, the entropy
increase due to (i) chemical reactions, (ii) conductivity, and (iii) viscosity.
The simplest way to ensure that the second law of thermodynamics is
satisfied is to make each  term positive definite.

At this point, the  formalism must be completed by some (suitably
simple) model for the various terms. A reasonable starting point would be
to assume that each term represents a linear deviation from equilibrium. For the chemical reactions this would
mean that we expand each $\Gamma_\x$ according to 
\begin{equation}
  \Gamma_\x = - \sum_{\y \neq s} {\cal C}_{\x \y} \mu^\y \ ,
  \label{bigGx}
\end{equation}
where ${\cal C}_{\x \y}$ is a positive definite (or indefinite) matrix
composed of the various reaction rates. Similarly, for the conductivity
term it is natural to consider ``standard'' resistivity such that
\begin{equation}
  \tilde{f}^\x_a =
  - \sum_{\y} {\cal R}^{\x \y}_{a b} v_\y^b \ .
\end{equation}
Finally, for the viscosity we can postulate a law of form
\begin{equation}
  \tau_\Sigma^{a b} =
  - \eta^{a b c d}_\Sigma {\cal L}_\Sigma \pi^\Sigma_{c d} \ ,
\end{equation}
where we would have, for an isotropic model,
\begin{equation}
  \eta^{a b c d}_\Sigma =
  \eta \perp_\Sigma^{a (c} \perp_\Sigma^{d) b} +
  \frac{1}{3} (\eta - \zeta)
  \perp_\Sigma^{a b} \perp_\Sigma^{c d} \ ,
\end{equation}
and the coefficients $\eta$ and $\zeta$ are identified as
representing shear and bulk viscosity, respectively.

A detailed comparison between Carter's formalism and the Israel--Stewart
framework has been carried out by \cite{priou91}. He concludes that
the two models, which are both members of a larger family of dissipative
models, have essentially the same degree of generality and that they are
equivalent in the limit of linear perturbations away from a thermal
equilibrium state. Providing explicit relations between the main parameters
in the two descriptions, he also emphasizes the key point that analogous
parameters may not have the same physical interpretation.

\subsection{Add a bit of chemistry...}

With the formal model development (at least at some level) in hand, it is natural to turn to the issue of the different dissipation coefficients. This effort has several different aspects. We may, for example, dig deeper and try to calculate the coefficients from some more fundamental---presumably microscopic---theory. At the same time, we may ask (still in the somehwat phenomenological vein) if we can make progress by considering the nature of the involved coefficient. Such questions inevitably takes us in the direction of chemistry, where the mechanics of mixtures and solvents tends to be explored in detail. The chemistry lab may seem a strange place to look for answers to astrophysics questions, but the problems we are interested in are truly interdisciplinary so it is perhaps not surprising that this is where we end up.

Central to any discussion of this kind is the Onsager symmetry principle \citep{1931PhRv...37..405O}, see \cite{andersson05:_flux_con,2012PhRvD..86f3002H} for relevant discussions. Focussing on the general idea--which is natural since the details depend on the application under consideration---we start by noting that,
for any system perturbations of the entropy density $s$ away from equilibrium must be represented by quadratic deviations. This allows us to write
\be
s\approx s_{\mathrm{eq}}-\frac{\Delta t}{2 T}\sum_{a,b} X_a L^{ab} X_b\ ,
\label{ds}\ee
or,  making use of the entropy creation rate $\Gamma_\s$:
\be
T \Gamma_\s = -\frac{1}{2}\sum_{a,b} X_a L^{ab} X_b=\sum_{a=1}^{N} J^a X_a\ ,
\label{entropy2} \ee
where the $X_a$ are known as ``thermodynamic forces''. They represent a measure of the departure of the system from equilibrium, while the ``thermodynamic fluxes'' 
\be
J^a=-{1\over 2}\sum_b L^{ab}X_b\ ,
\ee
 represent the response of the system. The Onsager symmetry principle simply states that microscopic reversibility implies that we should have $L^{ab}=L^{ba}$.
Comparing equation (\ref{entropy2}) to results like equation (\ref{2nd_final}) we can, by constructing the most general form for the tensor $L^{ab}$ in terms of the thermodynamical forces in the model, obtain the most general description of the dissipative terms in the  equations equations of motion. 

A key part of this construction is the observation that---because we are assuming an expansion away from equilibrium---we need the forces to vanish as thermodynamic equilibrium is reached. Hence, we should not work with the
chemical potentials, as in \eqref{bigGx}, because they  obviously  do not  vanish in equilibrium.
This point comes to the fore when we consider problems with reactions, as in the case of bulk viscosity.
We  need to replace the chemical potential with a more suitable ``force''. This leads us to introduce the affinity \citep{prigogine}. In the context of neutron stars, this point has been made in \cite{Carter04:_newtIII,2012PhRvD..86f3002H}. 

Suppose there are $N$ total reactions among $M$ various constituents
$\X$ of our multi-fluid system, to be characterized in the usual way as
stoichiometric relations between the particle number densities~\footnote{Technically speaking one should consider mole numbers in these relations. However, for the kind of reactions that we consider in neutron star cores there is no difference.} $\nu^\X=n^\X/\left(\sum_\X n^\X\right)$ ; i.e.
\be
   \sum_{\X }^M {\rm R}_\X^I~\nu^\X \to \sum_{\X }^M {\rm P}_\X^I~\nu^\X
           \quad , \quad I = 1,...,N\ , 
\ee
where ${\rm R}_\X^I$ and ${\rm P}_\X^I$ are, respectively, the reactant
and product stoichiometric coefficients. The affinity $A^I$ of the
$I^{\rm th}$ reaction is then defined as
\be
    A^I \equiv \sum_{\X }^M \left({\rm R}_\X^I - {\rm P}_\X^I\right)
        {\mu^\X} \ .
\ee
At thermodynamic equilibrium the affinities vanish, which is why they make
appropriate thermodynamic forces.

 It is intuitively clear that the affinities provide a natural description of the problem, but this does not mean that the 
formulation is complete at this point. In particular, it is worth noting that  the chemical potentials $\mu^\X$ 
become somewhat ambiguous in a multi-fluid context. Each
chemical potential should be defined as the energy per particle in the reference frame where the
chemical (or nuclear) reactions occur, but a multi-fluid mixture is characterized by
the presence of distinct velocity fields, neither of which represents the required frame. 
The relevant frame may, in fact, not be known a priori as the formulation we consider assumes an expansion away from ``equilibrium'', which ultimately involves
both dynamical and chemical considerations. The equilibrium frame may well depend on the dynamical evolution of the whole system. This complicates the issue, at least from the formal point of view. 

According to Hess's Law, for each chemical reaction there is only one
thermodynamic variable to track in order to determine the changes; namely,
the ``degree of advancement'' $\xi_I$ for the various reactants. For each of
the $I = 1...N$ reactions, a variation $\Delta \xi_I$ corresponds to a
variation $\Delta \nu^\X_I$ of the participating fluids:
\be
   \frac{\Delta \nu_I^{\rm r}}{{\rm R}_{\rm r}^I} = ... =
   \frac{\Delta \nu_I^{\rm s}}{{\rm R}_{\rm s}^I} =
  - \frac{\Delta \nu_I^{\rm u}}{ {\rm P}_{\rm u}^I} = ... =
  - \frac{\Delta \nu_I^{\rm v}}{ {\rm P}_{\rm v}^I} = \Delta \xi_I \ ,
\ee
where ${\rm r},...,{\rm s}$ and ${\rm u},...,{\rm v}$ represent the $\X$-components for which the
${\rm R}_\X^I$ and ${\rm P}_\X^I$ are non-zero. The (irreversible) change
$\Delta s$ in the entropy due to these reactions is given by
\be
   \Delta s = \frac{1}{T} \sum_{I = 1}^N A^I \Delta \xi_I \ .
\ee
By comparing with equation (\ref{ds}), we see that the $\Delta \xi_I$ represent the appropriate thermodynamic
``fluxes''.

The variations $\Delta \nu^\X$ of the individual number densities, in some time
interval $\Delta t$, can also be determined by
\be
    \Delta \nu^\X =  \Gamma_\X \Delta t \ ,
\ee
where $\Gamma_\X$ is the particle number creation rate.

Each of the $N$ reactions then has a corresponding change $\nu^\X_I$ that
contributes to $\Delta \nu^\X$, with the net result (as $\Delta t \to 0$)
\be
   \frac{d \nu^\X}{d t} = \sum_I \left({\rm R}_\X^I - {\rm P}_\X^I\right)
                          \frac{d \xi_I}{d t} \ .
\ee
Hence,
\be
 \Gamma_\X =  \sum_I \left({\rm R}_\X^I - {\rm P}_\X^I\right)
                    \frac{d \xi_I}{d t} \ .
\ee

If we take the reaction ``velocity'' $V^I\equiv \frac{d \xi_I}{d t}$ to be the thermodynamical flux,  then the change in entropy due to the reactions is
\be
   \Delta s =\sum_{\X \neq \s} \mu^\X \Gamma_\X=  \sum_{\X \neq \s} {\mu^\X} \left[\sum_I
                \left({\rm R}_\X^I - {\rm P}_\X^I\right)
                \frac{d \xi_I}{d t}\right]= \sum_{I} A^I V_I \ .
\ee
In the general framework the corresponding thermodynamic force will then be $A^I$ while the  flux is $-V_I$. Given this, we can  construct the fluxes out of the forces, limiting ourselves to  quadratic terms. An explicit example of such a construction can be found in \cite{2012PhRvD..86f3002H}.

\subsection{Towards a dissipative action principle}

Conventional wisdom suggests that an action principle---expressed as an integral of a 
Lagrangian, whose local extrema satisfy the equations of motion, subject to 
well-posed boundary constraints, see Sect.~\ref{sec:variational}---cannot exist for a dissipative system. However, this may be too dismissive. There have been a
number of (more or less successful) attempts to make progress on building dissipative variational models. A common approach has been to combine a variational model for the non-
dissipative aspects with an argument that constrains  the entropy production, often 
involving Lagrange multipliers (see \citealt{1994PhR...243..125I} for a review and 
\citealt{1975AcMec..23...17D,1980JPhA...13..431D,1982RSPSA.381..457M,1986AmJPh..54..997K,1986IJNLM..21..489V,1991PhLA..155..223H,Chien,2007EPJD...44..407N,2012PThPh.127..921F} 
for samples of the literature). The model we will consider is conceptually 
different. The conservative constraints on the system are built into the variation itself and 
the model does not involve (at least not in the first instance) an expansion away from 
equilibrium (in contrast to, for example, the model of Israel and Stewart or, 
indeed, any model that takes a  derivative expansion as its starting point). Formally, the new description remains valid  also for systems far away from 
equilibrium, and hence it provides a promising framework for the exploration of 
nonlinear thermodynamical evolution and associated irreversible phenomena---a 
problem area where a number of challenging issues remain to be resolved, involving for 
example maximum versus minimum entropy production for non-equilibrium systems 
\citep{1980ARPC...31..579J,2003JPhA...36..631D,2006PhR...426....1M,2010PhRvE..81d1137D}. 

Why should we expect a variational argument for non-equilibrium systems to 
exist?  The question is multi-faceted, but recall that one of the most topical 
problems in gravitational physics involves  two stars (or black holes) in a binary system, 
that lose orbital energy through the emission of gravitational waves. 
Gravitational-wave emission is a dissipative mechanism, yet the underlying 
theory is obtained from an action (see Sect.~\ref{sec:efevar}). This  tells us that you can, indeed, 
use a variational strategy for dissipative problems (a similar argument was  
made by \citealt{2013PhRvL.110q4301G,2014arXiv1412.3082G}). The key insight is that \emph{all the energy in 
the system must be accounted for}. In many ways this is trivial. If you account 
for all the energy in a given system, including the ``heat bath'', then there is no 
\emph{dissipation} as such. Rather, one tries to model the 
\emph{redistribution} of energy within the larger (now closed) system. 
This may be a natural \emph{logical} 
argument, but the question is if we can turn it into a \emph{practical} proposition.

The first step in this direction involves designing a variational 
argument that leads to the functional form of the dissipative fluid equations, adopting the attitude from classical mechanics 
where the equations of motion for a system can be written down without actual reference 
to a particular form for the energy. The completion of the model---fully specifying the various coefficients involved, which must draw on some level of microphysics understanding---is, of course, important 
but the problem is sufficiently complex that it is sensible to progress in manageable 
steps. 

The idea behind the new approach  is, conceptually, quite simple \citep{2015CQGra..32g5008A}. Recalling that the 
individual matter spaces (associated with the various fluid components) play a central 
role in the variational construction for a conservative system, let us consider the 
``physics'' of a dissipative system, e.g., with resistivity, shear or bulk viscosity. 
On the micro-scale dissipation arises due to particle interactions/reactions. On the fluid 
scale this naturally translates into an \emph{interaction between the matter spaces}. This interaction can be accounted for by letting each matter space be 
endowed with a volume form which depends on:
\begin{enumerate}
\item the coordinates of {\em all} the matter spaces, and 
\item the independent mappings of the spacetime metric into these spaces. 
\end{enumerate}
For example, if each $n^\x_{A B C}$ is no longer just a function of its own $X^A_\x$, the 
closure of $n^\x_{a b c}$ will be broken. As the fluxes are no longer  conserved, the 
formalism incorporates dissipation. Simple!

To see how this works, let us revisit the conservative problem from Sect.~\ref{sec:twofluids}. Recall that the scalar 
fields $X^A_\x$ label the (fluid) particles. If these are conserved, then the $X^A_\x$ must 
be constant along the relevant worldlines. That this is, indeed, the case is easy to 
demonstrate. Letting $\tau_\x$ be the proper time of each worldline, we have
\beq
             \frac{{ d} X^A_\x}{{ d} \tau_\x} = u^a_\x 
             \frac{\partial X^A_\x}{\partial x^a} = \frac{1}{n_\x} n^\x_{B C D} \epsilon^{a b c d} 
             \frac{\partial X^A_\x}{\partial x^{a}}
             \frac{\partial X^B_\x}{\partial x^b} 
             \frac{\partial X^C_\x}{\partial x^c} 
             \frac{\partial X^D_\x}{\partial x^{d}} =0  \ . \label{xAdrag}                   
\eeq
Since a fluid element's matter space coordinates $X^A_\x$ are constant along its 
worldline, it must also be the case that  
\beq
\frac{{\rm d} n^\x_{A B C}}{{\rm d} \tau_\x }= 0 \ .
\eeq  
In other words, the volume form $n^\x_{ABC}$ is fixed in the associated matter space. These steps demonstrate
that the key to non-conservation 
is to allow  $n^\x_{A B C}$ to be a function of more than the $X^A_\x$. This is quite 
intuitive. The worldlines of the various fluids will in general cut across each other, 
leading to interactions/reactions. A more general functional form for the matter space 
volume forms $n_{ABC}^\x$ may then be used to reflect this aspect of the physics. 
A schematic illustration of how this works is provided in Fig.~\ref{timeflow}.

\begin{figure} 
\centering
\includegraphics[width=0.5\textwidth]{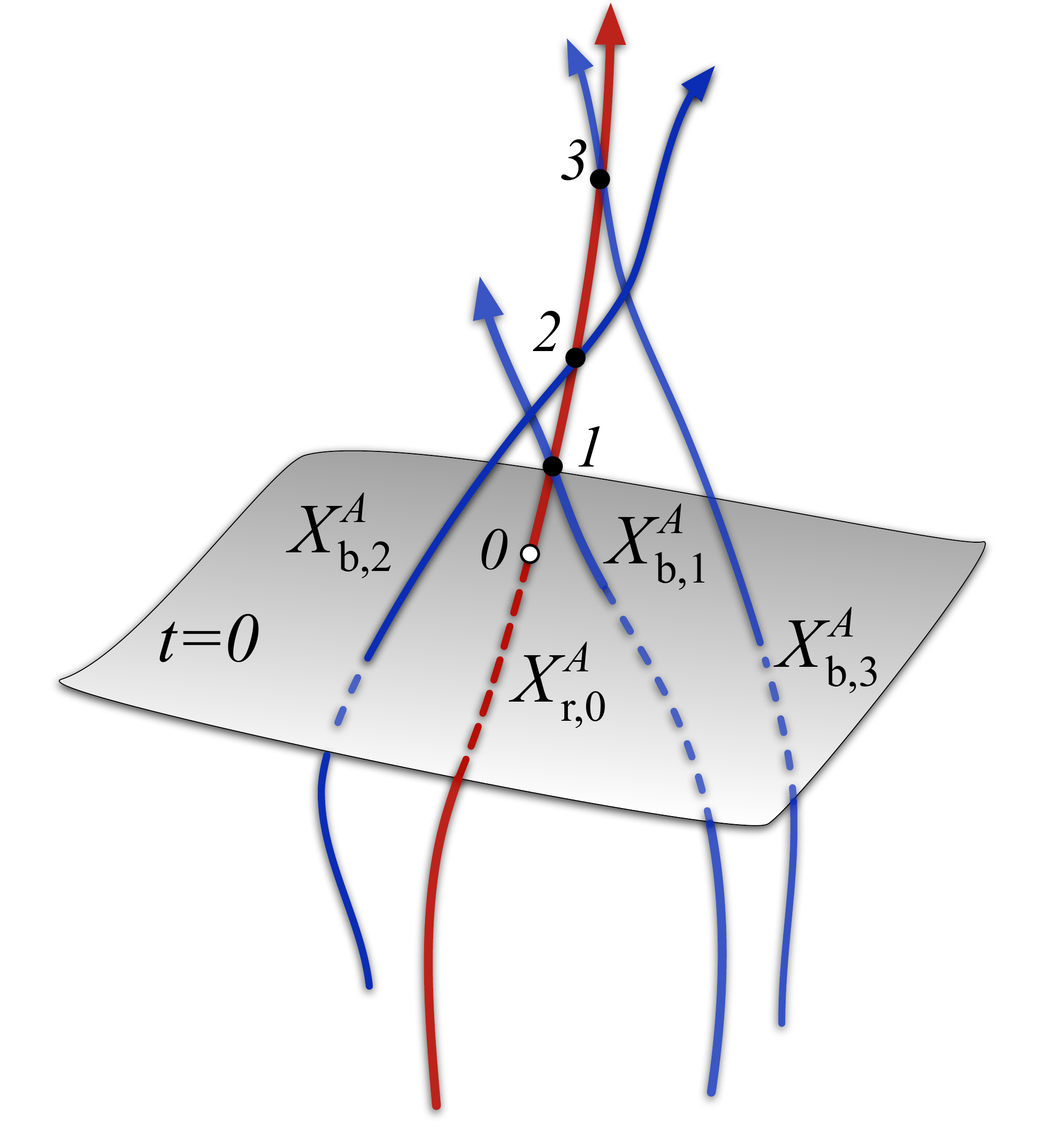}
\caption{An illustration of the notion that a coupling between  matter spaces may lead to 
dissipation. We consider the case of two fluids, labelled r and b (red and blue). The 
individual $X^A_\x$  do not vary along their own 
worldlines, even when the system is dissipative. By adding $X^A_\y$ ($\y\neq\x$) we 
get ``evolution'' since  the worldlines cut across each other. Let us choose a particular 
worldline of the r-fluid, say $X^A_\mathrm{r,0}$, meaning that $X^A_\mathrm{r}$ will 
take the same value at each spacetime point $x^a$along the worldline. At an 
intersection with a worldline of a fluid element of the b-fluid (the point labelled 1 in the 
figure, say) the other fluid's worldline will have its own label (in this case 
$X^A_\mathrm{b,1}$), which is the same  at every point on that worldline. At the next 
intersection (point 2), the worldline we are following has the same value for 
$X^A_\mathrm{r}$, but it is intersected by a different worldline from the other fluid 
($X^A_\mathrm{b,2}$), meaning that $X^A_\mathrm{b}$ at each intersection is different. 
Hence, $X^A_\mathrm{b}$, when considered as a field in spacetime, must vary along 
the r-fluid worldlines, and vice versa. This is how the closure of the individual volume 
three-forms is broken and ultimately why the model is dissipative.}
\label{timeflow}
\end{figure}

The seemingly simple step of enlarging the 
functional dependence of $n^\x_{ABC}$ allows us to build a variational 
model that incorporates a number of dissipative terms. However, in doing this we have to tread carefully. In particular, we must pay closer attention to 
the various matter space objects.  We are now dealing with geometric objects that 
actually live in the higher-dimensional combination of \underline{all} the matter spaces, 
e.g., we are dealing with an object of the form
\beq
n^\x_{ABC} \left( X_\x^D, X_\y^E \right) dX_\x^A \wedge dX_\x^B \wedge dX_\x^C \ , 
\qquad \y \neq \x \ .
\eeq
That is, a volume form in the x-matter space parameterised by points in the y-matter 
spaces. We can still pretend that the individual 
matter spaces (related to spacetime via the same maps as in the conserved case) 
remain somehow ``distinct'', but in reality this is not the case. 

When we allow $n^\x_{ABC}$ to be more complex we (inevitably) break some of the 
attractive features of the conservative model. Obviously, $n^\x_{ABC}$ is no longer a 
fixed matter space object. This has a number of repercussions, but we can still construct the action from matter space objects. 
To do this we need the map of the spacetime metric into the relevant matter space (as in the case of elasticity, see Sect.~\ref{sec:relastic}) 
\beq
    \hxab = \frac{\partial X^A_\x}{\partial x^a}  
                  \frac{\partial X^B_\x}{\partial x^b} g^{a b} 
              = \hxba \ . \label{gmap}
\eeq
Note that  $\hxab$ is not likely to be a tensor on matter space. In order for that to be the 
case, the corresponding spacetime tensor must satisfy two conditions: First, it must be 
flowline orthogonal (on each index).  This is true here since the 
operator which generates projections orthogonal to $\x$-fluid worldlines is  
\beq
\perp_\x^{ab} = g^{ab} + u_\x^a u_\x^b \ ,
\eeq
and because of Eq.~\eqref{xAdrag} we have
\beq
    \hxab = \frac{\partial X^A_\x}{\partial x^a}  
                  \frac{\partial X^B_\x}{\partial x^b} g^{a b} = \frac{\partial X^A_\x}{\partial x^a}  
                  \frac{\partial X^B_\x}{\partial x^b} \perp_\x^{a b} \ .
\eeq
The second condition that $\perp^{ab}_\x$ must satisfy so that $\hxab$ is a matter space 
tensor is \citep{2003CQGra..20..889B}
\beq
\mathcal{L}_{u_\x} \perp^{ab}_\x = 0 \ .
\eeq
This is not the case here. Indeed, this condition is too severe for most relevant applications.

Anyway,  it is easy to show that a scalar constructed from the contraction involving 
$g^{ab}$ and some tensor $t^\x_{a\ldots}$ is identical to the analogous contraction of 
the corresponding matter space objects \citep{Karlovini03:_elas_ns_1}. In particular, the 
number density follows from (as before)
\begin{multline}
n_\x^2 = - g_{ab} n_\x^a n_\x^b = \frac{1}{3!} g^{ad} g^{be} g^{cf} n^\x_{abc} n^\x_{def} \\
=  
\frac{1}{3!} g_\x^{AD} g_\x^{BE} g^{CF} n^\x_{ABC} n^\x_{DEF} \ , 
\end{multline}
while the chemical potential 
\beq
        \mu^\x = - u_\x^a \mu^\x_a 
\eeq
(according to an observer at rest in the respective fluid's frame) can be obtained from
\beq
n_\x \mu^\x = - n_\x^a \mu^\x_a = \frac{1}{3!} \mu_\x^{abc} n^\x_{abc} =  \frac{1}{3!} 
\mu_\x^{ABC} n^\x_{ABC} \ . 
\eeq
Here we have (as in Sect.~\ref{pullback}) introduced the dual to the momentum $\mu^\x_a$: 
\beq
    \mu_\x^{a b c} = \epsilon^{d a b c} \mu_d^\x 
                     \quad , \quad
    \mu_a^\x = \frac{1}{3!} \epsilon_{b c d a} \mu_\x^{b c d} 
               \ , \label{mu3form}
\eeq
and its matter space image;
\beq
     \mu^{A B C}_\x = \frac{\partial X^A_\x}{\partial x^{[a}} 
                                \frac{\partial X^B_\x}{\partial x^b} 
                                \frac{\partial X^C_\x}{\partial x^{c]}}  \mu_\x^{a b c} \ .
\eeq 

The key take-home message is that we can think of the matter  action as being 
constructed entirely from matter space quantities. In the simplest case of a single 
component one would have (see Sect.~\ref{sec:pullback})
\be
\Lambda \left( n_\x\right) = \Lambda \left( n^\x_{abc}, 
g^{ab}\right) \Leftrightarrow \Lambda\left( n^\x_{ABC}, g_\x^{AB} \right)\ .
\ee

\subsection{A reactive/resistive example}

Let us try to make the idea more concrete by working through the steps 
of the variational analysis, while allowing for general variations of the matter space 
density. Since the matter space coordinates  still vary according to \eqref{xlagfl} (this is 
essentially just the definition of the Lagrangian displacement) we easily arrive at the generic variation 
\beq
   \delta n^\x_{a b c} = - \lefx n^\x_{a b c} + 
                         \frac{\partial X^A_\x}{\partial x^{[a}} 
                         \frac{\partial X^B_\x}{\partial x^b} 
                         \frac{\partial X^C_\x}{\partial x^{c]}} 
                        \Delta_\x n^\x_{A B C} \ . \label{nxvargen}
\eeq
To make contact with \eqref{delnvec0} we need
\beq
\mu^\x_a \delta n^a_\x = \frac{1}{3!} \mu^\x_a \delta\left(  \epsilon^{bcda}
  n^\x_{bcd} \right) = - \frac{1}{3!} \mu_\x^{bcd} \delta n^\x_{bcd} + \frac{1}{3!} \mu^\x_a 
  n^\x_{bcd} \delta \epsilon^{bcda} \ ,
\eeq
where we recall \eqref{epsvar0}.
Hence,  we arrive at 
\beq
    \mu^\x_a \delta n^a_\x =  \frac{1}{3!} \mu^{a b c}_\x \lefx n^\x_{a b c} -  \frac{1}{2} 
    \mu^\x_a n^a_\x   g^{bc} \delta g_{bc} - 
     \frac{1}{3!} \mu^{A B C}_\x \Delta_\x n^\x_{A B C}  \ ,  \label{munxvar}
\eeq                 
and the ``final'' expression:                     
\begin{multline}
      \mu^\x_a \delta n^a_\x = \mu^\x_a \left(  n^b_\x \nabla_b \fd^a_\x - 
              \fd^b_\x  \nabla_b n^a_\x -  n^a_\x \nabla_b \fd^b_\x 
             - \frac{1}{2} n^a_\x g^{bc} \delta g_{bc}    \right) \\
             - 
                     \frac{1}{3!} \mu^{A B C}_\x \Delta_\x n^\x_{A B C}   \ .
              \label{mudgen} 
\end{multline}
The terms in the bracket are the same as in the conservative case, cf. \eqref{delnvec0}. The last term is new. 

The functional dependence of the volume form for a given fluid's matter space
is now the main input. Obviously,  $n^\x_{A B C}$ must depend on $X^A_\x$, 
the coordinates of the corresponding matter space, in order for us to retain the conservative 
dynamics. Adding to this,  let us include the coordinates $X^A_\y$ from the other, $\y 
\neq \x$, matter spaces. This breaks the closure of 
$n^\x_{a b c}$ and the model is no longer conservative.

The required variation of $n^\x_{A B C}$ becomes [in view of \eqref{DelX}]
\beq
  \Delta_\x n^\x_{A B C} = \sum_{\y \neq \x} 
         \frac{\partial n^\x_{A B C}}{\partial X^D_\y} \Delta_\x X^D_\y  = \sum_{\y \neq \x} 
         \frac{\partial n^\x_{A B C}}{\partial X^D_\y} \left(\xi_\x^a-\xi_\y^a\right) 
         \partial_a X^D_\y\ .
\label{DelNx}
\eeq
Comparing to \eqref{munxvar}, we see that it is natural to define
\beq
  \RXYa \equiv  \frac{1}{3!} \mu^{ABC}_\x  
           \frac{\partial n^\x_{ABC}}{\partial X^D_\y}  \partial_a X^D_\y \ . 
         \label{resist}   
\eeq
We then have
\begin{multline}
      \mu^\x_a \delta n^a_\x = \mu^\x_a \left(  n^b_\x \nabla_b \fd^a_\x - 
              \fd^b_\x  \nabla_b n^a_\x -  n^a_\x \nabla_b \fd^b_\x 
             - \frac{1}{2} n^a_\x g^{bc} \delta g_{bc}    \right) \\
             +  \sum_{\y \neq \x} \RXYa 
              \left(\fd_\y^a - \fd_\x^a \right)
                \ .
              \label{mudnxgen} 
\end{multline}

The final step involves writing down the variation of the matter 
Lagrangian, $\Lambda$. Starting from \eqref{dlamb}, we arrive at 
\begin{multline}
    \delta \left(\sqrt{- g} \Lambda\right) \\
    = - \sqrt{- g} \left\{\sum_{\x} \left( f^\x_a +\mu^\x_a 
    \Gamma_\x  -  R^{\x}_a\right)  \fd^a_\x - \frac{1}{2} \left(\Psi  g^{a b} + \sum_{\x} n^a_\x 
    \mu^b_\x  \right) \delta g_{a b}\right\} \\
    \\ +
  \nabla_a \left(\frac{1}{2} \sqrt{-g} \sum_{\x}
  \mu^{a b c}_\x n^\x_{b c d} \xi^d_\x \right)\ , \label{varladiss}
\end{multline}
where we have used 
\beq
  \sum_{\x} \sum_{\y \neq \x} \RXYa \fd^a_\y = 
  \sum_{\x} \sum_{\y \neq \x} \RYXa \fd^a_\x \ .
\label{RXYrel}\eeq
We have also defined
\beq
R^{\x}_a = \sum_{\y\neq\x} \left(  R^{\y\x}_a - R^{\x\y}_a \right) \ ,
\label{Rx}
\eeq
and
\beq
       \Gamma_\x = \nabla_a n^a_\x \ . \label{dualparc}
\eeq
Hence, the individual components are governed by the equations of motion
\beq
f^\x_a + \Gamma_\x \mu^\x_a  =  n^b_\x \omega^\x_{b a} +  \Gamma_\x 
\mu^\x_a = R^{\x}_a \ .
\label{force1}
\eeq
Since the force term $f^\x_a$ on the left-hand side is orthogonal to $n_\x^a$ (by the 
anti-symmetry of $\omega^\x_{a b}$) it is easy to see that this result implies that the 
particle creation/destruction rates are given by
\beq
\Gamma_\x = -  \frac{1}{\mu^\x} u_\x^a R^{\x}_a \ .
\eeq

Finally, an orthogonal projection of \eqref{force1} leads to
\beq
 2 n^a_\X \nabla_{[a} \mu^\x_{b]} 
+\Gamma_\x \perp_{\x b}^a \mu^\x_a  = \perp_{\x b}^a R^{\x}_a \ ,
\eeq
which provides the dissipative equations of motion for the system.

The bottomline is that, with Eq.~(\ref{varladiss}) we have a true action principle---in the sense that the 
field equations are extrema of the action---for a system of fluids that includes 
dissipation. It is also worth noting that the stress-energy tensor is still given by 
\begin{equation} 
     T^{a}{}_{b} = \Psi \delta^a{}_b + \sum_{\X} 
                       n^a_\X \mu^\X_b \ , 
\label{tab}
\end{equation}
 and we have
\beq 
      \nabla_b T^b{}_a =  \sum_{\x}\left(  f^\x_a + \mu_a^\x \Gamma_\x \right) = 0 \ ,
\eeq
since
\beq
   \sum_{\x}\rtotxa = 0 \ . \label{resvan}
\eeq
The requirement that the divergence of the stress-energy tensor 
vanish is automatically guaranteed by the dissipative fluid equations, in keeping with the 
diffeomorphism invariance of the theory.

As an immediate application of these relations, connecting with the discussion in Sect.~\ref{sec:heat}, let us consider the simplest relevant setting. Assume that we 
consider a system with two components; matter (labelled $\n$) and heat, represented by 
the entropy (labelled $\s$).  In principle, we need to provide an equation of state (that 
satisfies relevant physics constraints) in order to complete the model. Once this is 
provided we can calculate the resistivity coefficients from \eqref{resist} and then model 
the system using the momentum equations \eqref{force1}. However, let us consider the problem at the level of
phenomenology. We assume that the 
matter component is conserved, but the entropy does not need to be. 

First of all, given that we only have two components we must have
\beq
R^{\n}_a = R^{\s\n}_a - R^{\n\s}_a = - R^\s_a \ .
\eeq
Secondly, the conservation of the material component implies that 
\beq
\Gamma_\n = - \frac{1}{\mu^\n } u_\n^a R^{\n}_a =  \frac{1}{\mu^\n } u_\n^a R^{\n\s}_a 
= 0 \quad \Longrightarrow \quad   u_\n^a R^{\n\s}_a = 0 \ . 
\label{Gn}
\eeq
The upshot is that $R^{\n\s}_a$ must be orthogonal to \emph{both} $u_\n^a$ and 
$u_\s^a$. Meanwhile, the entropy change is constrained by the second law. That is, we have
\beq
\Gamma_\s = -  \frac{1}{T} u_\s^a R^{\s}_a =  \frac{1}{T} u_\s^a R^{\s\n}_a \ge 0  \ , 
\label{Gs}
\eeq
where we have introduced the temperature $T = \mu^\s$. Note that the constraints affect 
the two, likely independent, contributions to $R^\n_a$. We cannot infer a link between 
$R^{\n\s}_a$ and $R^{\s\n}_a$ at this point.

So far we have not introduced a privileged observer.  In order to facilitate a comparison with the discussion in Sect.~\ref{sec:heat}, let us  focus on an observer moving along with the 
matter flow. Then we have 
$u^a=u_\n^a$ and the relative flow required to express the entropy flux is defined such 
that 
\beq
u_\s^a = \gamma \left( u^a + w^a \right) \ , \eeq
where 
\beq
u^a w_a = 0\ , 
\quad \mbox{and} \quad \gamma = \left( 1 - w^2 \right)^{-1/2}  \ . 
\label{uxdecomp}
\eeq
The relative velocity $w^a$ is aligned with the heat flux vector (see, for example, Eq.~\eqref{entflux}).

Given \eqref{Gn} and \eqref{Gs} it  makes sense to introduce the decompositions
\beq
R^{\n\s}_a = \epsilon_{abcd} \phi_\n^b u^c w^d \ , 
\eeq
and 
\beq
R^{\s\n}_a = R_w w_a +  \epsilon_{abcd} \phi_\s^b u^c w^d \ , 
\label{Rsn1}\eeq
where $\phi_\n^a$ and $\phi_\s^a$ are unspecified vector fields.
We then see that \eqref{Gs} leads to
\beq
T \Gamma_\s =  \gamma R_w  w^2  \ge 0  \quad \longrightarrow \quad R_w > 0 \ . 
\label{second1}
\eeq
Meanwhile, the two components $\phi_\n^a$ and $\phi_\s^a$ are \emph{not} 
constrained by the thermodynamics. This leaves a degree of arbitrariness in the 
model. Should we be surprised by this? Probably not. A similar issue was 
discussed by \cite{2011RSPSA.467..738L} where it was demonstrated that the 
variational derivation leads to the presence of a number of terms in the heat equation that 
cannot be constrained by the second law. It was also pointed out that the difference 
between the model advocated by \cite{2011RSPSA.467..738L} and the 
second-order model of Israel and Stewart appeared at this level 
\citep{priou91}. It has not been established whether there are situations 
where these terms  have a notable effect on the dynamics.  This may be an interesting question.

\subsection{Adding dissipative stresses}

The previous example demonstrates how dissipation can be included in the variational multi-fluid 
formalism.  This is a positive step towards a better understanding of 
non-equilibrium systems in General Relativity. Dissipative contributions 
that tend to be  {\em postulated} can now be {\em derived} from first 
principles. Moreover, as the comparison with the problem of heat flow demonstrates, the model introduces new aspects of the problem. However, the example we 
provided only accounts for two particular non-equilibrium phenomena, particle 
non-conservation and resistivity. In order to argue that the model  
represents a credible alternative to established strategies, we need to demonstrate that the action principle 
generates terms of the tensorial form expected for more general 
processes. Thus, we consider the issue of dissipative stresses. 

The obvious starting point for an extension of the strategy is to ask what other 
quantities the matter space volume form, $n^\x_{ABC}$, may depend on. The natural 
object to consider is the mapping of the spacetime metric, $g_{a b}$, into the respective 
matter spaces.  As we will now demonstrate, this leads to a description that accounts for dissipative 
shear stresses. 

The mapping of the metric into the matter spaces introduces three independent 
possibilities. The most intuitive option involves allowing 
$n^\x_{ABC}$ to depend on $\hxab$, as defined in \eqref{gmap}.
Noting that Eq.~(\ref{xAdrag}) implies that the $X^A_\x$ will still be conserved 
along the associated flow, the variation of $n^\x_{A B C}$ is then such that
\beq
  \Delta_\x n^\x_{A B C} = \frac{\partial n^\x_{A B C}}{\partial \hxde}
         \Delta_\x \hxde       + \sum_{\y \neq \x}
         \frac{\partial n^\x_{A B C}}{\partial X^D_\y}  \Delta_\x X^D_\y  \ .
\label{dnnew}\eeq
The first term  in this expression is new, the second term is the same as in 
\eqref{DelNx} . The new term is easily worked out, following the steps from the simpler 
model. We find that
\beq 
      \Delta_\x \hxab =  \frac{\partial X^A_\x}{\partial x^a} 
                                      \frac{\partial X^B_\x}{\partial x^b} \Delta_\x g^{ab} =
                                      \frac{\partial X^A_\x}{\partial x^a} 
                                      \frac{\partial X^B_\x}{\partial x^b}
                                      \left[\delta g^{a b}- 2 \nabla^{(a} \fd^{b)}_\x \right] \ , 
\eeq
where we have used 
\beq
\Delta_\x g^{ab} = \delta g^{a b}- 2 \nabla^{(a} \fd^{b)}_\x \ , 
\eeq
(and round brackets indicate symmetrization, as usual.)

As in the previous example, the variation of the matter Lagrangian involves
$\mu^{A B C}_\x \Delta_\x n^\x_{A B C}$. The new contribution then takes the form
\begin{multline}
 \frac{1}{3!} \mu^{A B C}_\x  \frac{\partial n^\x_{A B C}}{\partial \hxde}
         \Delta_\x \hxde \\
         =  \frac{1}{3!} \mu^{A B C}_\x  
         \frac{\partial n^\x_{A B C}}{\partial \hxde}
         \frac{\partial X^D_\x}{\partial x^a} \frac{\partial X^E_\x}{\partial x^b}
         \left[\delta g^{a b} - 2 \nabla^{(a} \fd^{b)}_\x \right]  \\
         = - \frac{1}{2} \DXab \left[ g^{ac} g^{bd} \delta g_{cd} + 2 \nabla^{(a} \fd^{b)}_\x\right]  
         = - \frac{1}{2}  {S}_\x^{a b} \delta g_{ab} - \DXab \nabla^b  \xi_\x^a \ , 
\end{multline}
where we have defined
\beq
\DXab = \frac{1}{3}  \mu^{ABC}_\x \frac{\partial n^\x_{ABC}}{\partial \hxde}  
            \frac{\partial X^D_\x}{\partial x^a} \frac{\partial X^E_\x}{\partial x^b} = \DXba \ , 
            \label{dxab} 
\eeq
such that 
\beq
u_\x^a \DXba= 0   \ . 
\eeq

Combining the results, we arrive at 
\begin{multline}
      \mu^\x_a \delta n^a_\x = \mu^\x_a \left(  n^b_\x \nabla_b \fd^a_\x - 
              \fd^b_\x  \nabla_b n^a_\x -  n^a_\x \nabla_b \fd^b_\x   \right) + \DXab \nabla^b  
              \xi_\x^a \\
             + \sum_{\y \neq \x} \RXYa 
              \left(\fd_\y^a - \fd_\x^a \right)
             + \frac{1}{2} \left[ \mu^\x_c n^c_\x g^{ab} + {S}_\x^{a b} \right] \delta g_{ab}  \ .
\end{multline}

Introducing the total dissipative stresses, in this case trivially setting
\beq
\dtotxab = \DXab \ , 
\eeq
we see that  Eq.~(\ref{varladiss}) becomes
\begin{multline}
    \delta \left(\sqrt{- g} \Lambda\right) = - \sqrt{- g} \left\{\sum_{\x} \left(f^\x_a 
    + \Gamma_\x \mu^\x_a + \nabla^b\dtotxba  - \rtotxa\right)  \fd^a_\x \right.  \\
   \left.- \frac{1}{2} \left[\Psi  g^{a b} + \sum_{\x} \left(n^a_\x \mu^b_\x +  
    \dtotxab\right)\right] \delta g_{a b}\right\} \\
    + 
  \nabla_a \left[\sqrt{-g} \sum_{\x}
  \left(\frac{1}{2} \mu^{a b c}_\x n^\x_{b c d} + g^{ac} D^\x_{cd}\right) \xi^d_\x\right]\ , 
\end{multline}
where we have used  \eqref{RXYrel} and \eqref{Rx}  for the resistivity currents.

The  equations of motion now take the form
\beq
    f^\x_a + \Gamma_\x \mu^\x_a + \nabla^b \dtotxab =  \rtotxa 
             \ , \label{xfleqn}
\eeq
and the stress-energy tensor is 
\beq
       T^{a b} = \Psi  g^{a b} + \sum_{\x} \left(n^a_\x \mu^b_\x +  \dtotxabu\right) \ ,
                        \label{emstens}
\eeq
where the generalised pressure, $\Psi$, remains unchanged, cf. \eqref{seten2}.
As in the previous problem, it is  easy to show that 
\beq 
      \nabla_b T^b{}_a =  \sum_{\x} \left(f^\x_a + \Gamma_\x \mu^\x_a + 
                                          \nabla^b  \dtotxab\right) = 0 \ ,
\eeq
since \eqref{resvan} still holds. 

Finally, we can extract the various creation/destruction rates. We first contract 
Eq.~(\ref{xfleqn}) with $u^a_\x$, noting that $u^a_\x f^\x_a = 0$ and 
$u^a_\x \nabla^b \dtotxab = - \dtotxab \nabla^b u^a_\x$, to find 
\beq
       \mu^\x \Gamma_\x = - \rtotxa u^a_\x - \dtotxab \nabla^b u^a_\x    \ .
       \label{flxproj}
\eeq
When $\x=\s$ this gives the entropy creation rate which should be constrained by the 
second law.

Armed with the more general constraint \eqref{flxproj} for the dissipative terms, let us 
revisit the two-component model problem.  In particular, let us ask 
what  we can learn from the constraints that follow from the derivation.  As in the previous discussion of this 
problem we will use an observer moving along with the matter flow, such that 
$u^a = u_\n^a$ and $w^a$ represents the relative flow.

Let us first consider the matter component. Since we know that $R^{\n\s}$ should be 
orthogonal to $u_\s^a$ we introduce the decomposition
\beq
R^{\n\s}_a = R_u \left( w^2 u_a + w_a \right) + \epsilon_{abcd} \phi_\n^b u^c w^d \ .
\label{resistme} \eeq
Then \eqref{flxproj} implies that
\beq
D^\n_{ab} \nabla^b u^a = - R^{\n\s}_a u^a = R_u w^2  \ . 
\label{rel1} 
\eeq
Now, there are two possible cases to consider. In the general case, with a
distinct heat flow, we have $w^2>0$ which if we take $R_u >0$ implies that the left-hand 
side of \eqref{rel1} must be positive. To ensure that this is the case, we use the standard 
decomposition (with the same conventions as before, see \eqref{decomp})  
\beq
\nabla_a u^\x_b = \sigma^\x_{ab} +\varpi^\x_{ab} - u^\x_a \dot{u}^\x_b +\frac{1}{3} 
\theta^\x \perp^\x_{ab}\ , 
\label{deco}\eeq
where 
\beq
\sigma^\x_{ab} = D_{\langle a}u^\x_{b\rangle} \ , \qquad \mbox{with} \qquad D_a u^\x_b 
= \perp^\x_{ac} \perp^\x_{bd} \nabla^c u_\x^d \ , 
\eeq
where the angular brackets indicate symmetrization and trace removal (as in \eqref{angdef}),
\beq
\varpi^\x_{ab} = D_{[a} u^\x_{b]} \ , 
\eeq
\beq
\theta^\x = \nabla_a u_\x^a \ , 
\eeq
and 
\beq
\dot u^\x_a = u_\x^b \nabla_b u^\x_a \ . 
\eeq

With these definitions, each term in \eqref{deco} is orthogonal to $u_\x^b$. From the fact 
that $\DXab$ is symmetric and orthogonal to $u_\x^a$ it is easy to see that the condition 
inferred from  \eqref{rel1} is satisfied provided that we have
\beq
D^\n_{ab} = \eta^\n \sigma^\n_{ab} + \zeta^\n \theta^\n \perp^\n_{ab} \ , 
\eeq
with $\eta^\n>0$ and $\zeta^\n>0$. We recognise this as the dissipative (shear- and 
bulk viscosity) stresses expected in the Navier-Stokes equations. Interestingly, the 
second law of thermodynamics was not engaged in the derivation of this result. 

Finally, let us consider the entropy condition. Making use of the results from the simpler heat 
example, noting that we can still use \eqref{Rsn1} for 
$R^{\s\n}_a$, we see that \eqref{flxproj} leads to
\beq
T \Gamma_\s = \gamma R_w w^2 -  D^\s_{ab} \nabla^b u_\s^a \ge 0 \ , 
\eeq
as required by the second law. This suggests that, in addition to $R_w>0$ from before, 
we should have 
\beq
D^\s_{ab} =- \eta^\s \sigma^\s_{ab} - \zeta^\s \theta^\s \perp^\s_{ab} \ , 
\eeq
with $\eta^\s>0$ and $\zeta^\s>0$.

This example provides an indicative illustration, but it is (by no means) the most general model one may envisage, see \cite{2017CQGra..34l5001A}.

\subsection{A few comments}

The development of practical models---suitable for applications---for dissipative relativistic fluids remains very much a ``work in progress''. Having said that, there have been a number of recent potentially promising developments. We have covered the main ideas here, starting from   phenomenological models  constructed to
incorporate dissipative effects. The most ``obvious'' strategies---the ``text-book''
approach of \cite{eckart40:_rel_diss_fluid} and 
\cite{landau59:_fluid_mech}---fail completely, as they do not
respect causality and have  stability issues. Going further, we
described how the problems can be fixed by introducing additional
dynamical fields. We considered the formulations of 
\cite{stew77, Israel79:_kintheo2, Israel79:_kintheo1} and
\cite{carter91} in detail. From our discussion it should be
clear that these models are  examples of an extremely large family of
possible theories for dissipative relativistic fluids. Given this
wealth of possibilities, can we hope to find the ``correct'' model? To
some extent, the answer to this question relies on the extra
parameters one has introduced in the theory. Can they be constrained
by observations? This question has been discussed by
\cite{ger95} and \cite{lind96}. The answer seems to be
no, we should not expect to be able to use observations to single out
a preferred theoretical description. The reason for this is that the
different models relax to the Navier--Stokes form on very short
timescales. Hence, one will likely only be able to constrain the
standard shear and bulk viscosity coefficients, etc. Related questions
concern the practicality of the different proposed schemes. To a
certain extent, this is probably a matter of taste. Of course, it may
well be that the additional parameters required in a particular model
are easier to extract from microphysics arguments. With this in mind, we introduced a fairly recent development aimed at extending the variational approach to dissipative systems \citep{2015CQGra..32g5008A}.  This is conceptually interesting as it draws more directly of the matter space, but it is not yet clear how far this alternative strategy can be pushed. At the end of the day, it may well be that different circumstances require different logic. This would make the ``best'' formulation a matter of taste. Clearly,
there is scope for more thinking...

\section{Concluding remarks}

In writing (years ago) and updating (over several years) this review, we have tried to develop a coherent description of the diverse   building blocks required for fully relativistic 
fluid models. Although there are alternatives, we opted to base our
discussion of the fluid equations of motion on the variational approach
pioneered by \cite{taub54:_gr_variat_princ} and
developed further by \cite{carter83:_in_random_walk,
  carter89:_covar_theor_conduc, carter92:_brane}. This is an appealing
strategy because it leads to a natural formulation for multi-fluid
problems and there have been a number of extensions to cover (more or less) the full range of physics one may be interested in. This is reflected in the material that was added as the review was updated. We now go deeper into variational principles in relativity and consider applications ranging from superfluids with quantized vortices to elastic matter and electromagnetism. We also make contact with modern applications by discussing numerical implementations. Finally, the discussion of dissipative systems has been revised to reflect the ongoing discussion of this important, but still challenging problem. These changes are significant, but one could consider going further still. After all, fluids describe physics at many
different scales and there is a  lot of physics to discuss. The only
thing that is certain is that, whatever happens next, we expect to continue to enjoy the learning process!



\begin{acknowledgements}

Many colleagues have helped us develop our understanding of relativistic
fluid dynamics over the years. Instrumental to the developments described in this review have been the insights of Brandon Carter, David
Langlois, Reinhard Prix, Bernard Schutz and Lars Samuelsson, and for this we are particularly grateful.

NA acknowledges support from STFC via grant no.~ST/R00045X/1.

\end{acknowledgements}


\begin{appendix}
\normalsize

\section{The volume tensor in $n$-dimensions}
\label{appendix}

In this Appendix we provide a number of general identities for the completely antisymmetric volume tensor in 
$n$-dimensions. The most useful identities are those involving the tensor product (including, 
as needed, contractions over indices), of the volume tensor with itself \citep{wald84:_book}:
\begin{eqnarray}
  \epsilon^{a_1 \dots a_n} \epsilon_{b_1 \dots b_n} &=&
  \left( - 1 \right)^s n! \,
  \delta^{[a_1}{}_{b_1} \cdots \delta^{a_n}{}^{]}_{b_n},
  \label{epsum1}
  \\
  \epsilon^{a_1 \dots a_j a_{j+1} \dots a_n}
  \epsilon_{a_1 \dots a_j b_{j+1} \dots b_n} &=&
  \left( - 1 \right)^s \left( n - j \right)! \, j! \,
  \delta^{[a_{j + 1}}{}_{b_{j+1}}\cdots \delta^{a_n}{}^{]}_{b_n},
  \label{epsum2}
  \\
  \epsilon^{a_1 \dots a_n} \epsilon_{a_1 \dots a_n} &=&
  \left( - 1 \right)^s n!,
  \label{epsum3}
\end{eqnarray}
where $s$ is the number of minus signs in the metric (e.g., $s = 1$ for
spacetime). We have used the variation of the volume tensor with respect to
the metric in the actions principle presented in Sections~\ref{sec:pullback},
\ref{pbonemc}, and~\ref{sec:twofluids}. We will derive this variation here
using the identities above as applied to four-dimensional spacetime ($s = 1$
and $n = 4$).

Start by writing Eq.~(\ref{epsum1}) as
\begin{equation}
  g^{a_1 c_1}  g^{a_2 c_2} g^{a_3 c_3}
  g^{a_4 c_4} \epsilon_{c_1 c_2 c_3 c_4}
  \epsilon_{b_1 b_2 b_3 b_4} = \left( - 1 \right)^s
  n! \, \delta^{[a_1}{}_{b_1}\cdots \delta^{a_n}{}^{]}_{b_n},
\end{equation}
vary it with respect to the metric, and then contract the result with
$\epsilon_{a_1 a_2 a_3 a_4}$ to find
\begin{equation}
  \delta \epsilon_{b_1 b_2 b_3 b_4} =
  \frac{1}{4!} \epsilon_{b_1 b_2 b_3 b_4}
  \left( \epsilon^{a_1 a_2 a_3 a_4} \delta
  \epsilon_{a_1 a_2 a_3 a_4} + 4! \, g^{c d}
  \delta g_{c d}\right),
\end{equation}
where we have used
\begin{equation}
  0 = \delta \left( \delta^a{}_b \right) =
  \delta \left( g^{a c} g_{c b} \right)
  \qquad \Rightarrow \qquad
  \delta g^{a b} =
  - g^{a c} g^{b d} \delta g_{c d}\ .
\end{equation}
If we now contract with $\epsilon_{b_1 b_2 b_3 b_4}$ we find
\begin{equation}
  \epsilon^{a_1 a_2 a_3 a_4} \delta
  \epsilon_{a_1 a_2 a_3 a_4} =
  - \frac{4!}{2} g^{b c} \delta g_{b c}
\end{equation}
and thus
\begin{equation}
  \delta \epsilon_{a_1 a_2 a_3 a_4} =
  \frac{1}{2} \epsilon_{a_1 a_2 a_3 a_4}
  g^{b c} \delta g_{b c}.
  \label{epsvar}
\end{equation}

The last thing we need is the variation of the determinant of the metric,
since it enters directly in the integrals of the actions. Treating the
metric as a $4 \times 4$ matrix, and ``normalizing'' the $\epsilon$ by
dividing by its one independent component, the determinant is given by
\begin{equation}
  g = \frac{1}{4! \left( \epsilon^{0 1 2 3} \right)^2}
  \epsilon^{a_1 a_2 a_3 a_4} \epsilon^{b_1 b_2 b_3 b_4}
  g_{a_1 b_1} g_{a_2 b_2} g_{a_3 b_3} g_{a_4 b_4}.
\end{equation}
The right-hand-side is proportional to the left-hand-side of Eq.~(\ref{epsum3}) and thus
\begin{equation}
  \epsilon_{0 1 2 3} = \sqrt{- g},
  \qquad \epsilon^{0 1 2 3} = \frac{1}{\sqrt{- g}}.
  \label{normeps}
\end{equation}
It is not difficult to show
\begin{equation}
  \delta \sqrt{- g} =
  \frac{1}{2} \sqrt{- g} g^{a b} \delta g_{a b}.
  \label{deltadet}
\end{equation}



\section{The matter space Levi-Civita symbol}
\label{leviciv}

The pull-back formalism used in the variational approach builds on the
three-form densities $n^\x_{A B C}$. The associated matter-space analysis draws on basic facts from Linear Algebra
\citep{strang80:_lin_alg}, e.g., for constructing determinants and
matrix inverses to build the different $n^\x_{A B C}$ required
for fluids and solids. As it is helpful to understand the details, we summarize some of the key arguments here.

The first step is to introduce an arbitrary $3 \times 3$
matrix $M^{A B}$  ($A,B,C ... = 1,2,3$) and assume it has an
inverse $M_{A B}$, meaning
\beq
       M^{A C} M_{B C} = M_{C B} M^{C A} = \delta^A_B \ . \label{matinv}
\eeq
The first equality is the simple statement that left- and 
right-inverses must be equal for square matrices. 

The next step is to introduce the determinants of $M^{A B}$
and $M_{A B}$ --- $\det[M]$ and $\det[M^{- 1}]$, respectively.
In the same sense that $\sqrt{-g}$ is used to normalize
$\epsilon_{a b c d}$ (cf.~Eq.~\eqref{normeps} above),
$\det[M]$ and $\det[M^{- 1}]$ will serve as the normalizations
in their respective Levi-Civita symbols $\epsilon^M_{A B C}$
and $\epsilon_{M^{- 1}}^{A B C}$.

In an index form, where the Einstein summation convention is 
going to be used, determinants of $3 \times 3$ matrices
require completely antisymmetric three index objects, which
only take the values $\{\pm 1, 0\}$. These can be written in 
terms of standard matrix determinants with Kronecker-delta
symbols $\delta^A_B = \{1,0\}$ as the matrix entries:
\beq
      \left[A \ B \ C\right]^{\cal U} = \left|
                                             \begin{array}{ccc}
                                                  \delta^A_1 & \delta^A_2 & \delta^A_3 \cr 
                                                  \delta^B_1 & \delta^B_2 & \delta^B_3 \cr 
                                                  \delta^C_1 & \delta^C_2 & \delta^C_3  
                                             \end{array}
                                             \right| 
                                             = 3! \delta^{[A}_1 \delta^B_2 \delta^{C]}_3 = \{\pm 1, 0\} 
\eeq
and
\beq
     \left[D \ E \ F\right]_{\cal D} = \left|
                                             \begin{array}{ccc}
                                                  \delta^1_D & \delta^1_E & \delta^1_F \cr 
                                                  \delta^2_D & \delta^2_E & \delta^2_F \cr 
                                                  \delta^3_D  & \delta^3_E & \delta^3_F 
                                             \end{array}
                                             \right| 
                                             = 3! \delta^1_{[D} \delta^2_E \delta^3_{F]} = \{\pm 1, 0\} \ .
\eeq

When an Einstein summation on $B$ is performed for $\delta^A_B \delta^B_D$ it leads to the expression
\beq
       \delta^A_B \delta^B_D = \delta^A_1  \delta^1_D + \delta^A_2 \delta^2_D + \delta^A_3  
                               \delta^3_D \ .
\eeq
By working backwards on the indices with explicit $A = 1$, $B = 2$, etc., values, we can use the 
expression just above to show that
\beq
      \left[A \ B \ C\right]^{\cal U} \left[D \ E \ F\right]_{\cal D} = 3! \delta^{[A}_D \delta^B_E 
                      \delta^{C]}_F \ . \label{altsymproduct}
\eeq
This is the three-dimensional version of Eq.~\eqref{epsum1}
($n = 3$ and $s = 0$). 

The advantage of the $\left[A \ B \ C\right]_{\cal D}$ symbols
is that index notation can be used for the determinant of the 
matrix $M^{A B}$; namely, 
\begin{multline}
    \det[M] \equiv \left|
                                 \begin{array}{ccc}
                                        M^{1 1} & M^{1 2} & M^{1 3} \cr 
                                        M^{2 1} & M^{2 2} & M^{2 3} \cr 
                                        M^{3 1} & M^{3 2} & M^{3 3} 
                                  \end{array}
                           \right| \\
                           = \frac{1}{3!} \left[A \ B \ C\right]_{\cal D} \left[D \ E \ F\right]_{\cal D} 
                                        M^{A D} M^{B E} M^{C F} \ . \label{Mdet}
\end{multline}
Likewise, the determinant of the inverse matrix $M_{A B}$ is 
\begin{multline}
    \det[M^{- 1}] \equiv \left|
                                 \begin{array}{ccc}
                                        M_{1 1} & M_{2 1} & M_{3 1} \cr 
                                        M_{1 2} & M_{2 2} & M_{3 2} \cr 
                                        M_{1 3} & M_{2 3} & M_{3 3} 
                                  \end{array}
                           \right| \\ = \frac{1}{3!} \left[A \ B \ C\right]^{\cal U} \left[D \ E \ F\right]^{\cal U} 
                                        M_{A D} M_{B E} M_{C F} \ . \label{Minvdet}
\end{multline}

Now, we define the Levi-Civita symbols for $M^{A B}$ and $M_{A B}$ to be
\be
       \epsilon^M_{A B C} = \frac{1}{\sqrt{\det[M]}} \left[A \ B \ C\right]_{\cal D} \ , \label{lcM}
\ee
\be
       \epsilon_{M^{- 1}}^{A B C} \\
       =\frac{1}{\sqrt{\det[M^{- 1}]}} \left[A \ B \ C\right]^{\cal U} \ , 
                     \label{lcMinv}
\ee
and Eq.~\eqref{altsymproduct} takes the form
\beq
 \epsilon_{M^{- 1}}^{A B C} \epsilon^M_{D E F} = 3! \delta^{[A}_D \delta^B_E \delta^{C]}_F \ . 
 \label{epsMnorm}
\eeq
where we have used
\beq
       \det[M^{-1}] = \frac{1}{\det[M]} \ . \label{detinv}
\eeq
This follows simply from the determinant properties $\det[M^{- 1} M] = 1$ and 
$\det[M^{- 1} M] = \det[M^{- 1}] \det[M]$.

We end by noting that the determinants normalize the $\left[A \ B \ C\right]^{\cal U}$ and 
$\left[D \ E \ F\right]_{\cal D}$ symbols in the sense that Eqs.~\eqref{Mdet} and \eqref{Minvdet} become
\beq
       \epsilon^M_{A B C} \epsilon^M_{D E F} M^{A D} M^{B E} M^{C F} = 
       \epsilon_{M^{- 1}}^{A B C} \epsilon_{M^{- 1}}^{D E F} M_{A D} M_{B E} M_{C F} = 3! 
       \ . \label{Mnorm}
\eeq
Also, we can rewrite Cramer's Rule for obtaining the matrix
inverse $M_{A B}$ in and index form:
\beq
        M_{A B} = \frac{1}{2} \epsilon^M_{A C E} \epsilon^M_{B D F} M^{C D} M^{E F} \ . 
                         \label{Minv}
\eeq

\end{appendix}




\end{document}